\documentclass[aps,
english,preprintnumbers,nofootinbib,
twocolumn]{revtex4-1}

\usepackage{amsfonts,amsmath,amssymb}
\usepackage{graphicx}
\usepackage[utf8]{inputenc}
\usepackage{hyperref}
\usepackage{babel}
\usepackage{enumitem}

\newcommand\e{{\rm e}}

\newcommand\be{\begin{equation}}
\newcommand\ee{\end{equation}}
\newcommand\bea{\begin{eqnarray}}
\newcommand\eea{\end{eqnarray}}

\begin{document}

\def\rhoo{\rho_{_0}\!} 
\def\rhooo{\rho_{_{0,0}}\!} 

\begin{flushright}
\phantom{
{\tt arXiv:2006.$\_\_\_\_$}
}
\end{flushright}

{\flushleft\vskip-1.4cm\vbox{\includegraphics[width=1.15in]{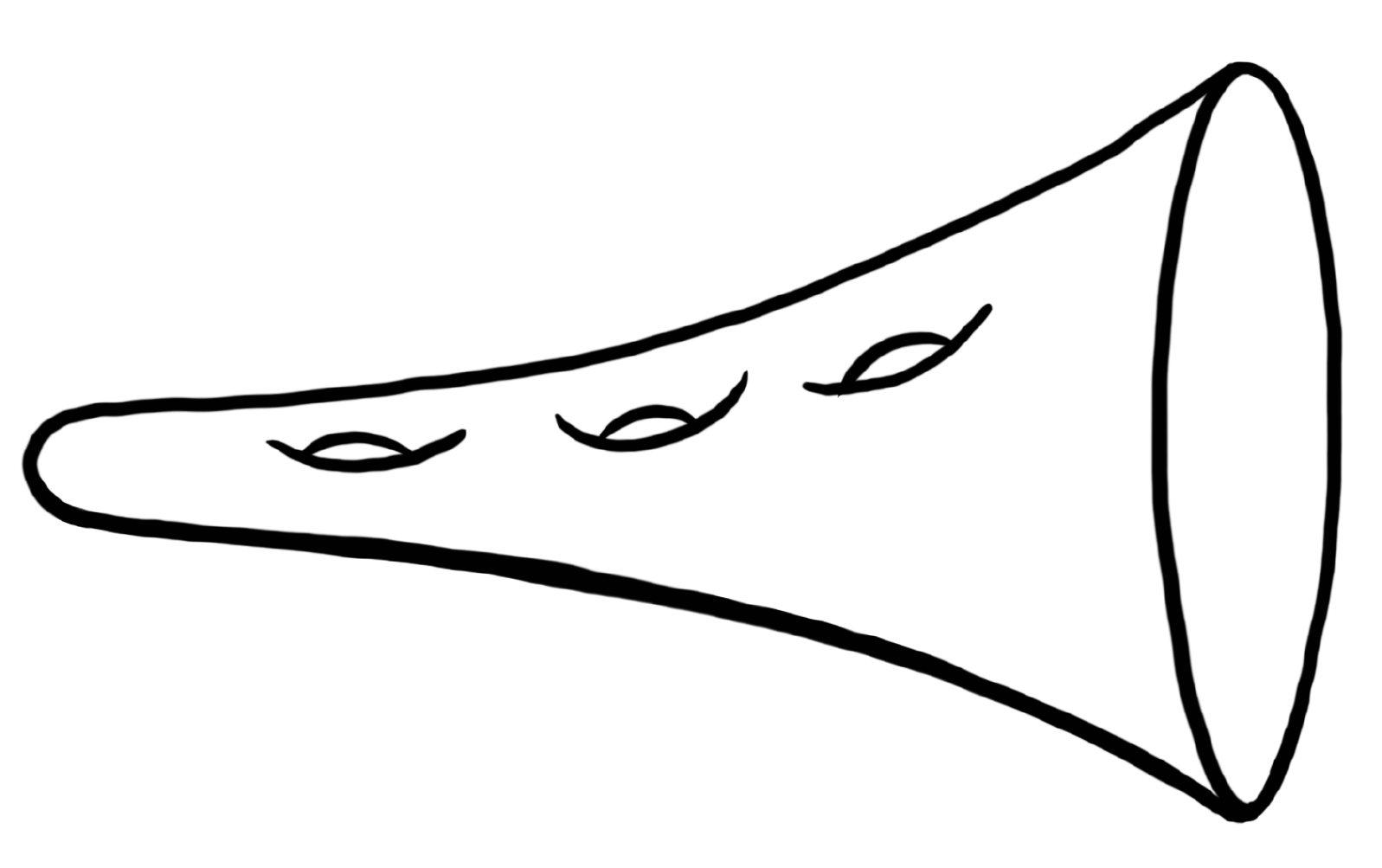}}}

\title{The Microstate Physics of  JT Gravity and Supergravity}
\author{Clifford V. Johnson}
\affiliation{\medskip\\ Department of Physics, Jadwin Hall,  Princeton University,
Princeton, NJ 08544-0708, U.S.A.\vskip0.15cm}
 \affiliation{\medskip\\ Department of Physics and Astronomy, University of
Southern California,
 Los Angeles, CA 90089-0484, U.S.A.}


\begin{abstract}
It is proposed that a complete understanding of two-dimensional quantum gravity  and its emergence in random matrix models requires fully embracing {\it both} Wigner (statistics) and 't Hooft (geometry). Using non-perturbative definitions of random matrix models that yield 
various JT gravity and JT supergravity models on Euclidean surfaces of arbitrary topology,  Fredholm determinants are used to extract precise information  about the  spectra of discrete microstates that underlie the physics.  A core result of the computations  is that in each case,  the (super) Schwarzian spectrum seen at leading order is only an approximation to a new kind of spectrum that is fundamentally discrete. 
It is further argued that the matrix models point to a 
{\it single} distinguished  copy of the discrete spectrum  that characterizes  the holographic dual of the Lorentzian (super) gravity theory. 
These facts suggest that the factorization puzzle is entirely resolved since the discrete spectrum underlying the Schwarzian can have a sensible quantum mechanical origin in a single theory,  without an appeal to an ensemble. It is argued that more generally, double-scaled matrix models contain information about both the Lorentzian and Euclidean approaches to quantum gravity. The former  comes from the Wignerian approach while the latter from the 't Hooftian $1/N$ topological expansion.
\end{abstract}


\maketitle

\section{Introduction}

\label{sec:introduction}

Random matrix models in the double-scaling limit of refs.~\cite{Gross:1990vs,Brezin:1990rb,Douglas:1990ve,Gross:1990aw} can be used to formulate a large number of two-dimensional (2D) Euclidean quantum gravity models on surfaces of arbitrary topology, at least perturbatively in the topological expansion parameter, ${\sim}1/N$. Here, $N$ sets the size of the matrices, and is taken to be large. The formulation can often  be extended to include non-perturbative physics too. In higher dimensions,  quantum gravity can also be formulated on manifolds with non-trivial topology, although it is a harder task. Nevertheless, it is sometimes   necessary to do so, because of certain physical questions that arise when studying for example, saddles of the Euclidean action that appear, such as  black holes and wormholes. It would be useful to know in general what the rules are for how and when to include other topologies, and how to interpret the results. The success with understanding the holographic nature of gravity~\cite{tHooft:1993dmi,Susskind:1994vu} has been a powerful guide.

On the one hand, the  ``traditional'' realization of  holography, as embodied in the AdS/CFT correspondence~\cite{Maldacena:1997re,Witten:1998qj,Gubser:1998bc,Witten:1998zw,Aharony:1999ti}, puts Lorentzian gravity (at least initially) on simpler surfaces, typically with one boundary and a bulk that is  topologically trivial. This follows in part from the process that motivated the correspondence in the first place, taking limits on  the gauge theory and gravity associated with  the string theory D--branes that gave rise to them. (For example, roughly speaking, Lorentzian global AdS$_5$  gravity is  dual to a system with a single Hamiltonian~$H$, which a $D{=}4$ Yang-Mills theory.) Later, after going to the Euclidean section~\cite{Gibbons:1976ue}, defining Euclidean time $\tau$  by taking  $t\to-i\tau$, ($\tau$ has  period $\beta=1/T$)  other topologies were seen to play a role, such as the large Euclidean AdS black holes that help capture, in gravitational language ({\it i.e.,} the Hawking-Page transition~\cite{Hawking:1982dh}), the  transition between the confined and deconfined phases of a dual Yang-Mills theory~\cite{Witten:1998zw}. The preferred phases are determined by summing over all  Euclidean manifolds ${\cal M}$, of both trivial and non-trivial topology $\partial {\cal M}$, that have the same boundary. (So $\partial{\cal M}=S^1_\tau\times S^3$  for the five dimensional case, and ${\cal M}$ can be thermal AdS$_5$ or the AdS$_5$-Schwarzschild solution.)

It was noticed by Maldacena and Maoz~\cite{Maldacena:2004rf} that creating wormholes by connecting  two copies of the boundary seems to present a challenge for holography, since now the gravity interpretation implies a failure of factorization in the holographic dual. This concern has recently been heightened by the landmark results of Saad, Shenker and Stanford~\cite{Saad:2019lba} (extended by Stanford and Witten~\cite{Stanford:2019vob}) that  made it clear that Euclidean Jackiw-Teitelboim (JT) gravity~\cite{Jackiw:1984je,Teitelboim:1983ux} and several variants of it  can be formulated as  random matrix models. At the time these matrix model results appeared, a  puzzle was that fact that the  leading result for JT gravity, captured by Schwarzian dynamics (see {\it e.g.,} refs.~\cite{Almheiri:2014cka,Maldacena:2016hyu,Jensen:2016pah,Maldacena:2016upp,Engelsoy:2016xyb}) yielded a spectral density that was continuous, and so could not be interpreted as the spectrum of a  holographic dual with a sensible Hilbert space~\cite{Stanford:2017thb,Harlow:2018tqv}.  The current view that followed from the appearance of the matrix model formulation of JT gravity  is that the 2D gravity model (and its variants) is {\it fundamentally equivalent} to the matrix ensemble that defines it on arbitrary topology. This idea fit nicely with the continuous density of the matrix model and moreover the presence of random matrix statistics gelled nicely with ideas about the black hole as a quantum chaotic system~\cite{Shenker:2013pqa,Maldacena:2015waa,Sachdev:1992fk,Kitaev:talks,Garcia-Garcia:2016mno,Cotler:2016fpe,Saad:2018bqo}. These beautiful results have been taken by many  to suggest that gravity is fundamentally an ensemble, which opens a can of worm(hole)s for higher dimensions, putting  factorization and traditional holography in direct tension with each other. This is the factorization puzzle, and it has resulted in attempts to either shore up higher dimensional holography, or  modify the matrix model in various ways to somehow preserve its gravitational content while recovering factorization. (For recent work see {\it e.g.} refs.~\cite{Saad:2021uzi,Blommaert:2021fob}.)

This paper suggests that there is another interpretation that entirely resolves factorization,  and  fits seamlessly with what is known about traditional holography.   Moreover it suggests a sharpening of the  meaning to  be given to the study of gravity on surfaces of higher topology  (involving wormholes, {\it etc.}) in any dimension. (At least when a holographic dual is available.) The point of view will strongly urge, with a great deal of evidence, a re-examination of the idea that gravity is  {\it necessarily} an ensemble. On the other hand, by examining in detail how  matrix models actually work {\it fully non-perturbatively} they will be seen to contain very powerful  (and beautiful) lessons for how gravity and topology work together with holography in all dimensions.

A hint of what is going in is already contained in the  higher dimensional AdS/CFT example mentioned above. Two very different possible phases (confined and deconfined) of the gauge theory, represented by sectors with different topologies, were included  in the Euclidean sum. Of course,  this does not mean that multiple variants of the theory are simultaneously at play. Multiple options is simply a part of the Euclidean apparatus--the path integral evaluating different possibilities. The journey back to a Lorentzian interpretation of course yields the physics of  the  dual gauge theory with a {\it single} Hamiltonian. 

The work of this paper will show that when fully non-perturbatively defined, the matrix model is an ensemble of discrete spectra that asymptotically resemble (in the perturbative limit) the leading (continuous) Schwarzian result. It immediately follows that any of them is therefore suitable to be the spectrum of a sensible quantum mechanical theory defining the holographic dual.  It will then be further argued that the job of learning what the matrix model has to say about the physics is not finished until the journey to the Lorentzian problem is complete, and there, a single particular spectrum emerges as the holographic dual.~\footnote{\label{fn:lorentzian-subtleties}There are subtleties in understanding the purely Lorentzian presentation of the quantization of JT gravity, as discussed in ref.~\cite{Harlow:2018tqv}, principally because it naturally has two boundaries. The approach taken here is that the leading result from the Euclidean approach, the effective Schwarzian action, is the beginning of a definition of the quantization. It suggests that there is a sensible Hilbert space, although incompletely described by the Schwarzian alone since the spectral density is continuous. Embedding it in the matrix model and carrying out the non-perturbative analysis of this paper shows how the discreteness arises, and equips the problem with a sensible Hilbert space. Interpreting the full matrix model correctly, it is argued, shows that a particular single Hamiltonian it picked out. Note that all of the structures involved in this  matrix model approach apply to a wider set of 2D gravity models than just JT gravity, as will be discussed in Section~\ref{sec:double-scaling}. Those models do not seem to have the Lorentzian puzzles presented by JT gravity, with its black hole and wormhole solutions. This is highly suggestive that this is a robust approach that gives a firm foundation for addressing the physics of JT.} This two-part demonstration entirely resolves the factorization issue, and in fact shows that 2D holography is quite similar to holography in higher dimensions.

Since the matrix model is able to capture the entirety of the Euclidean 2D gravity sum (because all the possible manifolds are fully classified), it is a particularly powerful and instructive laboratory. 
 A key lesson  is that the Euclidean gravity calculus is extremely resourceful in how it can capture the holographic physics using smooth geometry with various topologies. Sometimes in doing so it appeals to an  infinite ensemble of variants of the theory to achieve this.  This is {\it not} the same as saying that gravity is fundamentally an ensemble. It is just the power of the Euclidean approach. 
 
 The case of having two  boundaries to the gravity bulk is an excellent  illustration of this. Take the example of computing the spectral form factor~\cite{Saad:2018bqo,Saad:2019lba}:
\be
\label{eq:spectral-form-factor-C}
Z(\beta+it)Z(\beta-it)=\sum_{j,k}\e^{-\beta(E_j+E_k)}\e^{it(E_j-E_k)}\ ,
\ee
where $Z(\beta)$ is the partition function of the theory, and $\beta{=}1/T$ is the inverse temperature.  The energies $E_i$ are the spectrum of some Hamiltonian $H$.  Even if there is at early times $t$, a regime where the physics has a perfectly good description in terms of a smooth spacetime dual geometry (as is the case in the matrix model - the spectrum is approximately smooth),  at late $t$ the quantity will be driven to a regime where correlations between discrete ``microstate'' energies produce a wildly oscillating quantity. This is perfectly fine, as there is no mystery as to the oscillations' origin. However, this erratic behaviour cannot be captured in the  smooth geometrical language of a gravity theory and so is hard to model within the Euclidean sum over geometries.   Nevertheless, progress can be made by describing the {\it ensemble} of spectral form factors constructed from a family of Hamiltonians to which~$H$ belongs: $\langle Z(\beta+it)Z(\beta-it)\rangle$. The wild oscillations for each member of the ensemble are in slightly different places in~$t$, and so the average of all of them yields a smooth function of $t$, perhaps with broad features determined by the class of Hamiltonians under consideration. So now    a smooth gravity dual description can emerge. It is a wormhole through the bulk connecting two copies of the boundary where $Z(\beta)$ ``lives''.  This is precisely what happens in the matrix model.

It is natural to conjecture that  a version of this is what is at work in other examples of holography, in various dimensions. Ensembles allow smooth Euclidean geometry to arise as an ``average over ignorance", when necessary. The combination of  topology and smooth geometry will always  allow for a Euclidean calculation of  holographic dual physics. In this light, Eucildean wormholes should be considered no more mysterious than their black hole counterparts from the point of view of holography.

In short (!), the point of view of this paper is that rather than produce a factorization puzzle,  the matrix model  results have simply illustrated something profound about quantum gravity in all dimensions:  Gravity is fundamentally not an ensemble, but as a tool for Euclidean computations of properties of its holographic dual, it {\it can}  describe explorations of the ensemble through the use of smooth spacetimes with wormholes and other topologies.

Evidence for these assertions is needed, of course. It is best found by working in a situation where good control can be had over both gravity and the topology of the spacetimes it is placed upon. Hence the focus on the solvable case of two dimensions, through  double-scaled matrix models. The need to fully understand the content of the matrix models is paramount, and for that  non-perturbative methods are essential. Using them, it will be shown that there are features of the matrix model that fit extremely well with the above picture, and moreover that show that there are two {\it precise} mechanisms of just the kind described above for producing smooth spacetime geometry from the underlying matrix quantities. The first is the way familiar to most matrix model practitioners in the gravity context, and it is 't~Hooftian in spirit, fitting nicely with the intuition based on the topological organization of Feynman diagrams (and their dual tessellations) {\it via} the $1/N$  expansion~\cite{'tHooft:1973jz}. The other way is in the original Wignerian  spirit~\cite{10.2307/1970079}, and amounts to taking many copies of matrix spectra such that, even though they are discrete, the gaps between them get ``filled in'' to produce a smoothness that again yields a  geometrical spacetime description. This latter mechanism, which is more overtly statistical, has been less well appreciated by the gravity community, but a key point of this paper is that  {\it both} are at play when matrix models describe 2D gravity, whether it be for JT gravity and its variants, or for minimal strings and other applications.

\begin{figure}[t]
\centering
\includegraphics[width=0.48\textwidth]{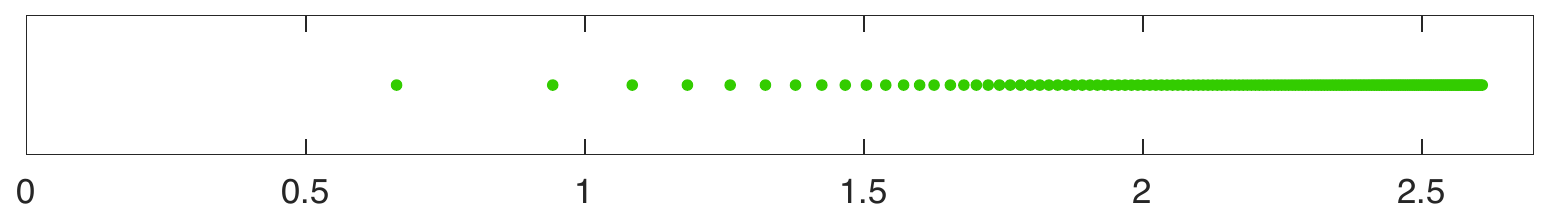}
\caption{\label{fig:JT-levels-150} An example of JT gravity's first 150 microstates, denoted  as the set $\{{\cal E}_n\}$ in the text. ($\hbar{=}1$).}
\end{figure}

Returning to the more controversial statement, the idea will be developed here, based on evidence to be discussed below,   that the matrix model strongly suggests (in a Wignerian manner)  the existence of an  holographic dual to JT gravity with a single Hamiltonian. The {\it ensemble}  of spectra appearing in the random matrix model, which undoubtedly precisely captures 2D gravity on Euclidean surfaces of general topology~\cite{Mirzakhani:2006fta,Eynard:2007fi,Saad:2019lba}, is  to be interpreted holographically not as JT gravity, but as an ensemble of the class of whatever {\it single} Hamiltonian  characterizes  JT gravity. Following the above discussion, it is entirely natural that they appear as part of the Euclidean framework. Specifically, everything will fit well with  factorization, and with traditional holography, if there exists  a dual for JT gravity (defined just on  disc topology and then Wick rotating back to the Lorentzian picture) that has a single Hamiltonian. It has a {\it discrete} energy spectrum, but  upon approaching energy regimes  where the familiar Schwarzian description in terms of a continuous spectrum should be  valid, it reproduces that physics.\footnote{In all likelihood this extends beyond JT gravity to the many other kinds of gravity that can be captured by random matrix models.} (See figure~\ref{fig:JT-levels-150} for an example, to be more thoroughly discussed shortly once conventions are established.) 

For some Readers this will be a bridge too far.  A fallback position is that somehow 2D gravity is a stark exception to how  holographic duality works in other dimensions, and that there is no definite Hamiltonian, as is commonly stated as the  current belief.  Intuitively, this seems less elegant than having holography work the same basic way in all dimensions. Moreover, the current view has no natural explanation for the discrete structures that are uncovered in this paper. Instead, the proposal will be made, based on evidence to be uncovered below, that  a definite dual exists, for every kind of JT gravity. 

\begin{figure*}
\centering
\includegraphics[width=0.90\textwidth]{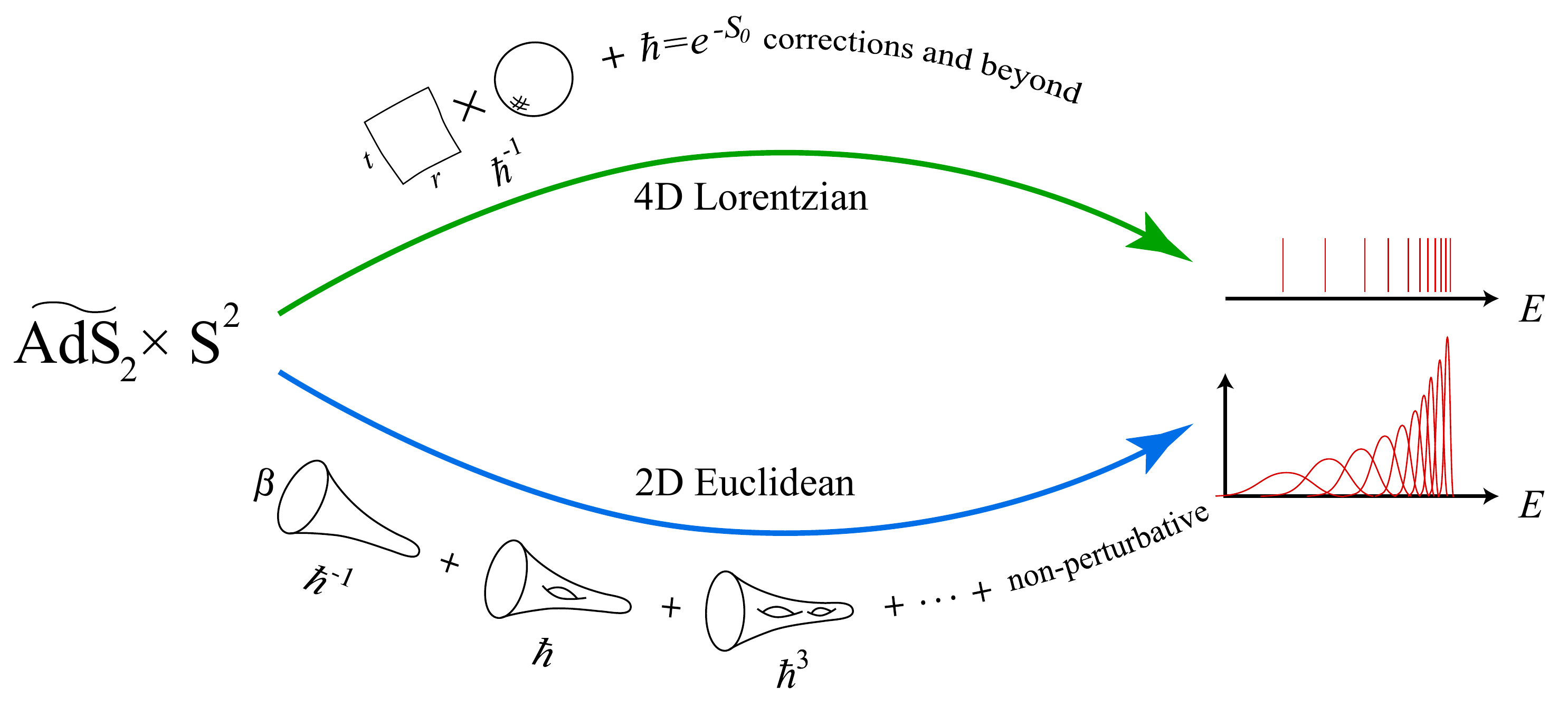} 
\caption{\label{fig:paths_to_spectrum} A schematic layout of the paths to finding the dual spectrum. The top path is four-dimensional, and Lorentzian. The bottom is two dimensional, and Euclidean. A Wignerian approach to the top path would study the ensemble of Hamiltonians with the correct asymptotic fixed by the Schwarzian result (see a  remark in footnote~\ref{fn:lorentzian-subtleties}), reading off the likely properties of the full spectrum at the end. This is a random matrix model. A 't Hooftian expansion of that same matrix model gives a topological expansion that coincides (as it must) with 2D Euclidean JT gravity and the perturbative matrix model of ref.~\cite{Saad:2019lba}.}
\end{figure*}

While this  assertion that there is a single dual spectrum cannot be proven within the random matrix model itself, the idea here is that it definitely points  to a potential loophole in the standard arguments underpinning the current belief that JT gravity cannot be dual to a system with a single Hamiltonian (see {\it e.g.} refs.~\cite{Stanford:2017thb,Harlow:2018tqv}.). A common feature of many  arguments seems to be the fact that the continuous spectrum of the Schwarzian dynamics that arises on the boundary (see below) is inconsistent with having a well-defined Hilbert space. However, as will be shown, the matrix model points to a definite {\it discrete} spectrum that only becomes effectively continuous at large $E$. So  this seems a good starting point for evading the usual arguments.

It is natural to wonder if  there is room for such a possibility in  non-matrix model approaches. Indeed,  ample opportunity seems available if the higher-dimensional origins of the JT gravity action are examined. The JT gravity action can be obtained by dimensional reduction from {\it e.g.,} a  four  dimensional Einstein-Maxwell system
with an ansatz $ds^2{=}g_{ab}(r,t)dx^adx^b{+}\Phi(r,t)^2d\Omega_2^2$, where $d\Omega_2^2$ is the metric on an $S^2$ with coordinates~$\theta$ and $\varphi$, and there is a magnetic flux $F_{\theta\varphi}{=}Q\sin\theta$. (See {\it e.g.,} ref.~\cite{Sarosi:2017ykf} for a review.) A $T{=}0$ solution is of course the extremal black hole throat AdS$_2{\times}S^2$ with $\Phi^2{=}Q^2 {=}A/4\pi {=} S_0/\pi$ where $S_0$ is the extremal Bekenstein-Hawking~\cite{Bekenstein:1973ur,Hawking:1974sw} entropy. An efficient way of handling this is to work with a $D{=}2$  dilaton gravity action, with $\Phi(r,t)$ as the dilaton and metric $g_{ab}(r,t)$. The choice $\Phi(r,t)^2{=}S_0{+}\phi(r,t)$ with $\phi$ small compared to a large $S_0$ is what gives rise to JT gravity. From there, in an Euclidean approach (see the action below in equation~(\ref{eq:JT-gravity-action})) $\phi$ is integrated out, yielding $R{=}{-}2$ and leaving the  Schwarzian action as the only dynamics on a fluctuating boundary of length~$\beta$. (Here $\beta{=}1/T$ is the period of Euclidean time $\tau=it$.) The spectral density of equation~(\ref{eq:leading-relation}) (below) emerges. Corrections to this  yields  perturbation theory organized in small $\hbar{=}\e^{-S_0}$ which can be interpreted as a topological expansion because $S_0$ multiplies $\chi$ (the Euler number of the two dimensional $(r,\tau)$ manifold) in the two-dimensional action.

The effective 2D gravitational coupling is related to the 4D one as follows: $G_N^{(2)}{=}G_N^{(4)}/S_0$, and so it is weak for large $S_0$. So to understand how to do more that just perturbation theory  about large $S_0$ requires  going beyond the safe regime of (semi-) classical gravity. The discreteness in the spectrum shows up precisely when away from the large $S_0$ limit, hence leaving room for additional physics to be present in the Schwarzian spectrum.

The core point here is that the $S^2$ origins of $\phi$ and its dynamics are important in the full story. Understanding the full spectrum of the (nearly) AdS$_2$ dynamics  needs a proper handling of quantum gravity on  the full ${\widetilde {\rm AdS}}_2{\times}S^2$, where the tilde is a reminder of the ``nearly'' aspect, that there is a  boundary of finite length $\beta$. This  is highly analogous to what happens in higher dimensions. Just doing semi-classical gravity on AdS$_5$ without  embedding it into a complete quantum gravity theory would never yield knowledge of  the ${\cal N}{=}4$ supersymmetric Yang--Mills holographic dual.  The complete physics required reference to the~$S^5$, yielding ten dimensional type~IIB supergravity,  then type~IIB string theory, and {\it then} its D3-branes and their world--volume dynamics. Without all that  (or something equivalent to it) 
 a sensible $D=4$ holographic dual cannot be discerned.

So there is room for a treatment of the ${\widetilde {\rm AdS}}_2{\times}S^2$ system that yields the discrete spectrum that might be expected on general grounds given the finite Bekenstein-Hawking entropy of the 4D black hole. Here is how it might have  happened in the  
spirit of a ``What Would Wigner Do?'' approach to the problem of finding the dual Hamiltonian,~$H$, (or at least its spectrum) of the Lorentzian  ${\widetilde {\rm AdS}}_2{\times}S^2$  system. Without the tools to solve for the desired Hamiltonian directly\footnote{\label{fn:d-brane-wrapping}There is presumably a non-matrix model way of constructing the right kind of  Hamiltonian directly, perhaps by building a near-extremal AdS$_2{\times}S^2$ system out of wrapped  intersecting higher-dimensional branes, probably  involving D5-branes, D1-branes, and Kaluza-Klein monopoles in IIB or D6-branes, D2-branes, and NS5--branes in IIA, or various S/T--dual variants thereof, and then looking at the effective Hamiltonian on their collective world-volume. See refs.~\cite{Johnson:1996ga,Maldacena:1996gb} for  early examples of  constructions, and ref.~\cite{Lozano:2021xxs} for a review of recent work, although variants without supersymmetry would be the preferred starting point.}  (following the upper path of   figure~\ref{fig:paths_to_spectrum}),  one might gather  as much  information as possible to constrain $H$'s properties, and then study the {\it ensemble} of Hamiltonians with those properties, to see if that approach might yield more information about  $H$'s expected  properties~\cite{10.2307/2331939,10.2307/1970079}. Here, the input information for such an approach  is the leading large $E$ behaviour, which is the Schwarzian spectral density (see equation~(\ref{eq:leading-relation}) below). It implicitly determines the potential (probability density function) of the matrix model.  The step that completes the story is then solving the model non-perturbatively~\cite{Johnson:2019eik,Johnson:2020exp}, and then (in Wignerian spirit) reading off the answer for what the dual Hamiltonian's spectrum most probably looks like. This last step  is what  will be discussed at length in this paper, building on work started in ref.~\cite{Johnson:2021zuo} that observed that the matrix model spectral density, although continuous, can {\it also} be written in as a sum of discrete peaks $p(n;E)$, each of which is in one-to-one correspondence with the energy spectrum $\{ {\cal E}_n\}$ shown in figure~\ref{fig:JT-levels-150}. It is key here that no reference to Euclidean physics was made. This was a search for an answer to a purely Lorentzian problem of determining the complete Hamiltonian of the theory, expected to have a discrete spectrum because of the finite entropy.

Now notice that there is another interpretation of this {\it same} matrix model. Its  't Hooftian $1/N{\sim}\e^{-S_0}$ expansion can be interpreted ({\it via} its Feynman diagrams in the usual way) as a topological expansion in terms of  {\it  Euclidean} ${\widetilde {\rm AdS}}_2$ geometries with handles. See the lower path of  figure~\ref{fig:paths_to_spectrum}. Of course, this is the content of the perturbative  results~\cite{Saad:2019lba} of Saad, Shenker and Stanford!  Now it is (hopefully) clear that it all arrives as part of a Wignerian toolbox for finding the single spectrum of a Lorentzian theory.

The primary purpose of this paper will be to thoroughly explore random matrix models (especially of JT gravity and JT supergravity, but using matrix model building blocks that individually also have independent lives as other kinds of 2D gravity theories) deploying non-perturbative tools\footnote{This exploration began in ref.~\cite{Johnson:2021zuo}, where Fredholm determinant techniques where first used, in this 2D gravity context, as a tool for examining the spectrum. The (successful) goal was to better understand methods for computing the quenched free energy of the system. The current paper started as a report on  generalizations of those results to JT supergravity, but it soon became clear  that they were a sign of a far deeper issue.} that are more forensically spectroscopic than usually used in the 2D gravity context (but are well-known in the broader random matrix community). The special kind of ``microstate'' spectrum $\{{\cal E}_n\}$ that emerges in each case is something that is present, incontrovertible,  and perhaps surprising to many. It does not currently have a good explanation in the standard paradigm that regards JT gravity as fundamentally an ensemble, and so it is not clear what its role is in that framework. That could be taken as a telling sign, but of course  it might be possible to give it a statistical interpretation and preserve the status quo. 

Nevertheless, the strategy to be taken here is that this special spectrum  is the matrix model's way of strongly signaling what the spectrum of the {\it single} dual Hamiltonian is for JT gravity. If this is true, holography for 2D gravity becomes again more of a piece with AdS/CFT and holography in other dimensions,  factorization is safe, and matrix models have given a new understanding  of the role of Euclidean quantum gravity on surfaces of higher topology, in any dimension.

\subsection{JT Gravity}
As already mentioned, Saad, Shenker and Stanford~\cite{Saad:2019lba} and Stanford and Witten~\cite{Stanford:2019vob} have shown that JT gravity (and several variants) on manifolds of arbitrary topology can be given a formulation as  double-scaled random matrix models
at least perturbatively in the topological expansion parameter $\hbar\equiv \e^{-S_0}$. The  Euclidean action for JT gravity on a two dimensional manifold ${\cal M}$ (with boundary $\partial{\cal M}$), which also includes a scalar $\phi$ to yield non-trivial dynamics, is:\begin{eqnarray}
\label{eq:JT-gravity-action}
I &=& - \frac12\int_{\cal M}\!\!\sqrt{g} \phi(R+2) -\int_{\partial \cal  M}\!\!\sqrt{h} \phi_b (K-1) \nonumber\\
&&\hskip 1.5cm -\frac{S_0}{2\pi}\left(\frac12\int_{\cal M} \!\!\sqrt{g} R +\int_{\partial {\cal M}}\sqrt{h}K\right)\ , 
\end{eqnarray}
where $R$ is the Ricci scalar and  in the  boundary terms,~$K$ is the trace of the extrinsic curvature for induced metric $h_{ij}$ and $\phi_b$ is the boundary value of $\phi$. The constant~$S_0$, which is in fact the $T{=}0$ entropy,  multiplies the Einstein-Hilbert action,  yielding the Euler characteristic $\chi({\cal M}){=}2{-}2g{-}b$ 
with $g$ handles and $b$ boundaries. 
Schematically, the partition function $Z(\beta)$ of a model (in an Euclidean formulation where $\beta{=}1/T$ is the period of $\tau{=}{-}it$) is:
\begin{equation}
\label{eq:partition-sum}
Z(\beta)=\sum_{g=0}^\infty Z_g(\beta)+\cdots = \int\! \rho(E) \e^{-\beta E}dE\ , 
\end{equation} 
where $Z_g(\beta)$ is the contribution from working on manifolds ${\cal M}$ with Euler number $\chi$ but with $b{=}1$. It contains a factor~$\hbar^{-\chi}$. The ellipses denote contributions beyond the perturbative expansion, which will be discussed considerably here. The continuous function $\rho(E)$ has a topological expansion too, and also non-perturbative parts. It is often referred to as the ``spectral density" in the JT gravity context, and one of the key points of this paper is that {\it this name should be exercised with considerable care,} because it is only partially true. 

Pulling on this thread will unravel much of the contemporary narrative tapestry about 2D gravity, but happily, the threads will be swiftly rewoven into a new narrative that fits all the known facts, and solves several puzzles.

\subsection{A New Arrangement for the Ensemble}

\begin{figure}[t]
\centering
\includegraphics[width=0.4\textwidth]{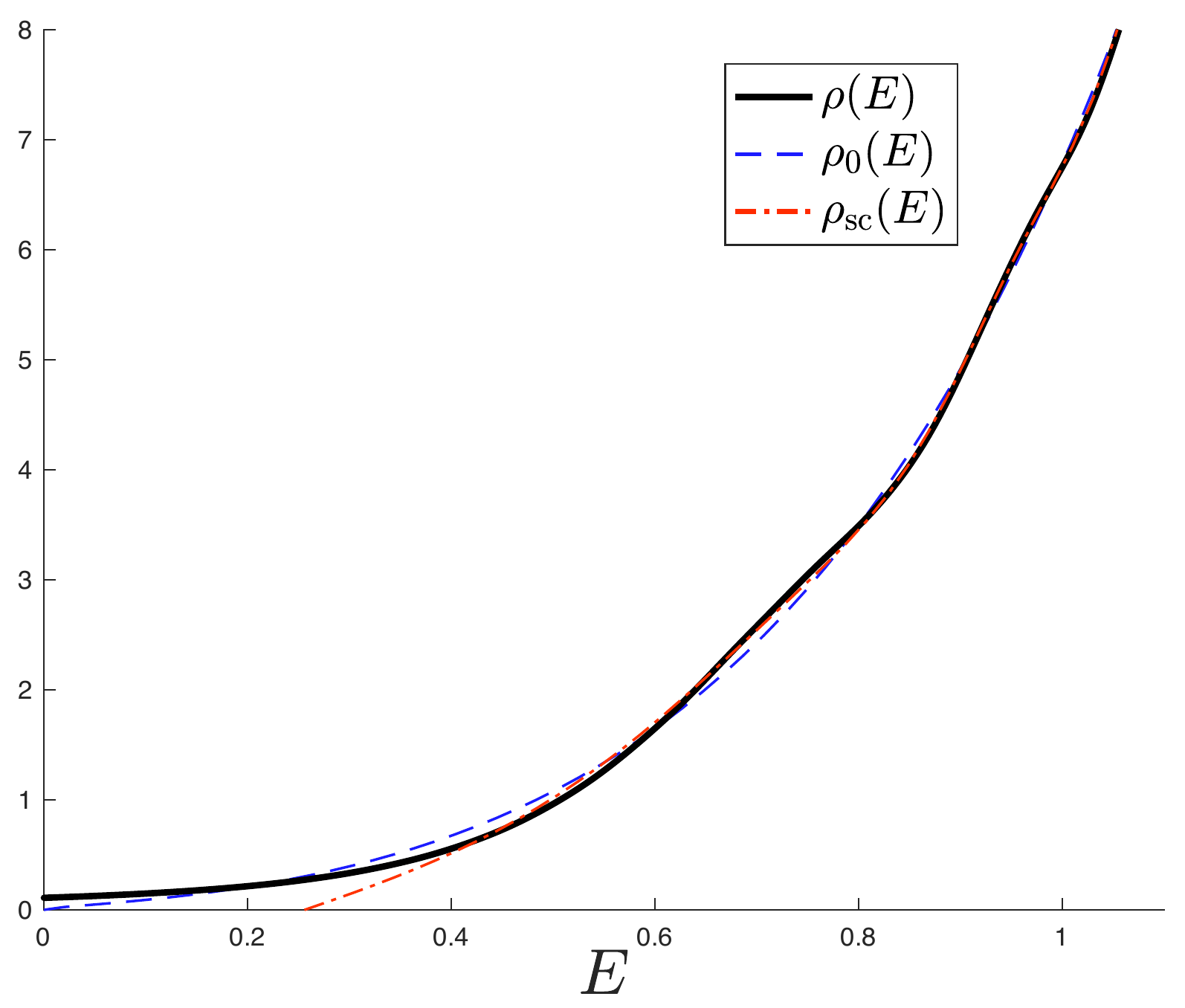}
\caption{\label{fig:full-density-vs-semi-classical} The full  spectral density $\rho(E)$ (solid black) for a non-perturbative completion of JT gravity, the leading disc result $\rho_0(E)$ (dashed), and the semi-classical estimate $\rho_{\rm sc}(E)$ (red, dash-dot). Here $\hbar{=}1$.}
\end{figure}

An oft-repeated piece of folklore in this context is that since it can be captured by a random matrix ensemble, JT gravity necessarily has a continuous spectrum. 
Nevertheless, a reconsideration of $\rho(E)$'s contents (begun in ref.~\cite{Johnson:2021zuo}) will be developed in this paper that brings to the fore 
 a very specific underlying {\it discrete} spectrum. As already stated above, it will be taken to be the {\it actual}   spectrum of quantized (Lorentzian) JT gravity. The first 150 levels have been computed to good accuracy  are displayed in figure~\ref{fig:JT-levels-150}, in the case of setting $\hbar=1$ (It can be done for smaller values too). Notice that it is only at large~$E$ (compared to~$\hbar{=}\e^{-S_0}$, which sets the size of the gaps in the spectrum) that an effectively continuous spectrum  emerges. This then coincides with the large $E$ limit of the function $\rho(E)$, indeed corresponding to  the  leading order  in topological perturbation theory  ({\it i.e.,} the disc), which for ordinary JT gravity  is: 
\be
\label{eq:leading-relation}
\rhoo(E) = \frac{\sinh{(2\pi\sqrt{E})}}{4\pi^2\hbar}\ ,
\ee
 which arises from Schwarzian dynamics~\cite{Maldacena:2016hyu,Stanford:2017thb}.
This function is plotted as the (blue) dashed line in figure~\ref{fig:full-density-vs-semi-classical}, where it becomes increasingly valid to the right.
Analogous spectra (and the accompanying proposal for their interpretation) will be made in several other JT models later in the paper.

For the evidence of this idea offered by the matrix model, return now to  the fully non-perturbative matrix model  function $\rho(E)$, and the note of caution sounded upon its first appearance above. For general~$E$ it will be seen to be a blurred version of the physical spectral density, where the word  ``physical'' privileges the perspective of the gravity theory, and not the matrix model. 
Specifically, it is a {\it fact} it can be written uniquely as:
\be
\label{eq:fattening}
\rho(E) = \sum_{n=0}^\infty p(n;E) \ ,
\ee
where $p(n;E)$ is an infinite family of specific peaked functions that are probability density functions of the $n$th energy level or microstate in the matrix ensemble, satisfying $\int \!  p(n;E)dE{=}1$. The mean energies of these distributions will be denoted as the set $\{ {\cal E}_n\}$, for $n{=}0, 1,\hdots\infty$, and it is this that is proposed as the JT gravity ``microstate'' spectrum, where~${\cal E}_0$ is the ground state.\footnote{\label{fn:peak-subtleties}Another natural possibility is to use the top of each peak $p(n,E)$, defining the most frequent energy at  level $n$. Either is a good choice, and it is not clear at present which prescription is correct. This requires further work. Other prescriptions for picking a representative energy at each level could emerge too. The point here is that there {\it is} some definite spectrum, and the peaks indicate its typical form. The choice used in this paper, the mean, seems appropriate since there is another natural object in the theory, a D-brane partition function, whose zeros represent the mean eigenvalues of the size $(N{-}1){\times}(N{-}1)$ matrix, as will be discussed in Section~\ref{sec:return-to-gravity}. That such an object appears  naturally in the theory is suggestive that mean values correlated across the entire matrix are a sensible notion. Ultimately, this detail matters less than the core idea that {\it some} definite discrete spectrum emerges at the end as the dual spectrum.} Strong evidence using matrix model analysis will be given below that the set $\{ {\cal E}_n\}$ does indeed have an interpretation as a very specific spectrum. In fact,  these points will be shown to be the fully non-perturbative ``locations'' of a kind of D-brane of the model. (Incidentally, whether the proposal is accepted or not, this paper uncovers new tools for understanding D-branes non-perturbatively in the matrix model that can be applied while motivated by  standard interpretations.)

\begin{figure}[b]
\centering
\includegraphics[width=0.48\textwidth]{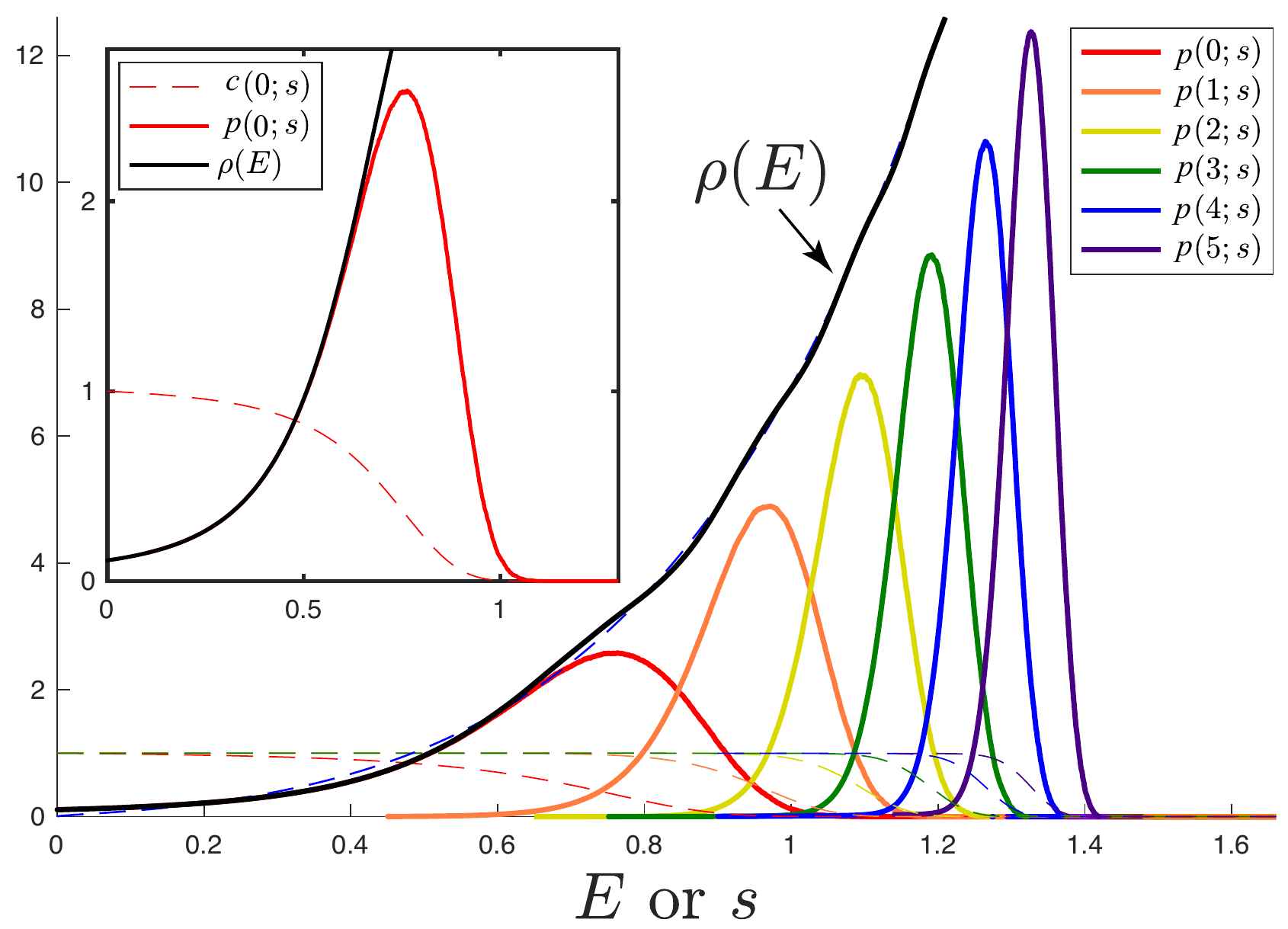}
\caption{\label{fig:JT-spectrum-from-fredholm} Full spectral density $\rho(E)$ (solid black), leading density $\rhoo(E)$ (blue dashed), and probability densities (also cumulative probabilities, dashed)  of the first 6  states of the  JT gravity microstate spectrum. ($s$ is another name for $E$ that will be explained later.)
Inset: Close-up of $\rho(E)$ and  distributions for the ground state, with ${\cal E}_0{\equiv}\langle  E_0\rangle{\simeq}0.663$. Note that $\hbar{=}{\rm e}^{-S_0}{=}1$ here. }
\end{figure}

The $p(n;E)$, many of which will be computed later in several examples, 
should be thought of as broadened $\delta$-functions located at the microstate energies. A figure from ref.~\cite{Johnson:2021zuo}, where they were first computed for a particular non-perturbative definition of JT gravity, is reproduced in figure~\ref{fig:JT-spectrum-from-fredholm}, showing the first six, for the case of $\hbar{=}1$ (results for other values of $\hbar$ can be readily computed--this choice makes them most manifest). The solid black curve is the fully non-perturbative~$\rho(E)$, and it is perhaps visible that the various peaks beneath actually sum to reproduce it, as will emerge later.
Crucially, their width and separation decrease as~$E$ increases, and  at large $E$ they  are well approximated by  a dense set of $\delta$-functions, forming a continuum. Compare to figure~\ref{fig:JT-levels-150} (on page~\pageref{fig:JT-levels-150}) where the green dots are the means of these peaks.

In fact, as will be explained later, at large level $n$, a useful asymptotic form for the $n$th energy ${\cal E}_n$ in the spectrum is:
\be
\label{eq:level-approximation}
\int^{{\cal E}_n}_0 \!\!\rhoo(E) dE = n\ ,
\ee
(this was used to compute the high levels in  figure~\ref{fig:JT-levels-150}), a result that applies beyond JT gravity to other gravity models, such as the  minimal series discussed in Section~\ref{sec:double-scaling}. From here it is easy to see that the density of these states  $\rho({\cal E}_n) {=} dn/d{\cal E}_n$ is the leading disc level density (the Schwarzian in the case of JT gravity).

All works to date on the matrix model description of JT gravity (and variants), even if using the full non-perturbative form of $\rho(E)$, have  assumed that it is the last word on the spectral content.  In fact it is merely   a rough draft of the spectral data: The $p(n;E)$ carry far more information, and the central point of this paper is that  they encode the true, {\it discrete}, quantum gravity spectrum of JT gravity.\footnote{Even in the very simplest, Gaussian, matrix models, the $p(n;E)$ have no known analytic form and must be computed by numerically evaluating certain Fredholm determinants. Later in this paper, more of these peaks will be computed for JT gravity, and for several JT supergravity examples, up to an energy  level where certain semi-classical techniques can be used to determine their location to arbitrarily high levels with extremely high accuracy.}

 Of course, it should be noted that a great deal has been learned about $\rho(E)$, and from it. For JT and several supersymmetric Jackiw-Teitelboim (SJT) gravity variants, it has been shown in a series of works (starting with refs.~\cite{Johnson:2019eik,Johnson:2020heh,Johnson:2020exp,Johnson:2020mwi}) that well-defined and computationally explicit non-perturbative definitions can be constructed\footnote{The existence of alternative non-perturbative approaches, and how to cross-compare them, has  discussed recently in ref.~\cite{Johnson:2021tnl}.}  that allow for the computation of the {\it full}~$\rho(E)$. (Figures~\ref{fig:full-density-vs-semi-classical} and~\ref{fig:JT-spectrum-from-fredholm} show it as the black curve, in the case of 
 JT gravity.)  
Such non-perturbative completions allow for the examination of key issues that cannot be fully addressed in perturbation theory such as stability, the nature of the low energy spectrum, and the late-time dependence of various useful diagnostic quantities such as   the spectral form factor. 

The non-perturbative  physics manifests as  undulations in  $\rho(E)$,  its two point correlator $\langle\rho(E)\rho(E^\prime)\rangle$, and higher point correlators. Much can be learned about various  phenomena from even just the leading form of these undulations, as is done in most of the literature on these matters (see {\it e.g.} the early work in refs.~\cite{Saad:2019lba,Blommaert:2019wfy,Saad:2019pqd}). The form can be deduced without full knowledge of the complete non-perturbative formulation, using methods~\cite{Saad:2019lba} equivalent to using~\cite{Johnson:2021tnl} a certain WKB approximation to be reviewed below. For JT gravity it is (for $E>0$):
\be
\label{eq:semi-classical-np-1}
\rho_{\rm sc}=\rhoo(E) - \frac{1}{4\pi E}\cos\left(2\pi\!\int^E_0\!\!\!\rhoo(E^\prime)dE^\prime\right)\ , 
\ee
where $\rhoo(E)$ is in equation~(\ref{eq:leading-relation}), and it is a good approximation toward larger $E$. At small enough $E$ it becomes inaccurate and eventually breaks down, as shown in figure~\ref{fig:full-density-vs-semi-classical} (red dash-dot line).

However, for many issues, there is no substitute for knowing the full non-perturbative physics. 
It is having access to it  that yields  the knowledge of the   ``microscopic''  physics underlying $\rho(E)$'s ubiquitous undulations. This informs equation~(\ref{eq:fattening}), leading to   the idea that it represents a discrete spectrum with a family of broadened peaks. The broadening is a consequence of the fact that a random matrix ensemble is being used to study the gravity theory. It is not a bad thing, but actually a rather powerful trick. As will be discussed in the next sections, it is  {\it precisely} this  broadening phenomenon that is responsible for the matrix model being able to yield smooth geometries of diverse topology, and  it has proven to be   a powerful tool for analysis of many key aspects of the physics. However, what the matrix model is actually doing  needs  to be re-assessed in the light of what has  been suggested above about the meaning of the spectrum. Even a single copy of  the full  $\rho(E)$ should instead  be regarded as an ensemble of  JT gravity-like spectra and quantities like $\langle \rho(E)\rho(E^\prime)\rangle$
are really spectral correlations averaged over {\it families} of JT theories. (A core point of this paper is that this is just fine in the Euclidean approach, and does not imply a fundamental ensemble dual.)

This will take some getting used to, but as mentioned above,  it will  have the pleasant consequence of placing the holographic discussion of JT gravity and its variants on a much more similar footing to that of higher dimensional AdS/CFT, and in the process  entirely resolve the  factorization puzzle in all dimensions.

\subsection{Rethinking Random Matrix Models}
 It is a common phenomenon in many walks of life that sometimes confusion  arises when  the instrument being used to study a thing gets mistaken  for the thing itself. The idea here is that perhaps this  has  happened with the matrix model, leading  to a sharp example of the factorization puzzle, and much understandable concern that gravity must   somehow necessarily always be an ensemble average of multiple variants of  its holographic dual. (See {\it e.g.,} discussion in refs.~\cite{Maldacena:2004rf,Saad:2019lba,Blommaert:2019wfy,Marolf:2020xie,Penington:2019kki} and {\it e.g.,} refs~\cite{Saad:2021uzi,Blommaert:2021fob} for recent attempts to resolve the puzzle by making modifications to the matrix model.)  
 
Well, the matrix model framework is a controllable model of emergent gravity, {\it i.e.,} it allows for complete control of the all the physics, and it is only in the  perturbative, regime that smooth Euclidean geometry arises. From the perspective of this paper (but also, operationally from how the integrals in ref.~\cite{Saad:2019lba} work at any order in perturbation theory), smooth geometry arises when the gaps in the spectrum can be ignored, effectively allowing a smooth spectral density to be substituted. This is a necessary (but not sufficient\footnote{Not all double-scaled matrix models with smooth $\rho(E)$ have a smooth spacetime interpretation. The Airy model, for example, since it arises as the double-scaling limit of the Gaussian model, has no surfaces at all. Its ``bulk'' spacetime physics is entirely topological.}) condition for a smooth spacetime geometry description. There are in fact two different ways this can happen, and it is worthwhile examining them in two separate types of scenario: Geometries with a single boundary, and geometries with multiple boundaries. The latter of course famously involves wormholes.

\subsubsection*{Single Boundary: Rethinking the Partition Function.} Consider geometries with a single boundary. They  contribute to the partition function $Z(\beta)$ in equation~(\ref{eq:partition-sum})--the disc, torus with a hole, double-torus with a hole, {\it etc}. As already seen, the gaps in the spectrum can be neglected by having large~$E$ compared to $\hbar{=}\e^{-S_0}$, (or large temperature $T{=}\beta^{-1}$). The smooth function $\rho_0(E)$ can be used, plus possible  corrections  perturbative about the large $E$ limit. This can then  be connected to path integrals over the appropriate smooth hyperbolic manifolds in the usual way, as elegantly described by ref.~\cite{Saad:2019lba}.  This is the first route to smoothness. 

A key point here is that even though one is in the matrix model, averaging over many spectra, all matrix spectra in this limit have the same large $E$ form, and so there is no interpretational worry about ensembles here. The average $\langle Z(\beta)\rangle$, which is {\it really} what the matrix model is computing, coincides well enough with $Z(\beta)$ for a single copy of the spectrum, as long as $E$ is large. To a good approximation then, the role of the matrix ensemble here is entirely in the 't Hooftian spirit of providing an elegant method for enumerating the geometry. 

As $E$ gets small enough (compared to $\hbar$) so that the continuum description of the spectra breaks down, things are somewhat more subtle. If there was just a single copy of the spectrum, with discreteness appearing, it would no longer be possible to ascribe a geometry this regime. This does not mean that access to the physics is lost, but just that a spacetime geometrical language is no longer the best tool. However, more can be done with a continuum language because a {\it  second} kind of filling of the gaps to obtain smoothness is possible. By combining  an entire ensemble of spectra (each looking a bit like figure~\ref{fig:JT-levels-150} but the locations of the discrete energies are different in each case), the gaps fill in to make the smooth $\rho(E)$--now the Wignerian spirit is in play! This smooth function  can be Laplace-transformed according to equation~(\ref{eq:partition-sum}) and used as the result for $Z(\beta)$. But now it is {\it really} important that it be understood as $\langle Z(\beta)\rangle$. This is usually referred to as the non-perturbative partition function of JT gravity but from the perspective of this paper, this is not really what it is. It is the average partition function of an ensemble of spectra, only {\it one} of which is the one identified as the JT gravity (dual) spectrum. This is the first example illustrating  the opening conjecture about how to reinterpret the matrix model's  Euclidean quantum gravity sum over surfaces of higher topology.

Simply put, this recasts the matrix model as an  ensemble of {\it  JT gravity-like duals}. One spectrum--the average or typical  one, in the spirit of Wigner--is JT gravity (or more precisely, the spectrum of its holographic dual). So what are all the others? Clearly they must correspond to deformations of JT gravity. This point will be returned to in a while. 

\subsubsection*{Clues from Thermodynamics} Before skepticism sweeps in, if it  hasn't yet, note that this perspective immediately solves a known puzzle about the thermodynamics: If $\langle Z(\beta)\rangle$ is the partition function, one should be able to compute the free energy from it, and hence the entropy,~$S(T)$. Starting with the simple Schwarzian result $Z_0(T)$, this fails to give a sensible answer~\cite{Engelhardt:2020qpv}. At some finite~$T$, $F(T){=}-T\log Z_0(T)$, which has negative slope, develops a maximum and then starts decreasing again. The entropy is therefore starting out positive, decreasing to zero at some finite $T$, and then going negative. In fact, it eventually diverges, negatively. Adding perturbative corrections does not help.\footnote{This was noticed in collaboration with Robie Hennigar, Felipe Rosso,  and Andrew Svesko  in March 2020, but left unpublished as it wasn't clear at the time what the meaning was.} Clearly this is not good behaviour for an entropy. Moreover, it never gets to $S_0$, which it is supposed to at $T=0$. It is natural to  assume  that this could be due to the fact that the realm of validilty in which  Schwarzian was derived breaks down at small enough $T$, and that it would be solved once  the full non-perturbative corrections are included. In fact, they do not help, as reported in ref.~\cite{Johnson:2020mwi}, with $S$ derived from the full $\langle Z(T)\rangle$ still going negative at some positive $T$. 

Ref.~\cite{Engelhardt:2020qpv}  attributed to the matter to the issue of computing $F_Q(T){=}{-}T\langle \log Z(\beta)\rangle$ {\it vs.} $F_A(T){=}{-}T\log\langle Z(T)\rangle $, {\it i.e.} the quenched free energy instead of the annealed free energy, in the language of disordered systems. Later, some techniques for computing $F_Q(T)$ in the matrix model  gave results in the ultra-low $T$ limit~\cite{Okuyama:2021pkf,Janssen:2021mek}, and in refs.~\cite{Johnson:2021rsh,Johnson:2021zuo} it was done successfully for complete matrix models using various methods, and in particular for JT gravity.

\label{sec:annealed}
However, the question of what it all {\it means} can still be asked. The answer is well known in the disordered systems literature, and it gels perfectly with the Wignerian perspective presented here. Generically, the annealed free energy of a system starts giving wrong answers at low temperatures because atypical configurations of the system are making large contributions to the sum. Such contributions  become more prevalent at low $T$. Notice that they simply add up in the average partition function, and then the log is taken. This will generically give a free energy that is smaller than the true value. On the other hand, the contributions of such configurations wash out in the quenched free energy upon averaging the logarithm of the partition sum. The final result for $F_Q(T)$   should therefore be attributed to the contribution of the most typical or average configurations of the system. 

Clearly a random matrix model is an excellent  testbed of this framework.   Now ``configurations'' are replaced by ``matrix spectra''. This was explored in ref.~\cite{Johnson:2021rsh} where a number of matrix models (with actual matrices) were explicitly queried by computing $F_Q(T)$  and $F_A(T)$. Some of the results will be recalled later.  In this paper, it will also be observed  that if one computes the mean spectrum $\{{\cal E}_n\}$ for a matrix model, and then directly evaluates the free energy $F_m(T){=}{-}T\log\left(\sum_{n=0}^\infty\e^{-{\cal E}_n/T}\right)$, the result (although not identical to it) correlates well with~$F_Q(T)$.

\begin{figure}[t]
\centering
\includegraphics[width=0.48\textwidth]{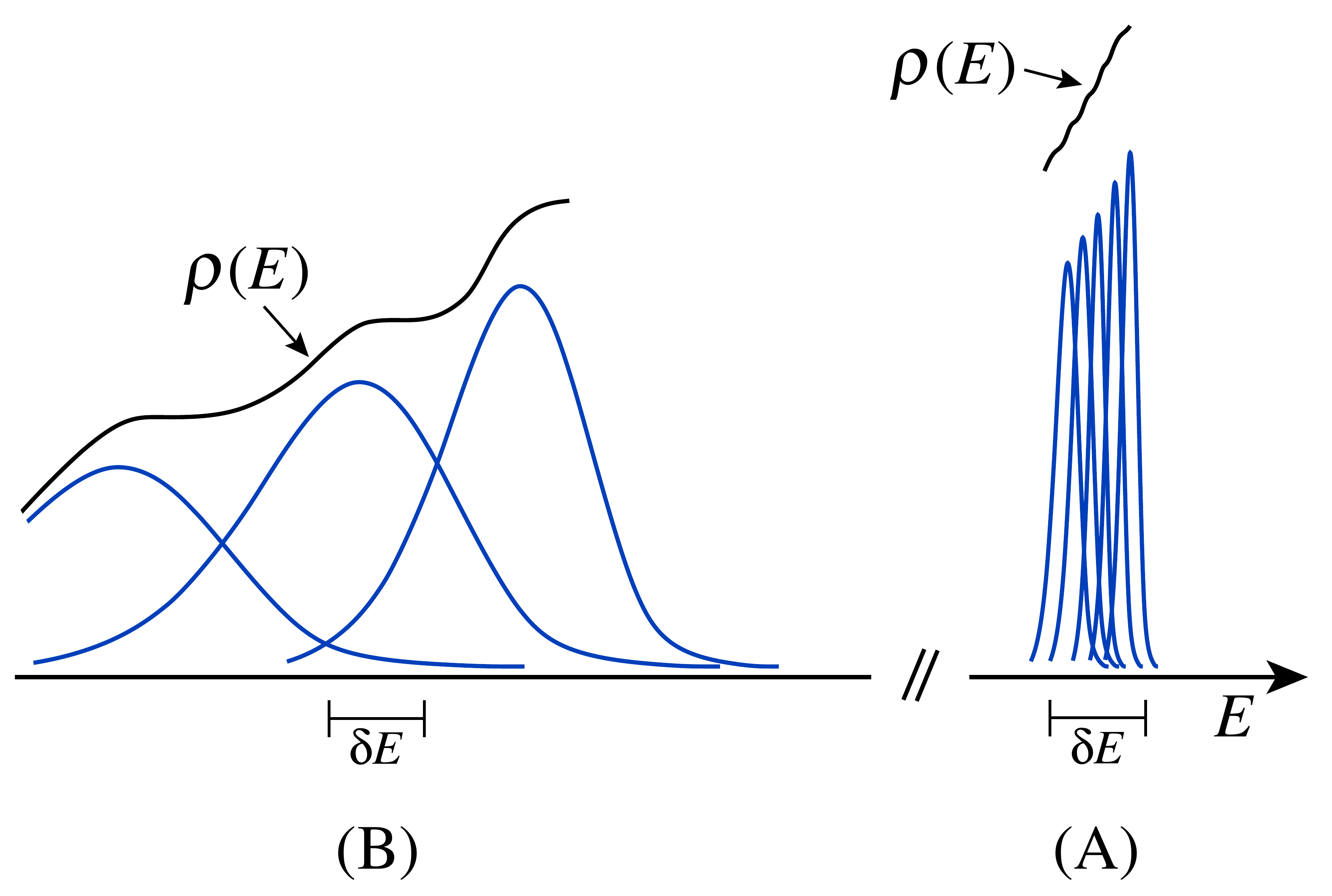}
\caption{\label{fig:energy-interval-scenarios} A sketch of low {\it vs.}  high energy regimes with the full spectral density $\rho(E)$ (solid black). The blue curves, whose sum gives $\rho(E)$, are the underlying microstate peak probability densities. In  the high energy scenario {\bf A} (right), for a given $\delta E$,  $\rho(E)$ captures more contributions from typical configurations than in the low energy scenario {\bf B} (left).}
\end{figure}

This also fits explicitly with the description above of how $\langle Z(\beta)\rangle$ is built by the matrix model. At successively lower $E$ (and hence $T$), in a given interval size $\delta E$, the smooth $\rho(E)$ it is made from includes successively  more contributions that are atypical, because the variance of the peaks $p(n;E)$ grows with lower energy. See figure~\ref{fig:energy-interval-scenarios}. So the $F_A(T)$ built from it will be increasingly afflicted by atypical configurations that skew the free energy result.

The suggestion is clear then. Correct and sensible thermodynamics derived by computing $F_Q(T)$ (averaging correctly over the entire ensemble of spectra) shares properties with   the free energy of a  {\it single} spectrum, taken here to be the mean spectrum. This lends a lot of support to the idea that the dual of JT gravity is has this (or similar)  single  spectrum.  So since it is  an ensemble average over a class of such spectra, each of which is in the class (defined by the matrix model) of Hamiltonians dual to a JT gravity:

\begin{itemize}[label={\raisebox{2pt}{\tiny\textbullet}},leftmargin=*,itemsep=-0.5ex]  
\item {\it  The random matrix model is  an ensemble of Hamiltonians dual to JT gravity-like theories.}
\end{itemize} 

Each different spectrum deserves to be thought of dual to a cousin of JT gravity, and it seems natural that they can be thought of as deformations of JT gravity. This immediately  follows from the fact that deformations of JT and SJT gravity correspond to deformations of the matrix model potential~\cite{Maxfield:2020ale,Witten:2020wvy,Rosso:2021orf}. This changes $\rho(E)$, and hence the pattern of peaks $p(n;E)$, and consequently the set $\{{\cal E}_n\}$ will change.

 \subsubsection*{Two Boundaries: Rethinking the Spectral Form Factor}
\begin{figure}[t]
\centering
\includegraphics[width=0.48\textwidth]{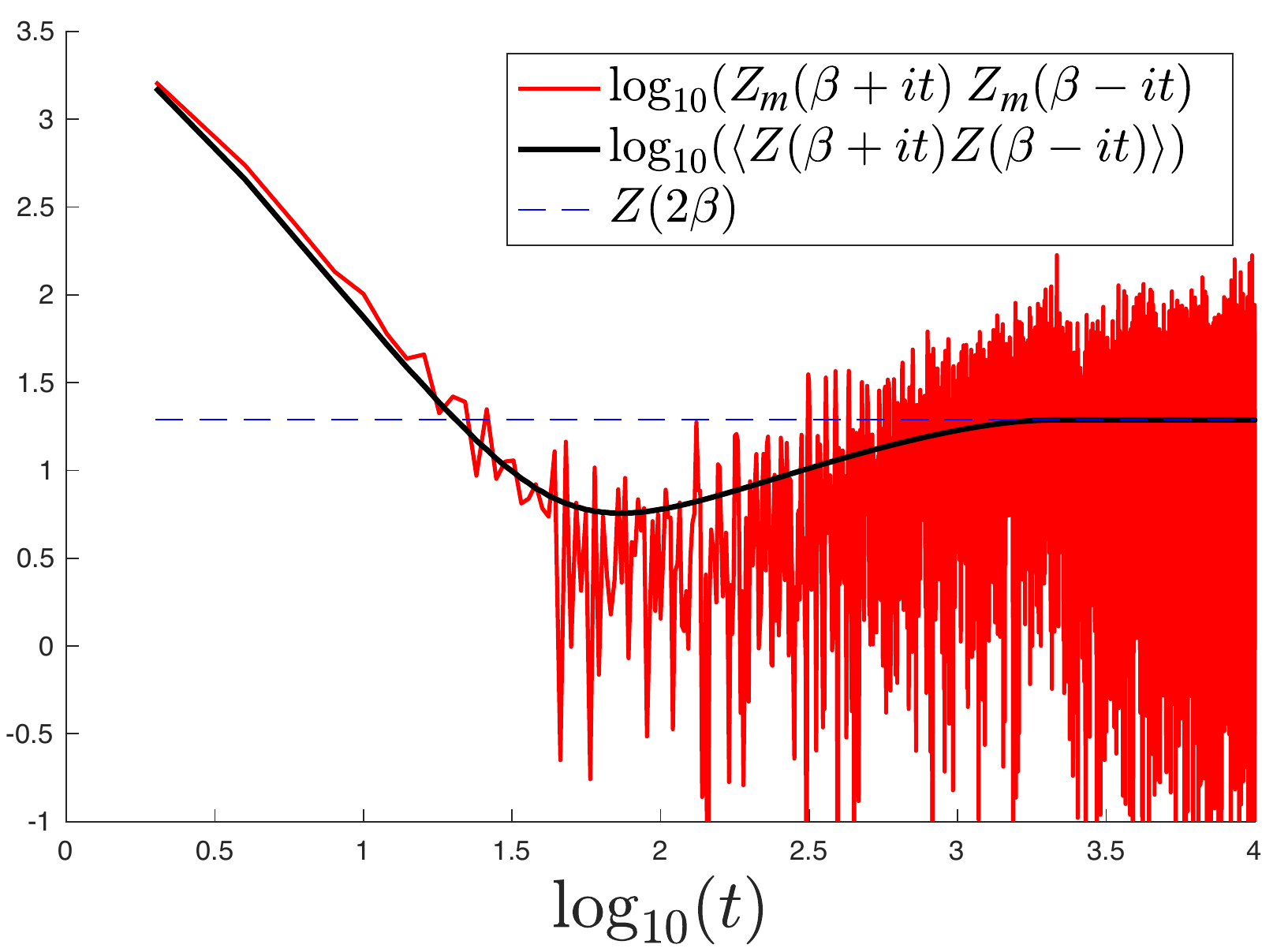}
\caption{\label{fig:JT_spectral_form_factor_compare_A} The spectral form factor for the mean spectrum $\{ {\cal E}_n\}$ (red, jagged) plotted against the ensemble averaged quantity, for JT gravity (black, smooth). The two curves follow each other closed at early times before the red curve begins fluctuations. The averaged quantity saturates to the plateau value $Z(2\beta)$. Here $\beta{=}1/2$, and $\hbar{=}1.$}
\end{figure}

Having two boundaries is necessary to describe, for example, the spectral form factor (see refs.~\cite{Garcia-Garcia:2016mno,Cotler:2016fpe,Saad:2018bqo} for early work in this context):
\be
\label{eq:spectral-form-factor-A}
Z_m(\beta+it)Z_m(\beta-it)=\sum_{j,k}\e^{-\beta(E_j+E_k)}\e^{it(E_j-E_k)}\ ,
\ee
where it is intentional that there are no averaging signs around this quantity. What is meant here by $Z_m(\beta)$ is the partition function built from a single copy of the spectrum, {\it i.e.} the  mean spectrum $\{{\cal E}_k\}$ that is a good representative of the proposed dual.   In such a case, it is inevitable that the discreteness of the underlying microstates must make their presences known in the physics. The limit of late time $t$ yields a particular pattern of (seemingly erratic) oscillations due to correlations between the discrete states. As a look ahead to later in the paper, see the red curve in figure~\ref{fig:JT_spectral_form_factor_compare_A}.  The first $175$ discrete states of the JT spectrum $\{{\cal E}_n\}$ were computed to good accuracy and used to construct the spectral form factor. 

To be clear, at  any $t$ the matrix model gives a specific understanding of what such states look like (in fact they they can be regarded as D-branes; See section~\ref{sec:return-to-gravity}), and so there is no mystery as  to their meaning. Their description, and how they play in the spectral form factor, does not involve a  smooth geometrical language however, due to the gaps in the spectrum as~$t$ becomes large enough. Again, the gaps can be filled in even at large $t$ by taking multiple copies  of an ensemble of 
spectra. This gives  an effective continuous spectral function, the matrix model $\rho(E)$, that has the gaps filled in, with only the bumps present.\footnote{Notice that this is highly analogous to smearing D-branes or fattening them into supergravity objects in higher dimensions.}  Now the resulting continuous spectrum allows a smooth geometrical description to emerge: It is the wormhole. 

The price paid for this process is a loss of the effects of the precise microstate spectrum. The oscillations of the spectral form factor are replaced instead by the softened aggregate behaviour of all the spectra, and the resulting ``ramp'' and ``plateau'' behaviours, nicely produced by the wormhole (see the solid black line in figure~\ref{fig:JT_spectral_form_factor_compare_A}, computed using the kind of matrix model studies of JT gravity presented in ref.~\cite{Johnson:2020exp}), give instead universal features of the  {\it class} of spectra to which the specific spectrum belongs. Of course, such universal information is nice to have, but it must not be forgotten that specific information about the {\it actual} spectrum is of interest to, for studying the specific quantum gravity of interest. 
(See Section~\ref{sec:spectral-form-factor-airy} for  further discussion of wormhole emergence, as well as the brief comment on implications for string theory a few paragraphs down from here.)

\subsubsection*{A Happy Implication: No Factorization Puzzle}
The overall point here is that the matrix model renders everything smooth by lumping discrete spectra into the simple object $\rho(E)$, making the passage to smooth geometry possible, in the spirit of 't Hooft. But it does this at the cost of involving all the spectra, in the spirit of Wigner. If the full matrix model spectrum is regarded as equivalent to that of JT gravity, as is currently conventional wisdom, then inevitably one must conclude that gravity is an ensemble, and all sorts of factorization issues arise (see earlier references), and questions about AdS/CFT in higher dimensions. But the   alternative interpretation given here, strongly suggested by the matrix model itself, and  the thermodynamics, and bolstered by the several steps of the discussion above, is that  JT gravity (defined on the disc topology and then Wick rotated to the Lorentzian signature) has a discrete spectrum (but becoming effectively continuous at large~$E$).  This is also  the spectrum of its holographic dual. The pleasingly simple logical conclusion, already made above, is that the matrix model has two complementary interpretations:  The Wignerian one picks out a single spectrum as the Lorentzian answer (the peaks show that there is a preferred one) while the 't Hooftian one gives the expansion over all Euclidean geometries, which are in a definite sense ``built'' from smooth data involving the whole set of spectra.  

At risk of seeming repetitive, this implies that  there is {\it no factorization puzzle} forced upon us by JT gravity after all. Also, it  makes the story of holography for JT gravity much more like higher dimensional AdS/CFT, and likely contains lessons for how to understand Euclidean computations in that context. (In case it is not clear, the analogue of the gauge theory dual in this story is the single copy of the spectrum, $\{ {\cal E}_n\}$.)

\subsubsection*{A Note on String Theory}

The above ideas and considerations about how wormholes appear through ensemble averaging the spectrum of 2D gravity theories applies to random matrix models more general than those that pertain to JT gravity.  Of course, random geometries with wormholes and higher topology are necessary for defining string theory world sheets, and the picture of this paper sheds some interesting light on how string theory works as well as it does. The limit of large~$E$ discussed here is equivalent to small loops. Translating that to string diagrams, one get small strings, in other words, the limit where $\alpha^\prime$ is small, corresponding to the point particle limit. There, perturbative string diagrams, such as the propagator for a closed string, again make sense as pertaining to mostly a single copy of a 2D gravity spectrum. 

Going to the limit where closed strings can become long is the analogue of the late $t$ limit of the spectral form factor discussed above. Using a single copy of the 2D gravity spectrum would mean that the physics will begin to enter a regime where the discreteness cannot be ignored, threatening a breakdown of worldsheet geometry. However, a smooth worldsheet description can still be obtained, by performing the averaging over an ensemble of different discrete 2D gravity spectra, resulting in the smooth $\rho(E)$ in a matrix model description. Now the emergence of the wormhole is possible, which in this case is the smooth closed string propagator, but now in the limit where strings are long,  bringing  into play all the essential  UV behaviour that makes string theory work so well. This picture is worth exploring further.

\subsection{Tour of the rest of this Paper}

The next task  is to survey and understand the tools needed to excavate the many results upon which the discussion above has been based. It is worth noting again that the tools and results to be discussed are useful regardless of whether one sits firmly on the Wignerian or 't Hooftian side of the aisle, accepting the new interpretation or not. 

At this point, it would normally  suffice to introduce the effective  one dimensional quantum mechanics problem that emerges from the double-scaled matrix models, in terms of which everything can be succinctly computed. However, it has become clear that this formalism  often seems to some Readers like a piece of magic, and so some attempt will be made to reach back a somewhat  further into the machinery in order to appreciate all the moving parts.  Another reason to do so is because presentations of matrix models in this context tend to keep in the foreground what might be called a t' Hooftian perspective, focussing  on continuum quantities that will yield smooth spacetime surfaces. The Wignerian perspective, much more statistical in character, is relatively neglected. Clearly from the results discussed already, {\it both} perspectives are needed in equal measure and so it is high time for this emphasis to change. 

Accordingly,  the long {\bf Section~\ref{sec:double-scaling}} that carefully describes the engine room of all the computations, will move back and forth between the perspectives, first ({\bf \ref{sec:search-gravity}}) developing the usual tools such as the Dyson gas picture, and the orthogonal polynomials, and  ({\bf \ref{sec:qm-toolbox}}) the curious one-dimensional model of quantum mechanics that emerges in the double scaling limit. This model (which is {\it not} the quantum mechanics dual to JT in any direct sense) is often regarded as confusing, not the least because of the way it is used, and so some key remarks about how it works are made, by connecting to a many-body fermion picture ({\bf \ref{sec:many-body}}) of the Dyson gas dynamics\footnote{The Author is  grateful to Robbert Dijkgraaf, Felipe Rosso, Edward Witten and Herman Verlinde for many questions about the formalism that made it clear that some exposition was needed. The act of writing it  was very useful.}. There is a nice dictionary between computations in the two systems that deserves emphasis (some of this material was recently put into ref.~\cite{Johnson:2021tnl}, but will be recalled and expanded upon here). {\bf Subsection~\ref{sec:Statistics}} then jogs back a bit to develop the statistical picture, which has a much older history than the methods described so far. From there, a core object, a certain ``self-reproducing'' kernel $K(E,E^\prime)$ will appear. It is a rather stunningly beautiful object in terms of which all statistical questions about the matrix model can be answered. It is the natural link between the two perspectives. On the one hand, its diagonal gives the smooth spectral density $\rho(E)$, and on the other hand, statistical results can be expressed in terms of determinants of it, bringing its off-diagonal elements into play.

But there is {\it different} determinant that is needed in order to derive the (proposed) JT gravity spectrum, discussed in {\bf subsection~\ref{sec:fredholm}}. It is a not a determinant of $K(E,E^\prime)$ but a much more subtle animal. Asking about the ``gap probabilities'' of {\it individual} eigenvalues in a given energy interval will have an answer in terms of the properties of an integral operator that lives on that interval, defining a Fredholm problem whose kernel is $K(E,E^\prime)$. The determinant of {\it that} integral operator (subtracted from the identity) is the Fredholm determinant. The matter of computing probability distributions from this object, and the fascinating new kinds of distributions that arise in various problems,  is a rich area of modern mathematical physics, statistical physics, and computational physics. Some key aspects are developed in this section, including some discussion of the (necessary) numerical computational techniques needed to extract answers.

{\bf Subsection~\ref{sec:return-to-gravity}} returns again to a more spacetime perspective, quickly discussing the insertion of loop operators and D-branes into the formalism, and re-interpreting a special case of the wavefunctions of the quantum mechanical problem of {\bf subsection~\ref{sec:qm-toolbox}} in this light. (This is often referred to as the FZZT D-brane partition function.) It is also a natural bridge between the two perspectives, since $K(E,E^\prime)$ can be built out of it. Its WKB form will turn out to be useful for certain computations. The free energy and spectral form factor computations for double-scaled matrix models are discussed in {\bf subsections~\ref{sec-free-energy}} and~{\bf \ref{sec:spectral-form-factor-airy}}, where they are compared to the same quantities computed for the single special copy of the (mean) spectrum. 

Throughout this long section, the tools and ideas are illustrated with the simplest matrix model of all (in this context), the Gaussian unitary ensemble, a model of $N{\times}N$ Hermitian matrices that at large $N$ yields the Airy model after ``double''-scaling (although here this really just means ``scaling'' with $N$. There are no higher order couplings to tune to justify the term ``double''). It is not a model of gravity that has smooth surfaces of finite area (it is a topological gravity model~\cite{Dijkgraaf:2018vnm}), but nevertheless serves as a good  testbed since on the one hand many results for it can be written analytically, and on the other hand, large Hermitian matrices with Gaussian probability distributions are readily 
simulated with a few lines of code on a modest computer. The results can be ``double-scaled'' too. So everything can be checked and illustrated explicitly against an ensemble of actual matrices, which is useful for checking intuition. 

{\bf Subsection~\ref{sec:bessel-models}} discusses some Bessel models as another set of models that can be simulated with actual matrices, extracting their ``microstates'' using the Fredholm techniques, and studying their thermodynamics (following on some work done in ref.~\cite{Johnson:2021rsh}). They help as a quite different setting for seeing in action some of the tools developed earlier, and also serve as precursors of important features that will appear in JT supergravity.

{{\bf Section~\ref{sec:jt-gravity}} begins the discussion of matrix models that yield  JT gravity, applying all the tools developed to extract the various results that underpin the perspective and proposal presented in the introduction. Some of these results were presented, briefly, in the Letter of ref.~\cite{Johnson:2021zuo}, but there will be much more to present here, along with the new perspective on it all. {\bf Section~\ref{sec:sjt-gravity}} presents the JT supergravity results for four models. 

{\bf Section~\ref{sec:concluding-remarks}} will end with some discussion, which will be mercifully brief since so much as already been said in the  Introduction.  

\section{Double-scaled Matrix Models}
\label{sec:double-scaling}
\subsection{A Search for Gravity}
\label{sec:search-gravity}
Central to the discussion are random matrix models in the ``double scaling limit''~\cite{Brezin:1990rb,Douglas:1990ve,Gross:1990vs}.\footnote{Some excellent reviews often mentioned are  refs.~\cite{Ginsparg:1993is,Eynard:2015aea}, but two excellent books with a non-gravitational perspective  that will help with the statistical view are refs.~\cite{Meh2004,ForresterBook}.}
 In the simplest matrix models, the probability distribution of the $N{\times}N$ Hermitian matrix $M$ is $p(M){=}{\rm e}^{-{\rm Tr} V(M)}$ with  Gaussian $V(M){=}\frac{1}{2}M^2$ being the most famous prototype {\it \'a la} Wishart and Wigner~\cite{10.2307/2331939,10.2307/1970079}. 
A more general polynomial  will be  of interest when trying to capture a theory of  gravity:
\be
\label{eq:raw-potential}
V(M){=}\sum_p g_p M^p\ .
\ee
    The matrix model partition function ${\widetilde Z} {=}\int {\e }^{-{\rm Tr}V(M)}dM $ can be thought of as a toy field theory, and given a  Feynman diagrammatic expansion. At large $N$, these diagrams can be viewed  (following 't Hooft~\cite{'tHooft:1973jz},  Brezin {\it et.~al.}, and Bessis  {\it et.~al.}~\cite{Brezin:1978sv,Bessis:1980ss}) 
as tessellations  of 2D Euclidean spacetimes,  each order in the $1/N$ expansion corresponding to the topology upon which the diagrams  can be drawn. So perturbatively the matrix model partition function can be written: ${\widetilde Z}=\sum_{g=0}^\infty {\widetilde Z}_g N^{2g-2}$, where ${\widetilde Z}_0$ is the sum of all diagrams that can be drawn on the sphere, ${\widetilde Z}_1$ the sum for the torus,  and so on. On any genus, one can compute the area $A$ of the dual surface using the number, $n$,  of faces in a given triangulation (each has on average an area ${\sim}\delta^2$ that can be set to unity in this initial discussion). In a study of random surfaces with some ``bare"  cosmological constant $\mu_B$, denoting the partition function (a bit unconventionally) by ${\widetilde F}(\mu_B)$,  the fixed area dependence can be isolated by a Laplace transform 
\be
{\widetilde F}_0(\mu_B)=\int_0^\infty\! dA \,\e^{-\mu_B A} {\widetilde F}_0(A)\ ,
\ee
(this is a sum over $n$ in a discrete model) and it is known~\cite{Knizhnik:1988ak,David:1988hj,Distler:1988jt} that 
$
{\widetilde F}_0(A)\sim \e^{\mu_c A}A^{\gamma_s - 3}+\cdots\ ,
$
where $\gamma_s<0$ is an exponent that depends upon whether there are additional ``matter'' degrees of freedom living on the random surface ($\gamma_s=-\frac12$ when there are none), 
and so 
\be
\label{eq:KPZ}
{\widetilde F}_0(\mu_B)\sim |\mu_B-\mu_c|^{2-\gamma_s}\ ,
\ee
which shows (upon computing the expectation value of area by differentiating with respect to $\mu_B$)  that there is a special point to tune to   ($\mu_B\to\mu_c$) where surfaces with large areas dominate the physics. In defining a theory of smooth surfaces (where details of the tessellations don't matter, {\it i.e.,} the typical triangle area,~$\delta^2$, is vanishingly small compared to the overall area), this is a good place to tune to, and this is what the double-scaling limit does. It is a combination of   sending $N{\to}\infty$ (to get the topological expansion) while  tuning the couplings~$g_p$ in the potential~(\ref{eq:raw-potential}) to critical values such that  ${\widetilde F}\sim\mu^{2-\gamma_s}$ where $\mu$ is the scaling piece of  $\mu_R=(\mu_B-\mu_c)=\mu\delta^2$ as $\delta{\to}0$.  How this works will be unpacked below.

Writing $M{=}U\Lambda U^{-1}$ where $U$ is a unitary matrix and $\Lambda{=}{\rm diag\{\lambda_1,\lambda_2,\cdots,\lambda_N\}}$, the matrix model can can be written as (up to a constant coming from integrating over the volume of the unitary group):
\begin{equation}
\label{eq:Dyson-gas}
 {\widetilde Z} {=}\int \prod_i d\lambda_i  \prod_{i<j}(\lambda_i-\lambda_j)^2 {\e }^{-\sum_iV(\lambda_i)}\ ,
\end{equation}
where the square of the Vandermonde determinant $\Delta(\Lambda)=\prod_{i<j}(\lambda_i-\lambda_j)={\rm det}\| \lambda_i^{j-1} \|$ is the Jacobian for the change of variables from $M$ to $\Lambda$.

The standard route taken at this point is to take a large $N$ limit and search for a smooth description, a certain leading saddle point solution,  which  will correspond to the leading contribution with spherical topology. Corrections to that are then developed in a $1/N$ expansion. This will be briefly reviewed in a moment. However, it should not be forgotten that at the heart of it all is a statistical system, with an underlying ensemble of random matrices. Expression~(\ref{eq:Dyson-gas}) says that the probability of finding the $N$ eigenvalues in positions/values  $\{\lambda_i\}$ is given by the integrand. Moreover, it is natural to ask about fewer groups of eigenvalues by simply integrating out as many as desired. This will be returned to in Section~\ref{sec:Statistics}.

It is useful to think of the system as a ``Dyson gas" of particles (at positions $\lambda_i$), and write it as:
\begin{equation}
\label{eq:Dyson-gas2}
 {\widetilde Z} {=}\int \prod_i d\lambda_i 
  \exp\Biggl\{{-N\sum_iV(\lambda_i)+\sum_{i\neq j }\log|\lambda_i-\lambda_j|}\Biggr\}\ ,
\end{equation}
making explicit an effective logarithmic repulsion in addition to the applied potential $V(\lambda)$. Additionally, the choice has been made to rescale $\lambda\to\lambda/\sqrt{N}$, which will ensure the the eigenvalues occupy, at large $N$, a {\it fixed} finite segment on the real $\lambda$ line as $N$ grows. A factor of~$N$ in front of $V(\lambda)$ then restores the canonical scaling that was started with. At large $N$  the matrix eigenvalues $\lambda_i$  can be well described with a function $\lambda(X)$, where $X{=}i/N$ is a continuous coordinate.  A spectral density
${\tilde\rho}_0(\lambda)=dX/d\lambda$ can be introduced, a smooth limit of 
\be
\label{eq:raw-density}
{\tilde\rho}_0(\lambda)=\frac{1}{N}\sum_i^N\delta(\lambda - \lambda_i)\ ,
\ee
and normalized so that $\int{\tilde\rho}(\lambda)d\lambda=1$.
Now the quantity inside the exponential can be written (remembering to use  $N^{-1}\sum_i\to\int dX$):
\be
\label{eq:dyson-action}
-N^2\Biggl[\int{\tilde\rho}(\lambda)V(\lambda) d\lambda-\int d\lambda\int d\mu {\tilde\rho}(\lambda){\tilde\rho}(\mu)\log|\lambda-\mu|\Biggl]\ ,
\ee
 and a large $N$
  saddle point solution results from the condition:
\be
V^\prime(\lambda) = 2P\int_{-a}^{a}\frac{{\tilde\rho}(\mu)}{\lambda-\mu}d\mu\ ,
\ee
where  $P$ means the principal part and $\lambda_c=a$ marks the end of the ``droplet''  saddle point solution, and it has been assumed (for simplicity) that the potential is even.\footnote{There are several excellent reviews of this. See {\it e.g.} refs.~\cite{Meh2004,ForresterBook,Eynard:2015aea}.} 
See figure~\ref{fig:dyson-gas}.

\begin{figure}[t]
\centering
\includegraphics[width=0.45\textwidth]{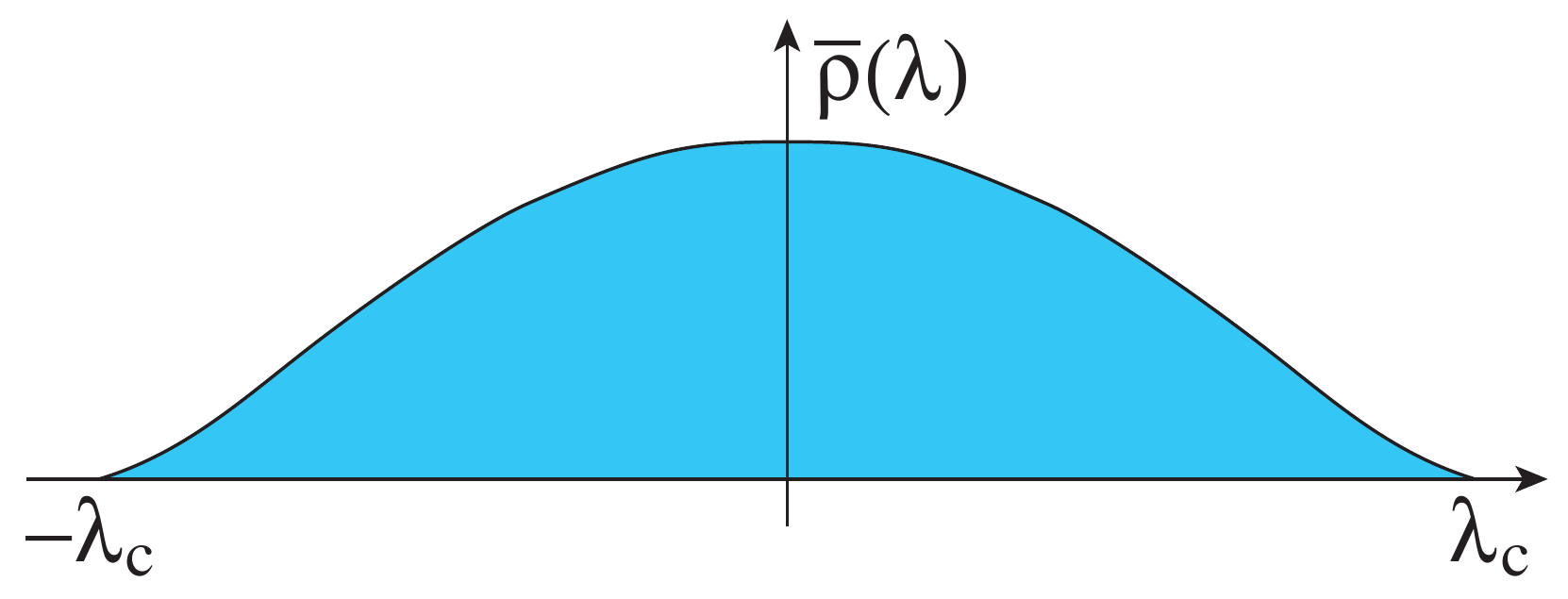}
\caption{\label{fig:dyson-gas} The Dyson gas at large $N$.}
\end{figure}

The next step is to introduce the analytic function $F(\lambda)=\int d\mu {\tilde\rho}(\mu)/(\lambda-\mu)$ that is real away from the region $-a<\lambda <a$, has leading behaviour of  $1/\lambda$ as $|\lambda|{\to}\infty$ (following from the density's normalization condition) and that has discontinuity $F(\lambda\pm i \epsilon)=\frac{V^\prime}{2}\mp\pi i{\tilde\rho}(\lambda)$.
A solution can be found by writing
\be
\label{eq:resolvent-saddle}
F(\lambda)=\frac{V^\prime(\lambda)}{2} - Q(\lambda)\sqrt{\lambda^2 - a^2}\ ,
\ee
where $Q(\lambda)$ is a polynomial (two orders fewer than $V(\lambda)$) whose coefficients are fixed by requiring the leading large~$\lambda$ behaviour. The density is then 
\be
\label{eq:tree-level-rho}
{\tilde\rho}_0(\lambda)=\frac{1}{\pi}Q(\lambda)\sqrt{a^2 - \lambda^2}\ .
\ee

The double-scaling limit ($N{\to}\infty$ with $g_p{\to}g^c_p$)  \cite{Gross:1990vs,Brezin:1990rb,Douglas:1990ve,Gross:1990aw}
focuses on the scaling region in the neighbourhood of an endpoint (see {\it e.g.,} Refs.~\cite{Kazakov:1989bc,Dalley:1991zs,*Bowick:1991ky} for discussion and classification of this behaviour), magnifying it while sending the other end to infinity,  as will be done shortly.

It might be useful to be very specific. Consider the  prototype  Gaussian case with $N{\times}N$ Hermitian matrices. Then it is easy to see that $a=2$ and $Q(\lambda)=\frac12$, and so   the result can be written as 
\be
\label{eq:wigner-semi-circle}
{\tilde\rho}_0(\lambda)=\frac{\sqrt{4-\lambda^2}}{2\pi}\ ,
\ee
 the famous Wigner semi-circle law. This can be checked explicitly experimentally by sampling an ensemble of 100${\times}$100 Hermitian matrices, randomly generated on a computer using Gaussian probability. See figure~\ref{fig:semi-circle}, and  footnote~\ref{fn:wigner} describes  how to make it.\footnote{\label{fn:wigner} A computational aside: This came from  writing a \textsf{MATLAB} loop that, in each iteration, generated a sample 100${\times}$100 Hermitian matrix {\tt M=(C+ctranspose(C))/2} with Gaussian probability through random complex matrix {\tt C=randn(N)+1i*randn(N)} in {\textsf{MATLAB}} where {\tt N=100}. Then the eigenvalues were gathered, using {\tt eigs(M)}, and dumped into a list (simply a \textsf{MATLAB} vector; call it {\tt alleigs} for argument's sake). This was done 10,000 times using the loop, taking about 3s. Then every element in {\tt alleigs} was divided by {\tt sqrt(N)}. A simple histogramming command ({\it e.g.,} {\tt histogram(alleigs)}) generated the semi-circle.}  

\begin{figure}[t]
\centering
\includegraphics[width=0.45\textwidth]{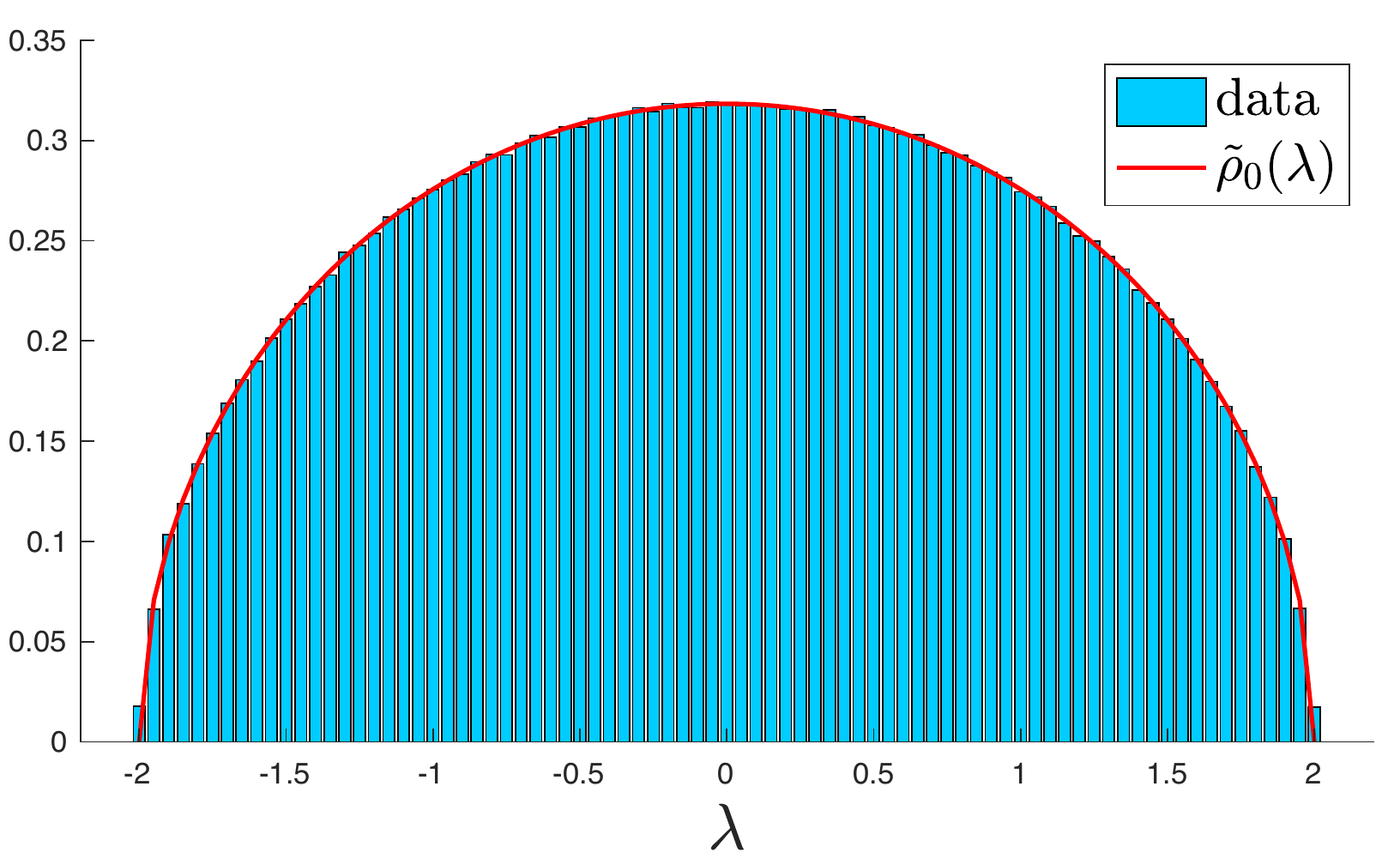}
\caption{\label{fig:semi-circle} 
Wigner's semi-circle law, using 10,000 samples of $100{\times}100$ Hermitian matrices.}
\end{figure}

Here is a precursor of the double-scaling limit: Scaling about the $\lambda_c{=}2$ end by writing $\lambda {=} 2{-}E\delta^2/N^{2/3}$, where $\delta\to0$ as $N\to\infty$, yields the spectral density in the neighbourhood of that end. It is going to be small compared to the ``bulk'' quantity, so writing ${\tilde\rho}_0(\lambda)d\lambda=C\rho_0(E)dE$, where $C$ scales inversely with $N$,  the finite piece $\rho_0(E)$ can be  extracted and called the scaled spectral density. It is:
\be
\label{eq:airy-tree}
\rho_0(E)=(\pi\hbar)^{-1}E^{\frac12}\ ,
\ee
 where it turned out that $C{=}N^{-2}$, and  the combination $\hbar=1/(N\delta^3)$ is held fixed as $N\to\infty$. So $\hbar$, (not really needed in this example, but it will lay the ground work for later more complicated models) is a  ``renormalized'' $1/N$ expansion parameter.  (This scaling can be done on the eigenvalue data in the computer code of footnote~\ref{fn:wigner}, and  will be discussed later.)

More complicated potentials $V(\lambda;\{g_p\})$ can be tuned to produce a family of generalizations of this behaviour by tuning to the critical point where  $Q(\lambda)$ in equation~(\ref{eq:tree-level-rho}) produces $k-1$ extra zeros  at the endpoint, which ultimately gives leading behaviour $\rho_0(E){\sim}\hbar^{-1}E^{k-\frac12}$ (where $\hbar=1/(N\delta^{2k+1})$.  There are corrections to this $E^{k-\frac12}$ behaviour, which can be written as a perturbative series in $\hbar$, plus non-perturbative pieces. The complete spectral density that includes them all will be written~$\rho(E)$.

For easy access to non-perturbative physics, it is prudent to work with a family of orthogonal polynomials, since in the double-scaling limit they will supply a powerful toolbox to use to study the physics. The  matrix model integral can be written entirely in terms of a system of polynomials of $n$th order,  $P_n(\lambda){=}\lambda^n+\cdots$ that are  orthogonal with respect to the measure $d\lambda {\rm e}^{-NV(\lambda)}$:
\be
\label{eq:orthogonality}
\int d\lambda  {\rm e}^{-NV(\lambda)}  P_n(\lambda)P_m(\lambda) = h_n\delta_{nm}\ ,
\ee
where $h_n$ is a normalization, and $h_0{=}\int d\lambda \e^{-NV(\lambda)}$. There is an infinite set of such polynomials $(n,m\in \mathbb{Z}^+)$. As will become clear in a moment,   the first $N$ of them (counting from $n{=}0$) will get used up in defining the large~$N$ solution.

For example, in the case of the Gaussian model, the $P_n$ are Hermite polynomials $H_n(\lambda){=}(-1)^n\e^{N\lambda^2/2}\frac{d^n}{d\lambda^n}[\e^{-N\lambda^2/2}]$. One intuitive way of seeing why is the fact that, for any particular one of the $\lambda_i$, the problem is simply the simple harmonic oscillator, for which the eigensolutions  are the  
Hermite functions ${\tilde \psi}_n(\lambda_i){=}\e^{-N\lambda^2/2}H_n(\lambda_i)$ with energies that have a dependence  ${\sim}(n+\frac12)$. Each of the $N$  problems with coordinate $\lambda_i$  can be in any of these quantum oscillator states. This is hugely important to appreciate for what is to come.

It is worth remarking that the $N$ different sectors are largely independent: The physics in one sector does not really affect the physics in another. This will mean that the collective description of the physics across the different sectors (a ``many body'' description) will be that of a {\it free} system. The logarithmic repulsion that keeps the $\lambda_i$'s from coinciding will instead be re-interpreted as having prepared  many-body states that are fermionic, as will be discussed further below.

 In fact, it is often useful to define a notation $|n\rangle{\equiv}\e^{-NV(\lambda)/2}P_n(\lambda)/\sqrt{h_n}$ so that the integral in equation~(\ref{eq:orthogonality}) is $\langle n|m\rangle=\delta_{nm}$. This sort of notation can be used for the entire matrix model problem, since in any of the $N$ sectors (with eigenvalue $\lambda_i$) one has the bra-ket basis to work with. So integrals involving insertions of any function of $\lambda$ can be recast as computing using the recursion relations and  using the bra-kets in each sector. This notation is also a precursor of the 1D quantum mechanics that will appear in the large $N$ limit shortly.
 
The first example of working in terms of the $P_n(\lambda)$ in the full matrix model is to solve for ${\widetilde Z}$ itself. Since (by taking linear combinations of rows or columns) the Vandermonde determinant can be written in terms of the first~$N$ of the $P_n(\lambda)$ as:
\be
\label{eq:vandermonde-polynomials}
\Delta=
{\rm det}\| \lambda_i^{j-1} \|_{i,j=1}^N ={\rm det}\| P_{j-i}(\lambda_i) \|_{i,j=1}^N \ ,
\ee
 it is easy to rewrite  ${\widetilde Z}$ in terms of properties of the polynomials, by expanding out the determinant factors and using the orthogonality relation. The outcome will simply be a combinatorial factor multiplied by a factor of $h_n$ from each of the $N$ sectors, and the answer is:
 $
 {\widetilde Z} = N!\prod_{n=0}^{N-1} h_n$. There's another way of writing this that is very illuminating.
The polynomials are related according to a recursion relation:
\be
\label{eq:recursion}
\lambda P_n(\lambda){=}P_{n+1}(\lambda) {+} R_nP_{n-1}(\lambda)\ ,
\ee
 for even $V(\lambda)$ (which will be used here for simplicity). In the case of Hermite polynomials, $R_n{=}n/N$, but in general the $R_n$ satisfy a more involved recursion relation  determined by knowledge of the potential $V(\lambda)$. Those can be worked out by studying the action of $d/d\lambda$ on the $P_n(\lambda)$. 
Since  $V(\lambda)$ has been chosen (for simplicity) to be even here, it is clear that
\be
\label{eq:identity1}
\int d\lambda   {\rm e}^{-NV(\lambda)}  P_n(\lambda)\frac{d}{d\lambda}P_n(\lambda)= 0\ ,
\ee
but something non-trivial arises if the derivative results in two polynomials of the same order being integrated, since using~(\ref{eq:orthogonality}):
\be
\label{eq:identity2}
\int d\lambda  {\rm e}^{-NV(\lambda)}  P_{n-1}(\lambda)\frac{d}{d\lambda}P_n(\lambda) = nh_{n-1}\ .
\ee
Integrating by parts and using~(\ref{eq:orthogonality}) again yields:
 \be
\label{eq:identity3}
N\!\!\int d\lambda {\rm e}^{-NV(\lambda)} V^\prime(\lambda)P_n(\lambda)P_{n-1}(\lambda)   = nh_{n-1}\ .
\ee
 This relation determines recursion relations for $R_n$, since the $V^\prime$ insertion can be simplified recursively using the relation~(\ref{eq:recursion}), and then orthogonality used to perform the resulting integrals. Their  non-linearity increases with the   order of the potential. For example for the Gaussian case, $V^\prime{=}\lambda$, and so  
 \bea
\label{eq:identity3}
&&N\!\!\int d\lambda {\rm e}^{-NV(\lambda)}(P_{n+1}(\lambda)+R_nP_{n-1}(\lambda)) P_{n-1}(\lambda) \nonumber\\
&&\hskip4.1cm  = NR_nh_{n-1} \ ,
\eea
confirming that $R_n=n/N$, as  stated above. The next non-trivial case has  $V=\lambda^2+g_4\lambda^4$,  (note the unit coefficient of the quadratic term), and the result is:
\be
\label{eq:recursionk2}
R_n\left[2+4g_4(R_{n-1}+R_n+R_{n+1})\right] = \frac{n}{N}\ .
\ee
This will  be used shortly.

 Knowing the coefficients $R_n$ turns out to be equivalent to solving the partition function integral or  any insertion into the integral  of traces of powers of $M$ 
{\it i.e.,} of $\lambda$.  This is helped by seeing that the $h_n$ and the $R_n$ are related according to $h_n=h_{n+1}/R_{n+1}$ (which can be proven by multiplying  the recursion relation~(\ref{eq:recursion}) by $P_{n+1}$ and integrating, using the orthogonality relation~(\ref{eq:orthogonality}) to write it in two ways). Hence,  
\bea
 \label{eq:matrix-simplified}
 {\widetilde Z} &=& N!h_0^N\prod_{n=1}^{N-1} R_n^{N-n}\\
 &=& N!h_0^N\exp\left(\sum_{n=1}^{N-1} (N-n)\log R_n\right)\nonumber\\
 &=& N!h_0^N\exp\left(N^2\cdot\frac{1}{N}\sum_{n=1}^{N-1} \left(1-\frac{n}{N}\right)\log R_n\right)
 \ . \nonumber
 \eea

At  large $N$, the index $n/N$ becomes a continuous coordinate $X$ that runs from 0 to~1. The orthogonal polynomials become functions of $X$, $P_n(\lambda){\to} P(X,\lambda)$, and so do the recursion coefficients: $R_n{\to}R(X)$. In terms of all these, the generator of the connected diagrams with no boundaries, ${\widetilde F}{=}{-}\log{\widetilde Z}$ is, after throwing away an additive constant:
\be
\label{eq:closed-free}
{\widetilde F} = -N^2 \int_0^1 (1-X)\log R(X)dX\ .
\ee

 There is  interesting behaviour for the integral in the neighbourhood of  $X=1$,  at which  $R(X)$ goes to~1, as can be seen in the Gaussian case given that $R(X){=}X$. (This is why this case isn't really a ``double'' scaling limit since large $N$ is all there is. There are no couplings in the potential.) For the quartic case it gets a bit more interesting: At large~$N$ the leading equation for $R(X)$ following from~(\ref{eq:recursionk2}) is $12gR^2+2R-1=0$, with solution
 \be
 R(X) = \frac{-1\pm\sqrt{1+12g_4}}{12g_4}\ .
 \ee
This can be expanded as a series in small $g_4$, yielding an enumeration of all  the connected diagrams that  can be drawn on surfaces of spherical topology. As $g_4\to g_4^{\rm c} = -1/12$, this series diverges---just the critical behaviour discussed at the beginning of this section---and $R(X)$ takes on the critical value $R_c{=}1$.
In Dyson droplet terms, this sets the endpoint of the distribution, which fits with the fact that the index $n$ has been correlated with the eigenvalue index $i$, so that $X$ running from~0 to 1  runs from one endpoint of the Dyson gas droplet to another.   
This endpoint is where the double scaled limit focuses. A direct connection between the two can  be seen from the following integral representation of  the leading density in terms of the function $R(X)$:
\be
\label{eq:sphere-integral-relation}
{\tilde\rho}_0(\lambda) = \frac{1}{\pi}\int_0^1\frac{dX}{\sqrt{4R(X)-\lambda^2}}\ ,
\ee
  where only contributions coming  from where the square root is real are kept.  A derivation may be found in ref.~\cite{Bessis:1980ss}, but it can directly  be thought of as another way of representing the result~(\ref{eq:tree-level-rho}) derived from saddle methods. Clearly the droplet endpoint position is $\lambda_c=2\sqrt{R_c}$. The simple Gaussian case of $R=X$  in this formula readily yields the semi-circle law~(\ref{eq:wigner-semi-circle}). 

In taking the limit, scaled versions of all the key quantities survive to define the continuum physics. For example, $X$ has a scaled piece~$x$,  {\it via}: $X{=}1{+}(x-\mu)\delta^{2k}$, where the parameter $\delta{\to}0$ in the limit, and~$k$ is some positive integer that will be determined by the order of the potential.    The  region $-\infty\leq x\leq \mu$ will have special significance shortly.   The topological expansion parameter $1/N$ also picks up a scaling piece, denoted~$\hbar$, {\it via}: $1/N {=} \hbar\delta^{2k+1}$. For the recursion coefficients $R(X) {=} 1{-}u(x)\delta^{2}$. 
 With these relations, in the limit $\delta\to0$ the surviving physics is
 \be
 \label{eq:connected-closed}
{\widetilde F} = \frac{1}{\hbar^2} \int_{-\infty}^\mu (x-\mu)u(x)dx\ ,
\ee
or alternatively 
\be
u(x) = \hbar^2\left.\frac{\partial^2 {\widetilde F}}{\partial \mu^2}\right|_{\mu=x}\ .
\ee
In the Gaussian case, $k{=}1$, and the relation $R_n {=} n/N$, which became $R(X){=}X$, simply yields $u(x){=}{-}x$, where~$\mu$ has been set to  zero, since the model has no parameter to which this quantity corresponds in the scaling limit. This will be different for higher order potentials, as will be clear in a moment. 

Another useful thing to do is apply these scalings to is the leading spectral density of (\ref{eq:sphere-integral-relation}). All the physics comes from the neighbourhood of the $X=1$ limit and as shown above, this is also the neighbourhood of an end $\lambda_c$ of the spectral density. So  generally the double-scaling limit ought to include scaling $\lambda$ away from $\lambda_c$ by a scaled amount: $\lambda=\lambda_c-E\delta^2$. (This is a slightly different approach than that used above~(\ref{eq:airy-tree}), but amounts to the same thing.) The power of $\delta$ is fixed by the denominator of the relation. The expectation should be that ${\tilde\rho}_0(\lambda)d\lambda\to\rho_0(E)dE$, once the scaling $\lambda=\lambda_c-E\delta^2$ is used. A factor $\delta^{2k-1}$ results from putting in the rest of the scaling relations, combining to give a factor $\delta^{2k+1}$. To get something finite at large $N$, a factor of $N$ should be multiplied in here, which gives the finite combination already identified:~$\hbar^{-1}$. Thus:
\be
\label{eq:leading-scaled-density}
\rho_0(E) = \frac{1}{2\pi\hbar}\int_{-\infty}^\mu\frac{dx}{\sqrt{E-u_0(x)}}\ ,
\ee
where $u_0(x)$ is the leading part of $u(x)$.  As a check, using $u_0(x)=u(x)=-x$ for the Gaussian case recovers~(\ref{eq:airy-tree}).
 This is an extremely useful formula that will also be derived in a different way later on.

More generally the recursion relations for $R_n$ at higher~$k$ become a {\it differential equation} for $u(x)$, which is often (for historical reasons) called a ``string equation''. (The trivial $k=1$ case had no such equation, just the exact relation~$u(x)=-x$.) For example, the quartic case gives an equation for $R(X)$ in terms of $X$ and $R(X\pm\epsilon)$, where $\epsilon=1/N$, which upon Taylor expanding gives:
\be
R(X)\left(2+4g_4\left[3R(X)+\epsilon^2\frac{\partial^2 R(X)}{\partial X^2}+\cdots\right]\right) = X\ ,
\ee
and now inserting the scaling relations $X{=}1{+}(x{-}\mu)\delta^{2k}$, $R(X) {=} 1{-}u(x)\delta^{2}$ and $\epsilon{=}\delta^{2k+1}\hbar$, and allowing the coupling to scale away from its critical value {\it via} $g_4{=}g_4^c{-}g_4^s\delta^{2k}$  yields that orders $\delta^0$ and $\delta^2$ cancel to zero, and a non-trivial equation appears at next order if  $k=2$, which is:
\be
\label{eq:painleveI}
-\frac{\hbar^2}{3}\frac{\partial^2u(x)}{\partial x^2}+u(x)^2 = -x\ .
\ee
It is here that the parameter $\mu$ now gets physical meaning (it could be scaled away by a coordinate shift in the $k=1$ case). It is actually the scaled value of the coupling $g_4$, which was set to $g_4^s=\frac{\mu}{12}$ in the above result. This fits nicely with the fact that  $g_4$ controlled the approach to the surfaces becoming critical. It is essentially the bare cosmological constant of the opening discussion, and $\mu$ is the renormalized value.

The ``string equation'' of equation~(\ref{eq:painleveI}) is in fact the Painlev\'{e}~I non-linear ODE. In principle, such equations yield both perturbative and non-perturbative information about the model, which was one of the great discoveries of the classic double-scaling limit papers~\cite{Brezin:1990rb,Douglas:1990ve,Gross:1990vs,Gross:1990aw}. Perturbation theory comes from expanding $u(x)$ in the $x{\leq}\mu$ region, giving for example:
\be
u(x) = \sqrt{-x}-\frac{1}{24}\frac{\hbar^2}{x^2} -\frac{49}{1152}\frac{\hbar^4}{x^4}+\cdots
\ee
which yields 
\be
{\tilde F} = \frac{4}{15}\frac{\mu^\frac52}{\hbar^2} + \frac{1}{24}\log|\mu|-\frac{7}{1440}\frac{\hbar^2} {\mu^\frac52}+\cdots
\ee
corresponding to the sphere, torus, and double torus orders in perturbation theory. The classic  models that can be obtained  with this simplest approach have for their leading behaviour at genus zero $u_0(x)=(-x)^{1/k}$ (for the Gaussian  case, $k{=}1$,  it is simply $u(x){=}-x$ to all orders.).   Note that putting this leading behaviour for $u_0(x)$ into the above confirms the $|\mu|^{2+\frac1k}$ form discussed beneath equation~(\ref{eq:KPZ}), with $\gamma_s{=}-\frac1k$. 

The ODEs that arise by following the steps from before to for higher $k$ can be written as:
\be
\label{eq:simple-string-equations}
{\cal R}=0\ , \quad {\rm where} \quad {\cal R}\equiv {\tilde R}_k[u]+x\ .
\ee
 and the ${\tilde R_k[u]}$ are $k$th order polynomials in $u(x)$ and its $x$ derivatives normalized such that their $u(x)^k$ term has coefficient unity. They have a pure derivative piece given by $2k-2$ derivatives acting on $u(x)$, and then various mixed non-linear pieces. The first three of these ``Gel'fand-Dikii''\cite{Gelfand:1975rn} polynomials are:
 \bea
 \label{eq:Gelfand-Dikii-polys}
 &&{\tilde R}_1[u]=u\ , \qquad   {\tilde R}_2[u]=u^2-\frac13u^{''}\ , \quad {\rm and }\nonumber\\
 &&{\tilde R}_3[u]=u^3-\frac12(u')^2-uu^{''}+\frac{1}{10}u^{''''}\ , 
\eea
where here a prime indicates an $x$-derivative times a factor of $\hbar$. The higher ones can be generated using a recursion relation. They will not be needed explicitly here.

Going beyond perturbation theory to define all the physics of the model  requires learning the full form of $u(x)$ in the $x{\leq}\mu$ region, which is where the integration takes place to define physical quantities. (This will be referred to as the ``Fermi sea" once a useful many-body picture  is developed in subsection~\ref{sec:many-body}. The location $x=\mu$ is the Fermi surface.)  This is in principle obtained from the full string equation, whose domain extends beyond  the Fermi sea to the ``trans-Fermi" region  $x>0$. This is natural since the $x$ is the scaled index of the infinite family of orthogonal polynomials. Only $N-1$ of them were used to define the Dyson gas, up to $x=\mu$. But consistency demands that data about the full family should inform the information about the first $N$. That is the role of the differential equation. The technique for establishing the non-perturbative physics is the matter of being able to find sensible solutions for $u(x)$ that extend to the whole real $x$ line, while preserving the perturbative form in the Fermi sea region. In fact, for the $k=2$ the simple string equation~(\ref{eq:painleveI}) fails to yield an appropriate solution. This is in fact true for all $k$ even. This is a direct result of a non-perturbative instability of the Dyson gas for these models. However, the instability can be removed, and a  more general string equation results (which has the same perturbative content), for which  physical  extensions of $u(x)$ to the trans-Fermi region are possible. The consistency conditions for constructing $u(x)$, and a full analysis of the string equation (both semi-classically and non-perturbatively) has been recently extensively discussed in the JT gravity context in ref.~\cite{Johnson:2021tnl}, and so to avoid repetition the Reader is referred there. Those better behaved string equations (the form of which was  discovered in refs.~\cite{Morris:1990bw,Dalley:1991qg,Dalley:1991vr}) will be used in refs.~\cite{Johnson:2019eik,Johnson:2020heh,Johnson:2020exp} (as will be recalled later in this paper) to examine the  fully non-perturbative physics of JT gravity, and (using a variant) also for the study of various JT supergravity models. 

Returning to the main story,  it is useful to look more closely at the organization of the physics the orthogonal polynomials provide. The normalized  polynomials, with an additional factor from the measure absorbed into them, $\e^{-NV(\lambda)/2}P_n(\lambda)/\sqrt{h_n}$, will become a  function denoted $\psi(x,E)$ in the double-scaling limit, about which more will be said  shortly. The basis denoted previously $|n\rangle$  becomes $|x\rangle$ in the limit. An important question to ask is what becomes, in the limit,  of the operation of multiplying the orthogonal polynomials by $\lambda$. This is interesting and important to work out, and follows from the recursion relation~(\ref{eq:recursion}). First, dividing throughout by $\sqrt{h_n}$ and then using the relation $h_n=h_{n+1}/R_{n+1}$ multiple times gives:
%
\bea
\label{eq:recursion3}
{ \lambda} |n\rangle{=} \sqrt{R_{n+1}}|n+1\rangle + \sqrt{R_n}|n-1\rangle\ ,
\eea
and preparing for the large $N$ limit this is:
\bea
\label{eq:recursion3}
{ \lambda} |X\rangle{=} \sqrt{R(X{+}\epsilon)}|X{+}\epsilon\rangle + \sqrt{R(X)}|X{-}\epsilon\rangle\ ,
\eea
and so multiplying by $\lambda$ can be written as an operator ${\hat \lambda} $ on the  $X$ dependence in the following way:
\bea
\label{eq:recursion3}
{\hat \lambda} {=} \sqrt{R(X{+}\epsilon)}\exp\left\{\epsilon\frac{\partial}{\partial X}\right\} + \sqrt{R(X)}\exp\left\{-\epsilon\frac{\partial}{\partial X}\right\} \ \nonumber\\
\eea
Taylor expanding and substituting the scaling relations from before gives ${\hat \lambda}{=}2{-}{\cal H}\delta^2+\cdots$, where: 
\be
\label{eq:hamiltonian}
 {\cal H}=-\hbar^2\frac{\partial^2}{\partial x^2} + u(x)\ ,
\ee
is the operator whose eigenvalue is the scaled energy $E$. The scaled wavefunctions in the limit are denoted $\psi(x,E)$ and are thus the eigenfunctions of ${\cal H}$.  With a known $u(x)$, ({\it that must extend over all of $x$}) this  then fully defines a quantum mechanics problem.  As emphasized already all the way back to  Ref.~\cite{Johnson:2019eik}, it  is this quantum mechanical system that naturally provides the tools with which to excavate the {\it fully  non-perturbative} physics of JT gravity and variants thereof. 

\subsection{A Quantum Mechanics  Toolbox}
\label{sec:qm-toolbox}

The  quantum mechanics' spectral problem, with   wavefunctions $\psi(E,x)$ and energies $E$: 
\be
\label{eq:eigenstates}
{\cal H}\psi(E,x)=E\psi(E,x)\ ,
\ee
(where ${\mathcal H}$ is given in equation~(\ref{eq:hamiltonian})) is now the focus. An important note of caution is required here, given the kind of confusion that sometimes arises about this  quantum mechanics. {\it It should not  be confused with a gravity dual.} It is simply an efficient organizing tool for describing the physics of the Dyson gas. Consequently, various parts of it  will be used  in unusual ways to answer questions about the parent system. Sometimes this will seem very odd from the point of view of the quantum mechanics, which can often lead to some initial confusion and disorientation. But this is simply because it is an auxiliary system.\footnote{These common (but mild) side effects will eventually pass. Practicing on the Gaussian case before operating the heavier machinery is worthwhile at this point, which is why it is often mentioned.}

A first taste of this is that, as already seen  energy $E$ of the quantum mechanical system  is actually  {\it position}  in the (scaled) Dyson gas problem. 
On the other hand  recall that the  position, $x$, in the quantum mechanics problem began life as the index of the orthogonal polynomials, which correlates with energy excitation of an oscillator problem.  This means  that $x$  labels  an {\it energy} in  the Dyson gas.  The Gaussian case is illustrative. Recall that the ${ \psi}_n(\lambda) =\e^{-NV(\lambda)/2}P_n(\lambda)/\sqrt{h_n}$ are Hermite functions, the classic harmonic oscillator wavefunctions. They have  energy $\left(n+\frac12\right)$ times a constant, confirming that $x$ (the piece of $n$ that survives the double scaled limit)  in fact labels an energy.  This generalizes to non-Gaussian cases too, as the index $n$ sets the  highest power  $\lambda^n$ in the polynomial factor of the wavefunction, controlling the number of nodes it has, and so it (and hence $x$) remains a good energy coordinate of the Dyson gas. 

In the double-scaling limit, for the Gaussian case $u(x){=}{-}x$ as  mentioned, and hence  the wavefunctions are simply Airy functions, written as:\footnote{A direct scaling limit  on the Hermite function shows that it indeed becomes the Airy function in the scaling limit, as first shown by Moore in ref.~\cite{Moore:1990cn}.}  
\be
\label{eq:airy-wavefunction}
\psi(E,x){=}\hbar^{-\frac23}{\rm }{\rm Ai}[-\hbar^{-\frac23}(E+x)]\ ,
\ee
where the normalization is chosen in a way that will match the leading result for the spectral density, as will be shown presently.
 In this special case (following from the form of ${\cal H}$), the objects $E$ and~$x$ come in the combination $(E+x)$ everywhere resulting in an  accidental symmetry on exchanging them, but in general this will not be the case.

The generating function~(\ref{eq:connected-closed}) for the closed universes that has appeared so far has not been the focus in the contemporary literature  on JT gravity and its variants. It does play a role though, since it is built from the function $u(x)$, which is a foundation for the object which {\it is} the focus:
The defining partition function of the JT gravity theory (once the matrix model is specified through the appropriate $u(x)$; see later) is  the double-scaled version of the ``loop operator''  $\langle \frac1N{\rm Tr} [{\rm e}^{\ell M}]\rangle$, which makes a loop of fixed length~$\ell$ in the surface. (It can be derived by taking the large $L$ limit, holding $\ell$ fixed,  of the matrix operator ${\rm Tr}[M^L]$ that generates (after going from vertices to tessellations in the t'Hooftian way)  a large loop.) 

Setting $\ell{=}\beta$, it  is, in the orthogonal polynomial language developed in the previous subsection~\cite{Gross:1990aw,Banks:1990df}:
\be
\label{eq:full-loop-expression}
Z(\beta)  = \int_{-\infty}^\mu \!\!dx \,\langle x| {\rm e}^{-\beta{\cal H}}|x\rangle\ ,
\ee  
(Later the constant $\mu$ will be fixed by matching to perturbation theory.  For JT gravity in the conventions of this paper, $\mu=0$.) 
A somewhat similar development can be done for the matrix models needed to treat  JT supergravity examples. the upshot will be (see later) that similar formulae will be useable for SJT, but with $\mu=1$. 

The integral in equation~(\ref{eq:full-loop-expression}) is a sort of projected trace over the exponentiated Hamiltonian, which it is useful to write for future reference as:
\be
\label{eq:define-projector}
Z(\beta) = {\rm Tr}[{\cal P} \e^{-\beta{\cal H}}] \quad{\rm where}\quad {\cal P}\equiv \int_{-\infty}^\mu \!\!dx |x\rangle\langle x| \ .
\ee
 Now,  exponentiating ${\cal H}$ and taking this trace are not intuitively obvious things to do from the perspective of  the simple quantum mechanics~(\ref{eq:eigenstates}). However, a few more steps show that the form is quite natural for the problem in hand. Putting $\int d\psi |\psi\rangle\langle\psi|{\equiv}\int dE |\psi_E\rangle\langle\psi_E| {\equiv}1$ into Eq.~(\ref{eq:full-loop-expression}) and using Eq.~(\ref{eq:eigenstates}) with $\psi(E,x){\equiv}\langle\psi_E|x\rangle$
gives the partition function as the Laplace transform:
\begin{equation}
\label{eq:partition-function}
Z(\beta){=}\int dE \rho(E) {\rm e}^{-\beta E}\ ,
\end{equation}
 of the  full spectral density: 
 \be
 \label{eq:full-density}
 \rho(E)=\int_{-\infty}^\mu|\psi(E,x)|^2dx\ .
 \ee
 More will be said about the meaning of this presently. First, consider again the example, the Airy case. Using the wavefunction~(\ref{eq:airy-wavefunction}) (and put $\mu{=}0$) gives the full non-perturbative density of the matrix model: 
\be
\label{eq:airy-density}
\rho(E){=}\hbar^{-\frac23}\left[ {\rm Ai}^\prime(\zeta)^2{-}\zeta {\rm Ai}(\zeta)^2\right] \,
\ee
 with $\zeta{\equiv}{-}\hbar^{-\frac23}E$,  where ${f}^\prime{\equiv}\partial f/ \partial\zeta$.  Note that its support includes  $E{<}0$. In contrast, taking  large $E$ gives  the classical (disc order) result obtained earlier in equation~(\ref{eq:airy-tree}): $\rhoo{=}E^\frac12/(\pi\hbar)$, which is  supported only on  $E{\ge}0$. The function is plotted in figure~\ref{fig:toy-densities-2}.\footnote{For the Reader following along with a \textsf{MATLAB} window open, this is a good point to return to footnote~\ref{fn:wigner}, and re-run the code but now with the shift and scaling given just before equation~(\ref{eq:airy-tree}), {\it i.e,} if {\tt e} is a list of raw matrix eigenvalues, accumulate   ${\tt E}{=}{\tt N}^{\tt 1/6}{*}{\tt (2{*}sqrt(N){+}e)}$ over the loop. On histogramming the new data, the bins  line up with the curve~(\ref{eq:airy-density}), for some way, although it  eventually begin to fall short due to the finiteness of $N$. This is also shown in figure~\ref{fig:toy-densities-2}. } 
  The non-perturbative undulations visible at lower energy have additional microscopic {\it statistical} information, not to be found in the gravity literature, but will be very important in what is to follow. This to be discussed in Section~\ref{sec:Statistics}.
  \begin{figure}[t]
\centering
\includegraphics[width=0.45\textwidth]{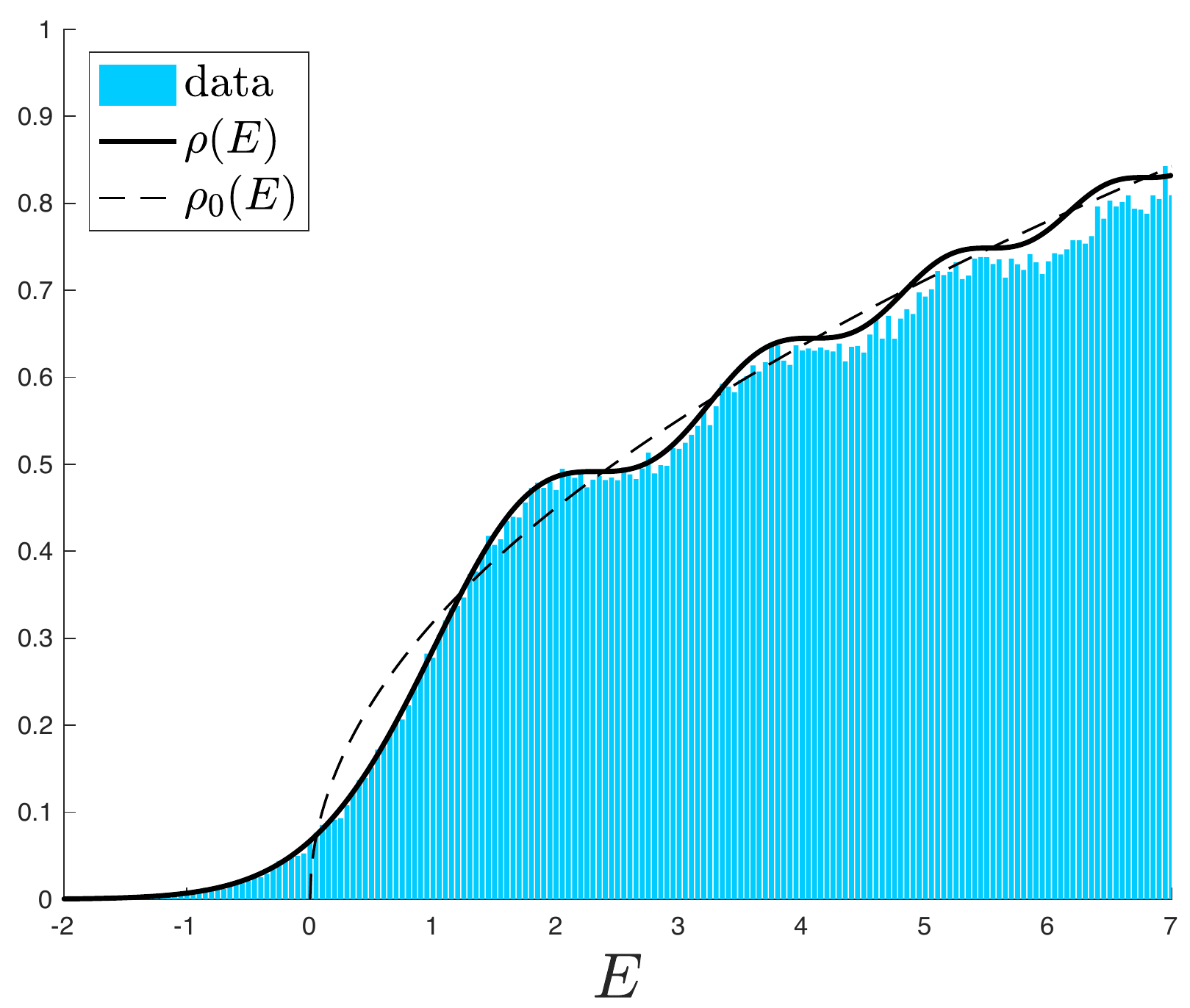}
\caption{\label{fig:toy-densities-2} The spectral density $\rho(E)$ for the Airy model (solid line). The rising dashed line is the leading result $\rhoo(E){=}\sqrt{E}/(\hbar\pi)$. Here, $\hbar{=}1$. Also shown are the histograms of data obtained from 100K scaled energy samples of 100${\times}$100 Hermitian matrices randomly generated with Gaussian probability. The non-perturbative undulations at lower energy hide additional microscopic information, to be discussed in Section~\ref{sec:Statistics}.}
\end{figure}

 For future reference, the Laplace transform of $\rho(E)$ can be computed exactly to give:
 \be
 \label{eq:airy-partition-function}
Z(\beta)=\int_{-\infty}^{+\infty} \rho(E)e^{-\beta E}dE = \frac{e^{\frac{\hbar^2}{12}\beta^3}}{2\pi^{1/2}\hbar\beta^{3/2}}\ .\qquad{\rm (Airy)}
\ee

Equation~(\ref{eq:full-density}), while operationally clear, is opaque from the perspective of the simple quantum mechanics. Tracing things back to its roots will nicely explain it in terms of a Fermionic many-body problem. 

\subsection{The Many-Body Picture}
\label{sec:many-body}
Recall that, from the point of view of the Dyson gas model, ~$E$ is not an energy but a position. Its precursors are the  positions $\lambda_i$. Moreover, notice that  $\rho(E)$ in equation~(\ref{eq:full-density}) is  built out of the square of an object that, at a given $E$, has  non-trivial structure in $x$.  The parameter $x$ was identified as an energy in the Dyson gas picture of the matrix model. Its ancestor before double scaling was the index $n$ on $\psi_n(\lambda_i)$. For every one of the discrete $\lambda_i$ there is a whole family of  the  wavefunctions  $\psi_n(\lambda_i)=\e^{-NV(\lambda)/2}P_n(\lambda_i)\sqrt{h_n}$. In the Gaussian case this is simply the presence of   a harmonic oscillator system at every position in space. This is the basis for a  many-body system  (with a continuum quantum field theory description) to emerge once the large $N$ limit is taken and the $\lambda_i$ merge into a continuum. Quantum fields or many-body wavefunctions with some given energy can be made by turning on modes (in each of the $N$ sectors at positions $\lambda_i$) built out of the the natural basis functions to hand, which are the $\psi_n(\lambda_i){\to}\psi(x,E)$. 

The nature of the matrix model problem to hand is such that these many body wavefunctions are in fact {\it fermionic}. Starting with the vacuum in all sectors (positions $\lambda_i$)  and turning on excitations (using the $P_n$ themselves), fermionic many body  states can be built as Slater determinants, {\it e.g.} a two body case involving two sites is $|\Psi_{12}\rangle{=}[\psi_1(\lambda_1)\psi_2(\lambda_2){-}\psi_1(\lambda_2)\psi_2(\lambda_1)]/\sqrt{2}$ which has energy indexed by 2 since $\psi_2$ is involved, and so forth. The fermionic nature is such that it is impossible to have two parts of the many-body state at the same energy, since the wavefunction will vanish.  In this way, it is natural to fill an incompressible ``Fermi sea'' up to some energy level  $n{=}N-1$. 

The question as to what state is natural for the matrix model is answered by equation~(\ref{eq:vandermonde-polynomials}), which rewrites the Vandermonde determinant (of which there are two copies in the defining model) in terms of the $P_n$, with a clear interpretation: The  Slater determinant uses all $N$ sites (positions $\lambda_i$) with  oscillators turned on all the way to energy $n{=}{N-1}$, and is manifestly fermionic since exchanging any two sites (rows) produces a minus sign. It is made by choosing  $N-1$ positions (so leaving out one--$N$ ways of doing this)  and then summing over all (signed) permutations of assignments of  the $P_1,P_2,\cdots,P_{N-1}$ to those sites ($(N{-}1)!$ terms). In other words, summing  kets of length $N$ of the form $|1\cdot 2\cdot 3\cdot 0\cdot 4\cdot \cdots |N{-}1\rangle\equiv |P_1\rangle | P_2\rangle| P_3\rangle|0\rangle |P_4\rangle\cdots |P_{N-1}\rangle$, where the  $0$ represents no excitement  in the omitted position, for a total of $N!$ terms. Denote this many-body state, after dividing it by $\sqrt{N!}$ for normalization,  as  $|\Psi\rangle$. As an example,  for $N=3$ it is:
\begin{widetext}
\bea
\label{eq:many-body-example}
|\Psi\rangle =\frac{1}{\sqrt{3!}}
\biggl( |0\cdot 1\cdot2\rangle - |0\cdot2\cdot1\rangle 
+|2\cdot0\cdot1\rangle - |1\cdot0\cdot2\rangle 
+ |1\cdot2\cdot0\rangle - |2\cdot1\cdot0\rangle\biggr)\ .
\eea 
%

\end{widetext}
So an important feature of this many-body state is that it is maximally spread across the entire site of  size $N$.

The special behaviour seen as $X\to1$ in expressions~(\ref{eq:matrix-simplified}) and~(\ref{eq:closed-free}) is essentially appearing as the Fermi level (which is at $X{=}1{-}\epsilon$) is approached.  The scaling limit that was taken was the process of magnifying the region at the edge of the Fermi sea, parameterizing it with~$x$, running from   $x{=}{-}\infty$ to the Fermi level at $x{=}\mu^-$. It is natural to ask what is the expectation value, or probability, of  a particular {\it position} $\lambda_j$ in this state. One must simply form $\langle\Psi|$ and combine it with $|\Psi\rangle$ to get, (using orthonormality, and  integrating over all $\lambda_i$ {\it except} $\lambda_j$), a sum $\sum_{i=0}^{N-1} |\psi_i(\lambda)|^2$ which in the double scaling limit becomes expression~(\ref{eq:full-density}).

This  explains, from the point of  view of the many body picture,  the presence of the integral over $x$ from $-\infty$ to~$\mu$ in several expressions. The integral is simply computing the expectation value of a certain object in the filled Fermi sea. The role of the quantum mechanics of equations~(\ref{eq:hamiltonian}) and~(\ref{eq:eigenstates}) is now extremely clear! The potential $u(x)$ contains the DNA (as it were) that ultimately determines, for each Dyson gas position $E$,  the precise amplitude of oscillation energy (labelled by $x$) that has been turned on at that level in the Fermi sea. The  DNA  is neatly expressed succinctly (through the Schrodinger equation~(\ref{eq:eigenstates}))  in the form of  the wavefunction $\psi(x,E)$. At a given $E$ therefore, it is very meaningful to sum up (integrate) $|\psi(x,E)|^2$ over the whole Fermi sea filling (out of which the state is defined). This is simply the density of  particles at position $E$ in the Dyson gas picture, formula~(\ref{eq:full-density}). Clearly, knowing $u(x)$ is therefore a means of fully specifying the desired matrix model.

With that dictionary in mind it is interesting to look afresh at the quantum mechanics~(\ref{eq:eigenstates}) in its own right.   There is of course a natural momentum to be identified,  $p{=}\pm\sqrt{E-u_0(x)}$.\footnote{Note  a subtlety here, reflected in the notation: As a solution to the string equations (to be discussed below), the full potential $u(x)$ of the problem  in general is of the form: $u(x)=\sum_{g=0}^\infty u_g(x)\hbar^{2g}+\cdots$ where the ellipsis denote non-perturbative contributions. The momentum $p$ in that potential can be discussed too, but it is already enough to look at the leading problem, the $g{=}0$ contribution $u_0(x)$,  to develop the key insights.} Now consider the  classical phase space with coordinates $(x,p)$.

For the Airy example, $p=\pm\sqrt{E+x}$, and so curves of constant $E$ are given by the parabolas $x=p^2-E$.  Figure~\ref{fig:phase_space_airy} shows the phase space (for positive $E$ values), with the Fermi level $x=\mu=0$ drawn in. Flying through the $x<\mu$ region (starting at the (red) dot), to the left with positive $p$ and then out again with negative $p$ has the simple interpretation as a tour through the many-body state that has been constructed.  
\begin{figure*}
\centering
\includegraphics[width=0.450\textwidth]{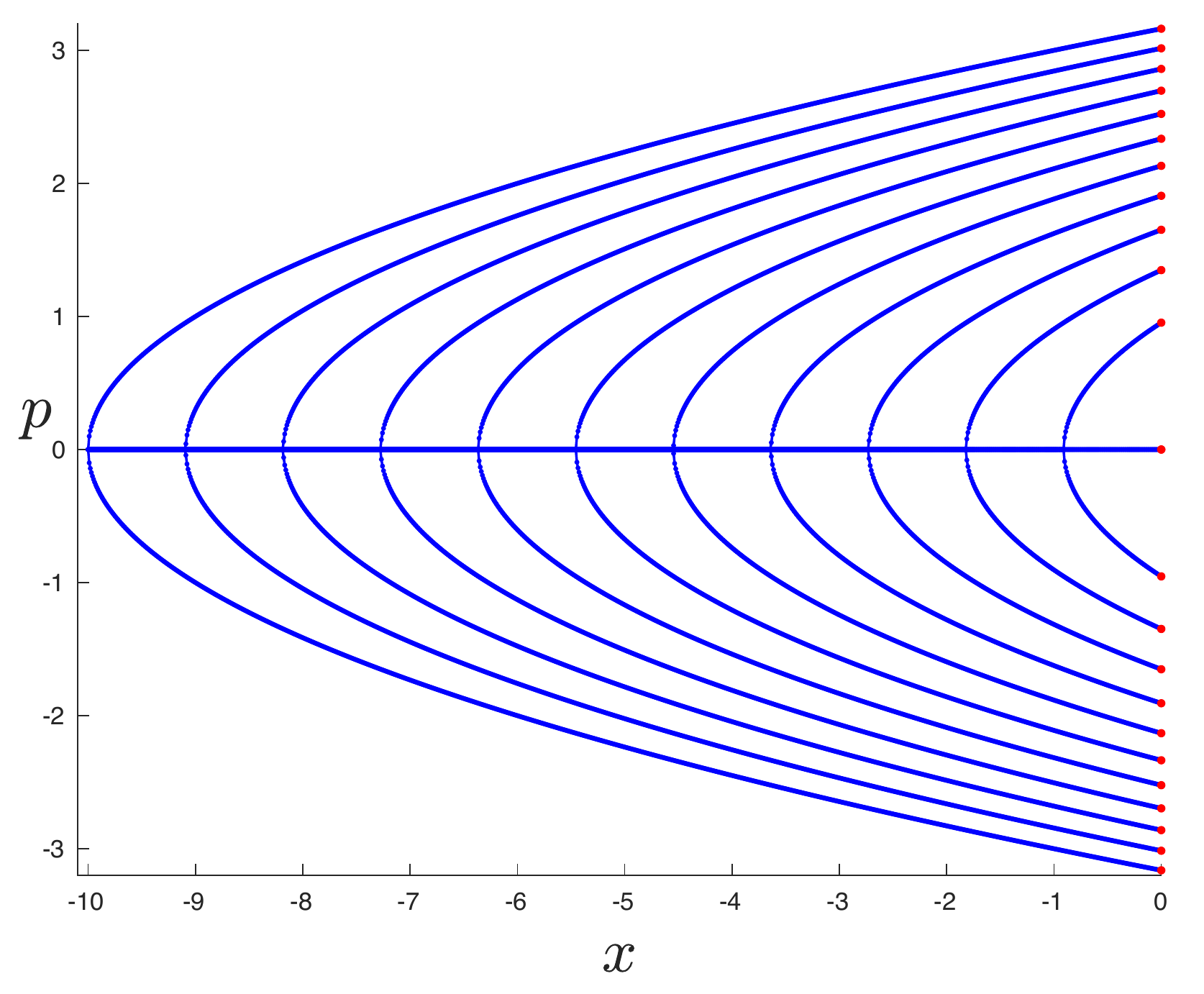} 
\hskip0.5cm
\includegraphics[width=0.450\textwidth]{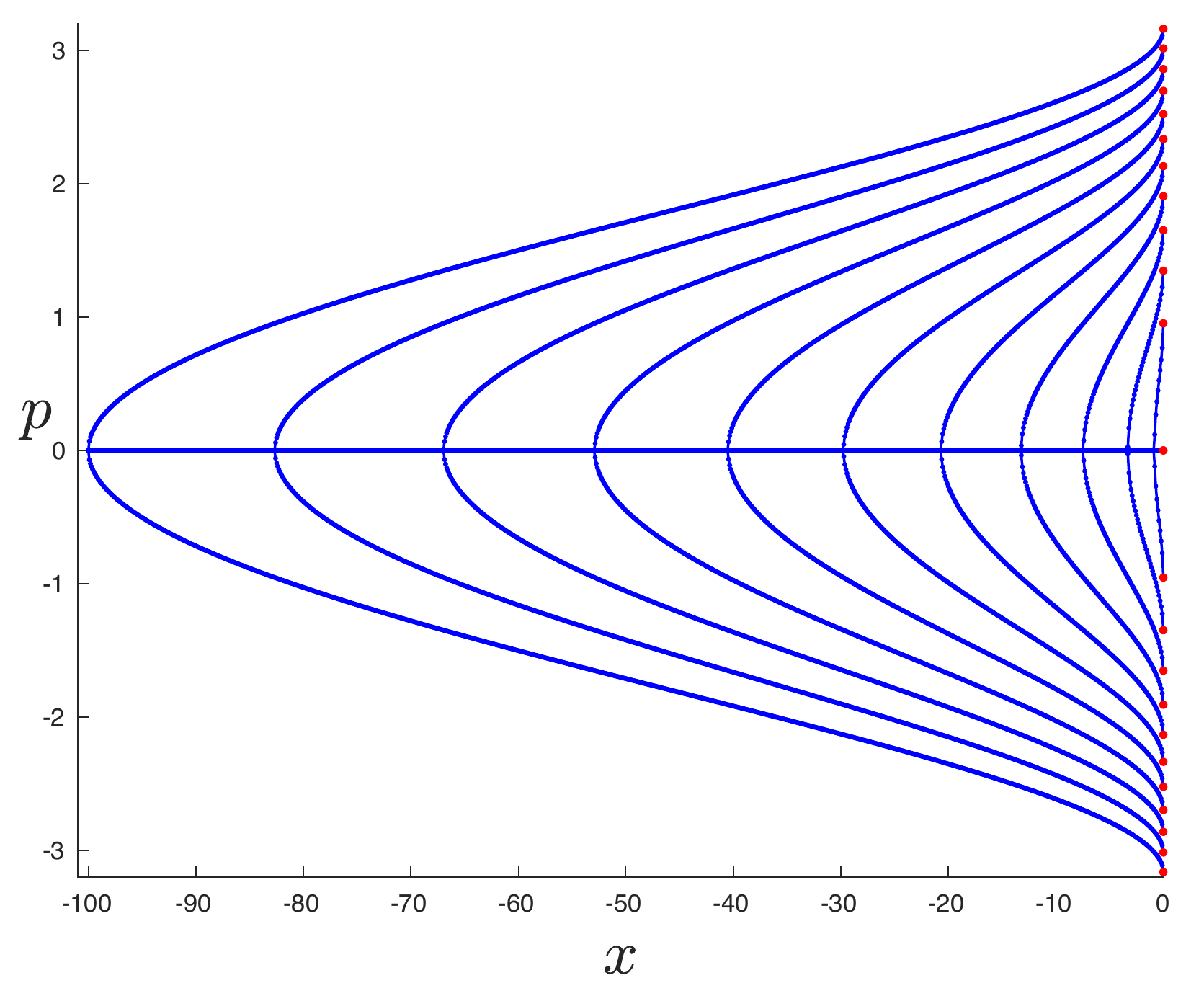} 
\caption{\label{fig:phase_space_airy} The left figure shows sample fixed $E$ trajectories in the phase space of the Airy model ($k=1$), which dives below the surface of the Fermi sea in the associated Dyson gas. On the right is the same thing for the $k=2$ model. Here the Fermi level is at $x=\mu=0$.}
\end{figure*}
For further illustration, the case of the (perturbative) $k=2$ minimal model is given (for the same sample energies) in Figure~\ref{fig:phase_space_airy}, on the right, where this time the classical string equation is $u_0(x)=(-x)^{1/2}$.
The notable new features that develop are the concavity of the curves as they leave the Fermi surface at $x=0$ before dipping into the bulk, and the fact that  they subsequently probe more deeply than for the $k=1$ case. With higher $k$ this becomes more pronounced.  
Looking ahead, the analogous pictures that can be drawn  for the JT and SJT cases later  combine the features of the various $k$ in an interesting way.

\begin{widetext}
Taking the WKB approximation of the wavefunctions in potential $u_0(x)$ is instructive. As usual, there are two pieces with general coefficients corresponding to a  normalisation and a phase:
\begin{equation}
\label{eq:WKB}
\psi(E,x)\simeq\frac{1}{[E-u_0(x)]^\frac14}\left(A_+\e^{{\frac{i}{\hbar}\int^x\!\sqrt{E-u_0(x')}dx'}}+A_-\e^{-{\frac{i}{\hbar}\int^x\!\sqrt{E-u_0(x')}dx'}}\right)\ ,
\end{equation}
These can be fixed by matching to the Airy case (the wavefunctions are Airy functions, with an asymptotic at large~$E$ that is precisely of WKB form):
\be
\label{eq:WKB3}
\psi(E,x) \simeq \frac{1}{\sqrt{\pi\hbar}}\frac{1}{(E+x)^{\frac14}}\cos\left(\frac{1}{\hbar}\frac23{(E+x)^\frac32}-\frac{\pi}{4}\right)\ ,
\ee
and so the  result in the classically allowed region is: 
\begin{equation}
\label{eq:WKB2}
\psi(E,x)\simeq\frac{1}{\sqrt{\pi\hbar}}\frac{1}{[E-u_0(x)]^\frac14} \cos\left({\frac{1}{\hbar}\int^x\!\sqrt{E-u_0(x')}dx'-\frac{\pi}{4}}\right)\ ,
\end{equation}
\end{widetext}
and the leading classical  expression for the  leading part of $\rho(E)$, using the density integral~(\ref{eq:full-density}):
\be
 \label{eq:leading-density}
 \rhoo(E)=\frac{1}{2\pi\hbar}\int_{-\infty}^\mu\frac{1}{\sqrt{E-u_0(x)}}dx\ .
 \ee
 There is an additional factor of $\frac12$ coming from the fact that, for large enough $E$, the frequency of the oscillations due to the cosine are fast enough to stand being integrated over.  Well, this is a result~(\ref{eq:leading-scaled-density}) obtained from a different perspective in subsection~\ref{sec:search-gravity}, but now here its all orders and beyond origins are now understood in terms of the complete wavefunctions~(\ref{eq:full-density}).  
 
 Of course, there is information in the semi-classical oscillations too, and it is sometimes useful to keep them. These are indeed the first rough draft of the non-perturbative information, very useful if $u_0(x)$ all that was available. They give an accurate depiction of the undulations in the spectral density for sufficiently large $E$. With a bit of care, the squared cosine  factor can be manipulated to yield equation~(\ref{eq:semi-classical-np-1}), where now $\rho_0(E)$ means expression~(\ref{eq:leading-density}) in general, and the rewriting:
 \be
\label{eq:integrate-momentum}
\int^\mu_{-\infty}\!\!\sqrt{E-u_0(x)}dx = \pi\hbar\!\int^E_0\!\!\!\rhoo(E^\prime)dE^\prime
\ee
was used. 

While the WKB form of the wavefunction will prove useful later, this WKB-corrected form of the density won't be needed here (given that complete non-perturbative data are available) and so much more will not be said about it than was already present in the Introduction. See a discussion in ref.~\cite{Johnson:2021tnl} for how to efficiently derive it from the perspective of wavefunctions, and more about how it fits into understanding non-perturbative physics. Ref.~\cite{Saad:2019lba}, where it was first derived (using a different perspective) has lots of interesting discussion about the semi-classical physics to be learned from it too.

\subsection{A Return to Statistics}
\label{sec:Statistics}
Much of the focus of the previous subsections was on getting access to physics that connects to properties of random surfaces of arbitrary topology, for the purposes of describing quantum gravity. The main (mostly) observable featured so far has been $\rho(E)$, which even after non-perturbative effects are included is a smooth function summarizing the collective or aggregate behaviour of the random matrices used to define the model. In fact because of this focus it is easy to forget that random matrix models began life as statistical endeavours! A main thrust of this paper is that this rich heritage can inform quantum gravity too, starting in two dimensions (and perhaps beyond), and that this is in fact necessary to fully understand it using matrix models. Indeed, there are several tools that are implicit in the previous that work, until recently, have not been fully brought into the spotlight in the gravity context. One such example is a Fredholm determinant, to appear in subsection~\ref{sec:fredholm},  that is built from an important object, the kernel $K(E,E^\prime)$, that organizes so much of random matrix theory. The kernel  in turn is built out of the wavefunctions $\psi(E,x)$, that have already been seen to be core elements of the construction. 

It all goes back to the Dyson gas starting point. The expression~(\ref{eq:Dyson-gas}) also has the interpretation as the integral over what is (up to a constant) the ``joint probability density'' $\textsf{P}_N(\lambda_1,\ldots,\lambda_N)$ for the eigenvalues to be at positions $\{\lambda_1,\ldots,\lambda_N\}$. From this one can derive the answers to questions about smaller groups of eigenvalues by simply integrating over the appropriate ranges over which ignorance is tolerated. As a familiar example, one can ask about just one eigenvalue position,~$\lambda$, and hence integrate over all the others, giving 
\be
\label{eq:one-point}
R_1(\lambda)={\tilde \rho}(\lambda) = N\int \prod_{i=2}^N d\lambda_i \textsf{P}_N(\lambda,\lambda_2,\ldots,\lambda_N)\ ,
\ee
and this is just the spectral density computed earlier. (The factor of $N$ is because any of the $N$ eigenvalues could have been the one Chosen  to be $\lambda$). However the answer for any number of them, $m$, can be computed to give the $m$-point correlation function:
\bea
\label{eq:em-point}
&&R_m(\lambda_1,\ldots,\lambda_m) = \\
&&\hskip1cm \frac{N!}{(N-m)!}\int \prod_{i=m+1}^N d\lambda_i {\textsf P}_N(\lambda_1,\lambda_2,\ldots,\lambda_N)\ ,\nonumber
\eea
with the straightforward combinatorial explanation for the factor.

\begin{figure*}
\centering
\includegraphics[width=0.48\textwidth]{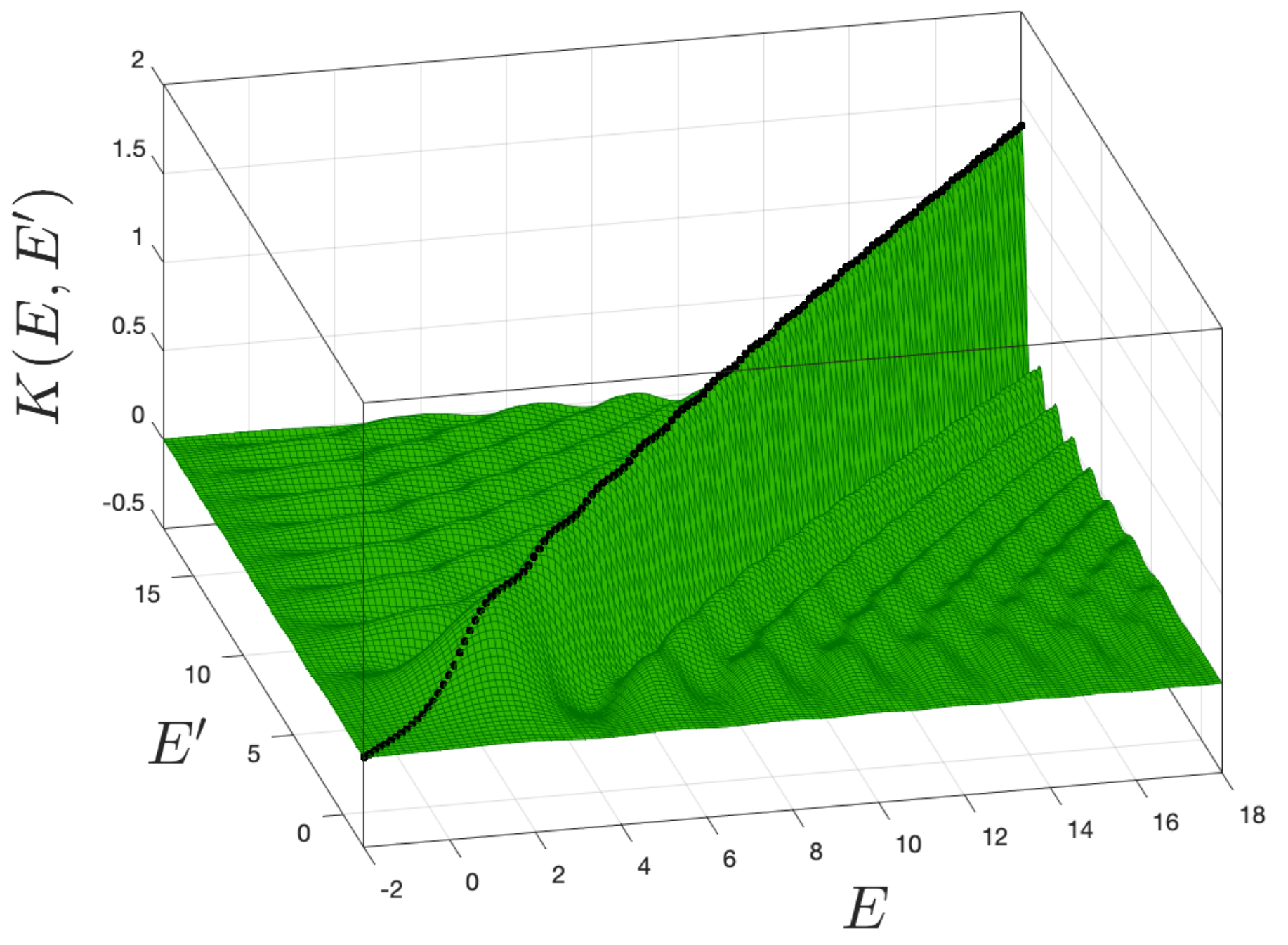}\hskip0.5cm\includegraphics[width=0.48\textwidth]{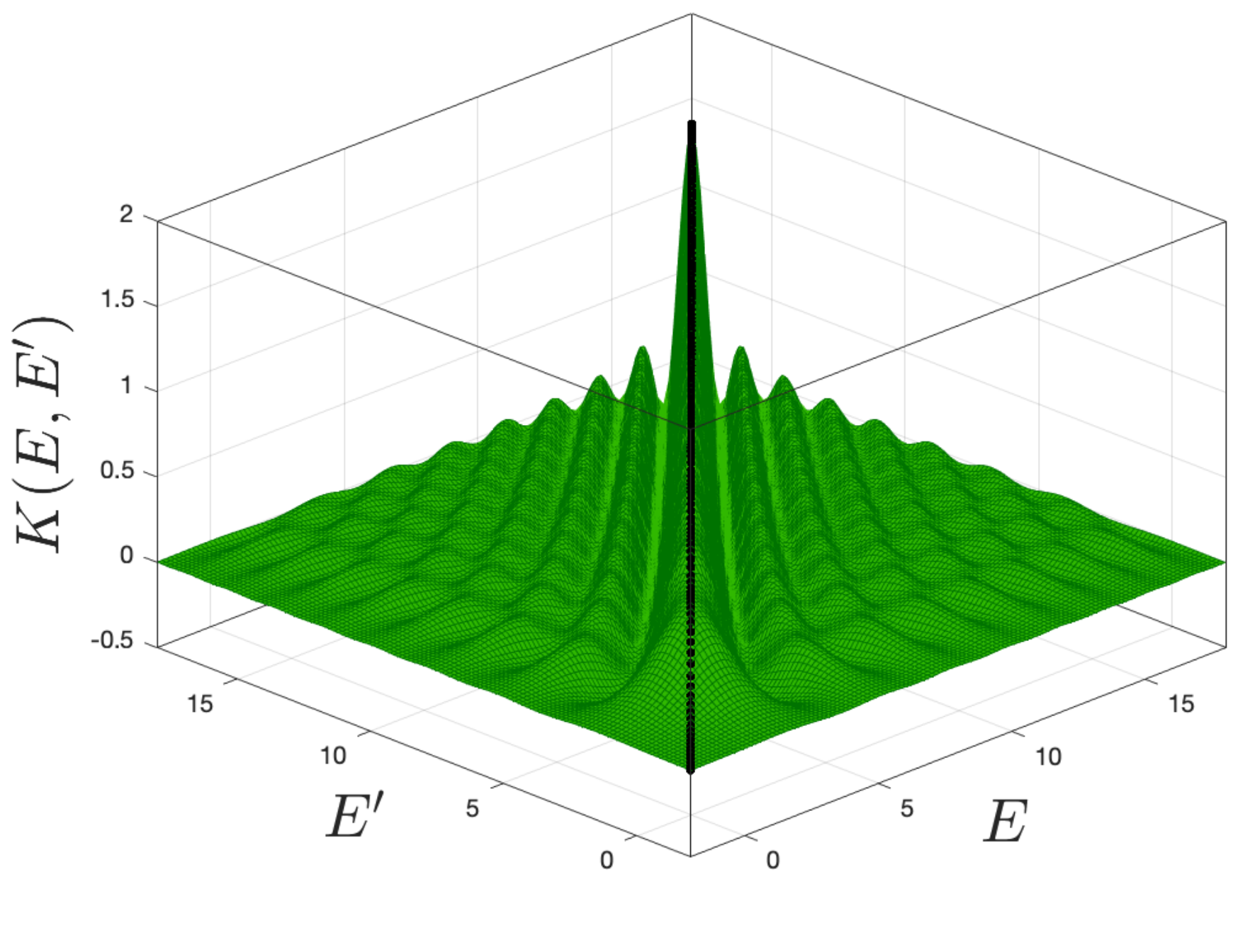}
\caption{\label{fig:Airy_kernel_3d}  Two views of the Airy kernel, showing the matrix model spectral density $\rho(E)$ along the diagonal dorsal ridge, with a pattern of ripples emanating outward into the off-diagonal regions. A great deal of microscopic physics can be extracted from this structure, and its analogues for other matrix models. See text.}
\end{figure*}
%

Some rather nice things happen when these quantities are written in terms of the normalized orthogonal polynomials, $\psi_n(\lambda)=\e^{-NV(\lambda)/2}P_n(\lambda)/\sqrt{h_n}$. Construct the matrix $(\mathbf{P})_{ni}=\psi_{n-1}(\lambda_i)$, in terms of whose determinant (up to a factor) the Vandermonde determinant was written in equation~(\ref{eq:vandermonde-polynomials}). In terms of $\mathbf{P}$, the master joint probability turns out to be\footnote{Apologies are offered  for there being three different P-quantities in these equations. $P_n(\lambda)$ are the orthogonal polynomials,~\textsf{P} is the joint probability density, and $\mathbf{P}$ is the matrix built from $P_n(\lambda)$s. It's really just for one paragraph.} $\textsf{P}_N={\rm det}(\mathbf{P}^{T}\mathbf{P})/N!$, where the $N{\times}N$ matrix
\be
\mathbf{P}^{T}\mathbf{P}(\lambda_i,\lambda_j)=K_N(\lambda_i,\lambda_j)\ ,
\ee 
is explicitly given as  
\be
\label{eq:kernel-discrete}
K_N(\lambda_i,\lambda_j)\equiv\sum_{n=0}^{N-1}\psi_n(\lambda_i)\psi_n(\lambda_j)\ .
\ee
This is a primary object of interest. In terms of submatrices of it,  formed in the obvious way by simply leaving eigenvalues out of the list, the $m$-point correlator can be succinctly written as: 
\be
\label{eq:m-correlator}
R_m(\lambda_1,\ldots,\lambda_m) ={\rm det}\biggl\|K_N(\lambda_i,\lambda_j)\biggr\|_{i,j=1}^{m}\ ,.
\ee
This  uses the nifty ``self-reproducing'' property of $K_N$ that follows from orthogonality of the $\psi_m(\lambda)$:
\be
\label{eq:self-reproducing}
\int d\kappa\,\, K_N(\lambda,\kappa) K_N(\kappa,\nu) = K_N(\lambda,\nu)\ .
\ee
The eigenvalue density seen earlier itself is just the determinant of the 1${\times}$1 matrix formed by leaving out all the $N-1$ possibilities.
\be
{\tilde \rho}(\lambda) = K_N(\lambda,\lambda)=\sum_{n=0}^{N-1} \psi_n^2(\lambda)\ .
\ee
This returns full circle to the double-scaled expression~(\ref{eq:full-density}). It in turn was also given an interpretation in terms of the many-body fermion wavefunction in the paragraph below~(\ref{eq:many-body-example}). This makes sense because the rewriting just performed of the probabilities in terms of orthogonal polynomials is again implementing the many-body picture. 
Just as the $\psi_n(\lambda_i)$ become the wavefunction $\psi(x,E)$ in the double scaling limit, correspondingly, the kernel, after double-scaling, is also a rather nice object:
\bea
\label{eq:edge-kernel}
K(E,E^{\prime}) &=& \int_{-\infty}^\mu \psi(E,x)\psi(E^{\prime},x) dx \\
&=& \frac{\psi(\mu, E)\psi^\prime(\mu,E^{\prime}) - \psi^\prime(\mu, E)\psi(\mu,E^{\prime})}{E-E^\prime}
\ , \nonumber
\eea
describing correlations at the (scaled) endpoint of the spectrum. The second line makes use of the continuum limit of the ``Christoffel-Darboux'' relation satisfied by orthogonal polynomials, and gives a useful form of the kernel in terms of just $\psi(E,\mu)$ and its first derivative, quantities that live entirely at the Fermi surface $x{=}\mu$. It is  natural to think of $K(E,E^\prime)$  as a sort of response function  encoding how two different parts of the Fermi surface, $E$ and $E^\prime$, talk to each other.  (The further significance of $\psi(E,\mu)$ will be discussed in Section~\ref{sec:return-to-gravity}.)

Many properties of region of the tail of the eigenvalue distribution, to which the quest for gravity has led,  can be expressed in terms of $K(E,E^\prime)$. In the last 30 years or so, there has been a great deal of interest in such regions, and many striking results, for a variety of random matrix (and related) problems, in the statistical physics community. A most celebrated example is in fact the Gaussian case being used for illustration here. The Kernel  the ``Airy Kernel'', which, after inserting the wavefunctions~(\ref{eq:airy-wavefunction}) into equation~(\ref{eq:edge-kernel}) and performing the $x$-integral,  can be written as:
\bea
\label{eq:Airy-kernel}
K(\zeta,\xi) &=&\frac{{\rm Ai}(\zeta){\rm Ai}^\prime(\xi)-{\rm Ai}(\xi){\rm Ai}^\prime(\zeta)}{\zeta-\xi}\ ,
\eea
where $\zeta = -\hbar^{-\frac23}E$ and $\xi = -\hbar^{-\frac23}E^\prime$, and in the above a prime denotes a $\zeta$ or $\xi$ derivative. For future reference, the wavefunction for Airy at the Fermi surface is:
\be
\label{eq:airy-wavefunction}
\psi(0,E) = \hbar^{-\frac23}{\rm Ai}(-\hbar^{-\frac23}E)\ .
\ee

%
It is highly worthwhile looking at this beautiful kernel object in three dimensions, since at a glance much can be appreciated about the information it encodes.
Figure~\ref{fig:Airy_kernel_3d} shows two views of a section of it.  
%
The black line delineating the top of what resembles the dorsal fin of some marine creature is the diagonal. This  is, as already noted, the full  spectral density $\rho(E) = K(E,E)$. Visible are the key non-perturbative undulations away from the simple $(\pi\hbar)^{-1}E^{\frac12}$ behaviour, as observed in figure~\ref{fig:toy-densities-2}. Moving off the diagonal  shows ripples developing, reflecting oscillations present in the wavefunctions from which it is made, and connecting smoothly to the undulations in the density. There is a rich amount of physics to be mined from this object (and its generalization to models of gravity to be discussed later) that will reveal a lot of interesting information.  {\it This information will include precise details of the microscopic spectrum.}


\subsection{More Statistics: The Fredholm Determinant}
\label{sec:fredholm}
The  properties of the Airy kernel were used by, {\it e.g.,} Forrester~\cite{FORRESTER1993709}, and Tracy and Widom~\cite{Tracy:1992rf} to characterize the probability distribution of the highest (or lowest) eigenvalue of the matrix distribution.  It is well worth seeing how it works more generally. 
The first or last eigenvalue problem is  computable in terms of $K(E,E^\prime)$ as follows. Returning to the full (unscaled, discrete) problem for a moment, the following question can be asked: What is the probability of finding no eigenvalue in the region $(a\leq\lambda\leq b)$? To work this out, as before  integrate~$\textsf{P}$  over all $\lambda_i$, over all allowed values but the region of interest.  After some manipulation~\cite{Gaudin1961SurLL,Meh2004}, it is:
\be
\label{eq:gap-probability}
E(0,(a,b))= {\rm det}\biggl\| \delta_{nm} - \int_a^b\psi_{n-1}(\lambda)\psi_{m-1}(\lambda)d\lambda\biggr\|_{n,m=1}^N\ .
\ee
(The unfortunate conventional name $E$ here has nothing to do with energy. Also, the zero in the brackets reminds that  it  is the probability of ``no eigenvalue''.)

Following Gaudin~\cite{Gaudin1961SurLL}, now consider the integral operator $\mathbf{K}|_{(a,b)}$ built from the kernel $K(\lambda,\kappa)$ (defined in equation~(\ref{eq:kernel-discrete})), acting on the space $a\leq \lambda\leq b$ on some eigenfunctions $f(\lambda)$ according to:
\be
\label{eq:eigensolutions-kernel}
\int_a^bK(\lambda,\kappa)f(\kappa) = \alpha f(\lambda)\ .
\ee
Decomposing $f(\lambda)$ in terms of the orthonormal wavefunction basis $\psi_n(\lambda)$ (used a lot in the previous subsection), a line or two of algebra shows that there are $N$ solutions~$\alpha_i$ defined by the characteristic equation:
\be 
{\rm det}\biggl\| \alpha\delta_{nm} - \int_a^b\psi_{n-1}(\lambda)\psi_{m-1}(\lambda)d\lambda \biggr\| = \prod_{n=0}^{N-1}(\alpha-\alpha_n)\ ,\ee
and hence the probability of interest~(\ref{eq:gap-probability}) is the Fredholm determinant~\cite{10.1007/BF02421317}:
\be
\label{eq:Fredholm-discrete}
E(0,(a,b)) = \prod_{n=0}^{N-1}(1-\alpha_n) = {\rm det}[\mathbb{I}-\mathbf{K}|_{(a,b)}]\ .
\ee 
Returning to the large $N$ limit and scaling to the endpoint as before (although it is also interesting to take the limit in the bulk of the distribution) gives the probability in the region $a\leq E\leq b$, now using the edge kernel~(\ref{eq:edge-kernel}).

As already seen in the  the prototype Gaussian/Airy case, although the Wigner semi-circle law put the eigenvalues (after scaling) on the positive real $E$ line ($\rho_0(E){=}(\pi\hbar)^{-1}E^\frac12$), this is just the leading (perturbative)  result: The eigenvalues are in principle distributed over the whole real line, with incursion into the $E<0$ region of magnitude exponentially smaller than the leading results.  (See equation~(\ref{eq:airy-density}) and figure~\ref{fig:toy-densities-2}.) To ask about the probability of the {\it first} energy, the interval of interest in the probability expression should be $a=-\infty$ and $b=s$, where the latter is some reference energy. The resulting determinant, and hence the probability sought, will be a function of $s$. For the Airy prototype it is (it is usually written with a conventional subscript of 2 to denote that it is from the $\boldsymbol{\beta}=2$ Dyson-Wigner ensemble):
\be
\label{eq:Airy-Fredholm}
E_2(-\infty,s)={\rm det}[\mathbb{I}-\mathbf{K_{\rm Ai}}|_{(-\infty,s)}]\ .
\ee
The second term inside the determinant means the Fredholm integral operator deployed on the real line from $-\infty$ to $s$, where the Airy Kernel in equation~(\ref{eq:Airy-kernel}) is used.  What should be expected is as follows. Far to the left of the interval, it is unlikely that there is a eigenvalue (since everything is mostly located in the region where~$s$ (or $E$) is positive), and hence the result (probability there is no eigenvalue) should be close to one. As one moves more the right, the likelihood increases  somewhat  that some outlier might have appeared, so the ``probability of none'' should decrease more. Moving closer to $s=0$ it will further decrease it since this is where the bulk of the eigenvalues starts. Moving well past $s=0$, eigenvalues almost certainly have already appeared, and so the probability should be falling rapidly to zero. This is in fact a cumulative probability density function (CDF), and so taking a derivative with respect to $s$ will yield the more usual probability density function (PDF) of the first eigenvalue, as will be shown below.

Tracy and Widom showed (following the methods of Jimbo, Miwa, M\^ori and Sato~\cite{Jimbo:1979rt}) that this can be characterized as a $\tau$-function, and can be defined in terms of the Hastings-McCleod~\cite{Hastings1980} solution of the Painlev\'e~II equation. This beautiful 
result represents considerable progress, since analytically computing the Fredholm determinant is possible only in the lucky circumstance when the eigenproblem~(\ref{eq:eigensolutions-kernel}) can be directly solved. The Painlev\'e approach gave a way of characterizing the result in terms of objects about which many results are known. However the full function can still only found by  numerically solving the (non-linear) Painlev\`e~II equation.

\begin{figure*}[t]
\centering
\includegraphics[width=0.48\textwidth]{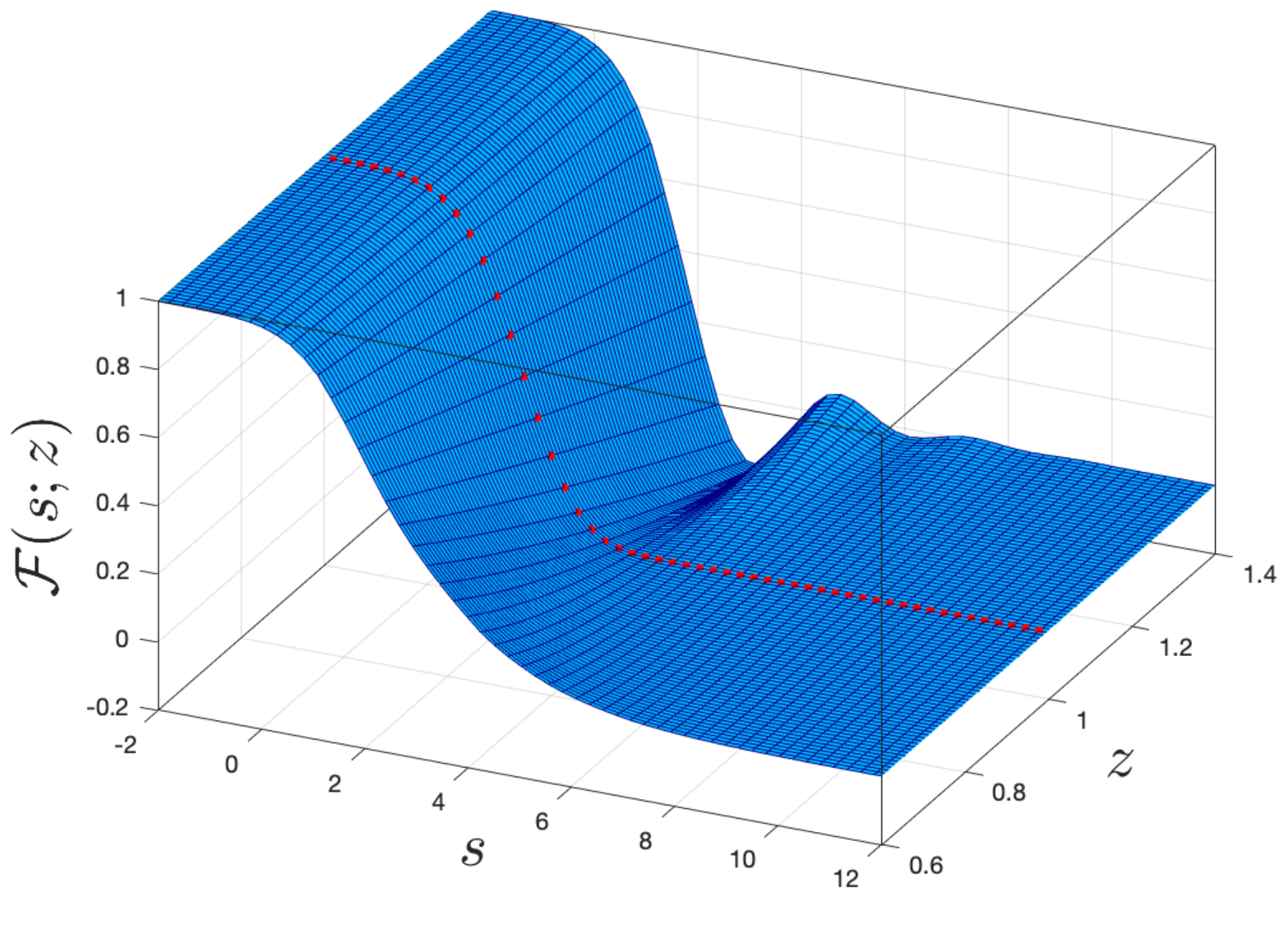}\hskip0.5cm \includegraphics[width=0.48\textwidth]{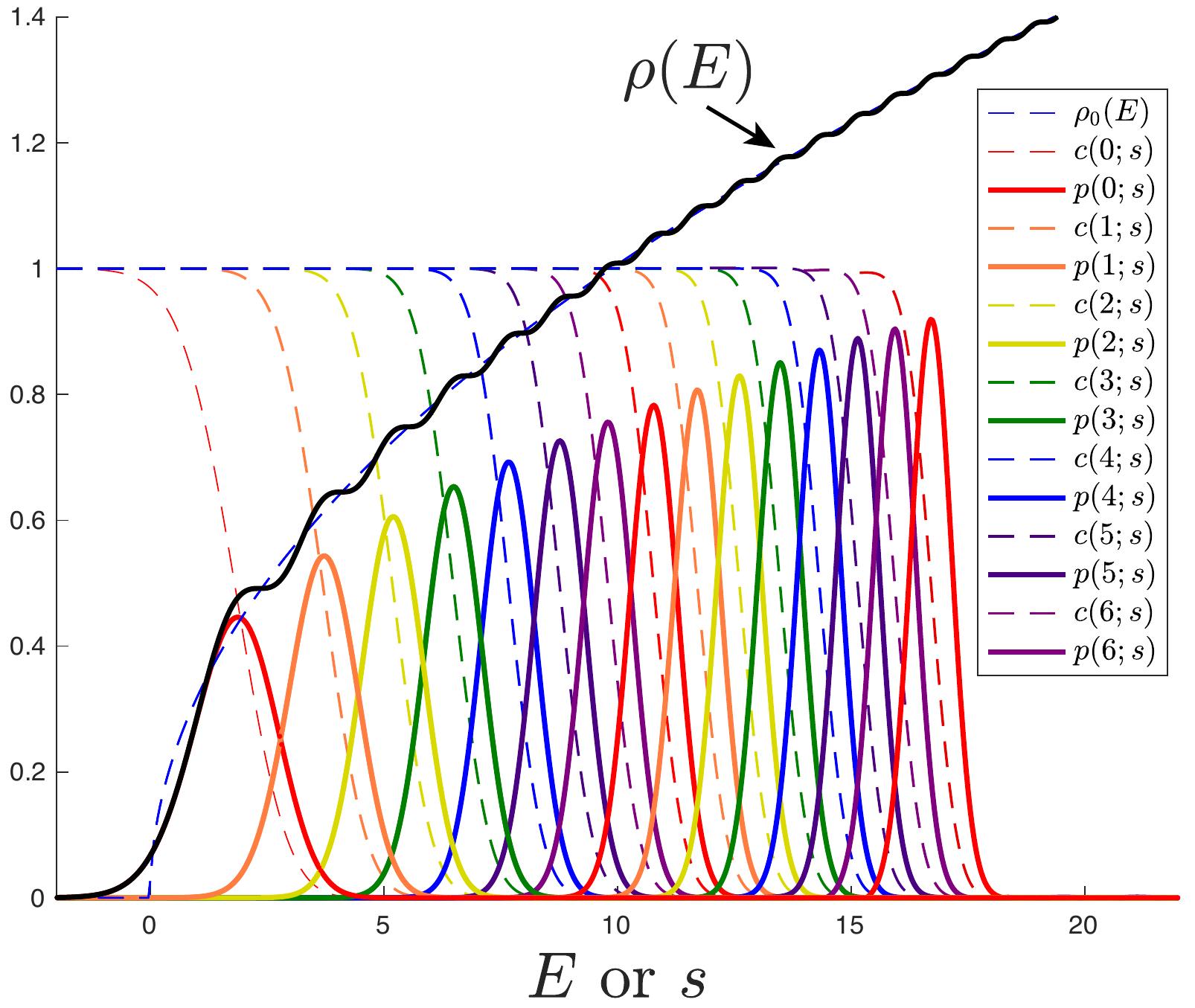}
\caption{\label{fig:Airy_fredholm_3d_and_peaks} On the left is the Fredholm determinant ${\cal F}((-\infty,s);z)$, from which all of the probabilities density functions on the right for the Airy model are derived. These are the first 15 energy levels.}
\end{figure*}

For many other problems, and certainly the gravity models to come, while it would be nice to have a generalization of this approach (defining the edge probability in terms of differential equations), the results needed are not known. Even if they were, a numerical approach to solving the equations would be needed to explicitly exhibit 
the whole result, as is clear from the Airy case. 

Happily, recent work by Bornemann~\cite{Bornemann_2009} entirely sidesteps the reliance on showing an equivalence to differential equations by showing  that Fredholm determinants can be very efficiently computed using quadrature methods, which in fact turn out to be more accurate and powerful than solving the associated non-linear equation ({\rm e.g.}, Painlev\'e~II), even when available. These are the methods that will be used in this paper. They are quite natural to use here, since fully defining  the gravity problems non-perturbatively {\it also} required numerically solving non-linear equations to obtain $u(x)$ for use as potential for the Hamiltonian~(\ref{eq:hamiltonian}), and then numerically solving the Schr\"odinger problem~(\ref{eq:eigenstates}) to get the $\psi(E,x)$ from which $K(E,E^\prime)$ will then be built. With three layers of numerical work and non-linearity there is the potential for the results to be badly afflicted by noise but it will turn out that very good progress can be made, as already shown in ref.~\cite{Johnson:2021zuo}. 

Before going on to do so, it is worth remarking that a slight generalization of the above allows for the computation of the probability distribution of the second eigenvalue, the third, and so on.  The notation will be as follows. Since the  first eigenvalue is the ground state energy, this will be denoted the 0th energy level. The second is then the 1st level, and so forth. Little $n$ will be the level label. Defining the object:
\be
\label{eq:fredholm-master}
{\cal F}((a,b);z)={\rm det}[\mathbb{I}-z\mathbf{K}|_{(a,b)}]\ ,
\ee
it turns out that the probability for $j$ eigenvalues to be in the interval is the result of  acting with $(-d/dz)^j$  on ${\cal F}([a,b];z)$, and then setting $z=1$.  The derivation of this is some straightforward (at least in the finite $N$ case)  fun with determinants involving writing  ${\cal F}(z)$ as $\prod_i(1-z\alpha_i)$, where ${\cal F}(z=1)$ is the original Fredholm determinant $E(0,(a,b))$ of equation~(\ref{eq:Fredholm-discrete}).  This will be omitted here, but see  Appendix A.7 of ref.\cite{Meh2004}. From here it is plain sailing, since the probability of having the $n$th level  appear  must come from adding the probabilities of the $(n-1)$th, $(n-2)$th, {\it etc.,} cases as well,  (as these are independent events that must have occurred for this to be the $n$th level) giving that 
 the cumulative distribution function of   the $n$th level    is:
\be
\label{eq:higher-levels}
c(n;(a,b)) = \sum_{j=0}^n \frac{(-1)^j}{j!}\frac{d^j}{dz^j} {\cal F}((a,b);z)\biggl.\biggr|_{z=1}\ .
\ee
and  the probability distribution $p(n;(a,b))$ for the $n$th level  can then be obtained from this by differentiation (see below). 

A brief summary of the methods of Bornemann~\cite{Bornemann_2009} are in order. The problem is to evaluate the determinant of an operator on the energy interval $(a,b)$. Thought of as a matrix, it is infinite dimensional, and so this is an additional challenge. 
The technique of quadrature  represents functions on an  interval in terms of a basis of special functions, and reduces the problem of integrating them to a sum. This is achieved by breaking up the  interval interval into~$m$ points $e_i$ and computing weights $w_i$ ($i=1\cdots m$) for each point. The values of the function at those points are weighted by the $w_i$ and the sum gives an approximation to the integral.   $\int_0^s f(E)dE \to \sum_i^m w_i f(e_i) $. The weights used depend upon the choice of special functions--each quadrature method has its own choices. The kind of quadrature that it is best to use depends upon the kinds of functions expected to be integrated on the interval. 
How well this works depends upon how well adapted the quadrature method is to the class of function being integrated, as well as the number of quadrature points~$m$. This is an ancient technique that works extremely well in a wide class of cases, and underlies much of the off-the-shelf numerical methods used by computers to solve integrals. Just as the  integral can be  reduced to a sum, an  integral operator becomes a finite matrix and so its determinant becomes a finite process that can be computed using the set of weights as used for ordinary quadrature:
 \be
 {\rm det}(\mathbf{I}-z\mathbf{K}|_{(-\infty,s)}) \to {\rm det}(\delta_{ij}-zw_i^\frac12 K(e_i,e_j)w_j^\frac12)\ .
 \ee

Bornemann has shown that this works extremely well (using {\it e.g.,} Clenshaw-Curtis and Gauss-Legendre quadrature) for many important Fredholm problems, including problems such as the Airy model. It is an interesting exercise to reproduce such results here, in preparation for  later adapting the methods for use with matrix models involving gravity. If the $\psi(E,x)$ are known analytically, as they are for Airy, it turns out that  the method  gives impressively accurate results for the first few levels with remarkably modest values of~$m$ such as~8, or 16, in matters of seconds.

Working on the interval $(-\infty,s)$, using  the Airy Kernel, the full ${\cal F}$ determinant of equation~(\ref{eq:fredholm-master}) can be constructed, in preparation for constructing information about higher levels as well. A portion of the result is displayed in figure~\ref{fig:Airy_fredholm_3d_and_peaks}, on the left. (It was made, with $m{=}64$, on a grid of $5000$ points for $s$ and $1000$ for $z$, and took~$\sim$81s to generate.\footnote{If the Reader is tempted to try this, it  requires only a little bit more coding than done for this section so far. Off the shelf quadrature weights can be used, but Bornemann actually supplies some lines of code for generating the quadrature weights, and some guidance as to how to use them. It is then simply a matter of writing some instructions to generate $K(E,E^\prime)$ out of Airy functions, and implementing the quadrature on $(0,s)$.}) The two-dimensional slice at $z{=}1$ (red line) shows the cumulative distribution function for the zeroth energy (ground state), $c(0;s){\equiv}E(0;(-\infty,s))$, and the more general $z$-dependence will produce, through $z$-derivatives, the CDFs for higher levels, $c(n;s)$, and their PDFs $p(n;s)=-d c(n;s)/ds$. (In the notation, dependence on $(-\infty,s)$ is simplified to just dependence on $s$.)

See the plot on the right of figure~\ref{fig:Airy_fredholm_3d_and_peaks} 
for the CDFs and PDFs for the first 15 levels. Key features are as follows:
\begin{itemize}
\item The rising dashed line is the perturbative result $\rhoo(E){=}(\pi\hbar)^{-1}E^{\frac12}$ obtained by scaling the Wigner semi-circle law.  See equation~(\ref{eq:airy-tree}).
\item The solid black line is the fully non-perturbative $\rho(E)$ given in equation~(\ref{eq:airy-density}).
\item The solid red line is the  PDF $p(0;s)$ of the lowest eigenvalue of each  matrix in  the ensemble, known as the Tracy-Widom distribution. The dashed red line is the cumulative CDF $c(0;s)$  for this lowest level. 
\item Successive peaks in solid lines show the PDFs for the next 14 energies of the ensemble, computed using~(\ref{eq:higher-levels}). Also shown are their CDFs.
\end{itemize}

%
%
As ``experimental'' confirmation of the results, figure~\ref{fig:numerics_for_airy} 
shows the histograms of the first six energy levels from 100K samples of  100$\times$100 randomly generated Hermitian matrices, where their energies $e_i$ have been scaled and shifted to give $E_i{=}N^{\frac16}(e_i{+}2\sqrt{N})$.\footnote{For the Reader with a \textsf{MATLAB} window open: This can be obtained straightforwardly from the earlier data by adding lines of code that simply compute the ordering of all the eigenvalues of each sample (use {\tt sort()}), and then creating separate lists that accumulate the eigenvalues according to which order they were in, per sample. Histogramming can then be done on each of these lists, and some colour coding added for good measure.}

\begin{figure}[t]
\centering
\includegraphics[width=0.450\textwidth]{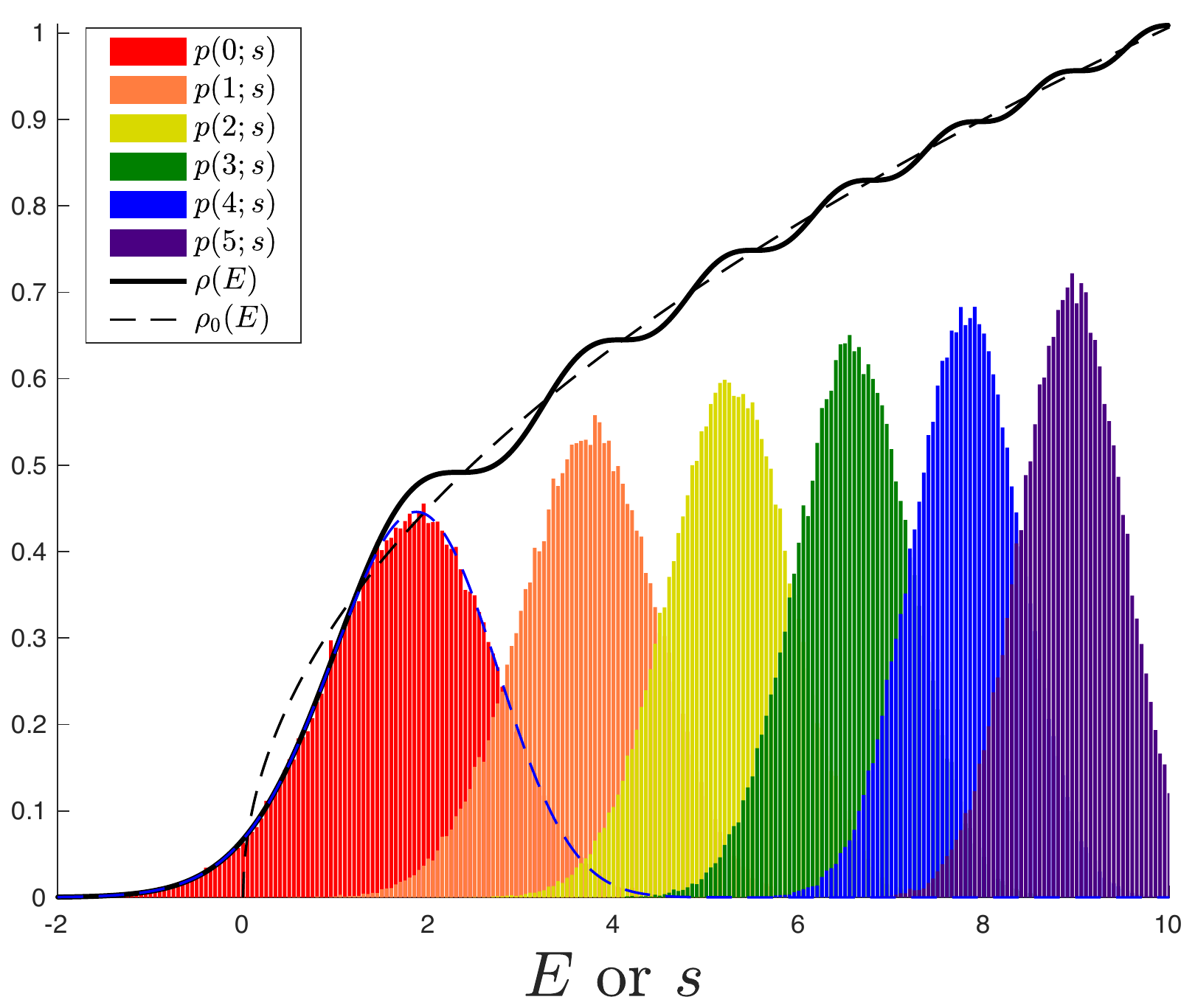} 
\caption{\label{fig:numerics_for_airy} The spectral density $\rho(E)$ for the Airy model (solid line). The rising dashed line is the leading result $\rhoo(E){=}\sqrt{E}/(\hbar\pi)$ for $\hbar=1$. The  histograms   $p(n;E)$ are frequencies of the $n$th energy level, extracted numerically from a  Gaussian random system of $100{\times}100$  Hermitian matrices, for $100K$ samples. Note the  correspondence with the undulations in~$\rho(E)$. The blue dashed peak  is the exact Tracy-Widom distribution~\cite{Tracy:1992rf} $p(0;E)$  for the ground state $E_0$, and  ${\cal E}_0\equiv\langle E_0\rangle\simeq1.77108$.}
\end{figure}

As emphasized in the Introduction it is now clear that the matrix model spectral density should be recast in the more {\it physical} manner as a sum (as given in equation~(\ref{eq:fattening})) over a {\it discrete} set of states, which happen to have broadened into probability distributions in the random matrix model's ensemble description of the physics. A representative choice   of this discrete spectrum is simply the set  of average energies ${\cal E}_n{=}\langle E_n\rangle$ of these distributions. The analogue of these in the full gravity theory (which will be discussed in the next section) is the proposed discrete  gravity spectrum of JT gravity. (See footnote~\ref{fn:peak-subtleties} for a note on whether the average of each peak is the definitive choice.) For later purposes it is useful to extract some data on the set $\{ {\cal E}_n\}$, for this value of $\hbar{=}1$. There's no closed form for these, so they are displayed in Table~\ref{tab:mean-levels-fredholm} for the first 15 levels. The variance of the PDFs from which they are drawn is also displayed, showing it decrease as energy increases. A plot at the bottom shows the steady narrowing of the gaps between the states.

\begingroup
\begin{table}[t]
\begin{center}
\begin{tabular}{|c|c|c||c||c|}
\hline
\textbf{energy} &\textbf{value}&\textbf{variance}&\textbf{value}&\textbf{Airy}\\
\textbf{level} &\textbf{(Fredholm)}&\textbf{(Fredholm)}&\textbf{(sampling)}&\textbf{zeros}\\
\hline\hline
\textrm{${\cal E}_0$}&\textrm{1.771087}&\textrm{0.813192}&\textrm{1.778}&\textrm{2.3381}\\
\hline
\textrm{${\cal E}_1$}&\textrm{3.675436}&\textrm{0.540525}&\textrm{3.707}&\textrm{4.0880}\\
\hline
\textrm{${\cal E}_2$}&\textrm{5.171345}&\textrm{0.433677}&\textrm{5.236}&\textrm{5.5206}\\
\hline 
\textrm{${\cal E}_3$}&\textrm{6.474627}&\textrm{0.372766}&\textrm{6.574}&\textrm{6.7867}\\
\hline
\textrm{${\cal E}_4$}&\textrm{7.656820}&\textrm{0.328587}&\textrm{7.801}&\textrm{7.9441}\\
\hline
\textrm{${\cal E}_5$}&\textrm{8.753390}&\textrm{0.295781}&\textrm{8.942}&\textrm{9.0227}\\
\hline
\textrm{${\cal E}_6$}&\textrm{9.789991}&\textrm{0.289320}&\textrm{10.025}&\textrm{10.0402}\\
\hline
\textrm{${\cal E}_7$}&\textrm{10.772994}&\textrm{0.275458}&\textrm{11.059}&\textrm{11.0085}\\
\hline
\textrm{${\cal E}_8$}&\textrm{11.691298}&\textrm{0.216385}&\textrm{12.051}&\textrm{11.9360}\\
\hline
\textrm{${\cal E}_9$}&\textrm{12.590839}&\textrm{0.208200}&\textrm{13.011}&\textrm{12.8288}\\
\hline
\textrm{${\cal E}_{10}$}&\textrm{13.508176}&\textrm{0.253614}&\textrm{13.944}&\textrm{13.6915}\\
\hline
\textrm{${\cal E}_{11}$}&\textrm{14.336514}&\textrm{0.221976}&\textrm{14.851}&\textrm{14.5278}\\
\hline
\textrm{${\cal E}_{12}$}&\textrm{15.089604}&\textrm{0.183207}&\textrm{15.738}&\textrm{15.3408}\\
\hline
\textrm{${\cal E}_{13}$}&\textrm{15.930288}&\textrm{0.192524}&\textrm{16.604}&\textrm{16.1327}\\
\hline
\textrm{${\cal E}_{14}$}&\textrm{16.753589}&\textrm{0.191854}&\textrm{17.454}&\textrm{16.9056}\\
\hline
\end{tabular}
\includegraphics[width=0.485\textwidth]{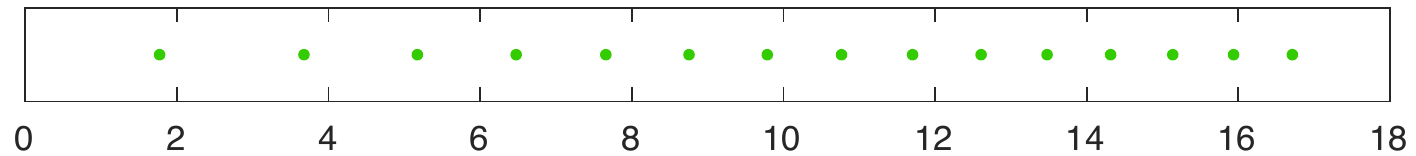}
\end{center}
\caption{The first 15 mean levels ${\cal E}_n=\langle E_n\rangle$ for the Airy model at $\hbar=1$, computed using  the Fredholm determinant method. The compute values are shown along with the (decreasing with energy) variance. The 4th column shows the first 15 levels  computed  by numerically generating 100K samples of  $100{\times}100$  random Hermitian matrices in a Gaussian ensemble and scaling. The 5th column shows the vanishing of the wavefunction~(\ref{eq:airy-wavefunction}), giving an approximation to the peaks' locations that improves as $n$ grows larger (see section~\ref{sec:return-to-gravity}).}
\label{tab:mean-levels-fredholm}
\end{table}
\endgroup

As a comparison, some ``experimental'' data mentioned above are included in the last column of the table. While the matrices are large, they are not infinite and so discrepancies begin to show up at the 1\% level for the leading energies. Moreover, since there are only 100 energies for each member of the ensemble, the first 15 reach a significant way into the ``bulk" and indeed they begin to deviate from the pure Airy case as energy is increased. A final   data set is in the last column. These are the zeros of the wavefunction of the Schr\"odinger problem  at $x{=}0$,  the Fermi surface, simply the Airy function in equation~(\ref{eq:airy-wavefunction}). Their full significance is discussed in Section~\ref{sec:return-to-gravity}, where it will be shown that they give an increasingly good approximation (as level $n$ increases)  to the  peaks' means and  (since they narrow) locations. 
Using the WKB form in equation~(\ref{eq:WKB3}), with $x{=}0$, these approximate values are given by the vanishing of the cosine, and at high  $n$ this gives a useful approximate formula:
\be
\label{eq:approximate-WKB-energies-Airy}
{\cal E}_n\simeq\left(\frac32 \left[n-\frac14\right]\pi\hbar\right)^{\frac23}\sim \left(\frac32 n\pi\hbar\right)^{\frac23}\ .
\ee
Notably, from this the density $\rho({\cal E}_n){\equiv}dn/d{\cal E}_n$ of the peaks can be computed at high $n$, and a short computation gives: 
\be
\label{eq:approximate-peak-density-Airy}
\rho({\cal E}_n)\sim\frac{{\cal E}_n^{\frac12}}{\pi\hbar}=\rhoo({\cal E}_n)\ ,
\ee
 the leading density of the Airy model derived in equation~(\ref{eq:airy-tree}). (In fact, this is all a special case of the more general formula~(\ref{eq:level-approximation}), which in turn is derived from the $x{=}0$ case of the  WKB formula~(\ref{eq:WKB2}), with~(\ref{eq:integrate-momentum}). See the next subsection for more on this.) 
 
 Result~(\ref{eq:approximate-peak-density-Airy}) is an instructive consequence of  equation~(\ref{eq:fattening})  (showing that that the peaks provide a discretization of the matrix model's continuous spectral density $\rho(E)$) combined with the additional observation that they become narrower and more closely placed at higher energy. The result is a density that approximates well the disc-level density in the large $E$ (high $n$)  limit.
%

Overall, these  results (well, their analogues when the full JT gravity matrix models are written down) will be used shortly to understand JT gravity, deep in an energy regime where its underlying quantum character is manifest. Note that the discrete spectrum $\{ {\cal E}_n\}$ goes hand-in-hand with a sensible finite entropy $S(T)$. The low temperature thermodynamics can be explicitly described in terms of populating these discrete states. This utterly fails if the continuum of states represented by $\rho(E)$ is used. (See discussion on page~\pageref{sec:annealed}.) 

\subsection{A Return to Geometry: D-Branes} 
\label{sec:return-to-gravity}
There is a smooth  spacetime interpretation of some of the key tools introduced before, in terms of spacetime boundaries (equivalently, open string sectors, in the context of the models describing minimal string world sheets). Some of the earliest  matrix model constructions of open strings in   minimal string theories were in {\it e.g.} refs.~\cite{Kazakov:1990cq,Kostov:1990nf,Minahan:1991pv,Gross:1991aj,Dalley:1992br,Johnson:1994vk}. Explorations using the conformal field theory approach were carried out in {\it e.g.,} refs.~\cite{Zamolodchikov:2001ah,Fateev:2000ik,Teschner:2000md}, with early synthesis of the two approaches set out in refs~\cite{Sen:2002nu,McGreevy:2003kb,Klebanov:2003km,McGreevy:2003ep,Martinec:2003ka,Alexandrov:2003nn,Maldacena:2004sn}. Here are some of the key elements (that we need for this paper) with some well known elements as well as some aspects that do not seem to be in the gravity/string literature. Recall that expression~(\ref{eq:full-loop-expression}) corresponds to inserting a loop of fixed length $\beta$ into the system. It is the double-scaled limit of the  quantity  $\langle W(\beta)\rangle=\langle \frac{1}{N} {\rm Tr} [{\rm e}^{\beta M}]\rangle$. The Laplace transform of this with respect to $\lambda$  is the quantity:
 \be
 \langle {\widehat W}(\lambda)\rangle= \frac{1}{N}\left\langle{\rm Tr}\left[\frac{1}{\lambda-M}\right]\right\rangle= \frac{1}{N}\frac{\partial}{\partial\lambda}\langle {\rm Tr}[\log (\lambda-M)]\rangle \ ,
 \ee the matrix resolvent.  
Because of the factor $\e^{-\lambda\beta}$, from this perspective  $\lambda$ measures the cost of making a loop of length $\beta$ --- It is a ``boundary cosmological constant''. To compactly study the insertion of loops of all lengths, it is useful to construct the exponentiated version of the loop insertion, which is simply $\langle {\rm det}[\lambda-M] \rangle$. From a statistical perspective, this is in fact the ensemble average of the characteristic polynomial of a  matrix $M$ of the random ensemble, a very natural quantity to compute. Note that it is this is a polynomial with leading term at order  $\lambda^N$, with  unit coefficient  (there's a division by~$\tilde Z$ in computing the insertion). Well, it is an  elegant old result (see {\it e.g.,} an appendix of ref.~\cite{Bessis:1980ss}) that the average characteristic polynomial of the $N{\times}N$ matrix ensemble is none other than the $N$th orthogonal polynomial, $P_N(\lambda)$. Recall that in the double-scaling limit, $x{=}\mu$ is equivalent to orthogonal polynomial index $n$  landing on~$N{-}1$.  At naive large~$N$ this  is indistinguishable from $N$, but it makes a difference in the complete scaling limit, as will become clear shortly. Recalling the factors that scale nicely together, the natural object representing the insertion of loops of arbitrary length is the largest orthogonal polynomial involved in defining the physics, which becomes 
the wavefunction of the quantum mechanics~(\ref{eq:eigenstates}) evaluated at the Fermi level: 
\be
\frac{1}{\sqrt{h_N}}\langle \e^{-NV(\lambda)/2}{\rm det}[\lambda-M]\rangle\,\, \stackrel{DSL}{\longrightarrow}\,\, \psi(E,\mu)   \ .
\ee
Here $E$ is now the  boundary cosmological constant (loops of length $\ell$ come weighted by $\e^{-\ell E}$). 

A place where the 2D spacetimes can form boundaries of arbitrary length is simply a D-brane, and such objects---a natural component in a generic string theory--are seen to also emerge naturally in the formalism here. Naively,  they are to be interpreted as places where the world living on the 2D universe can come to an end, prompting the moniker ``end-of-the-world'' branes. Since they are also associated with a particular (scaled) eigenvalue $E$ (which simply sets their boundary cosmological constant), they have also been called ``eigenbranes'' in the literature~\cite{Blommaert:2019wfy}. In a fully string theory context, they are also known as ``FZZT'' branes.\footnote{This multiplicity of names make them, perhaps appropriately given their legendary magical properties, the Gandalf of this adventure story.} To pick one name, since  ``D-brane'' has worked well for 30 years,  that will be used here.

There are many notable features of the wave function $\psi(\mu,E)$ and its dual D-brane identity. Some, but not all, it seems, have been noted in the literature. Generically, it begins at small $E$ with a smooth positive part that transitions to  oscillatory behaviour that continues off to infinity. (For example, for the Airy/Gaussian prototype, it is just the Airy function, $\psi(E,0) = {\rm Ai}(-\hbar^{-\frac23}E)$.) Recall that in the unscaled model the determinant operator started life as   the average characteristic equation of the matrices, which means that its $N$ zeros mark the mean location of the eigenvalues of the $N{\times}N$ matrices of the ensemble. Something of this survives to the scaling limit of the matrix model physics, but parts are lost in translation. The matrix model is built from the first $N$ of the polynomials, starting with the zeroth order one, as can be seen from equation~(\ref{eq:vandermonde-polynomials}) and several that follow.  So the top wavefunction that survives to the limit is really built from the order $(N{-}1)$ orthogonal polynomial, and therefore its zeros should be located slightly more into the bulk than the true mean locations, (since the Dyson gas spreads out with $N$).  So the zeroes of $\psi(E,\mu)$  will give only an {\it approximate} correspondence to the mean eigenvalues, which makes sense since (as now hopefully clear from  last section) the precise information about the individual eigenvalues is extracted from data contained in the Fredholm determinant, built from the kernel $K(E,E^\prime)$. (On the other hand, the discrepancy becomes smaller as $E$ increases, and this will be useful later.) Recall from equation~(\ref{eq:edge-kernel}) that the kernel is in turn built from  {\it all} the $\psi(E,x)$ that spread across the whole Fermi sea $(-\infty\leq x\leq\mu)$, folded together with with $\psi(E^\prime,x)$ to make  and the Fredholm determinant built from it. Alternatively, $K(E,E^\prime)$ it is built from $\psi(E,\mu)$ {\it and} its derivative in a non-local (in~$E$) manner.  (In fact, the zeros of $\psi^\prime(E,\mu)$ are equally interesting
in this regard.) 

This is all consistent with the plot of figure~\ref{fig:D-brane-wavefunction}, where $\psi(E,\mu)$ is overlaid upon the density $\rho(E)$ for the prototype Airy case.
\begin{figure}[t]
\centering
\includegraphics[width=0.42\textwidth]{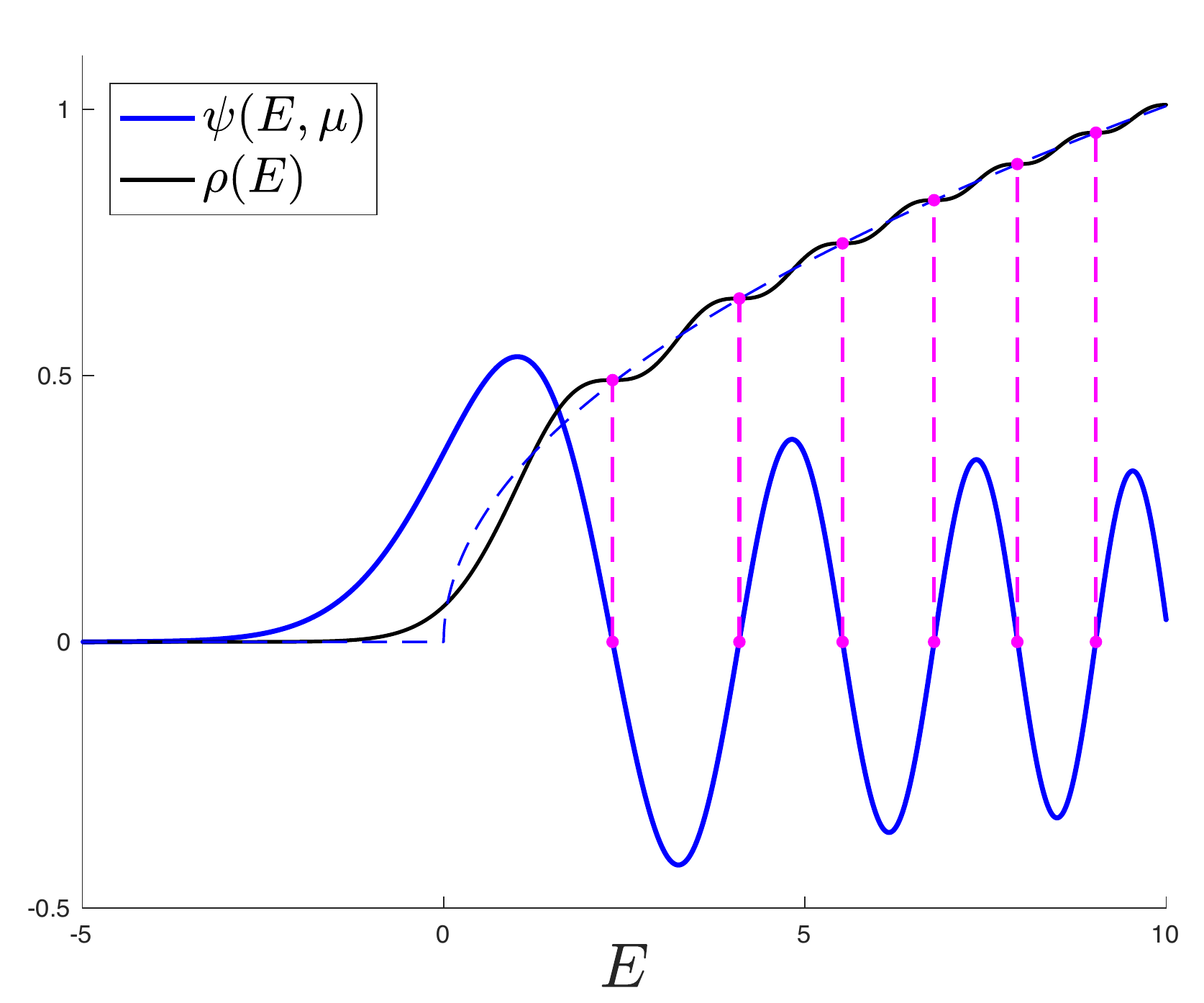}
\caption{\label{fig:D-brane-wavefunction} The D-brane partition function $\psi(E,\mu)$ plotted alongside the spectral density $\rho(E)$ for the Airy model, at $\hbar=1$. The zeros give a first draft of the mean values of the energy levels. They line up with the undulations of the spectral density.  The complete information about the locations of the mean energy levels requires the  Fredholm determinant~(\ref{eq:Airy-Fredholm}).} 
\end{figure}
The zeros correlate well with the dips or inflection points of the undulations, which are only an approximation to the location of the means of the individual underlying peaks (see figure~\ref{fig:Airy_fredholm_3d_and_peaks}).

This observation about $\psi(\mu,E)$ is an important clue about a deeper question. Because of its characteristic polynomial origins, its zeros {\it collectively} have meaning as an average spectrum of a matrix (albeit the size $(N-1)$ matrix). Their correlation with the peaks $p(n;E)$ lends support to the idea that their means represent the average spectrum of the full size $N$ system. So indeed, the matrix model does indeed strongly point to a meaningful specific spectrum, the $\{{\cal E}_n\}$, which as discussed in the Introduction is proposed as  the spectrum of the holographic dual to JT gravity. 

As already stated, since the peaks become narrower and more closely spaced at large $n$, the zeros of the wavefunction become a good approximation to the peaks' locations (whether it be the mean of the peaks or similar) and the WKB formula~(\ref{eq:WKB2}) at $x=\mu=0$, together with equation~(\ref{eq:integrate-momentum}), yields the result~(\ref{eq:level-approximation}), from which it follows readily that the density of the peaks asymptotes to the disc-level spectral density $\rhoo({\cal E}_n)$. This shows how (for JT gravity and for the wider class of gravity models the matrix model techniques can capture) the leading (continuous) perturbative  results for  the spectral  density  can be understood as an approximation to a discrete density that can arise from some sensible quantum theory.

As $\hbar$ becomes small, the WKB form~(\ref{eq:WKB}) can be used, and this is the limit where  $\psi(E,\mu)$ has been most used in the literature. It connects nicely to the observations above in the following way: For the form~(\ref{eq:WKB2}), increasingly smaller~$\hbar$  results in faster oscillations, and so the zeros of the wavefunction become increasingly closely spaced, until at $\hbar\to0$ they form a dense covering of the positive~$E$ line, as they should. This is the recovery of the smooth Dyson gas.  Meanwhile, for the situation where $E<0$, the behaviour is exponential, and 
\bea
\label{eq:semi-classical-np-2}
\psi_{\rm sc}&\simeq&  \frac{1}{(-E)^{\frac14}}\exp\left(-V_{\rm eff}(E)\right)\ ,
\eea
where $V_{\rm eff}(E)=2\pi\hbar^{-1}\int_0^{-E}\rho_0(-E^\prime)dE^\prime$ is the leading (saddle point) effective potential for one eigenvalue in the $E<0$ regime. 
To see that this is the case, return to the saddle-point large $N$ analysis that derived the Dyson gas droplet solution.  Once the saddle-point condition is met, the total energy ${\tilde E}_{\rm sad}$ of the configuration is in fact (minus) the order $N^2$ expression in equation~(\ref{eq:dyson-action}), since now ${\tilde Z}{=}\exp(-{\tilde E}_{\rm sad})$. 
Now the trick is to move one eigenvalue from position $\lambda_{\rm I}$ to position $\lambda_{\rm F}$, with result: 
\be
\Delta{\tilde\rho}_0(\lambda) = \frac{1}{N}\left[\delta(\lambda-\lambda_{\rm F})-\delta(\lambda-\lambda_{\rm I})\right]\ .
\ee
Inserting this into the saddle energy gives:
\begin{widetext}
\bea
V_{\rm eff} = -\Delta{\tilde E}_{\rm sad}&=&
-N\left\{V(\lambda_{\rm F}) -V(\lambda_{\rm I}) -2\int d\mu{\tilde\rho}_0(\mu)\left[\log(\lambda_{\rm F}-\mu)-\log(\lambda_{\rm I}-\mu)\right] \right\} \nonumber\\
&=&-N\int_{\lambda_{\rm I}}^{\lambda_{\rm F}} \left(V^\prime(\lambda)-2 F(\lambda)\right)d\lambda = -2\pi N \int_{\lambda_{\rm I}}^{\lambda_{\rm F}}\!\! {\rm Im}\,{\tilde\rho}_0(\lambda) d\lambda \nonumber\\
&\Rightarrow& -\frac{2\pi}{\hbar} \int_{E_{\rm I}}^{E_{\rm F}}\!\! {\rm Im}\,\rho_0(E) dE\ ,
\eea
where in the middle line, equations~(\ref{eq:resolvent-saddle}) and~(\ref{eq:tree-level-rho}) were used, and the final result after the double-scaling limit has been indicated. 
\end{widetext}
Since ${\tilde\rho}_0$ is real  on the cut (where the Dyson gas lives), this expression vanishes if the endpoints $\lambda_{\rm I,F}$ are there, saying that there's no cost to moving an eigenvalue around within the saddle solution. The expression really comes into its own when seeing what the cost is for moving an eigenvalue out of the saddle.
This is a useful partial diagnostic of the stability of a model~\cite{David:1990ge,David:1990sk}. 

For example, for the Airy case, where $u_0(x){=}{-}x$ and $\mu{=}0$, $V_{\rm eff}(E)=\frac{2\pi}{\hbar}\cdot\frac23 (-E)^\frac32$,  a potential that rises toward more negative $E$, producing a force that pushes eigenvalues to positive $E$, which is where the classical result puts the Dyson gas.  Actually, for simple Hermitian matrix models, odd $k$ are distinguished from even $k$ by the fact that the former have this rising behaviour while for the latter the effective potential ultimately falls, signalling an instability coming from the eigenvalues wanting to tunnel away from the positive~$E$ regime. There are ways of avoiding this problem and providing stable definitions, however. Such matters of stability  have been discussed thoroughly in the JT gravity context starting with the identification of the issue in ref.~\cite{Saad:2019lba}, a solution in ref~\cite{Johnson:2019eik}, and a recent detailed discussion of  the matter (with more solutions) in ref.~\cite{Johnson:2021tnl}, and so will not be pursued here.

It is clear now that  objects  naturally built from $\psi(E,\mu)$ (or from  the whole family, $\psi(E,x)$) ought to have a D-brane interpretation. The ubiquitous kernel $K(E,E^\prime)$, defined in equation~(\ref{eq:edge-kernel}), is such an object. It has already been discussed from the statistical point of view as a two-point correlator from which many useful quantities can be built.  Now it has a new interpretation as a sort of composite D-brane probe that is well-suited to detecting the ``location'' of individual energies.\footnote{The equivalent construction in terms of the two more basic  functions, $\psi(E,x)$ and $\psi(E^\prime,x)$, which each contain the full information about the Fermi sea construction of the underlying many-body,  resembles a process whereby they are ``zipped" together to construct $K(E,E^\prime)$. It is somewhat analogous to making a matrix-valued Yang-Mills quantity by tensoring together the Chan-Paton indices that label a stack of D-branes, in more familiar string theory settings.}

\subsection{Free energy}
\label{sec-free-energy}
As pointed out in ref.~\cite{Johnson:2021zuo}, once the spectrum of the random matrix model ({\it i.e.,} the probability distribution of each energy level)  is known, there are several things that can be readily computed using it, and among those are the free energy of the system. In fact with the information extracted so far, {\it four} distinct free-energy-like quantities can be computed and compared. The matrix model free energy quantities computed by averaging over the ensemble are the most obvious: The annealed free energy:
\be
\label{eq:annealed-free-energy}
F_A(T) = -T\log\langle Z(T)\rangle
\ee and the quenched quantity: 
\be
\label{eq:quenched-free-energy}  
F_Q(T) =-T\langle \log Z(T)\rangle\ ,
\ee where the angle brackets denote the ensemble average over matrices.  These can be done {\it directly} on the matrix model if one can generate the actual matrices and sample them. For the kinds of application needed here, the matrix ensemble has highly non-trivial probability weight (the raw potential $V(M)$), that is hard to simulate on the computer directly. Gaussians are special.   The exciting point here is that having information about the  individual statistics of each level, as afforded by the Fredholm analysis,  allows for the ensemble averaging to be done another way, by simply reverse engineering the probability information obtained for each level in order to generate samples that can then be analyzed. This is in principle possible (and was done with some success in ref.~\cite{Johnson:2021zuo}) although somewhat clumsy to do well, partly because care must still be taken to make sure that the ensemble samples created are still representative. Nevertheless,  direct methods for computing $F_Q(T)$ are desirable since they avoid using replica methods, and all the potential computational and interpretational problems that they can entail. The bottom line is that there should be no need for replicas to understand what is going on if there is firm knowledge of the spectrum, which is the case here. There have been some computations directly on the matrix model that have analytically extracted some results in the low temperature limit in certain toy models~\cite{Okuyama:2020mhl,Janssen:2021mek,Okuyama:2021pkf}, but the direct sampling approach~\cite{Johnson:2021rsh,Johnson:2021zuo} has so far been best suited to analyzing the behaviour at arbitrary~$T$.

The point of view of this  paper brings forth a third quantity  that can be computed quite simply: 
\be
\label{eq:mean-free-energy}
F_m(T) =-T\log \left(\sum_{n=0}^\infty \e^{-{\cal E}_n/T}\right)\ ,
\ee which is the free energy of the mean spectrum. 
Following through the reasoning for what the difference is between the quenched and annealed free energies already described in the Introduction (there is no need to repeat it here), this ought to not deviate too far from  the quenched free energy of the ensemble.

The fourth quantity is simply the  free energy computed using the logarithm of the partition function $\langle Z(T)\rangle$ made from computing the Laplace transform of the {\it full} $\rho(E)$, according to equation~(\ref{eq:partition-function}). This should coincide with the $F_A(T)$ computed from the ensemble method, since this really is what is meant by the matrix model's definition of $\langle Z(\beta)\rangle$ at the outset. However it is sometimes used as a useful (partial) check on the sampling methods, serving as a first flag of whether the samples used are (at least in aggregate) indeed representative, bolstering confidence in results obtained for $F_Q(T)$. 

It is perhaps useful to explain a little more about how to compute $F_Q(T)$ and $F_A(T)$ by sampling from the Fredholm-obtained data. For the purposes of determining contributions to the partition sum,  what is needed to get good results for them is the statistics of the energy levels over a large number of samples generated by knowing the individual probability distributions for each level.  There is no off-the-shelf operation for using a computer to randomly generate numbers with a  probability distribution that is not uniform or Gaussian. On the other hand, a  uniformly distributed  random number, $p$, between 0 and 1 can be readily generated. For a given level~$n$, recall that the Fredholm techniques give the cumulative probability distribution function $c(n;s)$. The probability of level $n$ is then obtained by evaluating the inverse of this function at $p$.  So, samples that have, in aggregate, similar statistical properties to the {\it actual} matrix model (after scaling) can  be generated this way, but it relies on being able to compute the CDFs up to high enough $n$ to capture the regime where the quenched and annealed free energies depart from each other.\footnote{Note that these samples are not actually literal representative  samples, since they can include samples for which, e.g. the randomly generated  energy for the $i$th level is higher in value than the randomly generated $(i+1)$th energy level. This should not matter a great deal for the purposes of the computing the partition function however, which does not track ordering of energies, and only cares about how frequently they appear in the ensemble. For large number of samples, the statistical properties of the samples generated this way should yield similar   results as samples that also included ordering information. This is all borne out by the results.} It becomes numerically difficult, especially for the gravity related models, to extract accurate CDFs from Fredholm much after $n$ goes above 10--15 (as will be shown), but then two other facts can be used. The first is that the variance (and skewness) of the PDFs reduce as~$n$ increases. This means that they can be reasonably well approximated,  for larger $n$,  by Gaussians (of course using ${\rm erf}(s)$ as the CDF). This presents the problem of how to locate the positions and  standard deviations of such placeholder Gaussians. This is solved by the observation that as $n$ increases, the WKB form of the wavefunction $\psi(E,\mu)$ gives a rather good estimate of the position of the $n$th peak. In fact it overshoots, while the zeros of its derivative undershoots. So together they give a rough estimate of the mean and standard deviation of the Gaussians. After all this is input, it can  simply be done  for as many levels as desired, and then performing the partition sum. Computing the logarithm of the average of the sum or the average of the log of the sum is then readily done. In examples studied in this paper, typically 5000--10000 samples were constructed, and sometimes up to levels in the low hundreds (although for the temperatures considered the physics had already settled down for much smaller levels than that).

The four quantities described were computed for the Airy model, where the samples of the last paragraph were generated purely from the Fredholm data, up to 15 energy levels. 
Moreover in this case,  direct computation using genuine matrix samples can also be done as a test! That gives two sets of three  quantities (computed different ways) to be compared: $F_Q(T)$ and $F_A(T)$  made from samples generated from Fredholm data, or made from directly sampling (5-10K) actual matrices, $F_m(T)$ made from the mean spectrum, and $F_A(T)=-T \log(\langle Z(T)\rangle)$. For the latter, while equation~(\ref{eq:airy-partition-function}) can be used to get the exact result, instead the Laplace transform of $\rho$ was taken between $E=-3$ and $17$, the approximate value of the highest level, to match the cutoff on the energies in the sampling.  These are all shown in figure~\ref{fig:Airy_free_energy}, and the results are interesting. Happily, both of the  direct sampling methods of computing $F_A(T)$ match well with the result computed using the matrix model's $\rho(E)$. (There is some numerical variance at the lowest temperatures, to be expected from the fact that configurations with (mostly rare) negative energies are contributing to a diverging annealed quantity there.)  At higher temperatures the direct sampling $F_A(T)$ is slightly above the correct amount due to the matrices being of finite size producing a slight  falloff in the energy population (see figure~\ref{fig:toy-densities-2}).  At low temperatures all the $F_Q(T)$  coincide very well (and they must agree at $T=0$, where the result is simply the average energy of the ground state.) 
\begin{figure}[t]
\centering
\includegraphics[width=0.48\textwidth]{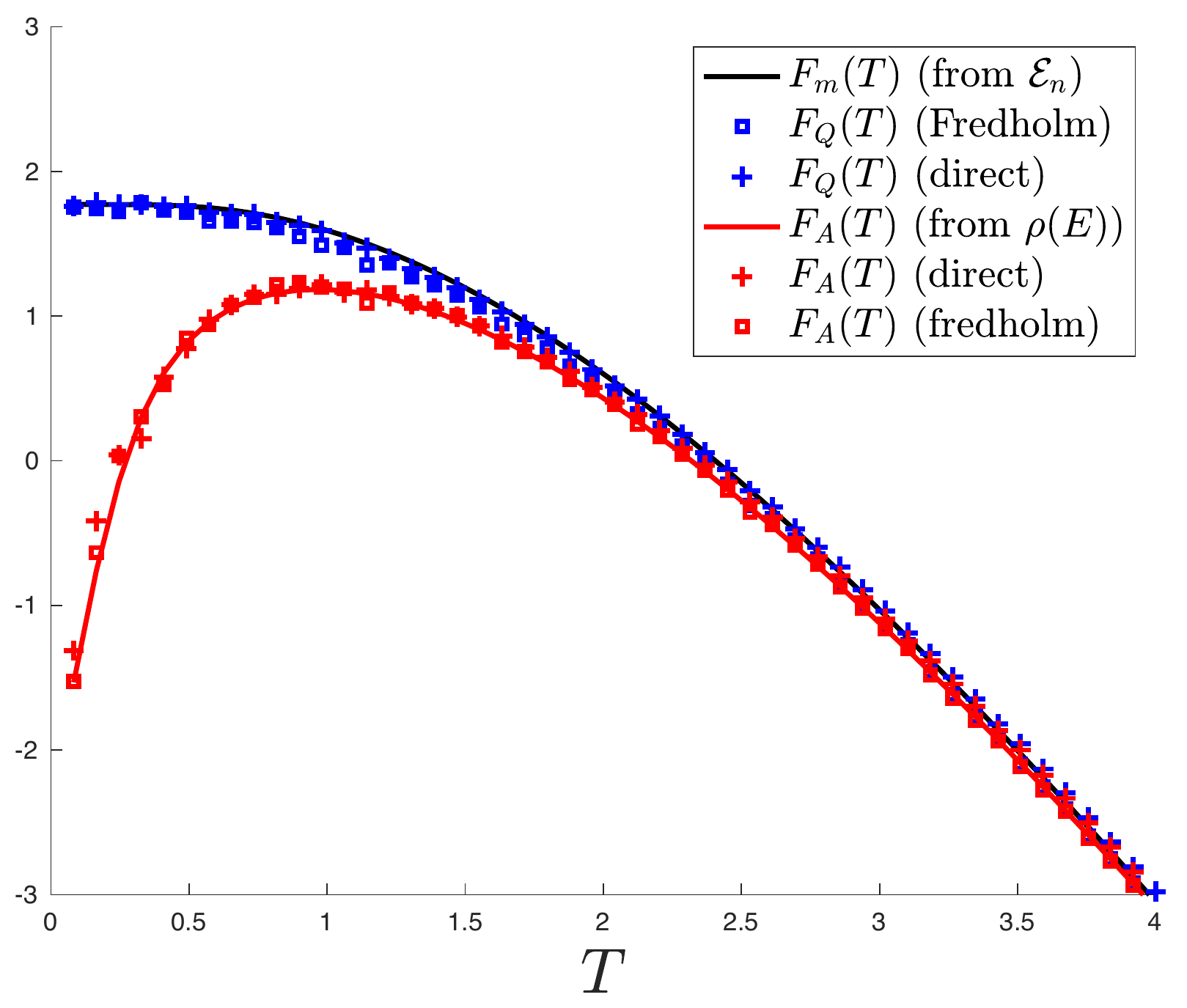}
\caption{\label{fig:Airy_free_energy} The Airy free energy, computed from the direct spectrum ${\cal E}_n$, compared to computing the  quenched quantity $F_Q(T)$  in the matrix model by two methods. The truncation level is $n=15$.  All three  cases become $\langle E_0\rangle\simeq 1.77$, and overall show good agreement. Also shown is the annealed quantity $F_A(T)$,computed three different ways. See text. }
\end{figure}
Notably, for some intermediate temperatures, $F_Q(T)$ computed the Fredholm way falls slightly below its counterpart from direct sampling, (the correct answer), while $F_m(T)$ follows well the direct result, but slightly above. Toward higher $T$, the quenched energies all agree very well, and merge with the annealed free energies as they should. Slight discrepancies among them can be fully accounted for by finite size effects, and the slightly differing methods  of implementing truncation to 15 levels in each case. This was confirmed by running versions of this same computation where the truncation level was much higher.

Overall, the discrepancies are slight enough to support the central idea: The quenched free energy using  the matrix ensemble is in strong agreement with the free energy computed using  the average spectrum. They needn't have matched exactly to support the idea, and it is already encouraging that they do so well for this exact model, where there is a direct computation to compare to using actual matrix samples. This gives confidence in the methods used, for JT applications (where there will not be matrix samples to compare to).

\bigskip

\subsection{Spectral form factor}
\label{sec:spectral-form-factor-airy}
Another interesting and important quantity to reconsider is the spectral form factor, a two-point function of the partition function whose late time behaviour is a useful diagnostic of the properties of the spectrum of the theory:
\be
\label{eq:sff-single}
Z(\beta+it)Z(\beta-it)=\sum_{j,k} \e^{-\beta(E_j + E_k)}\e^{it(E_j-E_k)}\ ,
\ee
for a given energy spectrum $\{ E_k\}$.
In particular, at late times, fluctuations begin to dominate the quantity, the nature of which  depend upon the  details of the low energy spectrum. The random matrix model readily yields the ensemble average of this quantity, and the result is naturally decomposed into the sum of a disconnected piece and a connected piece:
\bea
\label{eq:sff-averaged}
&&\langle Z(\beta+it)Z(\beta-it)\rangle =\\
&&\hskip 1.0cm \langle Z(\beta+it)\rangle\langle Z(\beta-it)\rangle+ \langle Z(\beta+it)Z(\beta-it)\rangle_{\rm c}\ . \nonumber
\eea This can all be written in terms of the free fermion language quite readily. The disconnected piece is made from two copies of equation~(\ref{eq:full-loop-expression}) while the connected piece is~\cite{Banks:1990df,Ginsparg:1993is} (remembering ${\cal P}$'s definition in equation (\ref{eq:define-projector})):
\begin{eqnarray}
\label{eq:correlator-connected}
&&\langle Z(\beta) Z(\beta^\prime)\rangle = {\rm Tr}(e^{-\beta{\cal H}}(1-{\cal P}) e^{-\beta^\prime{\cal H}}{\cal P}) \\
&&= {\rm Tr}(e^{-(\beta+\beta^\prime){\cal H}}) - {\rm Tr}(e^{-\beta{\cal H}}{\cal P} e^{-\beta^\prime{\cal H}}{\cal P})\nonumber \\ 
&&=  Z(\beta{+}\beta^\prime) -\!\! \int \!\!dE\!\int \!\!dE^\prime K(E,E^\prime) K^*\!(E^\prime,E)\e^{-\beta E-\beta^\prime E^\prime}
\ ,\nonumber
\end{eqnarray}
 which becomes, on setting $\beta\to\beta+it$ and $\beta^\prime=\beta-it$:
 \begin{widetext}
\bea
\label{eq:full-two-loops-expression}
\langle Z(\beta+it)Z(\beta-it)\rangle_{\rm c}  =
Z(2\beta) -\!\! \int \!\!dE\!\int \!\!dE^\prime K(E,E^\prime) K^*\!(E^\prime,E)\e^{-\beta (E+E^\prime)-it(E- E^\prime)}\ ,
\eea
which reveals a time dependent piece that  in fact goes to zero at large $t$, leaving a  time independent positive term, of magnitude $Z(2\beta)$, which is the ``plateau" to which the averaged quantity saturates. (This is also directly visible in the defining quantity~(\ref{eq:sff-single}) by looking at  the sum along the diagonal, {\it i.e.} $E_j=E_k$.)
In  the special case of Airy, the spectral form factor is computable in closed form, since the full spectral density can be Laplace transformed to give $Z(\beta)$ in equation~(\ref{eq:airy-partition-function}), and 
  implementing equation~(\ref{eq:correlator-connected}) yields the connected piece to be:
 \bea
 \label{eq:connected-airy}
 \langle Z(\beta)\rangle\langle Z(\beta^\prime)\rangle = \frac{e^{\frac{\hbar^2}{12}(\beta^3+{\beta^\prime}^3)}}{4\pi\hbar^2(\beta\beta^\prime)^{3/2}}\ ,\ \; \quad{\rm and}\quad
 \langle Z(\beta)Z(\beta^\prime)\rangle_{\rm c} =
 \frac{\e^{\frac{\hbar^2}{12}(\beta+\beta^\prime)^3}}{2\pi^{1/2}\hbar(\beta+\beta^\prime)^{3/2}}{\rm Erf}\left(\frac{1}{2}\hbar\sqrt{\beta\beta^\prime(\beta+\beta^\prime)}\right)\ ,
  \eea
  and so
  \bea
  \label{eq:sff-exact-airy}
  \langle Z(\beta)Z(\beta^\prime)\rangle = \frac{\e^{-\frac{1}{2}\hbar^2 \beta  t^{2}+\frac{1}{6} \hbar^2\beta^{3}}}{4\pi \hbar^2\left(\beta^{2}+t^{2}\right)^{\frac{3}{2}} } + \frac{\e^{\frac{2 \hbar^2\beta^3}{3}}  }{4 \sqrt{2\pi}\, \hbar\beta^{\frac{3}{2}}} \mathrm{Erf}\left(\hbar\sqrt{\frac{{\beta  \left(\beta^{2}+t^{2}\right)}}2}\right)\ .
  \eea
\end{widetext} Notice that at large $t$ the first term dies away to zero, and the last term indeed becomes $Z(2\beta) {=}\e^{\frac23\hbar^2\beta^3}/4\sqrt{2\pi}\hbar\beta^{3/2}$.

In general, the connected term, rather  famously,  can be interpreted  geometrically in terms of a spacetime wormhole {\it when the matrix model has a gravity dual}. That this is a smooth geometrical object is at the heart of the current discussions in the literature about gravity as an ensemble, so it is interesting to examine the matter here in the light of what has been uncovered.

It is important to understand that there are two separate issues at play, which are often conflated. One issues is that the average behaviour over the  ensemble of spectral  form factors for all samples  captures has the effect of smoothing out the late time fluctuations.  The point is that a single sample has wildly fluctuating behaviour, and every sample fluctuates differently. Summing over them smooths it all out. The precise nature of what emerges from the smoothing depends upon the model ({\it e.g.,} class of ensemble) under consideration.  Of course, a most distinguished single spectrum  is the set $\{ {\cal E}_n\}$, given by the mean of the Fredholm peaks uncovered here, the average spectrum. It is interesting to place this alongside the result~(\ref{eq:sff-exact-airy}) for the matrix model ensemble average in the case of Airy, and the result is shown in figure~\ref{fig:Airy_spectral_form_factor_compare}. %
\begin{figure}[b]
\centering
\includegraphics[width=0.48\textwidth]{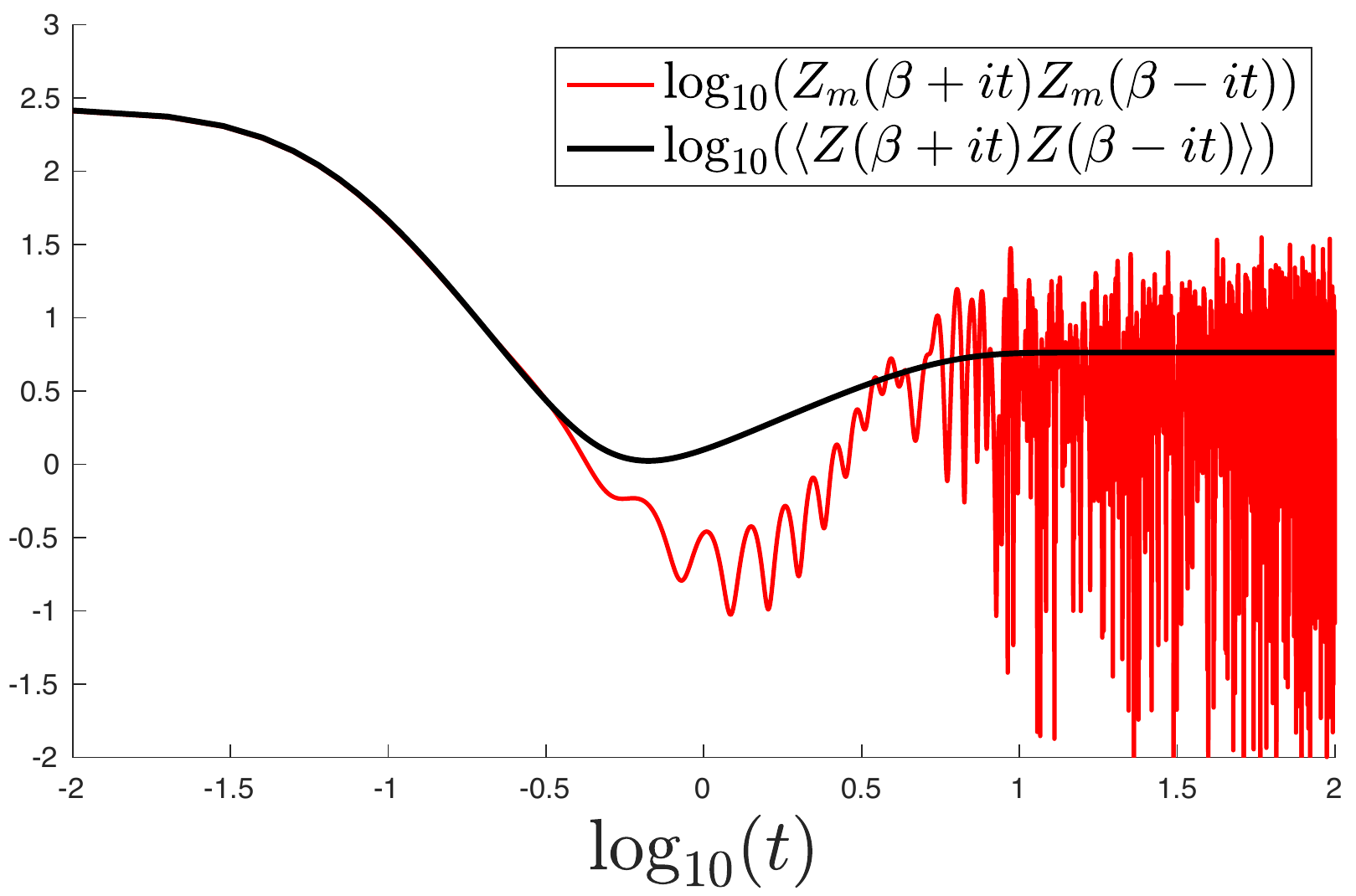}
\caption{\label{fig:Airy_spectral_form_factor_compare} The spectral form factor for the mean spectrum $\{ {\cal E}_n\}$ (red, jagged) plotted against the ensemble averaged quantity, for the Airy model (black, smooth). The two curves follow each other closed at early times before the red curve begins fluctuations. Here $\beta{=}\frac{1}{15}$, and $\hbar{=}1.$}
\end{figure}

So far this smoothing from averaging spectra has {\it nothing to do with gravity}. In fact, as already mentioned, Airy has no bulk gravity dual. So this makes it a useful testbed for separating out ideas of ensemble averaging, smoothing, and  gravity.  A separate issue  from the smoothing of the spectral form factor's erratic behaviour is whether a given ensemble has as gravity dual or not, (meaning a regime where the results can be written as a path integral over non-trivial smooth surfaces of different topologies). The matrix models for JT gravity and various variants of it  {\it do} have such a bulk gravity dual language. It is only in such cases that the extra step of calling the connected term a gravity wormhole is then meaningful. In such a case then there is a clear meaning to ascribing  Euclidean wormholes in gravity to an ensemble average.

One can go in the other direction and ask if there is any sense in ascribing a smooth wormhole geometry to the spectral form factor involving a single copy of the gravity spectrum. Are there circumstances under which a wormhole is a good description. The answer seems clear. It is never a good description of a single copy. The spectral form factor is a quantity that by definition has erratic behaviour at late enough time due to the low-lying discrete spectrum. In a model with a gravity counterpart, (such as JT gravity), the quantity is meaningful, but it is a separate question as to whether it has a description in terms of a path integral over a smooth Euclidean manifold (the wormhole). A careful point of this paper is that in the Euclidean theory, the matrix model achieves a smooth description by filling in the gaps between the  discrete microstates, doing so by averaging over the ensemble from which the dual spectrum is drawn.  In the Lorentzian picture where (it is proposed) there is a definite dual spectrum, the microstates are fully understood--they have a life as the D-branes associated to the $\{ {\cal E}_n\}$ for example--and the spectral form factor can be written in terms of them as done above. The ensemble averaging which is in the domain of the Euclidean description) effectively borrows D-branes from across all the members of the ensemble--smearing them out from their discrete positions--to give a smooth answer. The price one pays is the loss of the erratic behaviour of the spectral form factor that encodes a particular spectrum. Instead there is the smooth quantity, which gives useful universal information about the class of theory one is working in, but not the particular theory. That's of course ok.

The broad lesson is that wormholes (and other non-trivial topology) in gravity theories with a dual do imply that an ensemble average is taking place. But it is key to note that this is  the domain of the Euclidean description, which averages over microstates to produce smooth geometries.  Clearly, in the framework advocated in this paper,  the converse is not necessarily true: Gravity is {\it not fundamentally required} to be an ensemble average of anything. The averaging arises, when computing certain quantities in the Euclidean domain. In the Lorentzian picture, where a Hamiltonian is most meaningful, there is no ensemble of Hamiltonians defining the theory, and hence a single discrete spectrum.\footnote{There are some related discussions of the role of wormholes and ensembles in the Liouville theory context in {\it e.g.,} refs.~\cite{Betzios:2020nry,Betzios:2021fnm}. Reflections on how a unique boundary dual can be reconciled with the wormhole picture are also presented there.}

\subsection{Complex Matrices and Bessel Models}
\label{sec:bessel-models}
Before moving to JT gravity and supergravity, in this section an instructive side trip will be taken to study an entirely different family of exactly solvable toy models.  They arise by 
considering  $M$ as a Gaussian randomly generated {\it complex} matrix random matrix, but the combination of interest is  $H{=}MM^\dagger$. The eigenvalues of these (``Wishart''~\cite{10.2307/2331939} form) matrices  $H$ are manifestly positive. The system can be usefully thought of as rather like the   prototype Hermitian matrix model, but with a ``wall''   placed at $E{=}0$, stopping the eigenvalues  from flowing to negative values.  Again, in this Gaussian case it is straightforward and instructive to simulate these case, and it can be done with a simple  modification of the lines of code suggested in footnote~\ref{fn:wigner}.\footnote{\label{fn:marchenko} In \textsf{MATLAB}, generate complex {\tt C} as before but now form {\tt H=C*ctranspose(C)}, and study its eigenvalues over many samples, histogramming as before.}    This will yield a model system with label~$\Gamma{=}0$. Other integer~$\Gamma$ can easily be obtained by a further modification: The complex matrix~$M$ does not need to be square, but instead of size $N{\times}(N+\Gamma)$, or $(N+\Gamma){\times}N$. (This can be implemented by starting  with a random~$M$ that is $(N+\Gamma){\times}(N+\Gamma)$, and then  deleting either~$\Gamma$ rows or~$\Gamma$ columns before forming $H$ with give a system to be given label $\Gamma$. Notice that when rows are chosen, $H$ will have $\Gamma$ repeated zero eigenvalues. In fact, these have an interpretation as additional background D-branes in some models~\cite{Dalley:1992br,Johnson:1994vk,Johnson:2021tnl}, and R-R punctures in others~\cite{Stanford:2019vob}.) 

In the Altland-Zirnbauer~\cite{Altland:1997zz} classification of random matrix ensembles, this is the  $(\boldsymbol{\alpha},\boldsymbol{\beta}){=}(2\Gamma+1,2)$ system, where the case with rows deleted has $\Gamma$ negative and the case with columns has~$\Gamma$ positive. In fact, $\Gamma$ can also be half-integer in this classification scheme, but that isn't accessible here in terms of counting rows and columns of  matrices. Much can be learned  from just integer~$\Gamma$, and indeed ref.~\cite{Johnson:2021rsh} explored many new properties of the quenched free energy of such models by direct sampling, seeing how the resulting physics connects to many well known results in the random matrix literature. Here, the cases $\Gamma=0$ and $\Gamma=\pm\frac12$ will be briefly featured, as they contain many features that will be helpful for studying JT supergravity. 

The first of these cases is readily amenable to simulating with actual matrices. As before, large numbers of samples  of the case $N{=}100$ can be readily generated, and a pattern emerges. It  is the analogue of Wigner's semi-circle law, in this case the Mar\v{c}enko-Pastur law~\cite{zbMATH03259601}, a shape with the same~$\lambda^\frac12$ behaviour seen earlier at one end (away from the wall), but there is instead a~$\lambda^{-\frac12}$ divergence at the ``wall'' end. The end is at $4N$ this time, instead of $2\sqrt{N}$ for the Wigner case ($H$ being instead the product of two matrices). This means that the typical spacing between energy levels is of order 1, and so some of the scalings to follow will reflect this difference. Rescaling  to $\lambda^\prime{=}\lambda/4N$ and henceforh dropping the prime, the unit-normalized density is now 
\be
{\tilde\rho}(\lambda){=}\frac{\sqrt{2-\lambda}}{\sqrt{\lambda}\pi}\ .
\ee
It is shown in figure~\ref{fig:marchenko-pastur}.
\begin{figure}[t]
\centering
\includegraphics[width=0.49\textwidth]{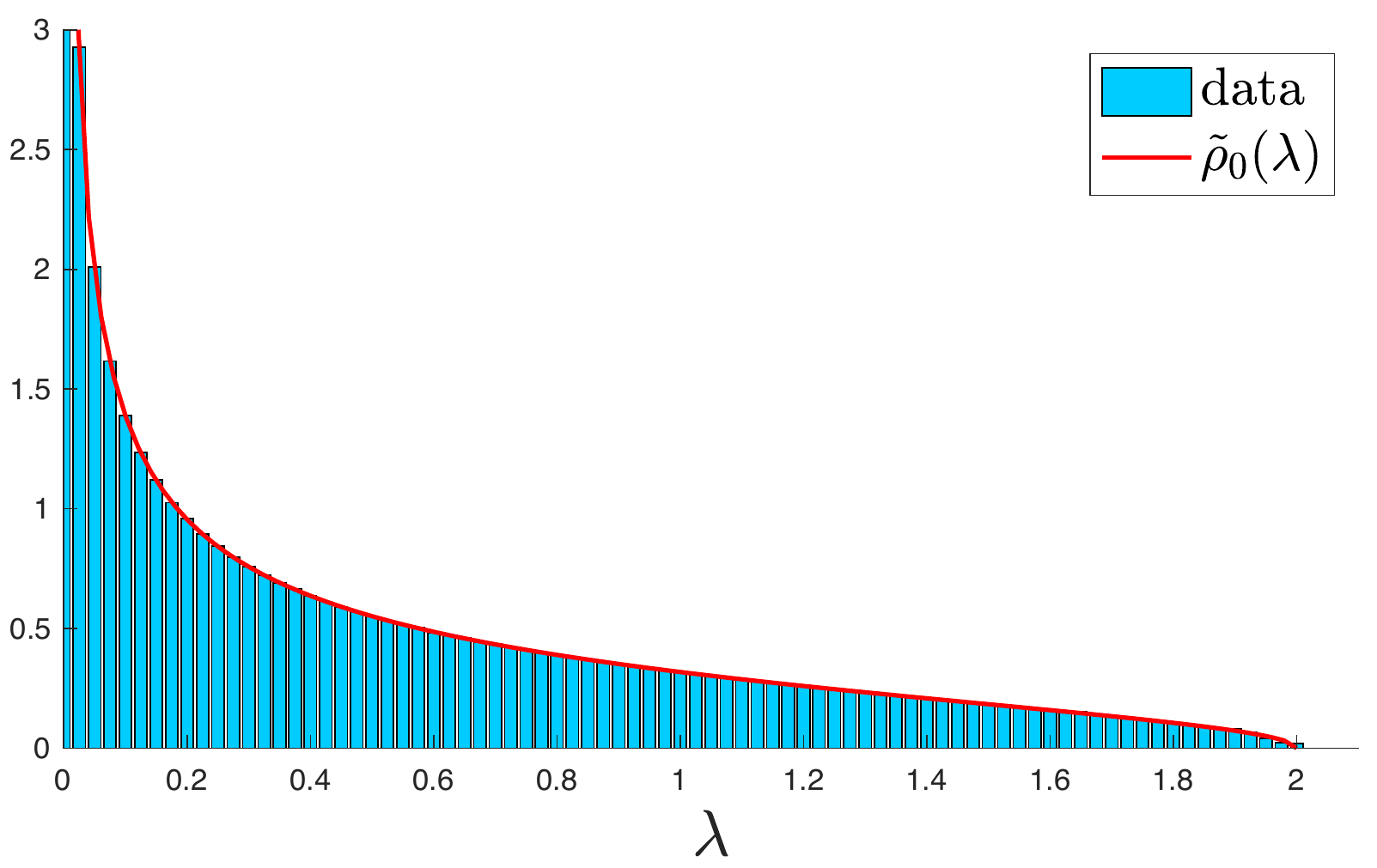} 
\caption{\label{fig:marchenko-pastur} The Mar\v{c}enko-Pastur distribution, generated from 10K samples of 100$\times$100 matrices $H$ generated according to footnote~\ref{fn:marchenko}.}
\end{figure}

 \begin{figure*}
\centering
\includegraphics[width=0.45\textwidth]{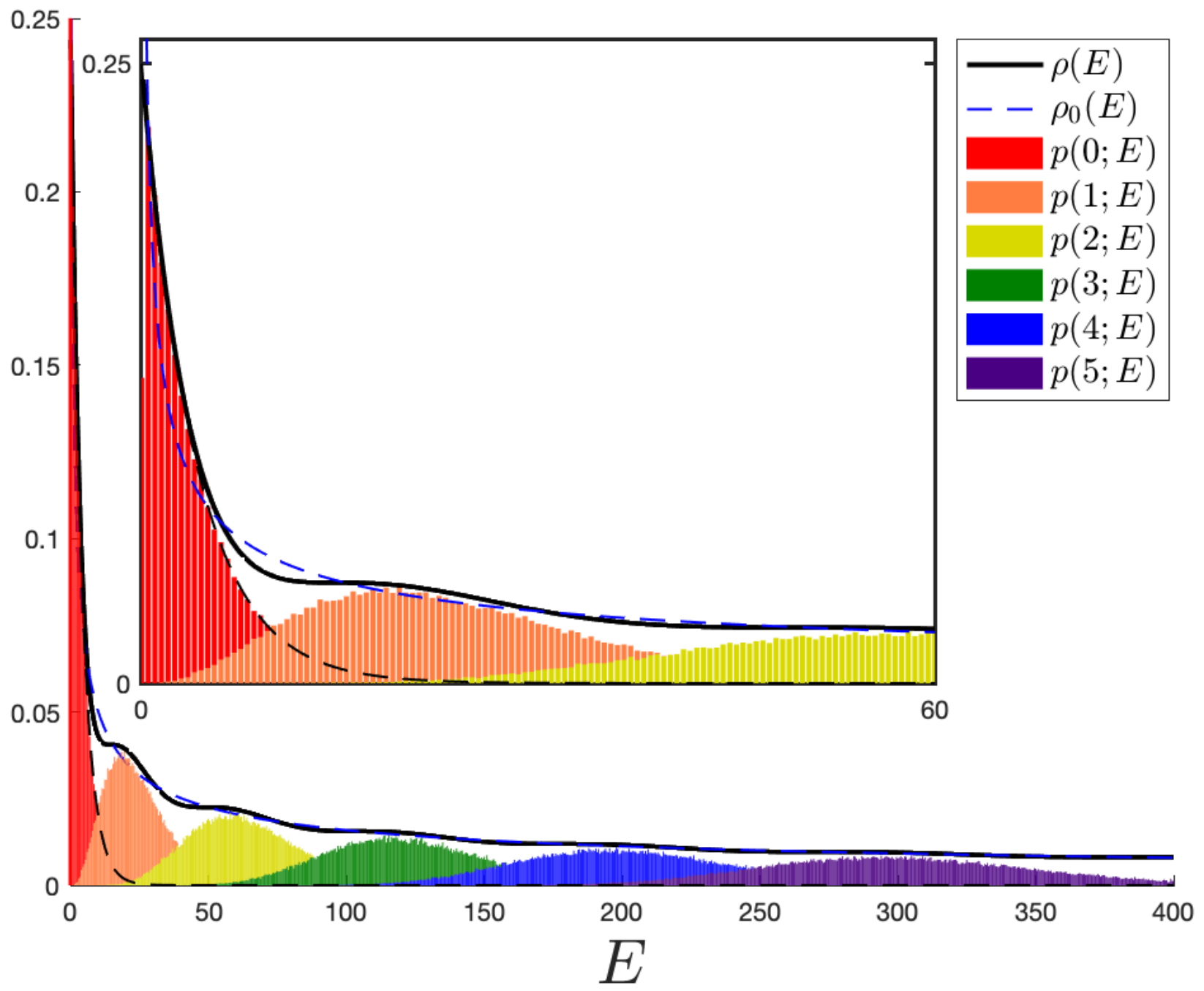} 
\hskip0.5cm
\includegraphics[width=0.45\textwidth]{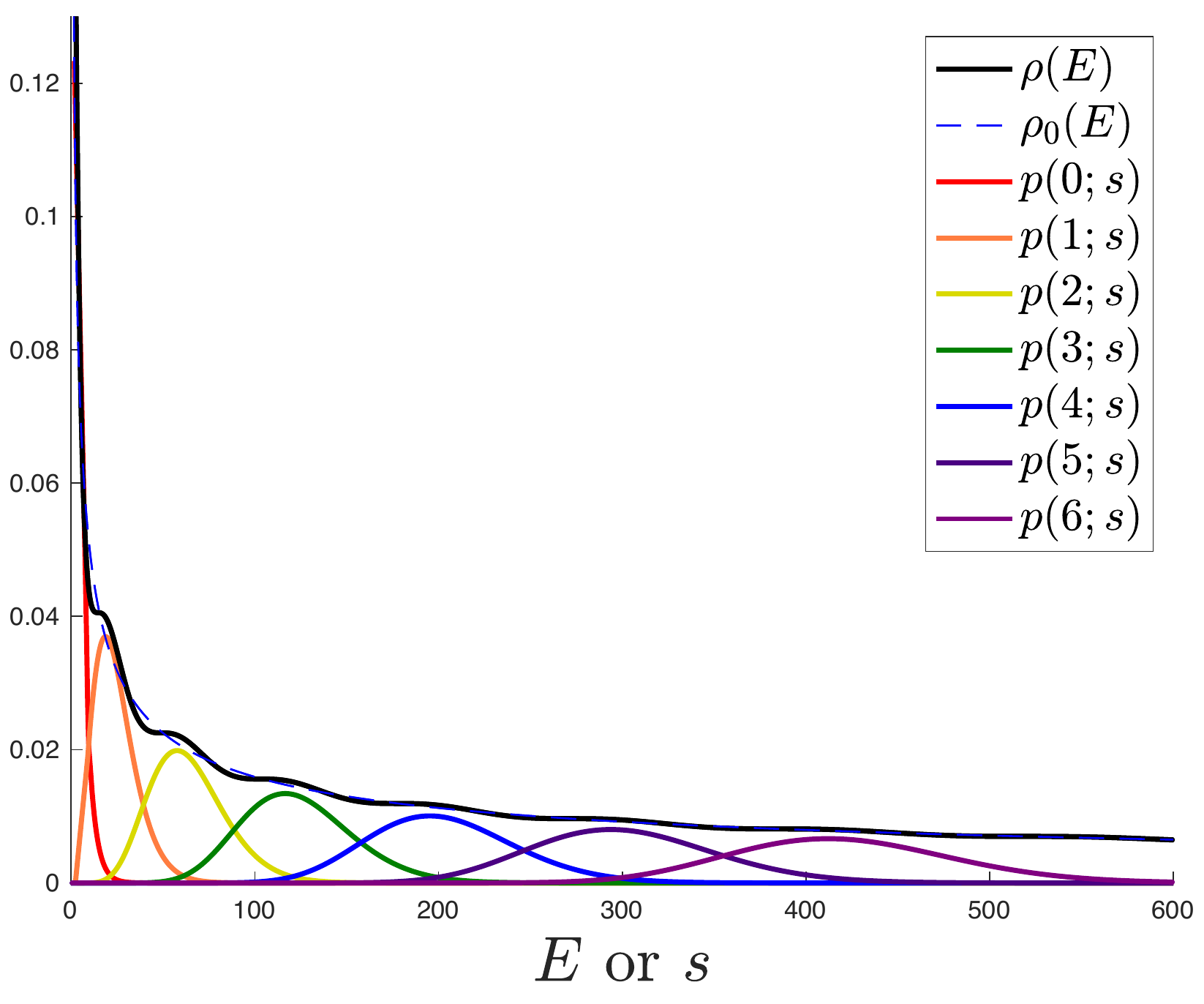}
\caption{\label{fig:bessel-simulation} Left: T spectral density $\rho(E)$ for the $\Gamma{=}0$ Bessel model (solid line). The  blue dashed line is the leading result $\rhoo(E){=}1/(2\pi\hbar\sqrt{E})$. The  histograms  of $p(n;E)$ are frequencies of the $n$th energy level, extracted from a system of $100{\times}100$   Gaussian random Hermitian matrices $H{=}MM^\dagger$, for $100K$ samples. Note the  correspondence with the undulations in~$\rho(E)$.   $p(0;E){=}\frac14{\rm e}^{-E/4}$, shown as a more swiftly falling dashed line (black) is the exact form  for the distribution of ground states, with mean   ${\cal E}_0{\equiv}\langle E_{\rm 0}\rangle{=}4$. Shown on the right are the first 7 microstate peaks for the (1,2) Bessel models computed using Fredholm determinant. (Not visible is that the first peak, like $\rho(E)$, stops ends at the origin at a height of $\frac{1}{4\hbar^2}$). Here, $\hbar{=}1$.}
\end{figure*}

In much of the random matrix model literature on  this type of system, the physics in the neighbourhood of the wall endpoint is usually referred to as  ``hard-edge'' (in contrast to the unconstrained ``soft-edge'' of the semi-circle distribution that gives rise to Airy after scaling). In this case,  Bessel models  arises, in the analogue of the ``double'' scaling limit~\cite{Morris:1991cq,Dalley:1991qg,doi:10.1063/1.530157,FORRESTER1993709,Tracy:1993xj}. The  zooming-in change of variables done in the previous section is simply $E=2N\lambda$. 

%

In fact, the spectral density for the matrix model is known exactly, and it is:
\be
\label{eq:bessel-density}
 \rho(E)\! = \frac{1}{4\hbar^2}\!\left[J_\Gamma^2(\xi)\!+\!J_{\Gamma+1}^2(\xi)\!-\!\frac{2\Gamma}{\xi}J_\Gamma(\xi)J_{\Gamma+1}(\xi)\right]\ ,
\ee
 (where $ \xi\equiv{\sqrt{E}}/{\hbar}$) and as noted above, when $\Gamma$ is a negative integer there are $\Gamma$ zero eigenstates, and so  $|\Gamma|\delta(E)$  should be added to the spectral density. The leading form is $\rhoo(E){=}1/(2\pi\hbar\sqrt{E})$, and the non-perturbative physics here consists of undulations around this. In contrast to the Airy case, the undulations broaden out and get further apart as they go to larger energy, so the underlying physics is quite different. It all follows from the presence of the hard wall causing a ``bunching up'' of the eigenvalues that propagates outwards, but becomes weaker farther away.

This shows another  interesting way that  these models are quite  different from the Airy case: In the  high energy limit,  no continuum  emerges for any individual matrix spectrum. Smoothness is achieved by combining discrete spectra at all energy scales. Aspects of these  models' behaviour will still play a crucial role in gravity models later, where
the large $E$ emergent continuum on any individual spectrum will be present (each one becoming a Schwarzian spectrum). This will be because the models will turn out to be a portmanteau of  features  from Bessel and the kinds of features seen in the models of the previous section.

 The case of $\Gamma=0$ is distinguished by having a non-zero and finite value of $\rho(0)=1/(4\hbar^2)$. This will be readily seen in the simulation in a moment. 
 In the $\Gamma=\pm\frac12$ cases, the Bessel functions reduce to trigonometric functions (divided by a square root), with the spectral densities becoming:
\be
\rho(E)_{\pm\frac12} = \frac{1}{2\pi\hbar \sqrt{E}}\mp\frac{1}{4\pi E}\sin\left(\frac{2\sqrt{E}}{\hbar}\right)\ .
\ee
Remarkably, the $\Gamma=\frac12$ case exactly cancels the classical divergence at $E=0$  to zero. This is a feature that will be present in JT supergravity results to come (see also ref.~\cite{Stanford:2019vob}).

As with the Airy case, histograms can be made of the statistics of the energy levels,  giving distributions called $p(n;E)$, $n=0,\cdots,\infty$. (They will be derived using Fredholm techniques shortly.) Again,  the peaks line up precisely with the familiar undulations  of the  Bessel models' spectral density  (with $\hbar{=}1$). See the left illustration in figure~\ref{fig:bessel-simulation}. In contrast to the Airy case, some exact results are known for some of the distributions in these sorts of models~\cite{10.1137/0609045,EDELMAN199155,FORRESTER1993709}. In particular, for this case,   the distribution of ground states is known exactly as:
\be
\label{eq:exact-ground-1}
p(0;E){=}\frac14{\rm e}^{-\frac{E}{4}}\ ,
\ee
and multiplying by $E$ and integrating gives  the mean value of the ground state as simply  ${\cal E}_0=\langle E_0\rangle{=}4$.
\begin{figure*}
\centering
\includegraphics[width=0.49\textwidth]{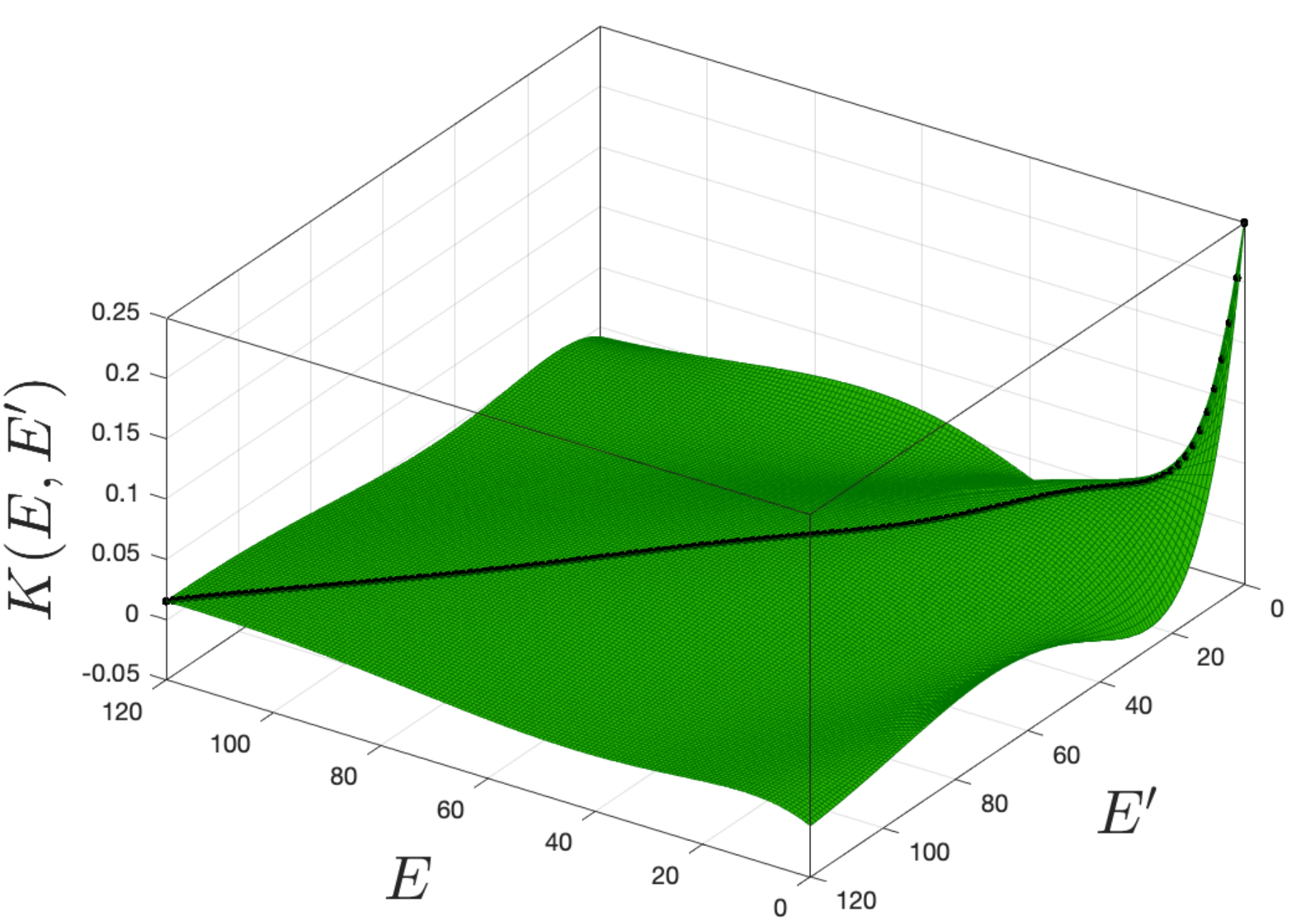}\hskip0.5cm\includegraphics[width=0.48\textwidth]{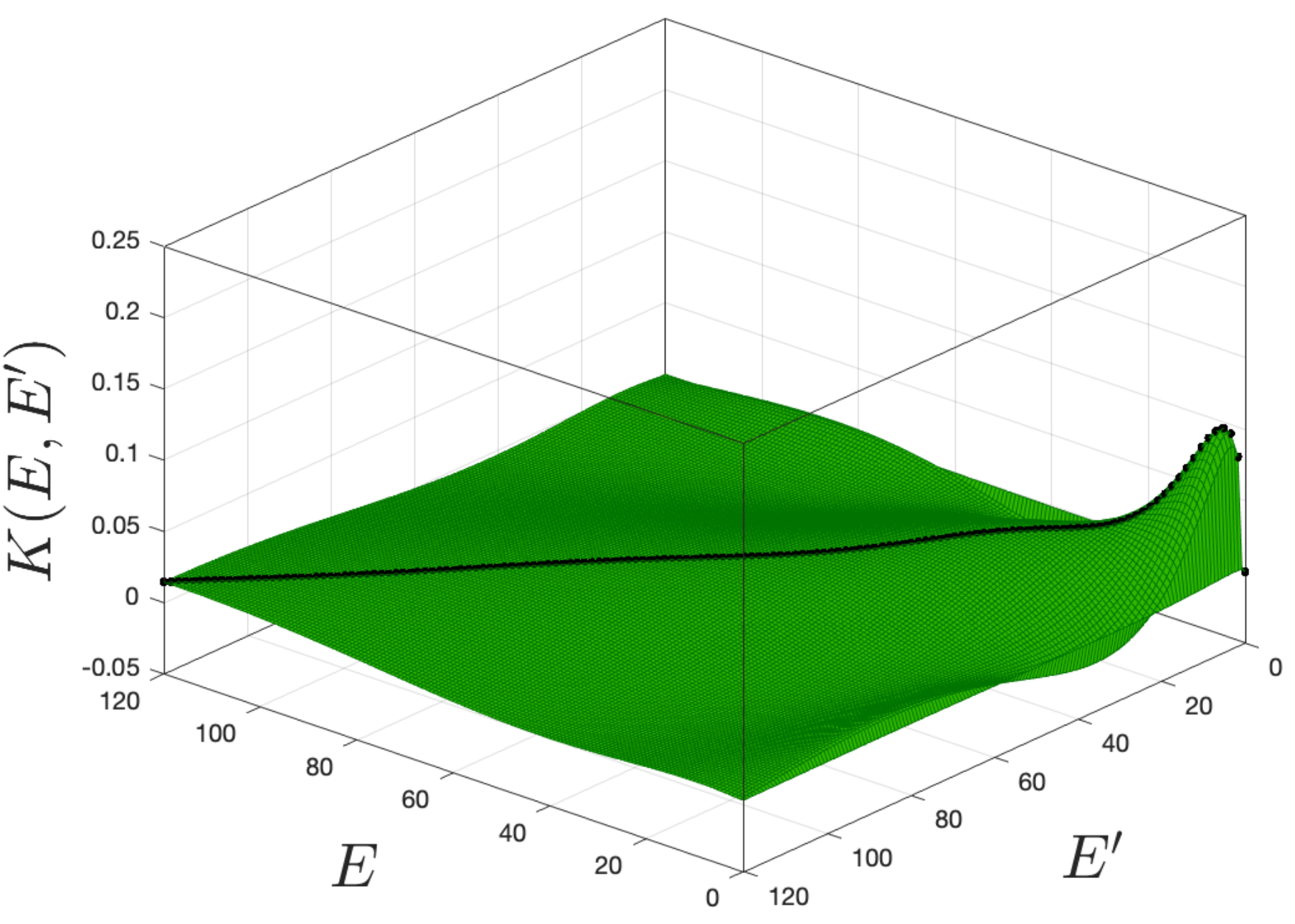}
\caption{\label{fig:Bessel_kernel_3d} Portions of the Bessel kernel for the cases (1,2) ($\Gamma{=}0$) on the left, showing the peak of height $\frac{1}{4\hbar^2}$ at the origin, and (2,2) ($\Gamma{=}\frac12$) on the right, where the function instead vanishes.}
\end{figure*}
%
\begin{figure*}
\centering
\includegraphics[width=0.48\textwidth]{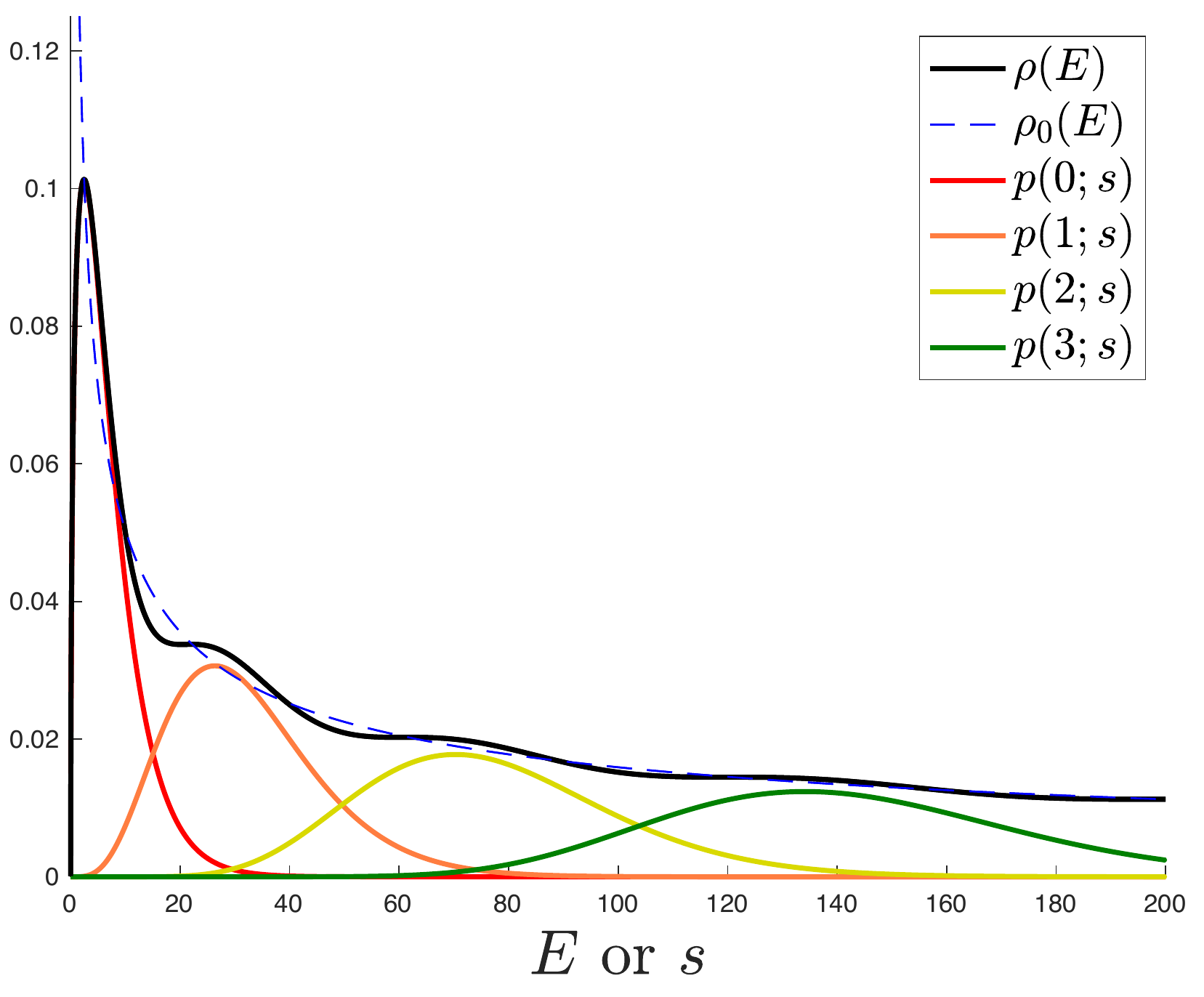}\hskip0.5cm\includegraphics[width=0.48\textwidth]{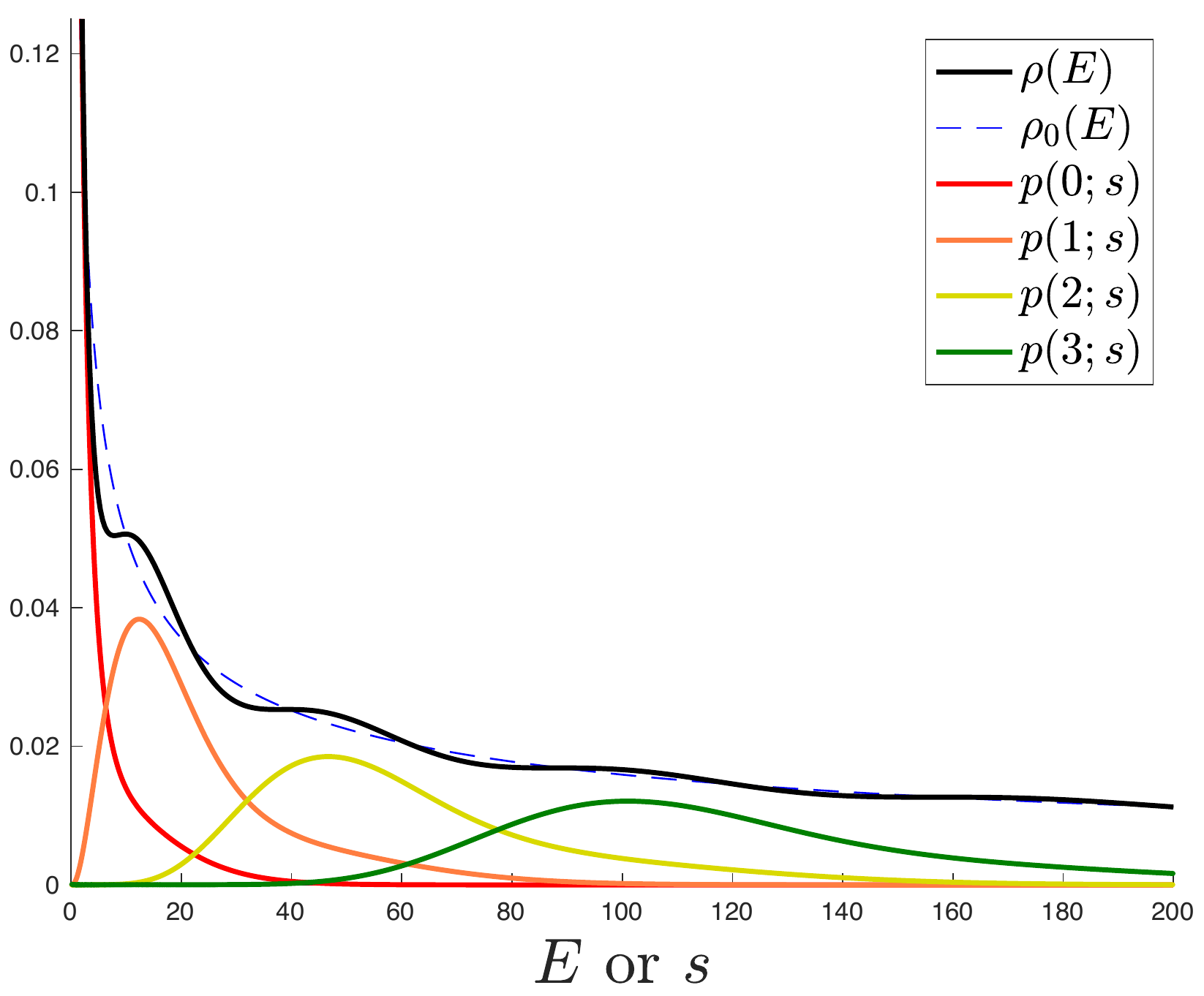}
\caption{\label{fig:Bessel_22_02_microstates} The matrix model spectral densities $\rho(E)$ and first four microstate peaks for the Bessel models, with (2,2) ($\Gamma{=}\frac12$) on the left and (0,2) ($\Gamma{=}{-}\frac12$) on the right.}
\end{figure*}
%
%

As might be expected, there is an orthogonal polynomial story for these matrix models too. The Hermite polynomials of the Wigner case are now exchanged for Laguerre polynomials~\cite{doi:10.1063/1.530157},  and in the scaling limit the wavefunctions (after scaling in a factor of the measure as before) become Bessel functions:
\be
\label{eq:Bessel-wavefunction}
\psi(E,x) = \frac{1}{\hbar} \sqrt{\frac{x}{2}}J_\Gamma\left(\frac{x\sqrt{E}}{\hbar}\right)\ .
\ee
They are in fact (with a specific normalization) wavefunctions of the Hamiltonian~\cite{Carlisle:2005wa,Johnson:2019eik}
\be
\label{eq:Bessel-Hamiltonian}
{\cal H} = -\hbar^2\frac{\partial^2}{\partial x^2}+\frac{\hbar^2\left(\Gamma^2-\frac14\right)}{x^2}\ .
\ee
The analogue of the Fermi sea   now runs from $x=0$ to $x(=\mu)=1$, which fits nicely with the fact that the scaling to the endpoint results {\it entirely} from a scaling with $N$, with no translational component. Looking ahead, the JT supergravity cases of Section~\ref{sec:sjt-gravity} will be seen to have an interpretation as having a Fermi sea where $x$ runs from $-\infty$ (like regular JT) up a the Fermi level at $\mu{=}1$, where the portion up to $x=0$ is like a regular JT Fermi sea (but filled differently so as to yield the super Schwarzian at leading order), and then the portion from $x=0$ to 1 has more of the Bessel character seen here. 

 From these wavefunctions  the ``Bessel kernel''  can be built  (using the analogue of equation~(\ref{eq:edge-kernel})):
\begin{widetext}
\be
K(E,E^\prime)=\int_0^1 \psi(E,x) \psi(E^\prime,x)dx =  \frac{1}{2\hbar}\left[\frac{\sqrt{E}J_{\Gamma-1}\left({\sqrt{E}}/{\hbar}\right) J_{\Gamma}\left({\sqrt{E^\prime}}/{\hbar}\right) - \sqrt{E^\prime}J_{\Gamma-1}\left({\sqrt{E^\prime}}/{\hbar}\right) J_{\Gamma}\left({\sqrt{E}}/{\hbar}\right)}{E-E^\prime}\right]\ .
\ee

\end{widetext}
Once again, this is  interesting  to look at in 3D, and the cases $\Gamma=0$ and $\Gamma=\frac12$ are shown in figure~\ref{fig:Bessel_kernel_3d}, with (again) some amusing resemblance to  aquatic creatures.

From here, the Fredholm technology of section~\ref{sec:fredholm} can be used wholesale for the Bessel models~\cite{Tracy:1993xj}, now on the interval $(0,s)$  (using the notation of that section) to derive the $p(n;s)$.  The computations done for Airy are readily adapted to this case, and some results follow. The first few peaks are shown for each of the cases $\Gamma=\pm\frac12$  in figure~\ref{fig:Bessel_22_02_microstates}. (The divergence in the (0,2) case presents considerable extra difficulty to control numerically. Such  numerical difficulty will also afflict some of the results of two JT supergravity models somewhat.)

\begin{figure}[t]
\centering
\includegraphics[width=0.45\textwidth]{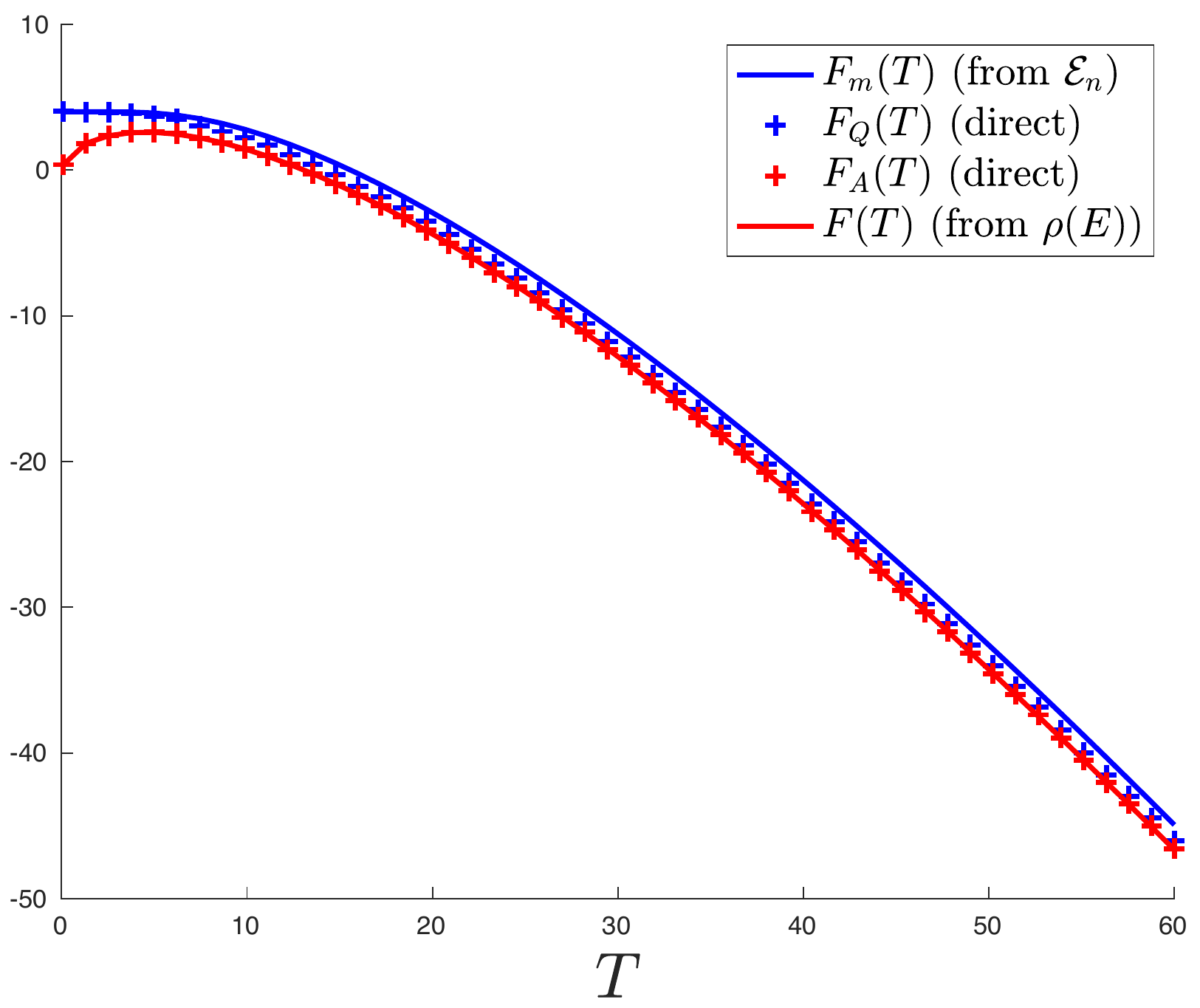} 
\caption{\label{fig:bessel-free-1} The blue crosses are the  quenched  free energy, $F_{Q}$, computed by direct sampling the matrices  for the (1,2) Bessel model, alongside the free energy computed from the mean spectrum (solid blue), including the first seven states.  They both  land at $\langle E_0\rangle{=}4$, the exactly known value of the averaged ground state. The annealed result from sampling, $F_A$, are red crosses (lower). They sit directly on top of the solid line  resulting from computing $F_A$ from transforming $\rho(E)$.}
\end{figure}

The first six peaks for $\Gamma=0$ are shown  on the right in  figure~\ref{fig:bessel-simulation}, and of course should be compared to the results from ``experiment'' on the left of the same figure.  The values of the first seven peaks for this case may be readily extracted as: ${\cal E}_0=4.00, {\cal E}_1=24.3, {\cal E}_2=64.11, {\cal E}_3=123.5, {\cal E}_4=202.6, {\cal E}_5=301.4, {\cal E}_6=417.9$, and they will be used shortly. In all cases, once again, the zeros of the wavefunctions evaluated at the Fermi surface ($x=1$) give a rough guide to the locations of the peaks, but unlike the Airy case (and other examples to come) the {\it spreading} of the peaks, and the gaps between them,  with  growing energy makes the wavefunctions a less useful guide.

 While it is certainly interesting to explore these models further, the main reason for bringing them into the discussion at this point (besides noting some features that will appear again in the JT supergravity cases) is that it serves as another testbed for the free energy computation: seeing how well the quenched free energy matches  with the free energy of the mean spectrum. Just as with the Airy example, the quenched and annealed free energies can be computed by direct sampling. The results are shown, using the first 7 levels (for $F_m(T)$), in  figure~\ref{fig:bessel-free-1}. (For the temperatures shown, the physics has already settled down to what it would look like if more levels were added.) The annealed energies agree very well. They go to zero here instead of diverging negatively (as they do for Airy) since there are no negative energy states involved in badly skewing the free energy as happens for Airy. Here the spectrum has $E=0$ as its lowest value.
 
As in the case of Airy, the quenched free energy of the ensemble and the quenched free energy of the mean spectrum follow each other well, with the ensemble's quantity lying slightly (on the scale of the typical energies of the states involved) below. They merge together as they approach the ground state at ${\cal E}_0=4$.

\medskip
\centerline{* * *}
\medskip

This is a good point to leave aside the development of the tools, and the study of toy models, and instead delve into matrix models that actually describe gravity. All the intuition developed so far will help fill out a rather satisfying story for several models of JT gravity.

\section{JT Gravity}
\label{sec:jt-gravity}
The string equation approach to formulating non-perturbative matrix model descriptions of JT gravity was developed in refs.~\cite{Johnson:2019eik,Johnson:2020exp} (and extensions to include deformations in ref.~\cite{Johnson:2020lns}).  Recently, ref.~\cite{Johnson:2021tnl} clarified several aspects of the non-perturbative definition for JT, and used the many-body language reviewed in the previous Sections to show how to describe a family of non-perturbative completions. Clearly there is no need to review every detail of the construction here, but some key aspects will be highlighted in order to have a reasonably self-contained narrative flow. 

Recall from subsection~\ref{sec:search-gravity} the potential $u(x)$ arising from studying double-scaled Hermitian matrix models is governed by an ordinary differential equation (ODE) of the form ${\cal R}=0$. See equation~(\ref{eq:simple-string-equations}). In fact, a linear combination (at the level of building equations)  defines a more general model that can be considered as an interpolating model in the sense of a field theory flows, defining the couplings $t_k$:
\be
{\cal R}\equiv\sum_{k=1}^\infty t_kR_k[u]+x\ ,
\ee
and now ${\cal R}=0$ defines (at least perturbatively) a more general model of random surfaces. A given set of values for the $t_k$ will determine the precise model. JT gravity, for example, will correspond to a particular choice, to be reviewed below.  Before moving on it is worthwhile swiftly dealing with the instability that was also mentioned in that section: The $k$ even models are unstable non-perturbatively, corresponding to eigenvalues wishing to tunnel out of the Dyson Gas droplet to lower $E$'s disconnected from the main group. This is undesirable from the point of view of trying to find a theory where (at least perturbatively,  the smooth surfaces of interest should be present. It is intuitively clear what needs to be done. If instead an ensemble of Hermitian matrices is used that has some lowest (scaled) eigenvalue $\sigma$ after double scaling,  such that $\sigma$ is above the (fully non-perturbative) threshold where the instability develops, but not so high that it plays any role at any order in perturbation theory, then this constitutes a  non-perturbative completion of the physics. This was first done (for $\sigma=0$) in this context in ref.~\cite{Johnson:2019eik,Johnson:2020exp}, and the broader case fully discussed and analyzed recently in ref.~\cite{Johnson:2021tnl}, showing that for JT gravity a (narrow) range of allowed $\sigma$ is possible, and explaining it in terms of the dynamics of background D-branes. 

It is straightforward to see how the larger string equation that results from defining such an Hermitian matrix ensemble arises. This was done explicitly long ago in refs.~\cite{Dalley:1991xx} (the original equation was found by studying a certain model of complex matrices~\cite{Morris:1990bw,Dalley:1991qg,Dalley:1991vr}), and rather than reproduce all the steps here, the essential aspects are described in a slightly different manner that might be illuminating. The simple Hermitian matrix model string equation can be formulated in an illuminating (in some respects) language as follows. Notice that the two non-trivial identities, (\ref{eq:recursion}) and~(\ref{eq:identity2})  that gave rise to the recursion relations (and ultimately the string equations) were to do with the action of $\lambda$ and $\frac{d}{d\lambda}$ in the matrix model. Specifically, (\ref{eq:recursion}) led to the operator realization of $\lambda$ as $Q=-{\cal H}$. Similarly the action of  $\frac{d}{d\lambda}$ can be thought of as an operator, and it is the momentum counterpart to ``position'' $Q$, which deserves to be called $P$. Now it is easy to state the string equation. It is simply the canonical commutation relation~\cite{Douglas:1990dd}.
\be
[P,Q]=1\ , 
\ee
and in fact it fully non-perturbatively encodes the fact that the Dyson gas problem has (at the level of the action) a translation invariance along the spectral line. The instability of the even $k$ models is simply the failure to find good solutions that are consistent with that beyond perturbation theory. 

Deriving an equation for a model of random Hermitian matrices with eigenvalues restricted to be within the range $(-\Sigma,\Sigma)$ (indeed, the potential is still chosen to be even for simplicity) is simply a matter of writing an additional identity for the insertion of $\lambda\frac{d}{d\lambda}$. In integrating all identities by parts, there will also now be boundary terms, involving $P_{N-1}(\pm\Sigma)$ and $P_{N}(\pm\Sigma)$. Such terms can be eliminated between the two equations, giving a single, larger string equation upon using subsection~\ref{sec:search-gravity}'s double scaling relations, supplemented with scaling the ``wall'' position with $\Sigma=2-\sigma\delta^2$, and it is:
\be
\label{eq:string-equation-big}
(u-\sigma){\cal R}^2 -\frac{\hbar^2}{2}{\cal R}{\cal R}^{\prime\prime}+\frac{\hbar^2}{4}({\cal R}^\prime)^2=0\ . 
\ee
This is easy to state in the operator language~\cite{Dalley:1992yi}. If the operator corresponding to an insertion of  $\lambda\frac{d}{d\lambda}$ into the Dyson gas is denoted ${\widetilde P}$, then the string equation is equivalent to the canonical commutation relation:
\be
[{\widetilde P}+\sigma P,Q]= P+\sigma\ .
\ee
If $\sigma=0$ this is just the expression of the scale invariance left over after breaking translations, and non-zero $\sigma$ gives the full form showing that $\sigma$ can be translated.
For the purposes of this paper, it will be enough to set $\sigma=0$. The other allowed values of $\sigma$ that are allowed by the consistency conditions of ref.~\cite{Johnson:2021tnl} (for JT gravity) give physics that is extremely similar.\footnote{In fact, a slight generalization of the equation is possible, where the right hand side has the constant $\hbar^2\Gamma^2$. This corresponds, here, to the result of adding background D-braneswith cosmological constant $\sigma$~\cite{Dalley:1992br,Johnson:1994vk} , but this will not be pursued here. In the next sections, this same equation will appear, but used in a different manner, to study JT supergravity, and then $\Gamma$ non-zero will be essential.}

Recall that, as discussed in subsection~\ref{sec:search-gravity}, the perturbative description of the matrix model is  entirely done in terms of the Fermi sea regime, $x<\mu$, as an expansion in $\hbar/|x|$. This is equivalent to a large $|x|$ expansion in the $x<0$ region of the string equation.  Correspondingly, the solution for $u(x)$ in this regime is captured by the simpler ${\cal R}=0$ solution of the string equation~(\ref{eq:string-equation-big}). In other words, perturbation theory is (as it should be) identical to the content of the simpler Hermitian matrix model.  

Setting  $\hbar=0$ for the leading behaviour (the large~$N$ saddle) in the $x<0$ regime results in the algebraic equation 
\be
\label{eq:leading-R0}
{\cal R}_0[u]\equiv\sum_kt_ku_0^k+x=0\ ,
\ee
 where $u_0$ is the leading part of $u(x)$.  This will result in some leading spectral density $\rho_0(E)$. The one of interest for JT gravity is
  the familiar Laplace transform of the Schwarzian result, given in~(\ref{eq:leading-relation}). The next step is to use the relation~(\ref{eq:leading-density}) to see what combination of the $t_k$ can give the $u_0(x)$ that yields this $\rho_0(E)$, and it is uniquely picked out to be~\cite{Dijkgraaf:2018vnm,Okuyama:2019xbv,Johnson:2019eik}:
 \be
\label{eq:teekay}
t_k = \frac{\pi^{2k-2}}{2k!(k-1)!}\ ,
\ee
with the Fermi level at $\mu=0$. (So in this section, all remaining occurrences of $\mu$ will be understood to be using $\mu=0$, but sometimes $\mu$ will be left in place for clarity. It will turn out to be 1 in the sections on JT supergravity.) 

  The leading solution for $u_0(x)$ can be written as:
\be
\label{eq:u0-equation-JT}
\frac{\sqrt{u_0}}{\pi}I_1(2\pi\sqrt{u_0})+x=0\ ,
\ee
where $I_1$ is the first modified Bessel function. 
In this sense, JT gravity can be understood as a particular combination of an infinite number of minimal models in this ``KdV'' basis.\footnote{Many things are called minimal models, and $t_k$, in the literature, together with  and the basis of operators the $t_k$ couple to. Here, the minimal model controlled by $t_k$ simply yields a behaviour $E^{k-\frac12}$ in the spectral density. This is sometimes referred to as the KdV basis in the literature, since it is in that basis that the $t_k$ act as generalised times in an organizing Korteweg-de Vries integrable flow between models.  There are also different common normalizations for the $t_k$ arising from whether the coefficient of~$u^k$ in~(\ref{eq:leading-R0}) is chosen as unity or not, as it is here. The KdV basis   differs from the operator basis more naturally occurring in the conformal minimal modelsl~\cite{Moore:1991ir}, and so the $t_k$s in such models are a mixture of the $t_k$s  used here. See ref.~\cite{Mertens:2020hbs}. After translating, the approach used here  aligns nicely with the large~$k$  approach first suggested in ref.~\cite{Saad:2019lba}. } Higher order terms in perturbation theory come from expanding the string equation in parameter $\hbar/x$ where $x$ is negative.  This is the perturbative description of the corrections to the Fermi sea description of the basic matrix model construction of JT gravity.  This yields corrections to the potential that can be written as $u(x)=u_0(x)+\sum_{g=1}^\infty u_g(x)\hbar^{2g}$. The quantum mechanics can be solved perturbatively (in $\hbar$) on this potential, and together they yield the topological perturbation theory described (using a different approach)  by Saad, Shenker and Stanford~\cite{Saad:2019lba}.


The most interesting issue for the purposes of this paper is that staying in perturbation theory will never reveal the microstructure of the JT gravity. As seen in the prototype example of Airy, what is needed is the full set of wavefunctions $\psi(x,E)$  from which everything about the matrix model can be constructed. For Airy, the exact potential to all orders was $u(x)=-x$, and all the non-perturbative physics came from solving the Schrodinger problem with that potential. Happily this was just the Airy equation, which resulted in Airy functions for the complete wavefunctions. Already a great richness resulted (from computing the Fredholm determinant, difficult enough in this simple case) yielded a lovely underlying structure of states. Here, the task is complicated by the fact that $u(x)$ has both perturbative and non-perturbative corrections described by a highly non-linear equation. There will be no exact solution to display here for $u(x)$. Once it is known (numerically), the wavefunctions and energies must be constructed, again numerically. The techniques for this were developed and explained in ref.~\cite{Johnson:2020exp}.\footnote{A key aspect of that work was to show that although the string equation is technically of infinite order, as $k$ grows the relative influence of $t_k$ diminishes. Also, the solutions sought are such that they asymptote to perturbative behaviour for large positive and negative $x$, with the non-trivial behaviour in the interior. A truncation of the equation to some high enough value of $k$ can be done such that the error in doing so would only be significant at large enough $x$ where the perturbative solution for $u(x)$ is sufficient to describe the physics. For JT gravity it turns out that $k=7$ (resulting in needing to solve a 15th order ODE) works very well, for $\hbar=1$.} Hence, the kernel for JT gravity will also only be known numerically. {\it Then} the step to constructing the Fredholm determinant must be done.  The two additional layers of numerics before getting to the Fredholm computation result in  quite a challenge for controlling accuracy, but it can be done after considerably further refining the numerical methods.\footnote{\label{fn:more-tips}For the Reader interested in doing these computations, the first step is to read the suggestions for numerical work on solving string equations given in ref.~\cite{Johnson:2020exp} (Appendix B). These are good enough for solving the string equations for $u(x)$ and obtaining wavefunctions that yield accurate results for $\rho(E){=}K(E,E)$, although it has since been found that they can be much  improved in speed by passing vectorized  Jacobians to the chosen \textsf{MATLAB} ODE solver. However the spectral methods need improvement  for the task of computing the Fredholm determinant since the delicate interplay of the off-diagonal components of $K(E,E^\prime)$ require greater accuracy and handling of the wavefunctions $\psi(E,x)$. For this, it is was found that the {\tt chebfun} package was extremely useful, where whenever possible operations are performed on representations of all quantities as Chebychev polynomials.  It can be readily installed into  \textsf{MATLAB}.  It allowed for much cleaner wavefunctions, and hence a more accurate kernel. It is also recommended that if available, parallelization of the code be done as much as possible to improve speed of some operations that must be done for every computed energy.}

Turning to results, the spectral density $\rho(E)$ first found in ref.~\cite{Johnson:2020exp} was already displayed in figure~\ref{fig:full-density-vs-semi-classical}, showing the non-perturbative undulations. (Several others for different values of $\sigma$ were constructed recently in ref.~\cite{Johnson:2021tnl}, but for illustration purposes $\sigma=0$ will remain the focus here.) Given the $\psi(E,x)$ that were constructed to yield this result,  it is straightforward to extract the special quantity $\psi(E,\mu)$, which, as seen for the Airy case, has a lot of useful information. Its zeros should line up with the points of inflection of the undulations, and this is confirmed in figure~\ref{fig:JT-wavefunction-zeros}. 

\begin{figure}[t]
\centering
\includegraphics[width=0.42\textwidth]{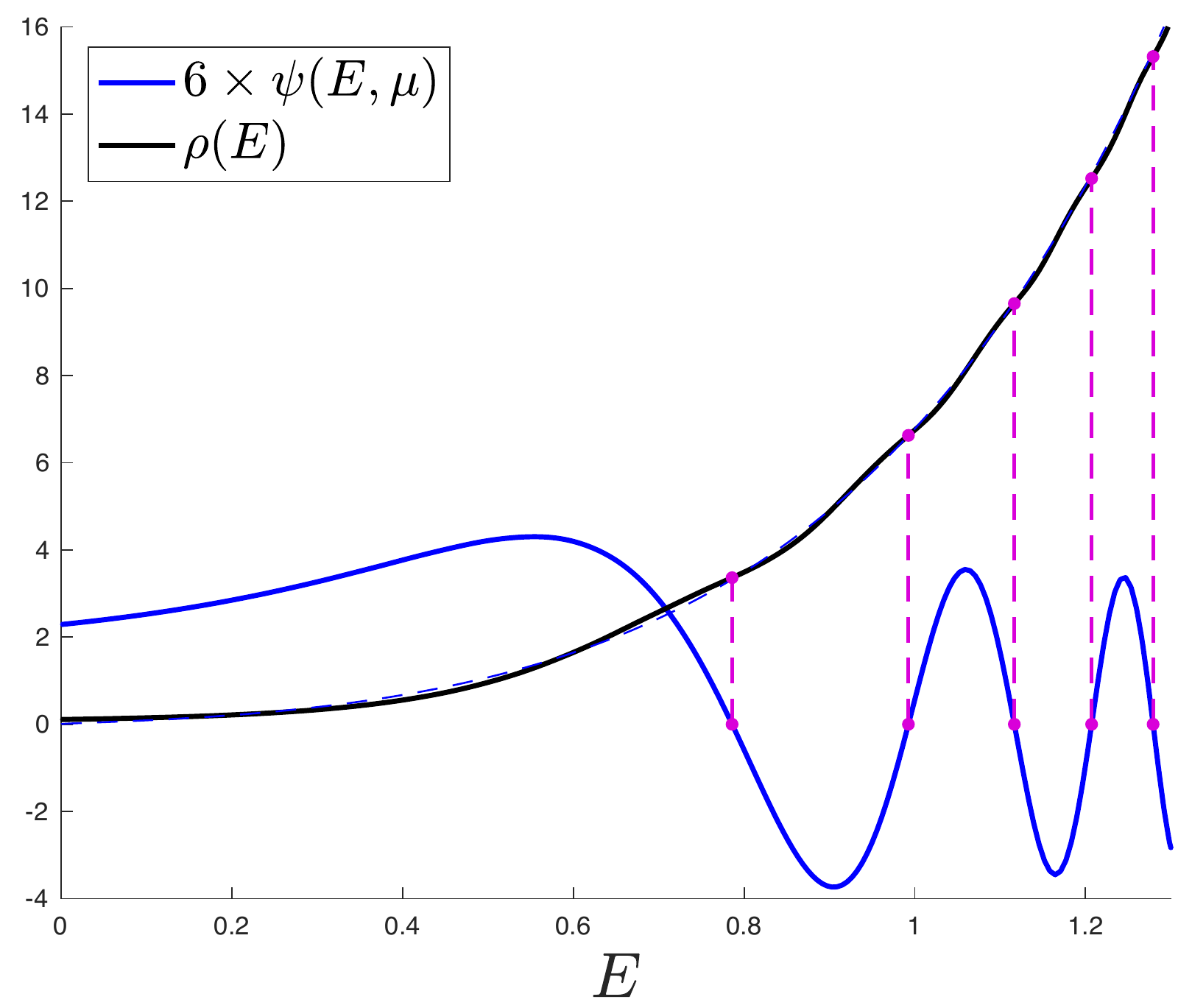}
\caption{\label{fig:JT-wavefunction-zeros} The D-brane partition function $\psi(E,\mu)$ plotted alongside the spectral density $\rho(E)$ for JT gravity. The zeros give a first draft of the mean values of the energy levels, lining up with the undulations of the spectral density.  The complete information about the locations of the mean energy levels requires the  Fredholm determinant to be computed. See text.}
\end{figure}

This gives a rough draft of the microstate spectrum, which becomes more accurate at higher energies, as already discussed in subsection~\ref{sec:return-to-gravity}.  As discussed there, crucially, it also confirmed that the $\{ {\cal E}_n\}$, as means of the $p(n;E)$, form a meaningful spectrum, which is  suggested as the JT dual.    At high enough energies, the truncation and numerical methods can be supplemented by the WKB form of the wavefunction~(\ref{eq:WKB2}). Figure~\ref{fig:JT-wavefunction-compare} shows the exact (numerical) $\psi(E,0)$ superimposed with the WKB form, the two together allowing the entire energy range to be covered. A hybrid wavefunction can be constructed by joining the two at large enough $E$ 
and it is displayed in figure~\ref{fig:JT-wavefunction-compare}. Its zeros will be discussed shortly.

\begin{figure}[t]
\centering
\includegraphics[width=0.42\textwidth]{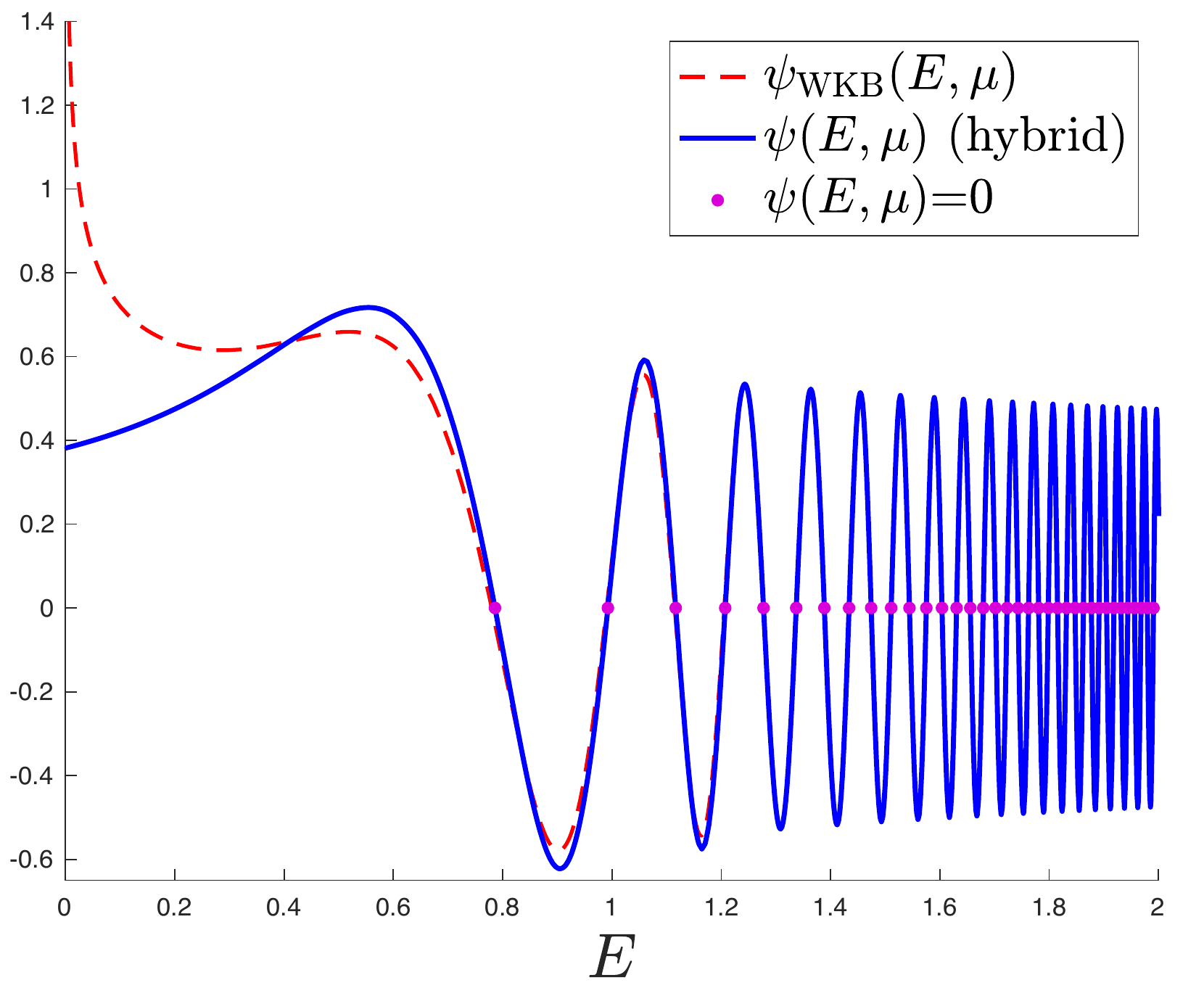}
\caption{\label{fig:JT-wavefunction-compare} The D-brane partition function $\psi(E,\mu)$ for JT gravity plotted alongside the WKB approximation of equation~(\ref{eq:WKB2}).}
\end{figure}

\begin{figure}[b]
\centering
\includegraphics[width=0.48\textwidth]{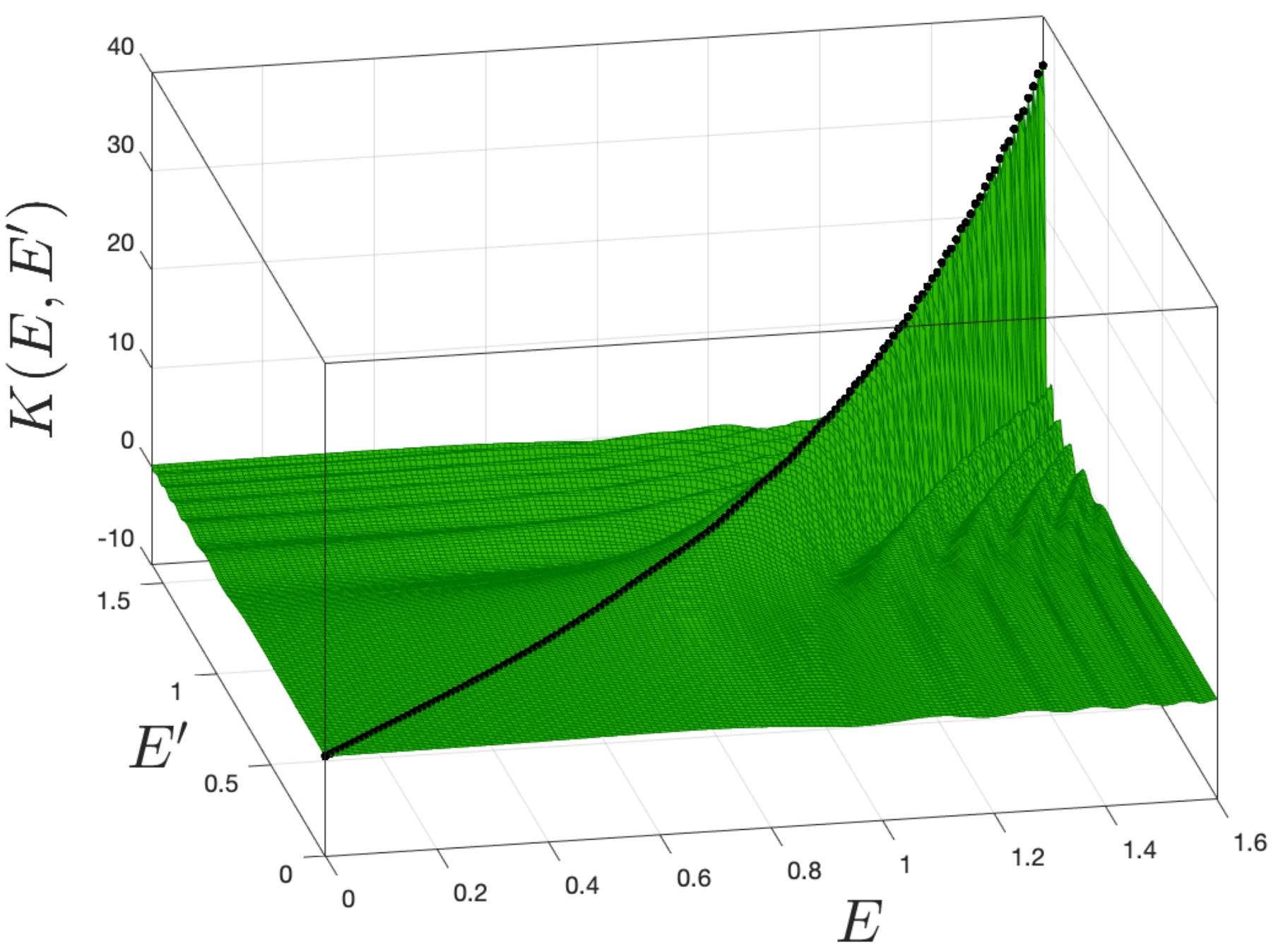}
\caption{\label{fig:JT_gravity_kernel_3d} A 3D view of the JT gravity kernel, showing {\it e.g.,} the  density $\rho(E)$ along the diagonal. ($\hbar{=}1$.)}
\end{figure}
\begin{figure*}
\centering
\includegraphics[width=0.48\textwidth]{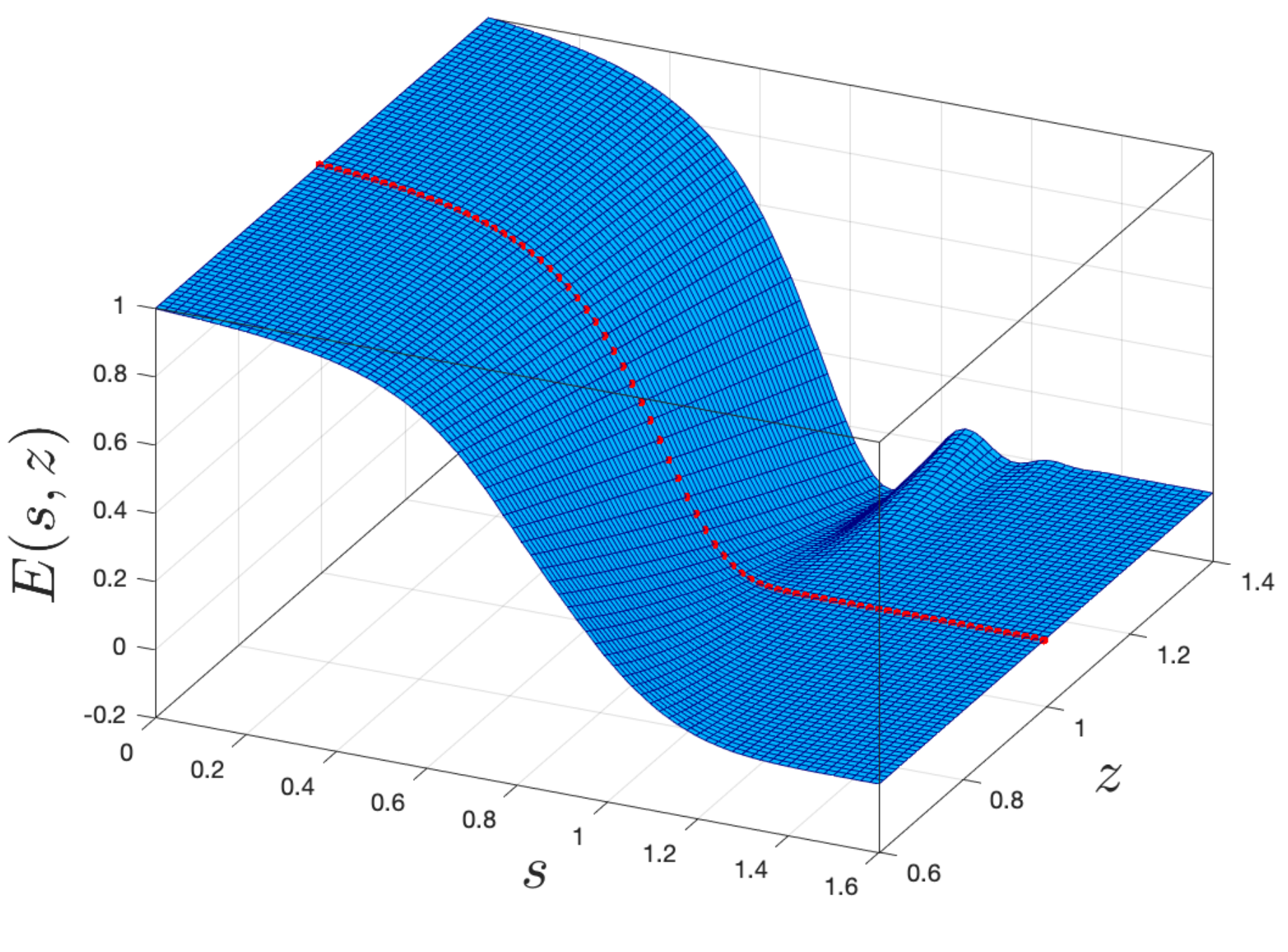}\hskip0.5cm
\includegraphics[width=0.48\textwidth]{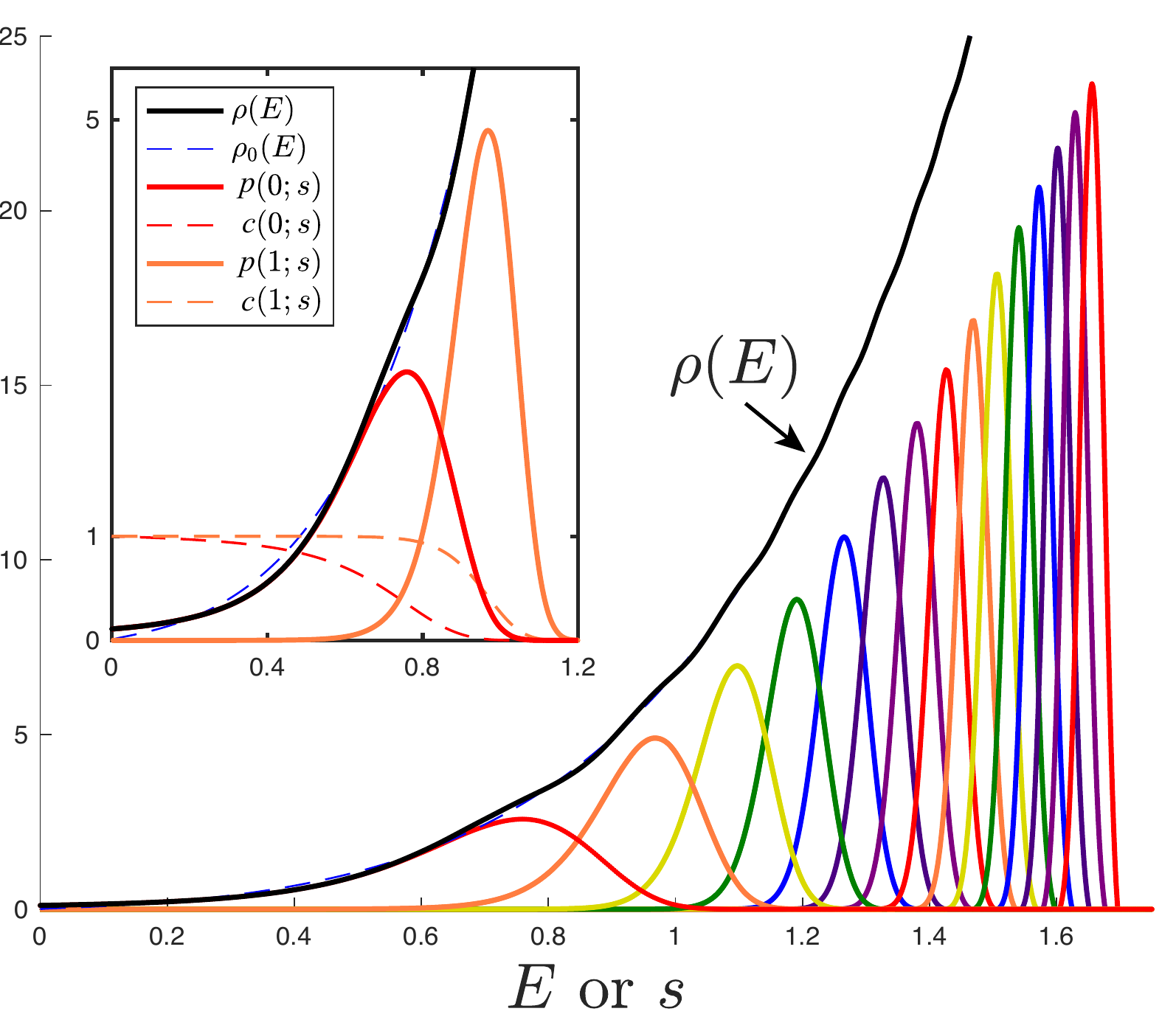}
\caption{\label{fig:JT_gravity_microstates_15} The Fredholm determinant ${\cal F}(s;z)$ (left) and the statistical distribution of the first fifteen energy levels (right) derived from it for JT gravity, shown alongside the leading ($\rho_0(E)$) and full ($\rho(E)$) spectral density. Here $\hbar{=}1$.}
\end{figure*}

Knowing $\psi(E,0)$ and its derivative is in principle enough to  compute the Kernel $K(E,E^\prime)$ (because of the Christoffel-Darboux form), but  numerically, it is not as useful as knowing the full set $\psi(x,E)$, because of the denominator $E-E^\prime$.  
The kernel can be constructed  numerically using the wavefunctions and as before it is illuminating to look at it, so a three-dimensional portion is shown in figure~\ref{fig:JT_gravity_kernel_3d}. 

With $K(E,E^\prime)$ in hand, the same procedures as done for Airy can be carried out in order to construct the Fredholm  determinant. A key difference here is that the interval upon which the computation is done begins at zero (more generally, $\sigma$) instead of $-\infty$. The considerable additional numerical difficulty resulted in extra struggles with noisy results but they were not insurmountable. (See footnote~\ref{fn:more-tips}.)  

A three-dimensional portion of ${\cal F}(s,z)$ is shown on the left in figure~\ref{fig:JT_gravity_microstates_15}.
The two-dimensional slice at $z=1$ (red line) shows the cumulative distribution function for the zeroth energy (ground state), $c(0,s){\equiv}E(0;s)$, and the more general $z$-dependence will produce, through $z$-derivatives, the CDFs for higher levels, $c(n;s)$, and their PDFs $p(n;s)=-d c(n;s)/ds$.
Despite appearances, numerical noise is still potentially a problem here, since the curve needs to be known to very good accuracy to increasingly high values of $s$, for an increasingly high number of $z$-derivatives to be taken, with the results from all preceding  derivatives added together. While smoothing algorithms can be helpful, they must be used sparingly lest the physics get erased, obscuring the properties of each level. Despite all this, it was possible to obtain again 15 energy levels for which good accuracy could be obtained for the statistics. (Several more levels could be discerned, but with less good confidence about the numbers.) 
They are shown, on the right, in figure~\ref{fig:JT_gravity_microstates_15}.

Table~\ref{tab:mean-levels-fredholm} lists the mean values ${\cal E}_n=\langle E_n\rangle$ of the peaks found and their variance, while the final column gives the values of the zeros of $\psi(E,\mu)$. As stated earlier, the latter become an increasingly good estimate of the location of the peaks (and their mean) as energy increases. In fact, by the 10th level it can be seen that they are already within $1\%$ of each other. So for later applications to be discussed shortly, it will suffice to use these zeros of $\psi(E,\mu)$ above level 15 as a good estimate of ${\cal E}_n$. It must be repeated that for low energies there is no substitute for the correct results obtained from the Fredholm determinant. Results from $\psi(E,\mu)$ are incorrect for this purpose.

\begingroup
\begin{table}[b]
\begin{center}
\begin{tabular}{|c|c|c||c|}
\hline
\textbf{energy} &\textbf{value}&\textbf{variance}&\textbf{Wavefunction}\\
\textbf{level} &\textbf{(Fredholm)}&\textbf{(Fredholm)}&\textbf{zeros}\\
\hline\hline
\textrm{${\cal E}_0$}&\textrm{0.662873}&\textrm{0.035562}&\textrm{0.785914}\\
\hline
\textrm{${\cal E}_1$}&\textrm{0.942654}&\textrm{0.007705}&\textrm{0.992520}\\
\hline
\textrm{${\cal E}_2$}&\textrm{1.085112}&\textrm{0.003494}&\textrm{1.116604}\\
\hline 
\textrm{${\cal E}_3$}&\textrm{1.183948}&\textrm{0.002086}&\textrm{1.207184}\\
\hline
\textrm{${\cal E}_4$}&\textrm{1.260617}&\textrm{0.001412}&\textrm{1.277322}\\
\hline
\textrm{${\cal E}_5$}&\textrm{1.323784}&\textrm{0.001034}&\textrm{1.337170}\\
\hline
\textrm{${\cal E}_6$}&\textrm{1.377675}&\textrm{0.000795}&\textrm{1.388628}\\
\hline
\textrm{${\cal E}_7$}&\textrm{1.424785}&\textrm{0.000632}&\textrm{1.433868}\\
\hline
\textrm{${\cal E}_8$}&\textrm{1.466769}&\textrm{0.000515}&\textrm{1.474301}\\
\hline
\textrm{${\cal E}_9$}&\textrm{1.504666}&\textrm{0.000438}&\textrm{1.510901}\\
\hline
\textrm{${\cal E}_{10}$}&\textrm{1.539360}&\textrm{0.000360}&\textrm{1.544367}\\
\hline
\textrm{${\cal E}_{11}$}&\textrm{1.571591}&\textrm{0.000112}&\textrm{1.575221}\\
\hline
\textrm{${\cal E}_{12}$}&\textrm{1.599964}&\textrm{$<$0.000112}&\textrm{1.603861}\\
\hline
\textrm{${\cal E}_{13}$}&\textrm{1.623555}&\textrm{$<$0.000112}&\textrm{1.630601}\\
\hline
\textrm{${\cal E}_{14}$}&\textrm{1.655204}&\textrm{$<$0.000112}&\textrm{1.655689}\\
\hline
\end{tabular}
\includegraphics[width=0.485\textwidth]{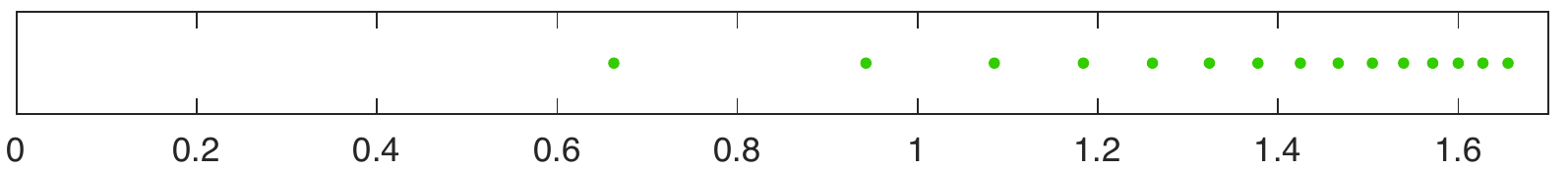}
\end{center}
\caption{The first 15 mean levels ${\cal E}_n=\langle E_n\rangle$ for JT gravity at $\hbar=1$, computed using  the Fredholm determinant method. The computed values are shown along with the (decreasing with energy) variance. The 4th column  shows the vanishing of the wavefunction $\psi(E,\mu)$, giving an approximation to the peaks' locations that improves as $n$ grows larger.}
\label{tab:mean-levels-fredholm}
\end{table}
\endgroup

As was discussed in section~\ref{sec-free-energy}, and tested on the example of the Airy model, the free energy of the model can be constructed from the knowledge of the spectrum's statistics. The quenched free energy from direct sampling of reconstructed samples was first computed in this manner in ref.~\cite{Johnson:2021zuo}, and as with the Airy case, it comes in slightly below the result for computing the free energy of the mean spectrum. They both asymptote to the annealed free energy as they should.  See  figure~\ref{fig:JT_gravity_free_energy}.  Henceforth, the free energy of the mean spectrum will be taken as a guide to the quenched free energy, and that will be displayed in further examples to come, alongside the annealed quantity computed from $\rho(E)$.  
\begin{figure}[b]
\centering
\includegraphics[width=0.48\textwidth]{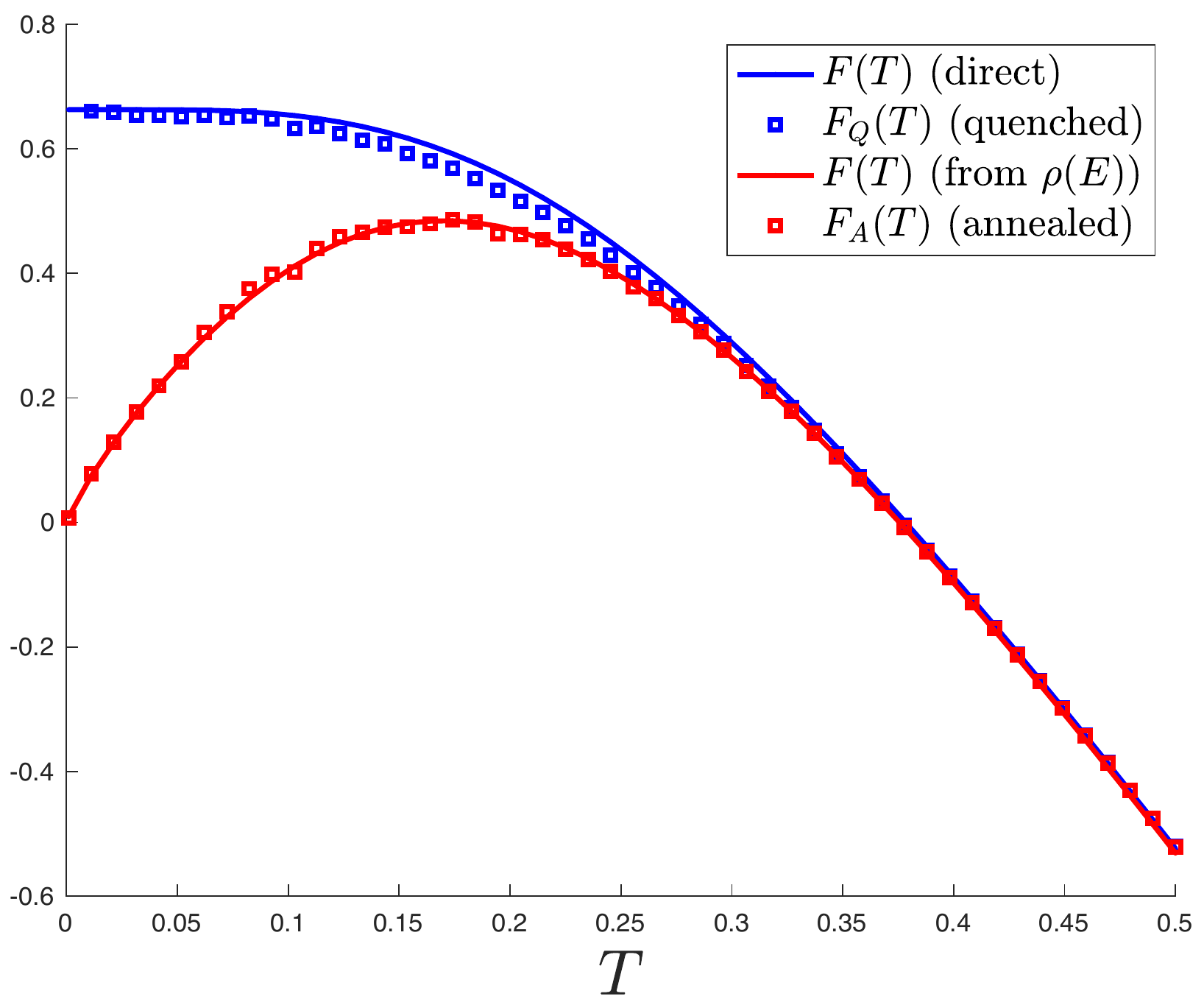}
\caption{\label{fig:JT_gravity_free_energy} The JT gravity free energy, computed from the direct spectrum ${\cal E}_n$, or as a quenched quantity $F_Q(T)$  in the matrix model. The truncation level is $n=155$.  Both cases become $\langle E_0\rangle\simeq 0.6628$. Also shown is the annealed quantity $F_A(T)$, as well as the naive free energy computed using the log of the partition function computed from $\rho(E)$ (integrated up to a similar energy).  Up to this $T$ there is hardly any further change to the quantities if higher energy levels are included.}
\end{figure}

Finally, it is informative to turn to the spectral form factor. What was done for the simple Airy model in section~\ref{sec:spectral-form-factor-airy} can also be done here. On the one hand, the Kernel can be used (see equations~(\ref{eq:sff-averaged}),~(\ref{eq:full-two-loops-expression}) and~(\ref{eq:correlator-connected})) to construct the smooth ensemble-averaged spectral form factor for JT gravity (this was first done fully non-perturbatively in ref.~\cite{Johnson:2020exp}), and on the other hand, the spectral form factor for the distinguished spectrum $\{ {\cal E}_n \}$ can be constructed. The two are shown in figure~\ref{fig:JT_spectral_form_factor_compare} for $\beta=\frac12$.\footnote{Notice that in ref.~\cite{Johnson:2020exp}, many of the results for the spectral form factor were presented for higher values of $\beta$, even up to 50. Since this corresponds to temperatures well below the mean of the first microstate peak, it is clear--in retrospect--that the resulting behaviour probed extremely low energy features of the behaviour of the ensemble well below the typical ground state energy. $\beta=0.5$ allows for the interplay of several of the lowest excited states to imprint strongly upon the behaviour of the spectral form factor, as can be seen by comparing the results. It  has become considerably easier to obtain the numerical control needed to obtain clean results at small $\beta$ since that work was published.}  

\begin{figure}[t]
\centering
\includegraphics[width=0.48\textwidth]{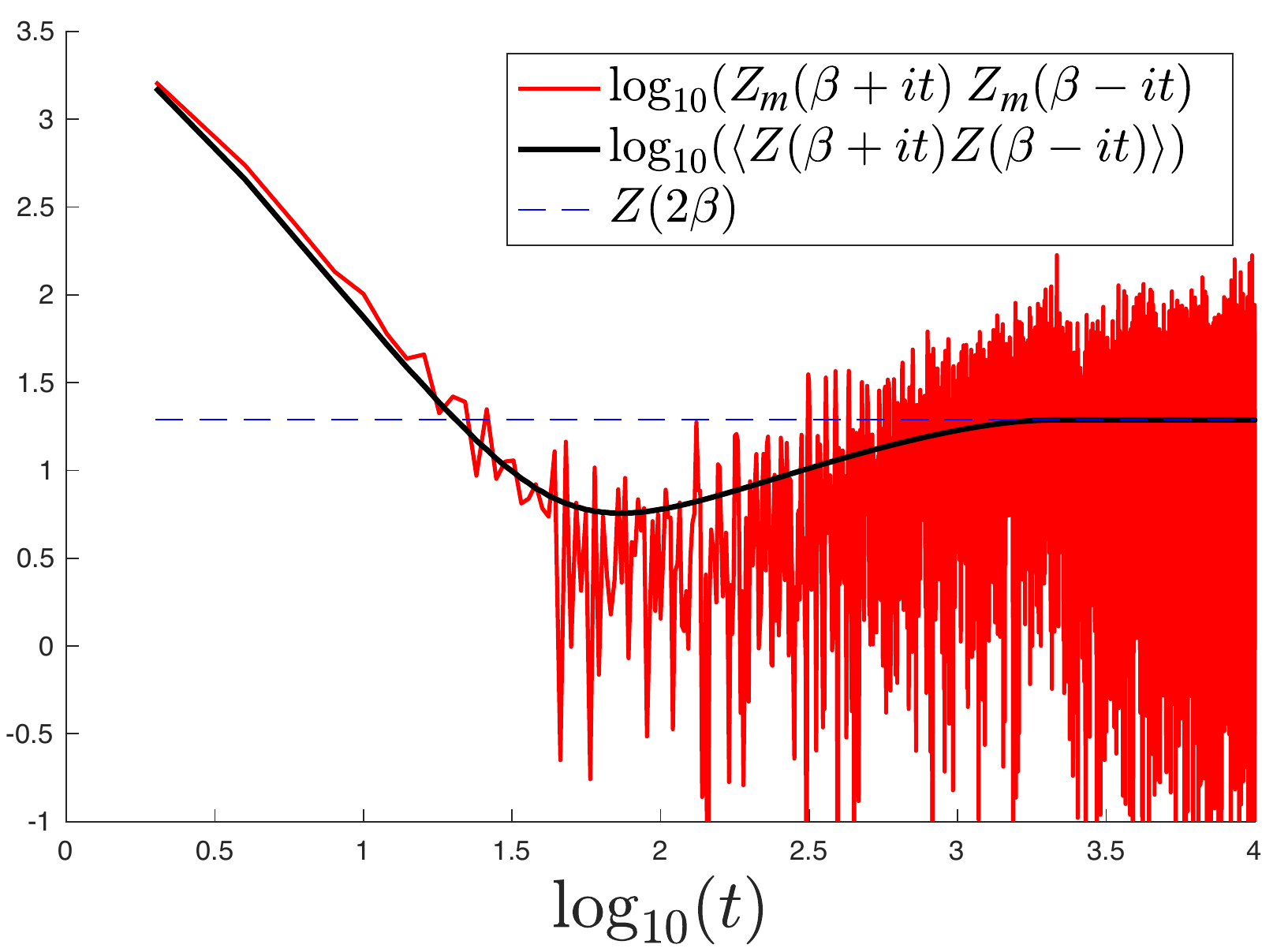}
\caption{\label{fig:JT_spectral_form_factor_compare} The spectral form factor for the mean spectrum $\{ {\cal E}_n\}$ (red, jagged) plotted against the ensemble averaged quantity, for JT gravity (black, smooth). The two curves follow each other closed at early times before the red curve begins fluctuations. The averaged quantity saturates to the plateau value $Z(2\beta)$. Here $\beta{=}1/2$, and $\hbar{=}1.$}
\end{figure}

 Again, this amounts to a lovely interplay between geometry and statistics. The connected contributions that produce the ramp and plateau features that result from averaging over the spectral form factors of the ensemble can indeed be attributed to a gravity wormhole, {\it via} the usual 't Hooftian interpretation of the double-scaled matrix model. This is makes sense now that there is an actual gravity counterpart of the matrix model. In the case of Airy, there's none of the {\it geometry} of gravity, just  topological counting, as can be seen by expanding the result~(\ref{eq:sff-exact-airy}) in  powers of $\hbar$, which are all even, equivalent to adding handles. On the other hand, it is averaging (in the sense of Wigner) over the ensemble of possible spectral form factors.

Again, it is worth repeating that there is no sense in which this is a requirement of the theory. It's just that a single spectral form factor requires knowledge of the microstate spectrum that makes it up. They are D-brane like objects associated with the definite energies that can be identified using the techniques see here, and they have a perfectly good description, but it is not in terms of smooth gravity. Averaging over the ensemble of possibilities smears them out,  sidestepping having to specify them, and allows for a smooth gravity description.

\section{JT Supergravity}
\label{sec:sjt-gravity}
Many of the observations made in the previous section have counterparts in various models of JT supergravity described perturbatively to all orders in the work of Stanford and Witten~\cite{Stanford:2019vob}, and it is possible to be just as explicit in constructing their non-perturbative content, as shown in refs.~\cite{Johnson:2020heh,Johnson:2020exp,Johnson:2020mwi} for the ``Type 0A'' family in the $(\mathbf{\alpha},\mathbf{\beta})=(2\Gamma+1,2)$ Altland-Zirnbauer classification, with  $\Gamma=0,+\frac12,-\frac12$, and in ref.~\cite{Johnson:2021owr} for the ``Type 0B'' family (a merging two-cut symmetric Hermitian matrix model).

The distinction between the Type~0A and Type~0B, as the name suggests, mirrors the distinction between the two types of string theory that bear the same name. Indeed  the minimal strings connected to  their matrix model description, generalizing the bosonic minimal string picture of the previous section, are of Type 0A and 0B. Denote the random matrix of one of these models as $H$. In these supersymmetric cases it can be written as the square of a supercharge, ${\cal Q}$, {\it i.e.,} $H={\cal Q}^2$. In the case of Type~0A, ${\cal Q}$ anticommutes with the operator $(-1)^{\textsf F}$, where ${\textsf F}$ is fermion number, and can be written as:
\be
{\cal Q}=\left(
\begin{matrix}
0 & M\cr
M^\dagger & 0
\end{matrix}
\right)\ ,
\ee
in a basis where $(-1)^{\textsf F}$ is block diagonal. Here $M$ is a complex matrix and $M^\dagger$ is its Hermitian conjugate. (See ref.~\cite{Stanford:2019vob} for more discussion of this.) So the type of random matrix model needed to describe such models should be of a complex matrix model $M$, but in the combination $MM^\dagger$. This is precisely the kind of matrix model whose double-scaling limit was studied long ago~\cite{Morris:1990bw,Dalley:1991qg}, and toy prototypes are the Bessel models presented in Section~\ref{sec:bessel-models}. As mentioned above, they are  notated $(\boldsymbol{\alpha},\boldsymbol{\beta})=(2\Gamma+1,2)$ in the Altland-Zirnbauer classification, although sometimes just the~$\Gamma$ values will be used.  In the case of Type~0B, ${\cal Q}$ does not respect $(-1)^{\textsf F}$, and so it is an ordinary Hermitian matrix. The matrix model of relevance isin the $\boldsymbol{\beta}{=}2$ Dyson-Wigner classification.  Ref.~\cite{Stanford:2019vob} conjectured that this should be a symmetric two-cut model, and this was recently confirmed in ref.~\cite{Johnson:2021owr}, where it was constructed fully non-perturbatively.  The matrix model of ${\cal Q}$, with eigenvalues $q$, has a spectrum symmetric in $q\to-q$, precisely what would arise from the merging of two cuts in a symmetric potential. The spectrum of $H={\cal Q}^2$ lies on the real positive line.

\begin{figure*}[t]
\centering
\includegraphics[width=0.45\textwidth]{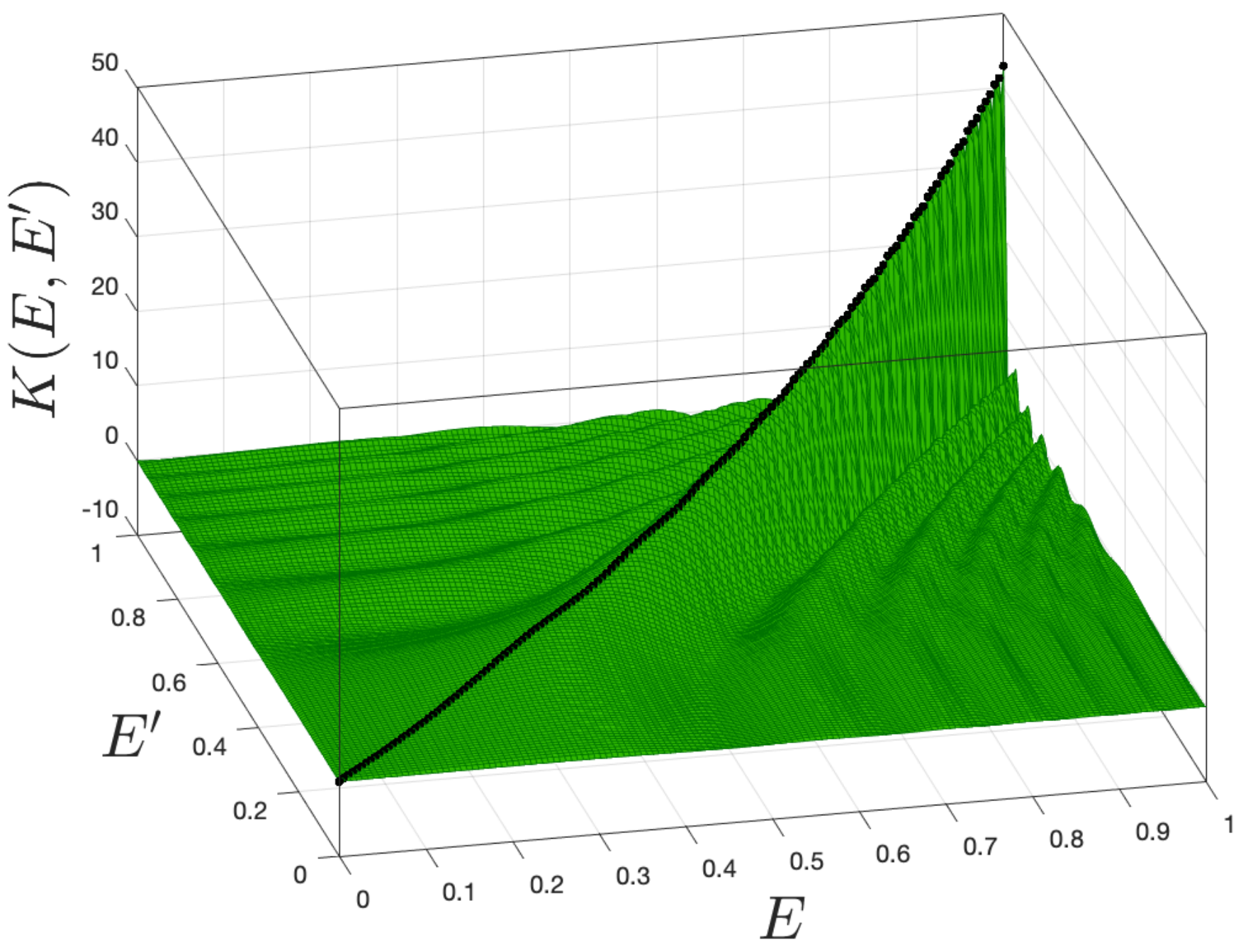}
\hskip1.5cm 
\includegraphics[width=0.45\textwidth]{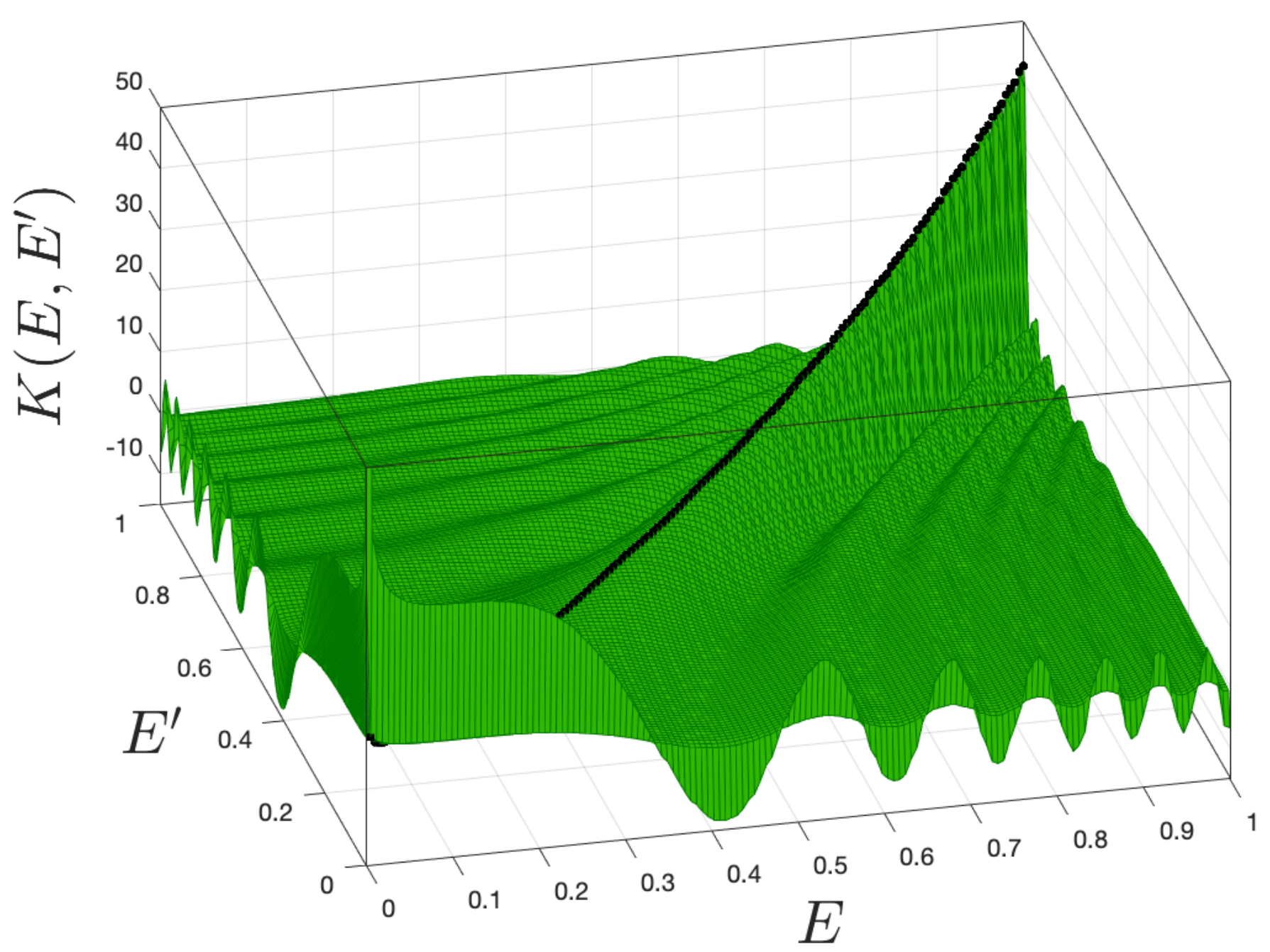}
\caption{\label{fig:JT_supergravity_22_kernel_3d} The kernel $K(E,E^\prime)$ for  $(2,2)$ JT supergravity (left) and $(0,2)$ JT supergravity (right). A notable feature of this model is an exact vanishing at~$E=0$. On the other hand, there is a divergence there for the (0,2) case. Here $\hbar=1$.}
\end{figure*}

Rather conveniently, the string equation that describes $u(x)$ for these supergravity cases has the same structure as equation~(\ref{eq:string-equation-big}) , but now has non-zero $\Gamma$ turned on.\footnote{It can be thought of as an integration constant in certain approaches to the derivation of the string equation, but from a direct matrix model perspective, for integer values it can be derived as resulting from using double scaled rectangular $N\times(N{+}\Gamma)$ or $(N{+}\Gamma)\times N$ matrices~\cite{Klebanov:2003wg,Myers:1991akt,Lafrance:1993wy}, depending upon the sign of $\Gamma$. It was observed in ref.~\cite{Carlisle:2005mk} that certain B\"acklund transformations change $\Gamma$ by an integer, but here, $\frac12$-integer values will be important too.} Writing it out again, it is:
\be
\label{eq:string-equation-big-2}
(u-\sigma){\cal R}^2 -\frac{\hbar^2}{2}{\cal R}{\cal R}^{\prime\prime}+\frac{\hbar^2}{4}({\cal R}^\prime)^2=\hbar^2\Gamma^2\ ,
\ee
but it is important to appreciate that the details of how the equation (and its solutions) is used are quite different in each case.  

The first key difference is that for all the supergravity cases, $\mu$ is a positive non-zero number, and  the conventions of this paper will have $\mu=1$. So the Fermi sea is now $-\infty\leq x \leq 1$. See also a comment on this after equation~(\ref{eq:Bessel-Hamiltonian}). The disc level solution $u_0(x)$ of the equation  still comes from  solving ${\cal R}_0=0$, but comes in two parts. For $x<0$ there is a non-trivial solution for $u_0(x)$, which arrives at $u_0{=}0$ when $x{=}0$. Between $x{=}0$ and $x{=}\mu({=}1)$, the leading solution is $u_0(x)=0$. The non-trivial part comes from matching to the JT supergravity disc solution for the spectral density, which is 
\be
\label{eq:leading-rho-SJT}
\rhoo(E) = \frac{\cosh(2\pi\sqrt{E})}{2\pi\sqrt{E}}\ ,
\ee
from which it can be seen from equation~(\ref{eq:leading-density}) that the $t_k$ are given in this case by~\cite{Johnson:2020heh}:
\be
\label{eq:teekay-super}
t_k = \frac{\pi^{2k}}{(k!)^2}\ ,
\ee
a different choice than that used for regular JT (see equation~(\ref{eq:teekay})). Correspondingly, $u_0(x)$ is defined by: 
\be
\label{eq:u0-equation-SJT}
I_0(2\pi\sqrt{u_0})-1+x=0\ ,
\ee
in contrast to  equation~(\ref{eq:u0-equation-JT}) which instead involves $I_1(2\pi\sqrt{u_0})$. The fact that $\mu=1$ has not made much difference so far (since at leading order $u_0(x)=0$ between $x=0$ and $x=1$, but this changes when developing perturbative corrections in $\hbar$ to get contributions from higher topologies, and of course its role is key non-perturbatively too. This makes sense in view of what emerged in  the review of subsection~\ref{sec:search-gravity}, because perturbation theory in $\hbar/x$ is now for {\it positive} $x$ whereas it was for negative~$x$ before. The leading correction to $u_0(x)$ in this regime is, for the $(2\Gamma+1,2)$ models: 
\be
 u_1(x)= \left(\Gamma^2-\frac14\right)\frac{\hbar^2}{x^2}\ ,
\ee
which was in fact the potential for the Bessel models in equation~(\ref{eq:Bessel-Hamiltonian}).
This becomes the seed, upon substitution, for the next order in perturbation theory and  it is straightforward to see with some simple algebra that  a factor $(\Gamma^2-\frac14)$ multiples every term at higher order in perturbation theory. This leads to the fact that for the cases $\Gamma=\pm\frac12$,  topological perturbation theory beyond the disc vanishes to all orders. As noted in ref.~\cite{Johnson:2020heh}, this  nicely matches what was found using the direct supergravity approach in ref.~\cite{Stanford:2019vob}. Additional special features of these two cases will emerge in what is to follow, as the microstate structure of all three $(2\Gamma+1,2)$ supergravities are uncovered in subsections~\ref{sec:Type-0A-22}~and some of subsection~\ref{sec:Type-0A-12}. 

\begin{figure*}
\centering
\includegraphics[width=0.38\textwidth]{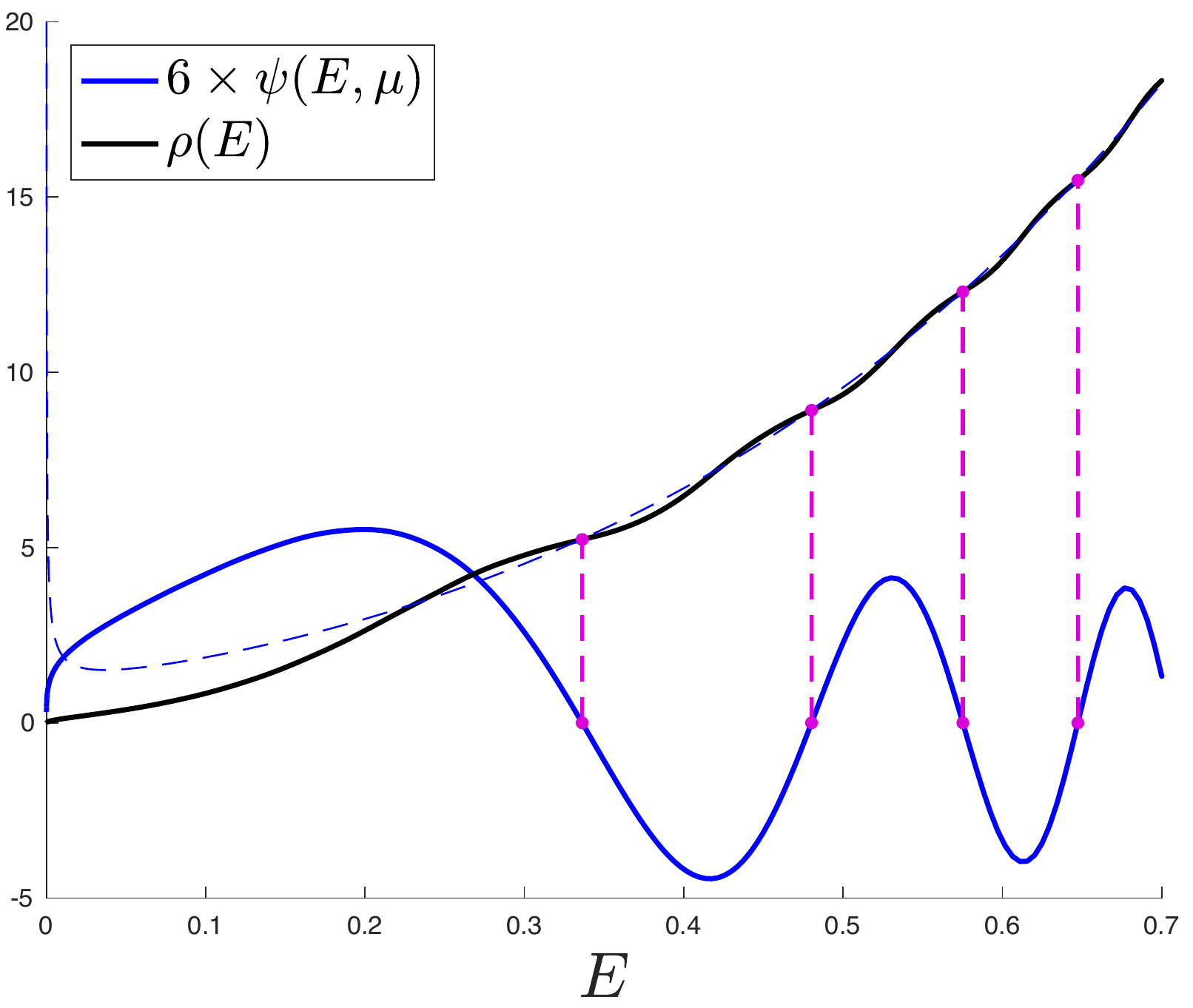} 
\hskip1.5cm
\includegraphics[width=0.38\textwidth]{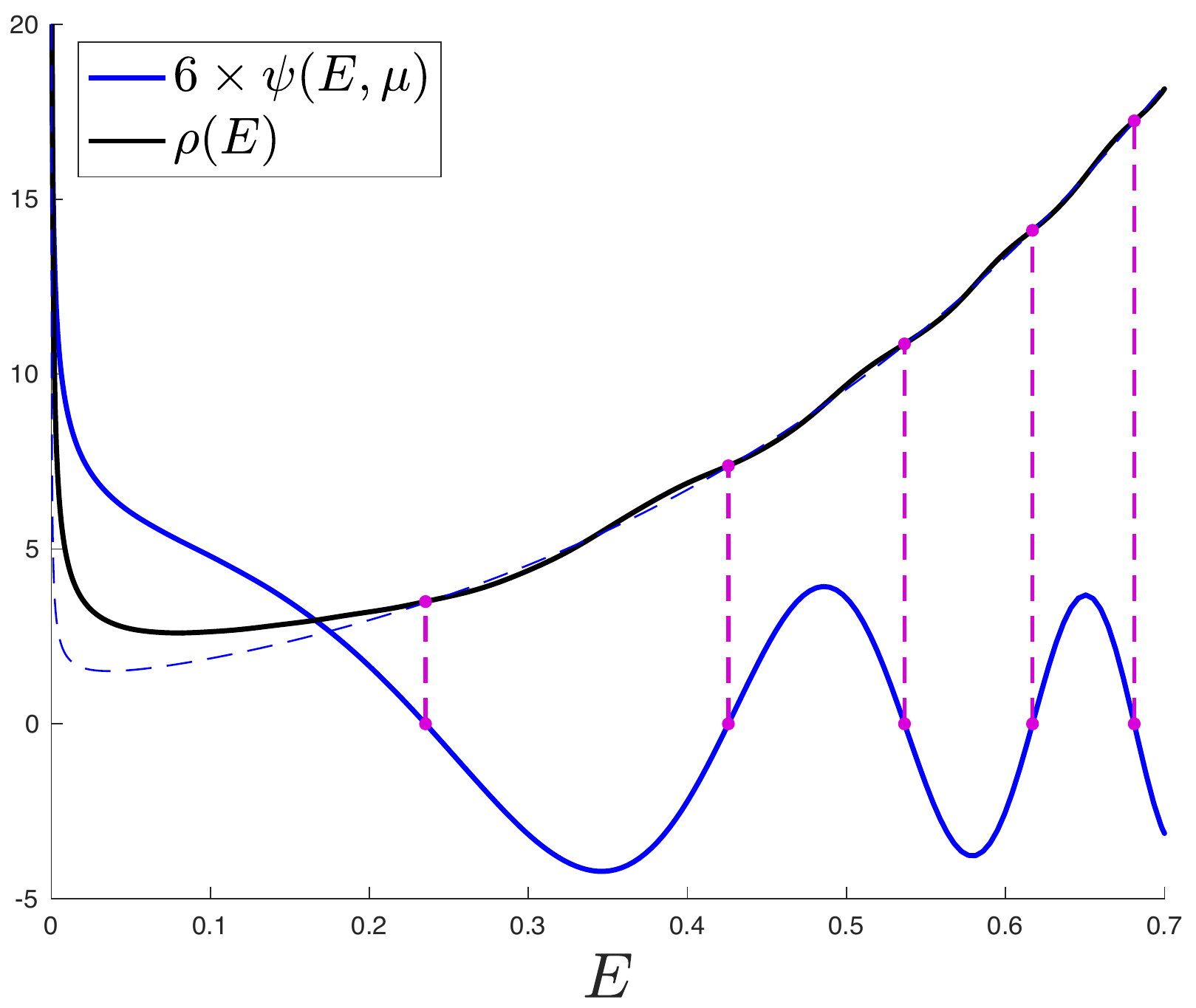}
\caption{\label{fig:JT-supergravity-22-wavefunction-zeros} The D-brane function $\psi(E,\mu)$ plotted alongside the spectral density $\rho(E)$ for (2,2) JT supergravity (left), and the (0,2) case  (right). The zeros give a first draft of the mean values of the energy levels. See text. 
}
\end{figure*}

As a bonus, these same special features of the string equation for $\Gamma=\pm\frac12$ play a role in the type~0B supergravity model described as a two-cut Hermitian matrix model. 
(This will be discussed more in subsection~\ref{sec:Type-0A-12}.) In that case, it was shown in ref.~\cite{Johnson:2021owr} that the physics can be described as a sum of two sectors, one based on a $u(x)$ solving the string equation~(\ref{eq:string-equation-big}) with $\Gamma=+\frac12$, and one based on a  $\Gamma=-\frac12$ solution. As a result, topological perturbation theory again vanishes to all orders, once again reproducing  observations made using supergravity techniques in ref.~\cite{Stanford:2019vob}. The microstate analysis for this  theory will be explored in subsection~\ref{sec:Type-0A-12}, 

As with the case of ordinary JT gravity, it is also useful to have the WKB form of the wavefunction $\psi(E,\mu)$. Using similar techniques to those used earlier (around equation~(\ref{eq:WKB2})), the appropriate form can be written as:
\bea
\label{eq:WKB-JT-supergravity}
&&\psi_{\rm WKB}(E,\mu) \simeq \\ 
&&\hskip1.0cm \frac{1}{\sqrt{\pi\hbar}}\frac{1}{E^\frac14} \cos\left(\pi\int^E \rho_0(E^\prime)dE^\prime - \frac{\pi}{4}(2\Gamma+1)\right)\ ,\nonumber
\eea
where $\rho_0(E)$ is given in equation~(\ref{eq:leading-rho-SJT}).
The normalization is such that it yields the leading spectral density {\it via} equation~(\ref{eq:leading-density}), and the phases match those of the corresponding toy Bessel models, equation~(\ref{eq:Bessel-wavefunction}).

For the Type~0A $(2\Gamma+1,2)$ theories, this form of the wavefunction will be used to complement the computation of the full wavefunction that was done deep in the non-perturbative regime, allowing it to be extended to high energies. As before, the zeros of $\psi(E,\mu)$ will turn out to be  a rough guide to the location of the mean microstates of the ensemble, increasing in accuracy with energy.  The story is more subtle for  Type~0B, and will be discussed  in section~\ref{sec:Type-0A-12}.

\begin{figure*}
\centering
\includegraphics[width=0.38\textwidth]{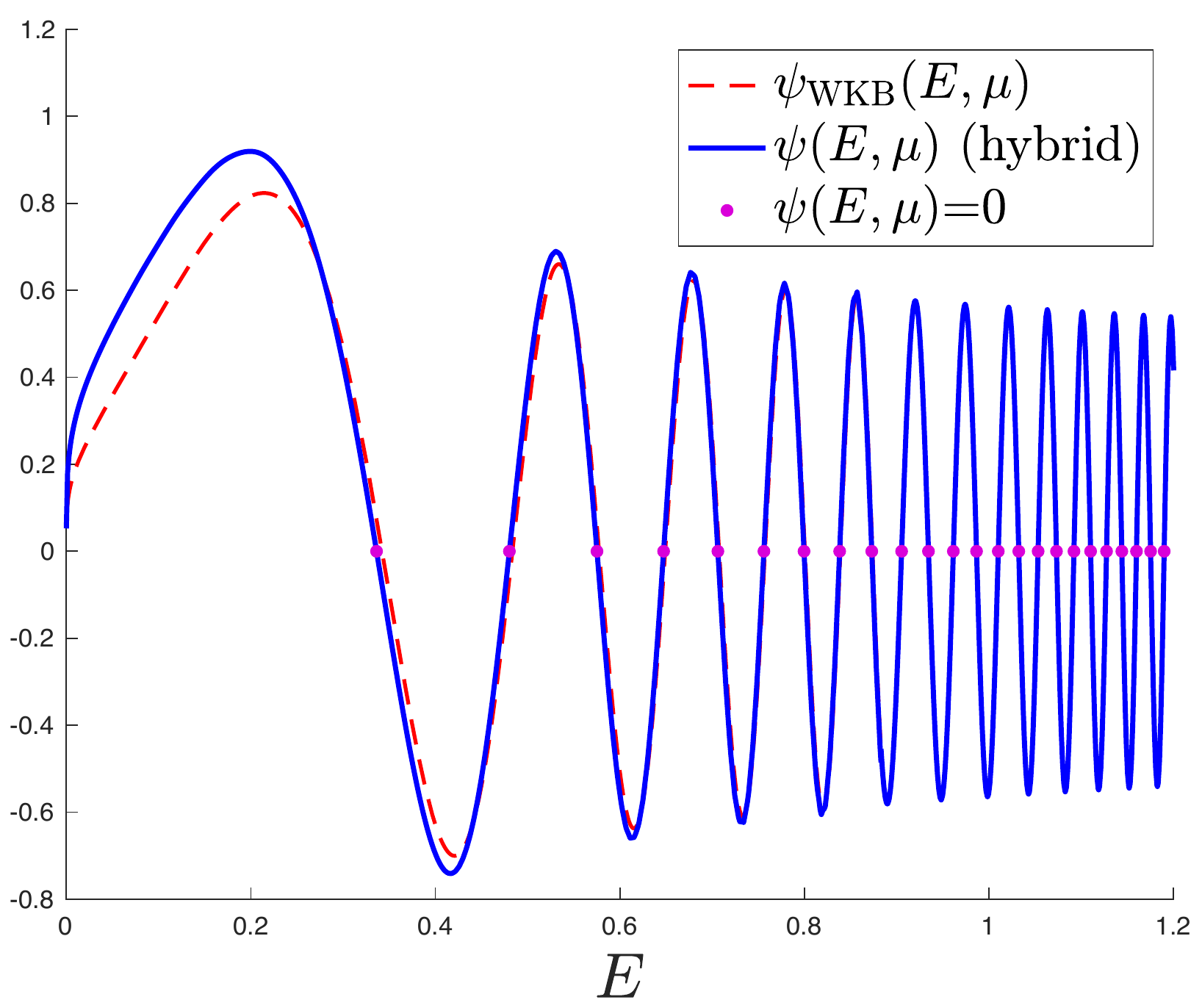}
\hskip1.5cm
\includegraphics[width=0.38\textwidth]{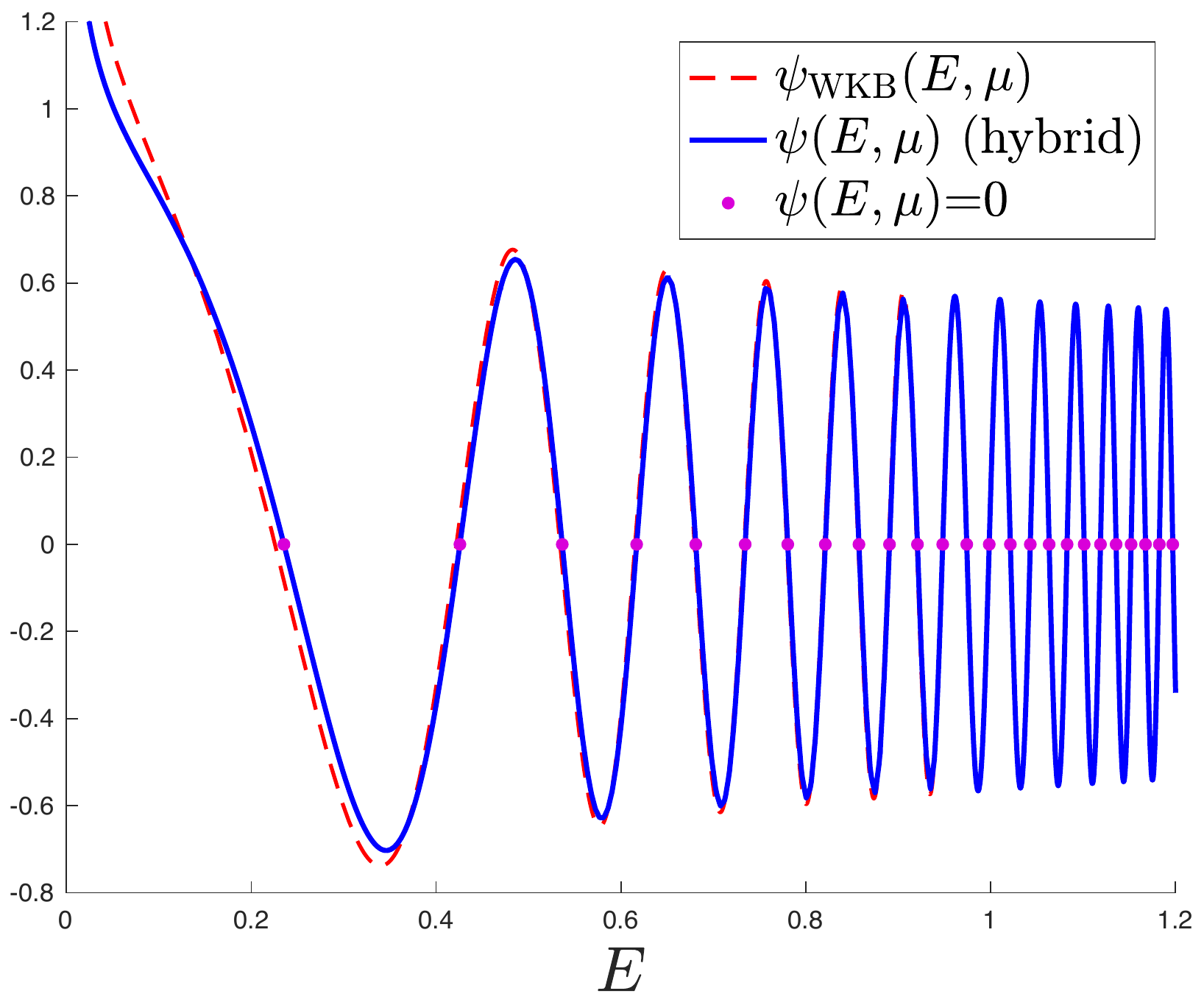}
\caption{\label{fig:JT-wavefunction-compare-22} The function $\psi(E,\mu)$ plotted alongside the WKB approximation of equation~(\ref{eq:WKB-JT-supergravity}), for (2,2) JT supergravity ($\Gamma{=}\frac12$) on the left and (0,2) JT supergravity ($\Gamma{=}{-}\frac12$) on the right.}
\end{figure*}

\subsection{$(2,2)$ and $(0,2)$ JT Supergravity\\(Type 0A, $\Gamma=\pm\frac12$)}
\label{sec:Type-0A-22}
The business of numerically solving the string equation for $u(x)$ for all the supergravity theories is much the same as it is for the ordinary JT gravity case. The truncation level chosen was, as before, at $k=7$, after which (for $\hbar=1$) the solution merges well with the classical solution at energies high enough for analytic methods to take over.  (Just as before, smaller values of $\hbar$ can readily be studied. Choosing $\hbar=1$ shows off all the key non-perturbative features to their greatest extent.) The wavefunctions are constructed as before, for 800 states, from which the kernel $K(E,E^\prime)$ defined in equation~(\ref{eq:edge-kernel}) can be constructed. 
  
Figure~\ref{fig:JT_supergravity_22_kernel_3d} shows (on the left) a 3D rendering of the kernel for the (2,2) theory. Again the black line along  the dorsal fin is the  full non-perturbative spectral density $\rho(E) = K(E,E)$. In this example, the non-perturbative physics corrects the classical divergence  at $E=0$ to zero, and correspondingly all other features off the diagonal smooth out to zero along the $E=0$ edges as well.  
The (0,2) case (shown on the right in the same figure) is quite different, with  divergences at $E=0$.

\begin{figure*}
\centering
\includegraphics[width=0.45\textwidth]{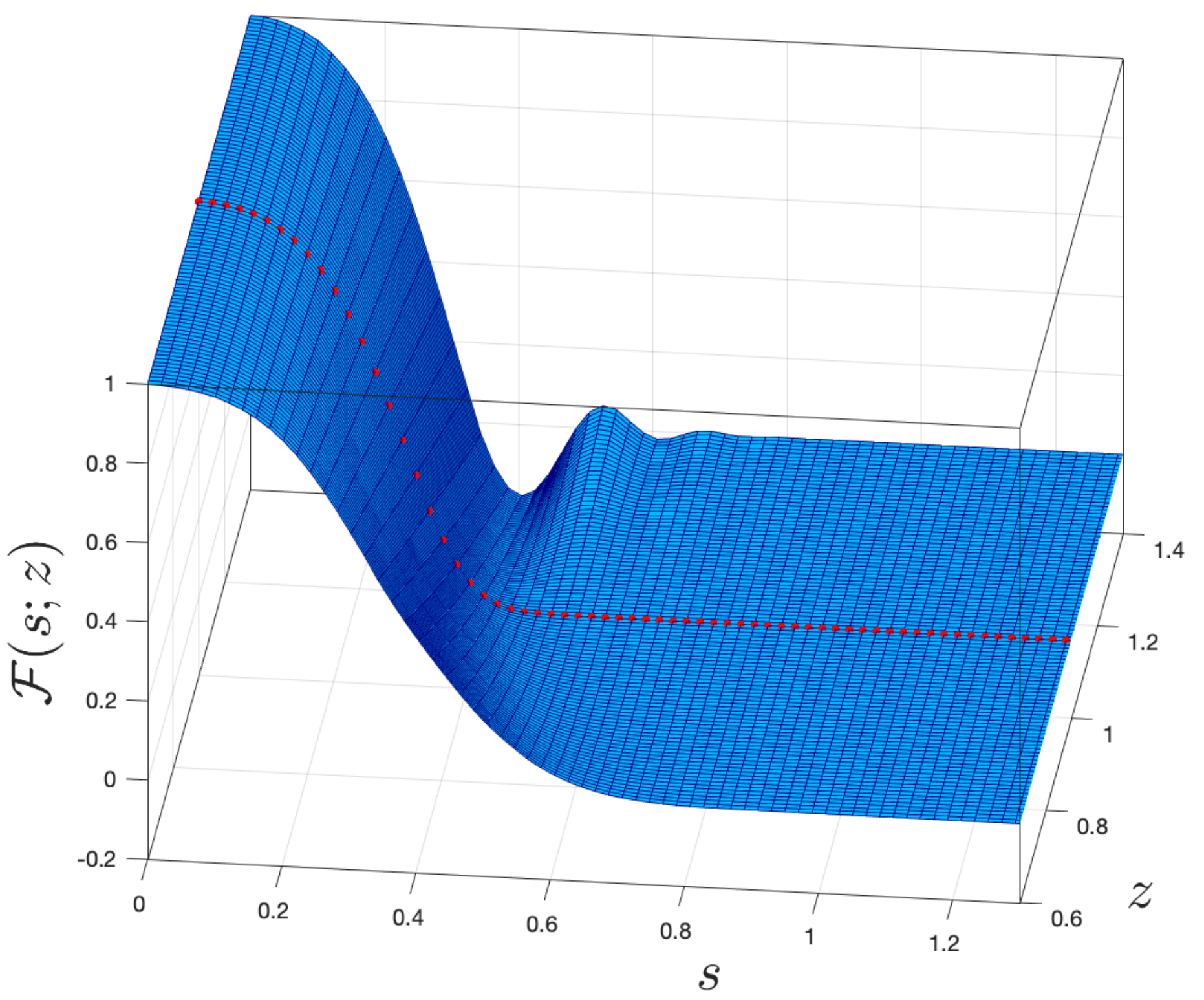}
\hskip0.5cm 
\includegraphics[width=0.48\textwidth]{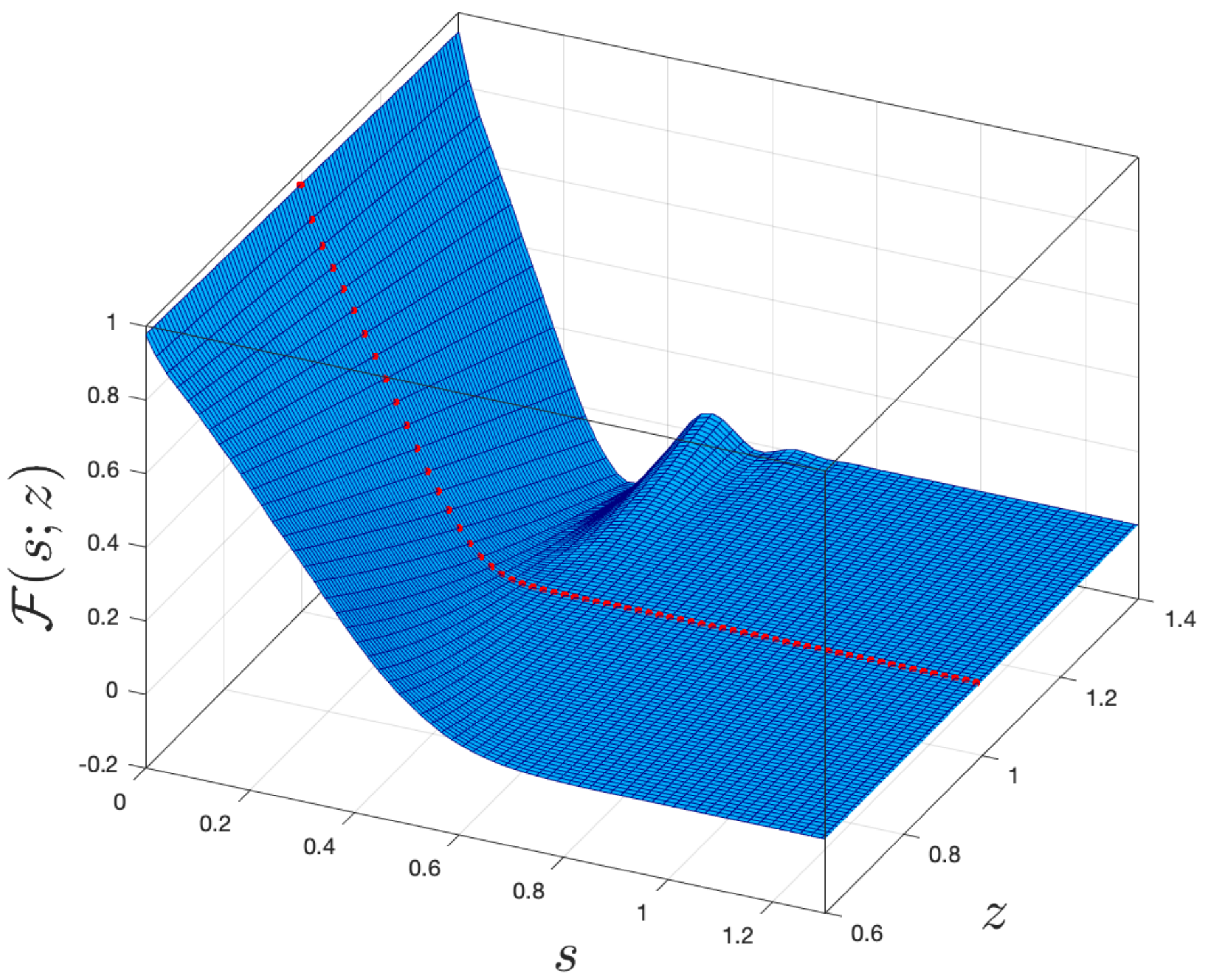}
\caption{\label{fig:JT_supergravity_22_fredholm_3d} Fredholm determinants computed for (2,2) JT supergravity (left) and (0,2) JT supergravity  on the right.}
\end{figure*}
\begin{figure*}
\centering
\includegraphics[width=0.48\textwidth]{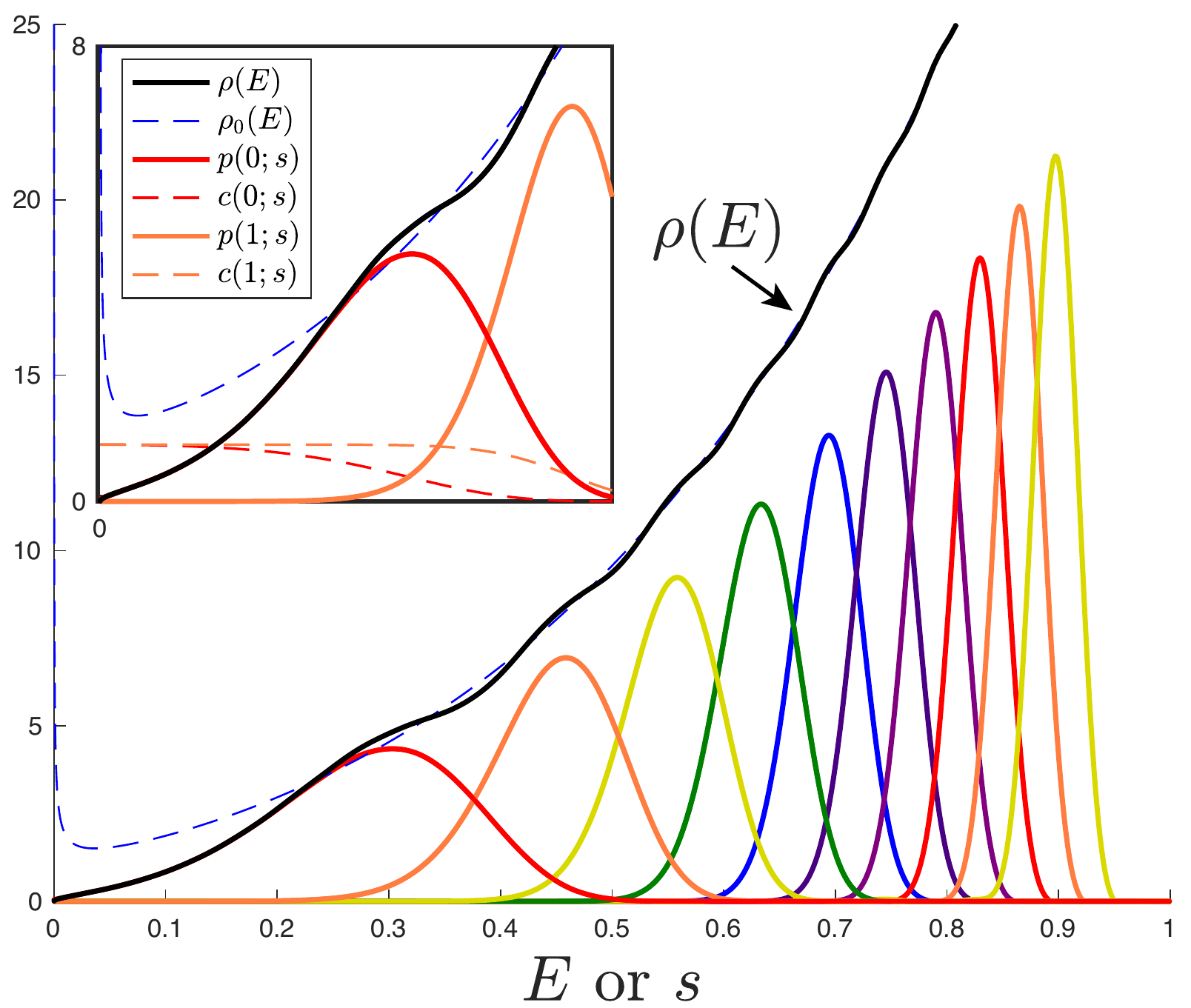}
\hskip0.5cm
\includegraphics[width=0.48\textwidth]{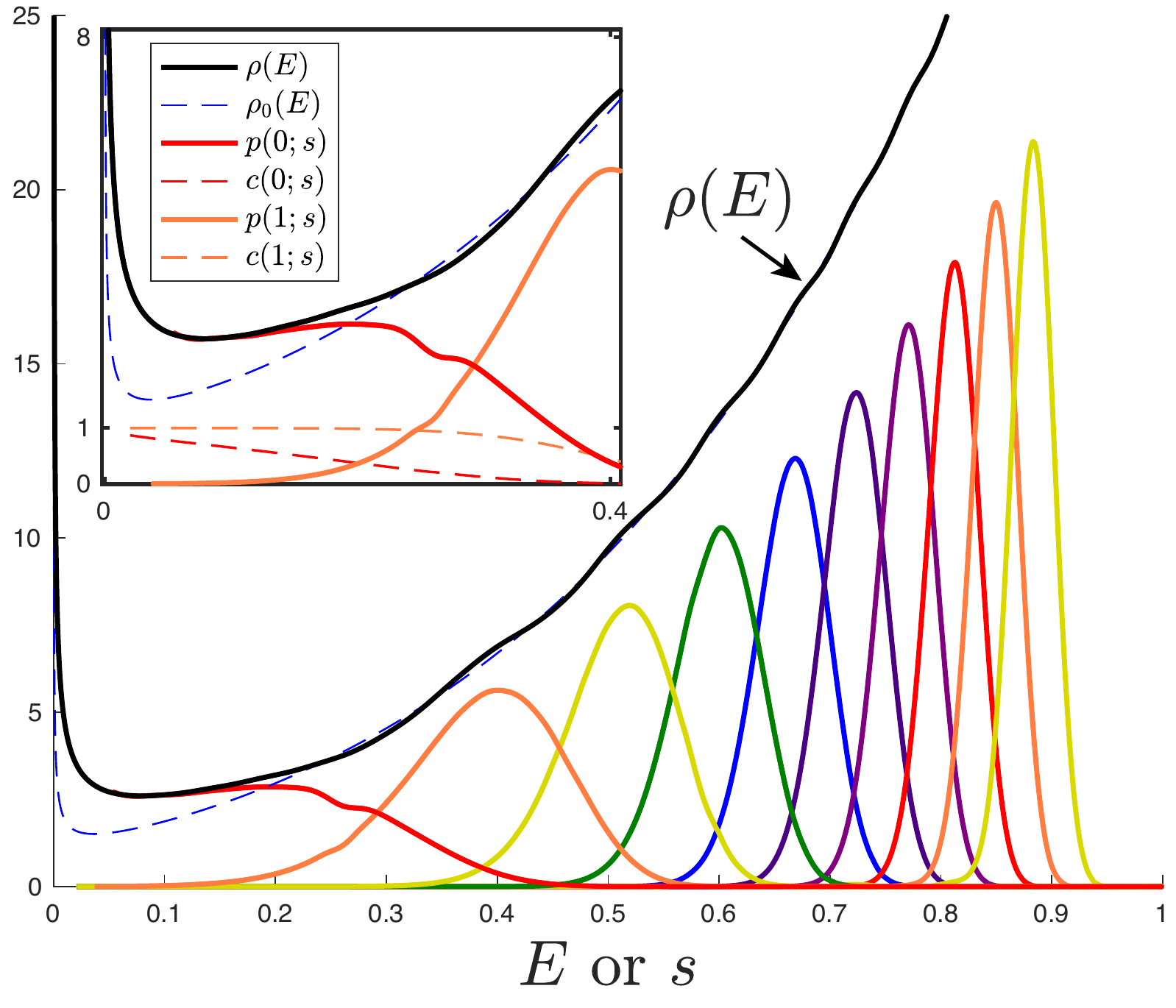}
\caption{\label{fig:JT_supergravity_22_microstates_10} The statistical distribution of the first ten energy levels  for  (left) (2,2) JT supergravity, and (right) (0,2) JT supergravity  shown alongside the leading ($\rho_0(E)$) and full ($\rho(E)$) spectral density. Here $\hbar{=}1$..}
\end{figure*}
%
 
\begin{table*}
\begin{center}
\begin{tabular}{|c|c|c||c|}
\hline
\textbf{energy} &\textbf{value}&\textbf{variance}&\textbf{Wavefunction}\\
\textbf{level} &\textbf{(Fredholm)}&\textbf{(Fredholm)}&\textbf{zeros}\\
\hline\hline
\textrm{${\cal E}_0$}&\textrm{0.276334}&\textrm{0.008374}&\textrm{0.336230}\\
\hline
\textrm{${\cal E}_1$}&\textrm{0.446558}&\textrm{0.003475}&\textrm{0.480204}\\
\hline
\textrm{${\cal E}_2$}&\textrm{0.551743}&\textrm{0.001915}&\textrm{0.575141}\\
\hline 
\textrm{${\cal E}_3$}&\textrm{0.629266}&\textrm{0.001250}&\textrm{0.647355}\\
\hline
\textrm{${\cal E}_4$}&\textrm{0.691330}&\textrm{0.000895}&\textrm{0.706188}\\
\hline
\textrm{${\cal E}_5$}&\textrm{0.743441}&\textrm{0.000681}&\textrm{0.756067}\\
\hline
\textrm{${\cal E}_6$}&\textrm{0.788520}&\textrm{0.000540}&\textrm{0.799550}\\
\hline
\textrm{${\cal E}_7$}&\textrm{0.828300}&\textrm{0.000443}&\textrm{0.838171}\\
\hline
\textrm{${\cal E}_8$}&\textrm{0.864245}&\textrm{0.000346}&\textrm{0.872967}\\
\hline
\textrm{${\cal E}_9$}&\textrm{0.897853}&\textrm{0.000045}&\textrm{0.904685}\\
\hline
\end{tabular}
\hskip1cm
\begin{tabular}{|c|c|c||c|}
\hline
\textbf{energy} &\textbf{value}&\textbf{variance}&\textbf{Wavefunction}\\
\textbf{level} &\textbf{(Fredholm)}&\textbf{(Fredholm)}&\textbf{zeros}\\
\hline\hline
\textrm{${\cal E}_0$}&\textrm{0.159610}&\textrm{0.009118}&\textrm{0.235327}\\
\hline
\textrm{${\cal E}_1$}&\textrm{0.384126}&\textrm{0.005324}&\textrm{0.425747}\\
\hline
\textrm{${\cal E}_2$}&\textrm{0.509673}&\textrm{0.002554}&\textrm{0.536497}\\
\hline 
\textrm{${\cal E}_3$}&\textrm{0.597023}&\textrm{0.001554}&\textrm{0.616981}\\
\hline
\textrm{${\cal E}_4$}&\textrm{0.664937}&\textrm{0.001068}&\textrm{0.680966}\\
\hline
\textrm{${\cal E}_5$}&\textrm{0.720957}&\textrm{0.000790}&\textrm{0.734411}\\
\hline
\textrm{${\cal E}_6$}&\textrm{0.768853}&\textrm{0.000614}&\textrm{0.780500}\\
\hline
\textrm{${\cal E}_7$}&\textrm{0.810801}&\textrm{0.000498}&\textrm{0.821139}\\
\hline
\textrm{${\cal E}_8$}&\textrm{0.848416}&\textrm{0.000382}&\textrm{0.857523}\\
\hline
\textrm{${\cal E}_9$}&\textrm{0.883475}&\textrm{0.000042}&\textrm{0.890567}\\
\hline
\end{tabular}
\includegraphics[width=0.485\textwidth]{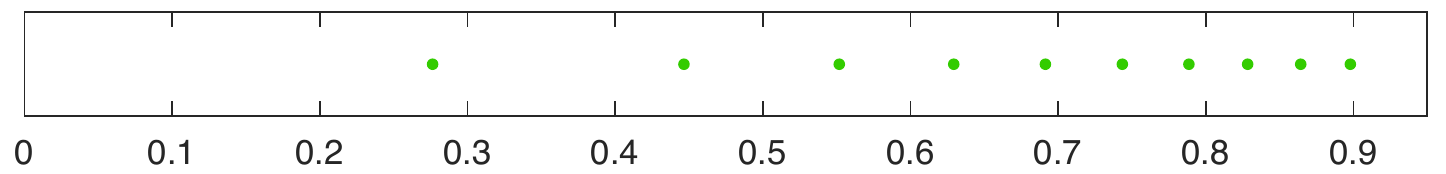}
\includegraphics[width=0.485\textwidth]{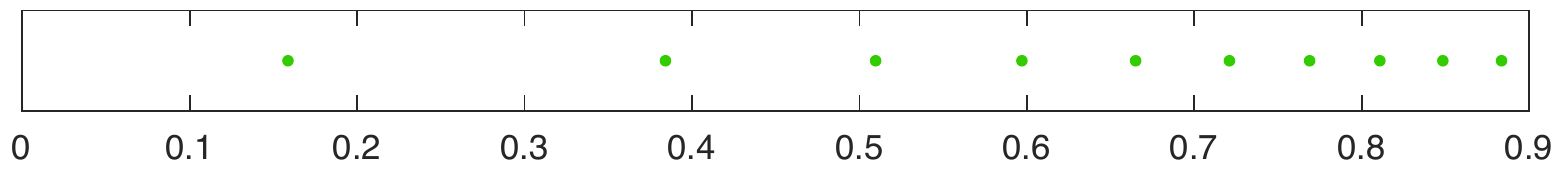}
\end{center}
\caption{The table on the left shows the first 10 mean levels ${\cal E}_n=\langle E_n\rangle$ for the (2,2) JT supergravity at $\hbar=1$, computed using  the Fredholm determinant method. The computed values are shown along with the (decreasing with energy) variance. The 4th column  shows the vanishing of the D-brane wavefunction, giving an approximation to the peaks' locations that improves as $n$ grows larger. The table on the right shows the corresponding quantities for the (0,2) case.}
\label{tab:mean-levels-fredholm-JT-supergravity-22}
\end{table*}
%

%
\begin{figure*}
\centering
\includegraphics[width=0.45\textwidth]{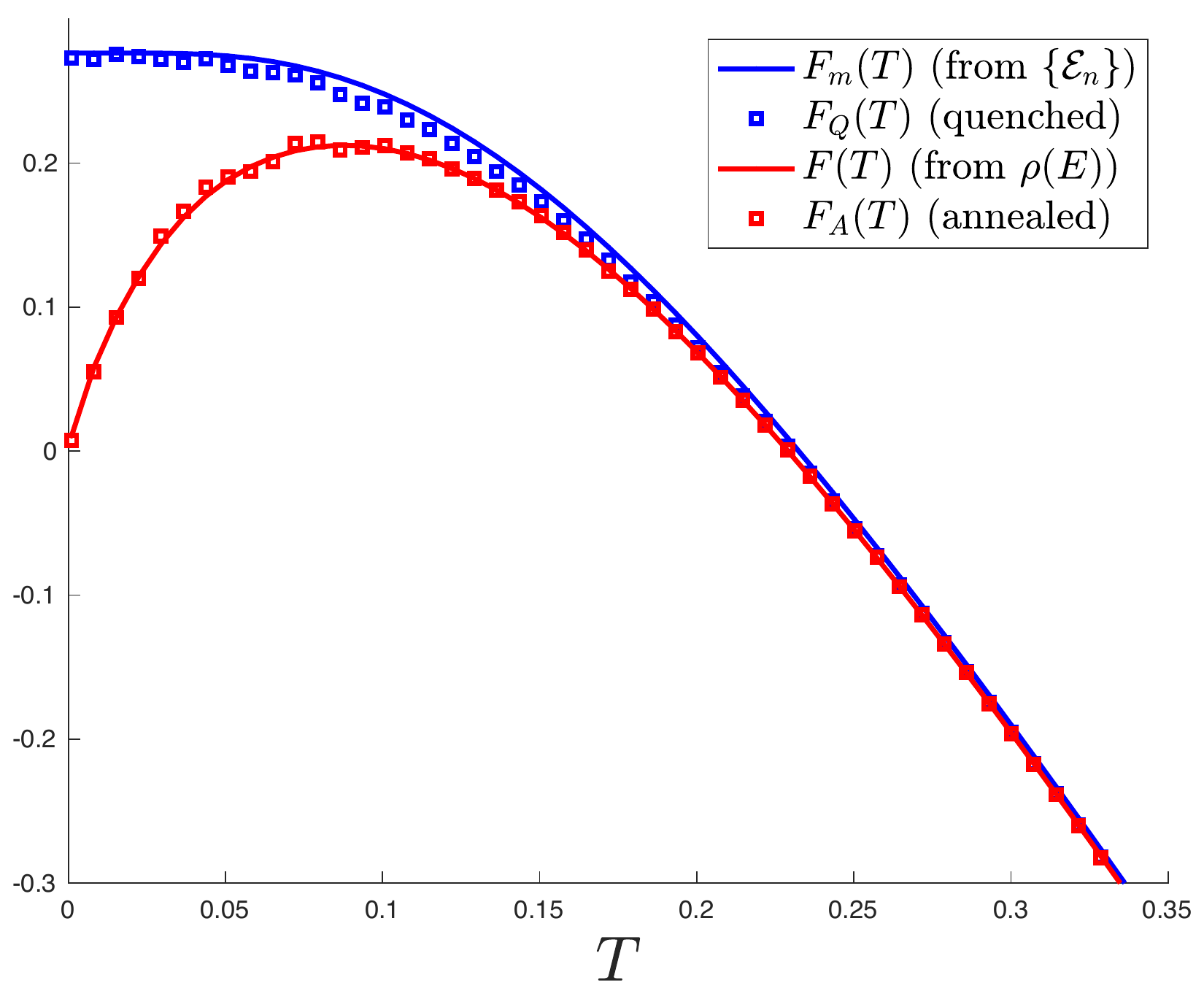}
\hskip1.5 cm
\includegraphics[width=0.45\textwidth]{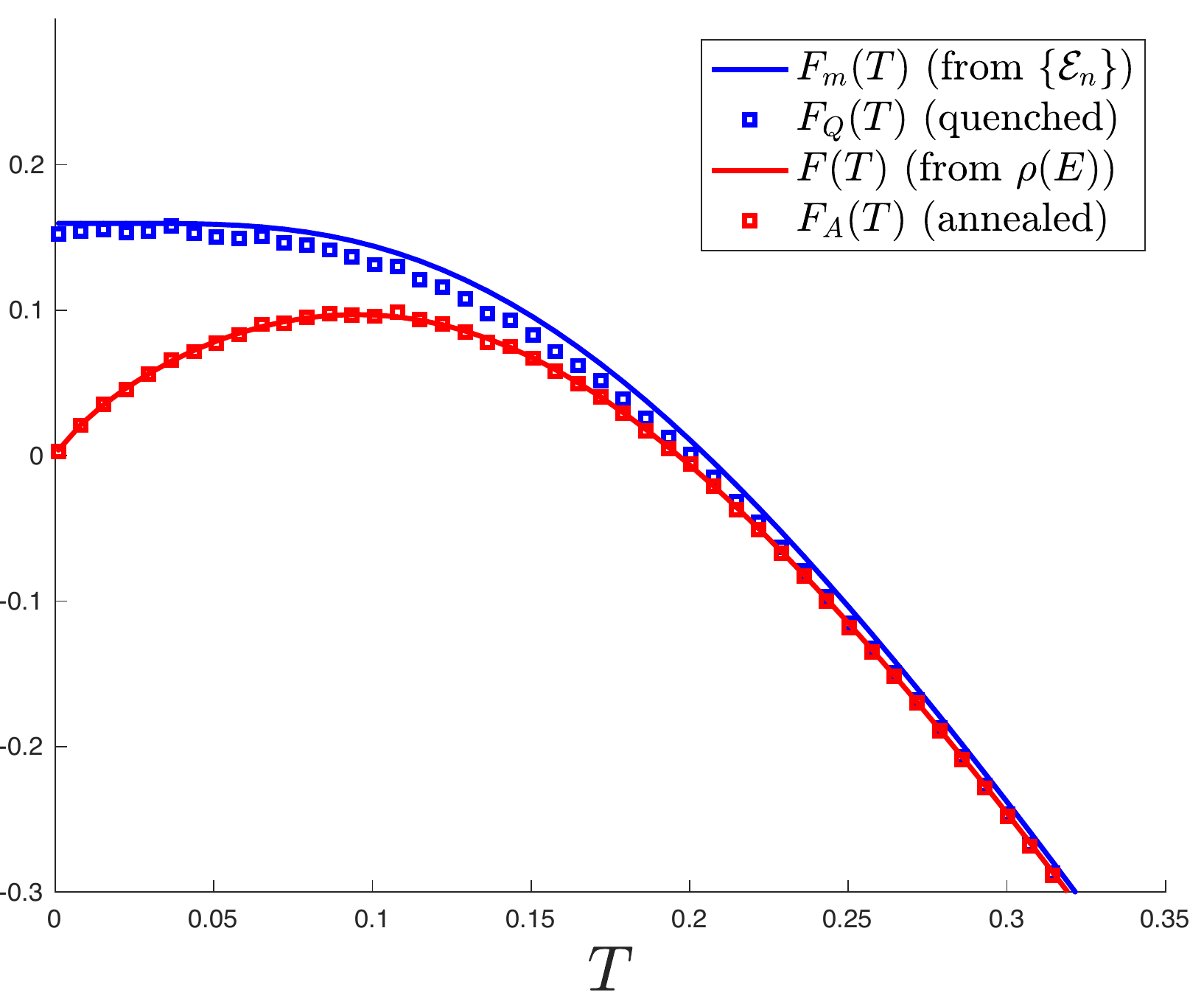}
\caption{\label{fig:JT_supergravity_22_free_energy} The (2,2) JT supergravity free energy (left), computed from the direct spectrum ${\cal E}_n$, or as a quenched quantity $F_Q(T)$  in the matrix model. The truncation level is $n=34$.  Both cases become $\langle E_0\rangle\simeq 0.2763$. Also shown is the annealed quantity $F_A(T)$, as well as the naive free energy computed using the log of the partition function computed from $\rho(E)$ (integrated up to a similar energy). On the right is the (0,2) case, with   $\langle E_0\rangle\simeq 0.1596$.}
\end{figure*}

\begin{figure*}
\centering
\includegraphics[width=0.48\textwidth]{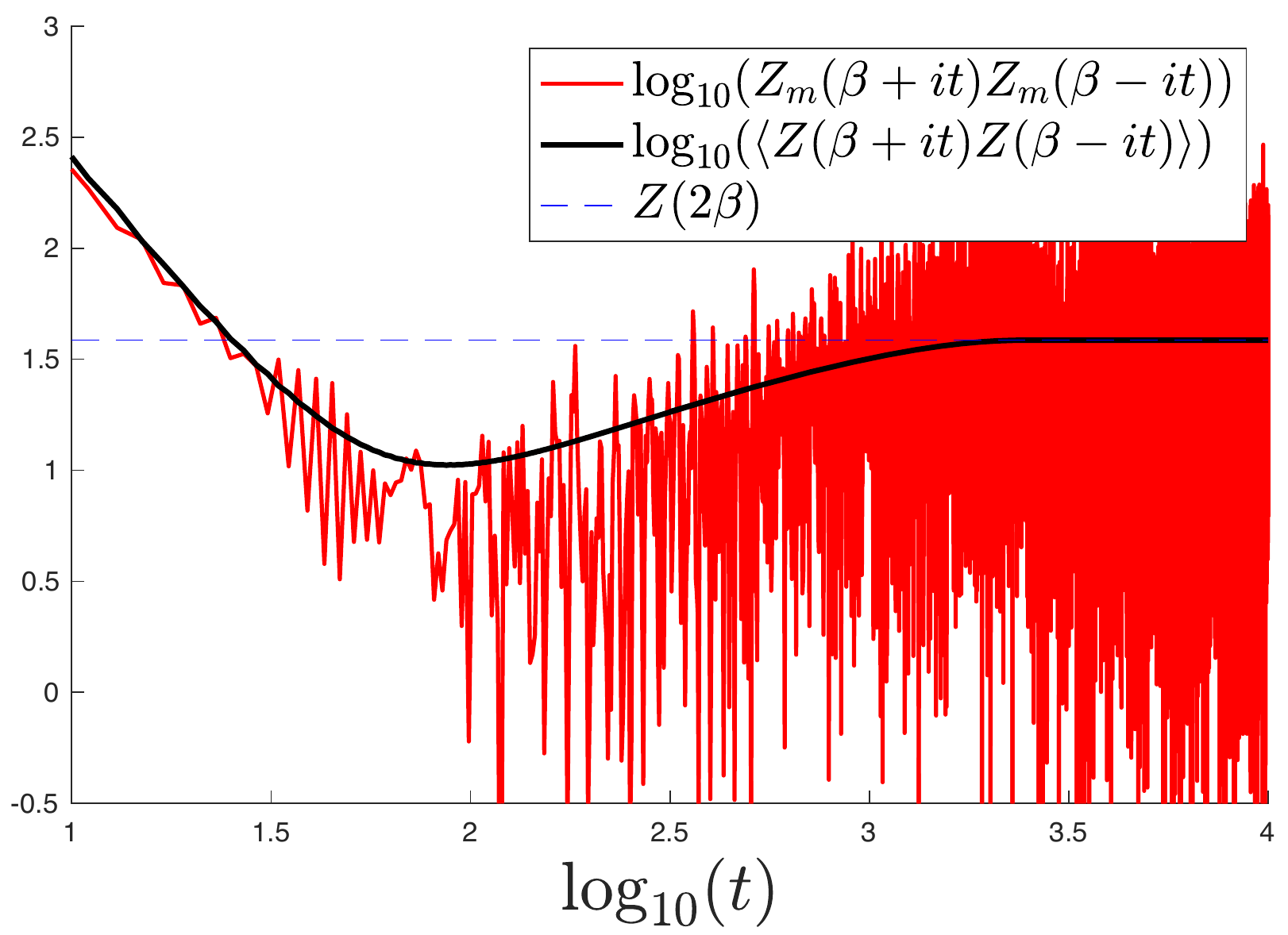}
\hskip0.5cm
\includegraphics[width=0.48\textwidth]{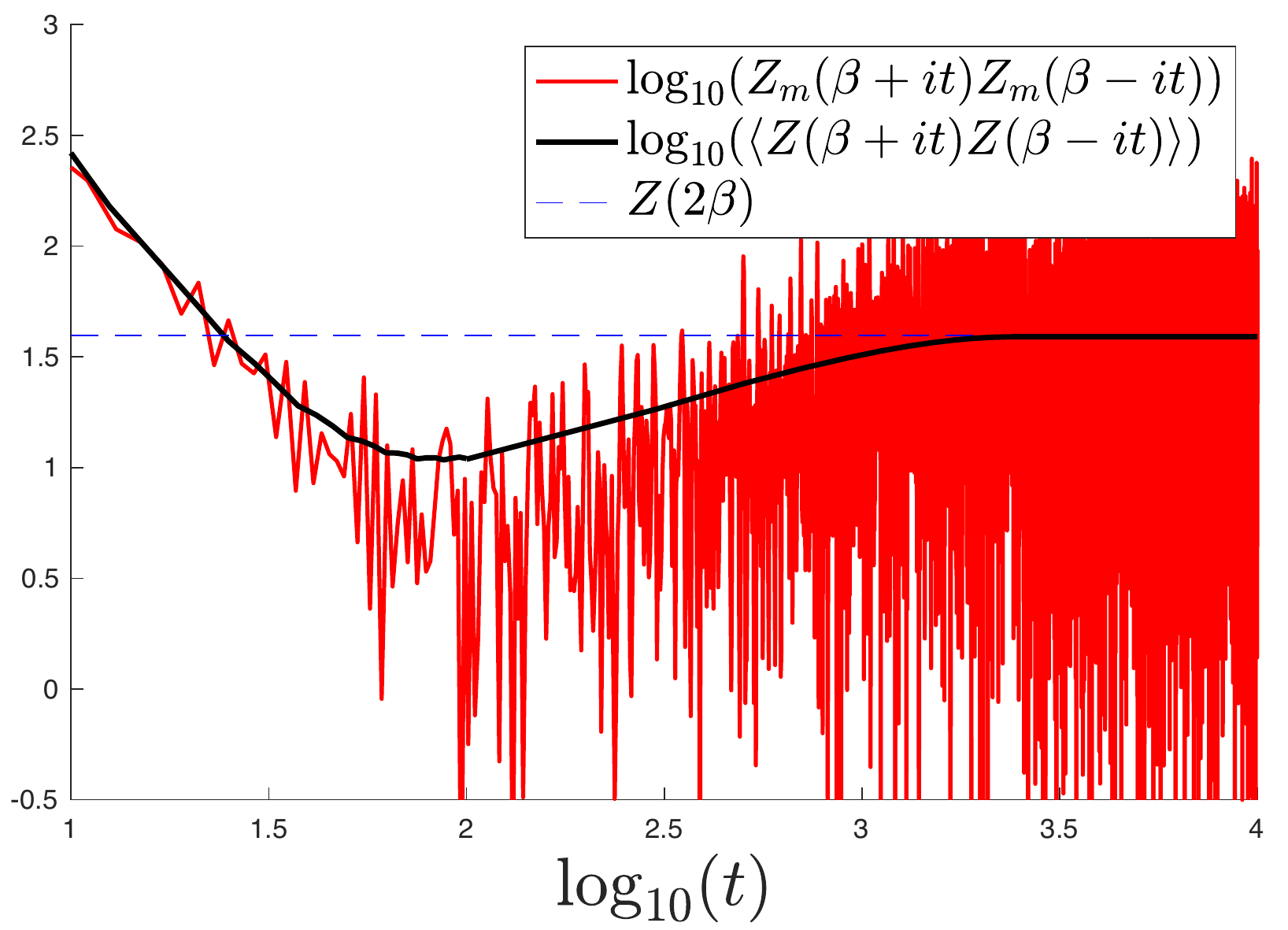}
\caption{\label{fig:JT_supergravity_22_spectral_form_factor_compare}  The spectral form factor for the mean spectrum $\{ {\cal E}_n\}$ (red, jagged) plotted against the ensemble averaged quantity, for JT supergravity (black, smooth). The (2,2) case is on the left and the (0,2) on the right. The two curves follow each other closed at early times before the red curve begins fluctuations. The averaged quantity saturates to the plateau value $Z(2\beta)$. Here $\beta{=}1/2$, and $\hbar{=}1.$}
\end{figure*}

These behaviours are reflected in the wavefunction $\psi(E,\mu)$ for each case, in both the full result computed from the quantum mechanics and in the WKB result. For the (2,2) the vanishing of the WKB result follows from the fact that  putting $\Gamma=\frac12$ into equation~(\ref{eq:WKB-JT-supergravity}) results in  a leading small $E$ behaviour of the cosine of $E^\frac12$, giving an $E^\frac14$ vanishing overall. The vanishing happens for the full (not just WKB) wavefunction as well. On the other hand, the (0,2) case is divergent near $E=0$, both for WKB and the full result. This low energy (qualitative) agreement near $E=0$ is a special case for these models.

 As they should, the zeros of the full wavefunction line up with the inflection points of the full density, as shown in figure~\ref{fig:JT-supergravity-22-wavefunction-zeros}. While the full and WKB wavefunctions  only qualitatively match  at low energies, the WKB form eventually begins to match the exact result well, and can henceforth be used out to arbitrarily  high energy. Figure~\ref{fig:JT-wavefunction-compare-22} shows a hybrid formed from the exact form and the WKB result, with the pure WKB form deviating as the red dashed line at low enough energies. Also highlighted are the zeroes, which will shortly be of use.

The full collection of $\psi(E,x)$ obtained from solving the spectral problem can be used to construct the kernel $K(E,E^\prime)$, using similar techniques to the ordinary JT gravity case. (See footnote~\ref{fn:more-tips} for suggestions on numerical techniques for computing these results.) The behaviour of the Fredholm determinant that is made from $K(E,E^\prime)$ is shown in figure~\ref{fig:JT_supergravity_22_fredholm_3d} for each case. 
 A two dimensional slice  along $z=1$ (shown in red) shows the cumulative probability $E(0;s)$ for the first energy level (ground state) of the system. Derivatives of the $z$-dependence away from this slice  generates the probability distributions for higher energy levels, and the first ten of these are shown for the (2,2) and (0,2) cases in  figure~\ref{fig:JT_supergravity_22_microstates_10}. (Note that there is small, but notable, bite taken out of the probability density function of the ground state in the (0,2) case, as a result of having considerably more noise than the (2,2) case,  because of the singular behaviour at $E=0$.)


 As with ordinary JT gravity, the peaks can be seen to grow together as energy grows, while also becoming sharper. As $E$ grows,  as discussed in the Introduction, they can eventually be well approximated by a dense set of delta-functions, and a safe cross-over to the smooth language of gravity can be made. Table~\ref{tab:mean-levels-fredholm-JT-supergravity-22} lists the details of the first ten peaks, confirming this observation. Also listed are the first ten zeros of the wavefunction $\psi(E,\mu)$. For larger~$E$  they become a more accurate locator of the mean energy, and can be used as such for the higher levels in  computations below.

The free energy can be computed for these energy levels, as before, and the results shown in figure~\ref{fig:JT_supergravity_22_free_energy}. This time, the free energy associated to the mean spectrum is shown, alongside the annealed result from using the matrix model partition function. 

Finally, the spectral form factor for the mean spectra can be computed in each case, and compared with the corresponding matrix model computation. These are shown in figure~\ref{fig:JT_supergravity_22_spectral_form_factor_compare}. To repeat the point of view expressed in the Introduction, this shows the matrix model creating what might be considered a {\it non-perturbative} spacetime wormhole by averaging over an ensemble of dual (2,2) (or (0,2)) JT supergravity spectra. The proposed {\it actual} dual in each case is a single copy of the spectrum, the mean one, and is shown in red.

\subsection{$(1,2)$ JT Supergravity  (Type 0A, $\Gamma=0$) and $\mathbf{\beta}=2$ JT Supergravity (Type 0B)}
\label{sec:Type-0A-12}

The (1,2) JT supergravity, while formulated in a way that is similar to the (2,2) and (0,2) cases of the previous subsection, has results that are qualitatively more akin to the case of the type~0B supergravity if the latter's results are presented for ${\cal Q}^2$'s spectrum instead of ${\cal Q}$'s. In view of that, and in the interest of using space efficiently, the two models' results are presented together. (In fact, another reason to group things this way is that the (2,2) and (0,2) theories preserve time-reversal invariance while the others do not.)

The construction of the type~0B case needs some additional description. The fact it {\it again} uses the string equation~(\ref{eq:string-equation-big-2}) is interesting, and goes back to properties of the equation that were discovered long ago by Periwal and Shevitz.~\cite{Periwal:1990gf,Periwal:1990qb}. There, it was noticed that the well-known Miura map of KdV integrable systems, $u(x) = r(x)^2\pm\hbar r^\prime(x)$ maps solutions $u(x)$ of the string equation to solutions of the Painlev\'e~II hierarchy of equations. These had already been identified as arising from doudle scaling limits of  unitary matrix models, which obtain critical behaviour when the ends of two cuts meet. (The prototype is the Gross-Witten transition~\cite{Gross:1980he}, also studied by Wadia~\cite{Wadia:1981rb}.)
That hierarchy was also connected to double cut Hermitian matrix models by Crnkovi\'c, Douglas, and Moore~\cite{Crnkovic:1990ms}, and shown by  Hollowood et al.~\cite{Hollowood:1992xq}, to embed into the larger Zakharov-Shabat and non-linear Schr\"odinger hierarchies. Later, Klebanov et. al.~\cite{Klebanov:2003wg} connected some of these systems to type~0A and 0B minimal strings. The recent work~\cite{Johnson:2021owr} then connected the symmetric case back to the string equation~(\ref{eq:string-equation-big-2}), and worked out some extra details of the loop equations in order to then apply it to the study of the 0B JT supergravity.

The loop operator is built by summing two sectors. In the ${\cal Q}^2$ picture, one is a copy of the $\Gamma=+\frac12$ solution of the string equation and the other is a copy of the solution at $\Gamma=-\frac12$. Together they build the physics of a symmetric double-cut Hermitian matrix model:
\begin{widetext}
\be
\label{eq:full-loop-expression-0B}
\langle {\rm Tr}(\e^{\beta{\cal Q}^2})\rangle = \int_{-\infty}^1 \!\!dx \,\langle x| {\rm e}^{-\beta{\cal H}_+}|x\rangle+\int_{-\infty}^1 \!\!dx \,\langle x| {\rm e}^{-\beta{\cal H}_-}|x\rangle \ .
\ee
Here, 
\be
{\cal H}_+ = -\hbar^2\frac{\partial^2}{\partial x^2}+[r(x)^2+\hbar r^\prime(x)]\ , \quad {\rm and} \quad {\cal H}_- = -\hbar^2\frac{\partial^2}{\partial x^2}+[r(x)^2-\hbar r^\prime(x)]\ ,
\ee
Calling $q$ the energy associated to ${\cal Q}$, which runs over the real line (where the two cuts live)  the Laplace transform in~$q$ space defines the spectral density $\rho(q)$ {\it via}
\be
\langle {\rm Tr}(\e^{\beta{\cal Q}^2})\rangle = \int_{-\infty}^{+\infty} dq \rho(q) \e^{-\beta{\cal Q}^2}\ ,
\quad
{\rm with}
\quad
\rho(q) = |q|\int_{-\infty}^1\!dx\left[|\psi(q^2,x)|_{+}^2+\psi(q^2,x)|_{-}^2\right]\ ,
\ee
\end{widetext}
where the wavefunctions $\psi(q^2,x)$ come from solving the spectral problems of ${\cal H}_\pm$. Using $E=q^2$ and so $dE=2qdq$, this identifies the $E=q^2$ density as:

\be
\rho(E) = \frac12 \int_{-\infty}^1\!dx\left[|\psi(E,x)|_{+}^2+\psi(E,x)|_{-}^2\right]\ .
\ee 
This all motivates the following form of the kernel of the two-cut system  (in $E$ variables) as simply the sum\footnote{This conclusion was reached in conversation with Felipe Rosso.}
\be
K(E,E^\prime) = \frac{1}{2}\left[K_+(E,E^\prime)+K_-(E,E^\prime) \right]\ ,
\ee
and from here  the Fredholm story of subsection~\ref{sec:fredholm} then goes through rather straightforwardly, on the interval~$(0,s)$. 

Moving on to various results, first some for the (1,2) JT supergravity that was introduced in the last section. The wavefunction along with the spectral density is shown in figure~\ref{fig:JT-12-wavefunction-zeros}. 
\begin{figure}[b]
\centering
\includegraphics[width=0.42\textwidth]{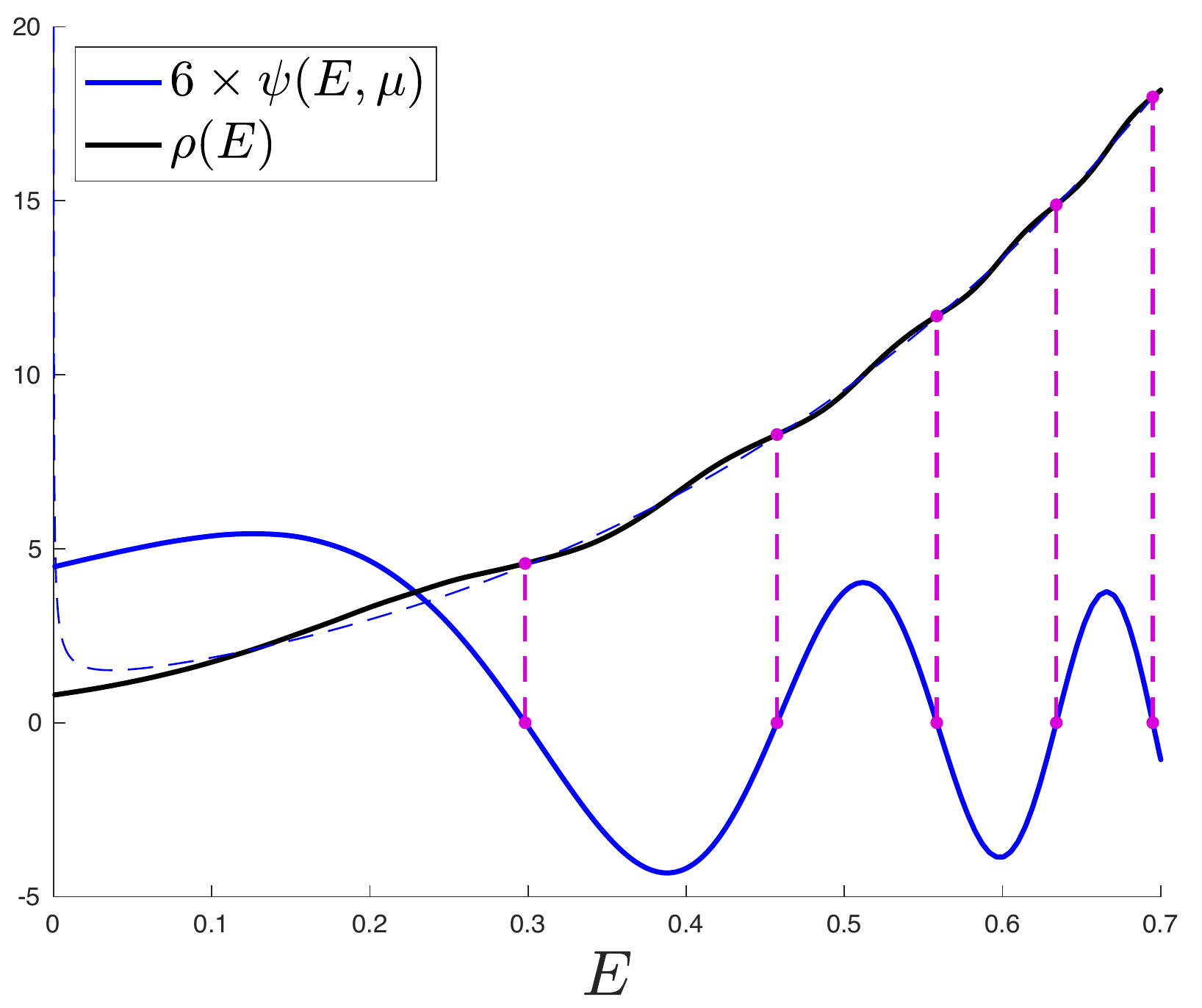}
\caption{\label{fig:JT-12-wavefunction-zeros} The $\psi(E,\mu)$ plotted alongside the spectral density $\rho(E)$ for (1,2) JT supergravity. The zeros give a first draft of the mean values of the energy levels, lining up with the undulations of the spectral density.  The complete information about the locations of the mean energy levels requires the  Fredholm determinant to be computed. See text.}
\end{figure}
Once again, at large enough $E$, it is well approximated by the WKB wavefunction given by putting 
$\Gamma{=}0$ into equation~(\ref{eq:WKB-JT-supergravity}), and the two sets of results can be conjoined to make wavefunction shown in figure~\ref{fig:JT-wavefunction-compare-12}. Its wavefunctions can then be used to construct the large~$E$ approximation to the spectrum of the gravity dual, just as before.
\begin{figure}[b]
\centering
\includegraphics[width=0.42\textwidth]{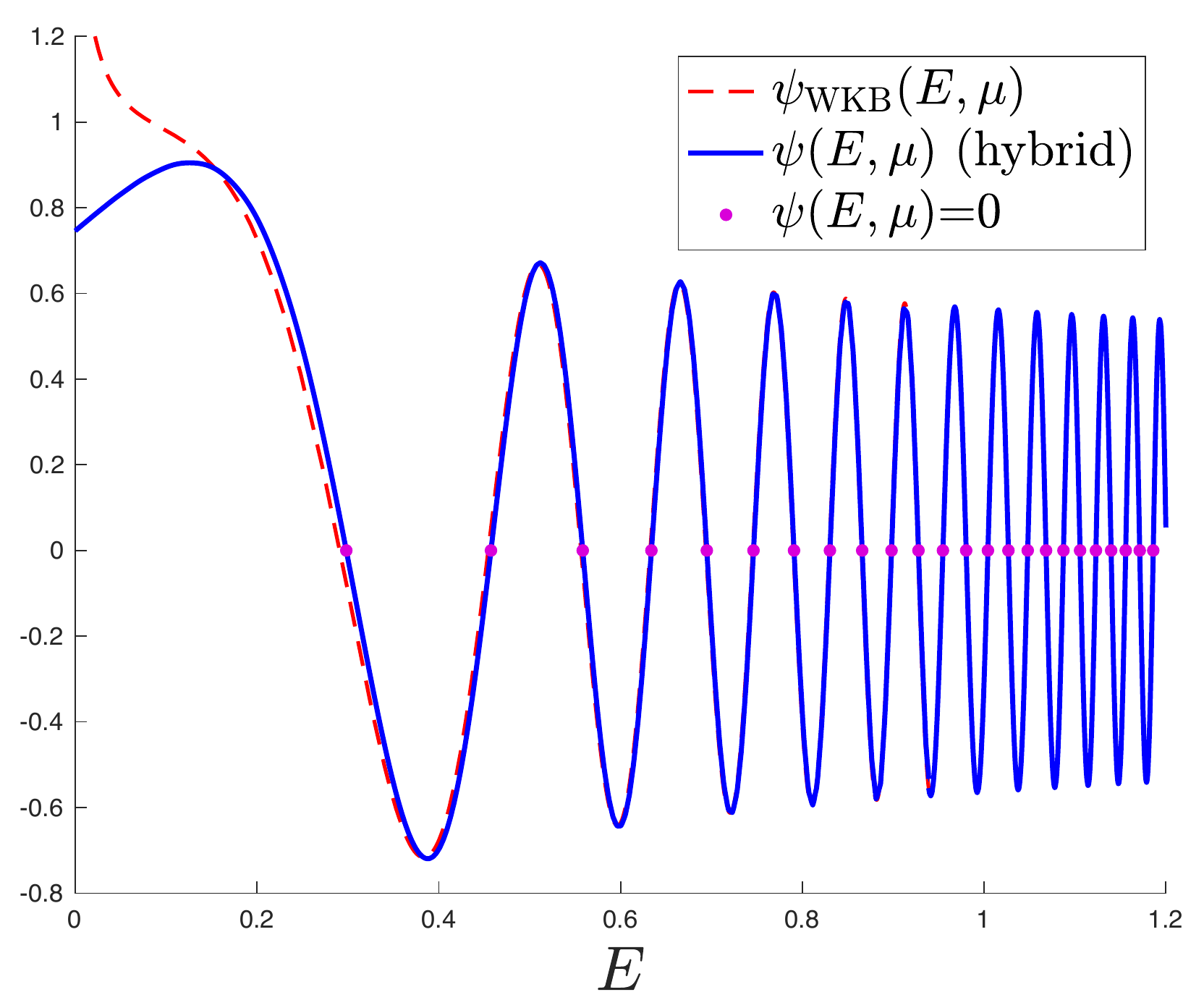}
\caption{\label{fig:JT-wavefunction-compare-12} The function $\psi(E,\mu)$ plotted alongside the WKB approximation, for (1,2) JT supergravity.}
\end{figure}

\begin{figure*}
\centering
\includegraphics[width=0.48\textwidth]{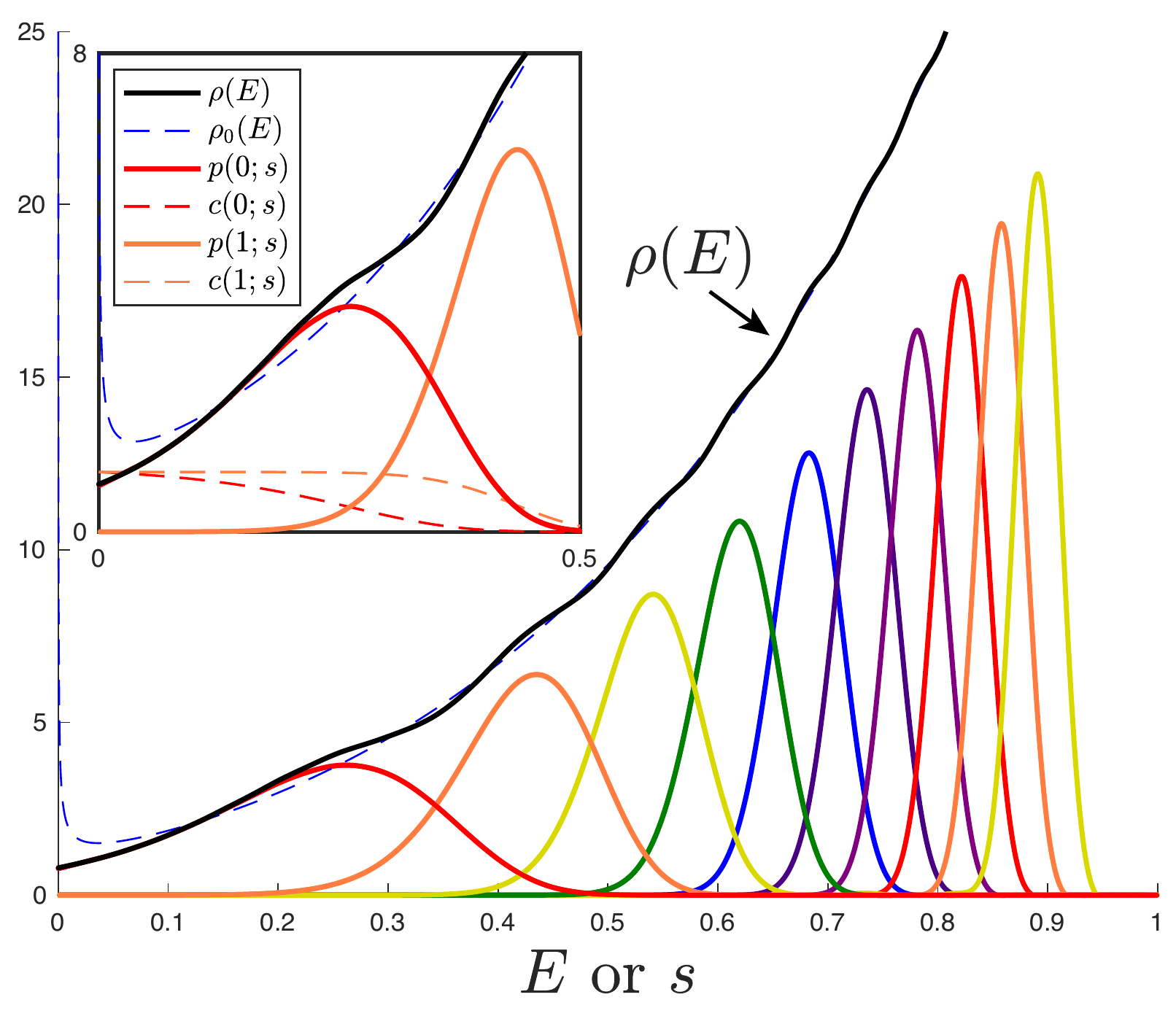}
\hskip0.5cm
\includegraphics[width=0.48\textwidth]{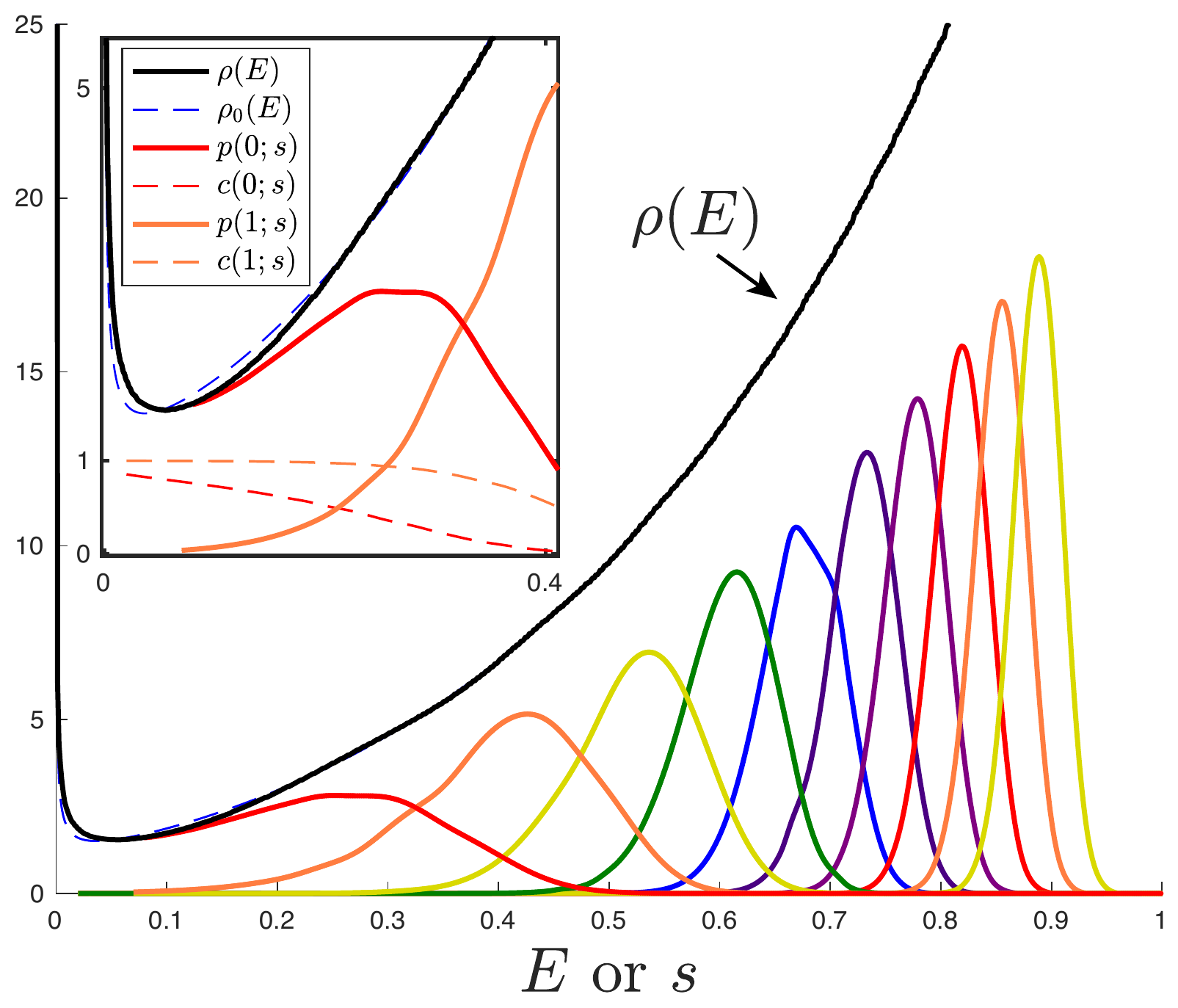}
\caption{\label{fig:JT_supergravity_12_microstates_10} The statistical distribution of the first ten energy levels  for  (left) (1,2) JT supergravity, and (right) 0B JT supergravity (in the $E$ basis) shown alongside the leading ($\rho_0(E)$) and full ($\rho(E)$) spectral density. The peaks are affected by more numerical noise in the latter case due to the behaviour near the origin. Here $\hbar{=}1$.\\\\}
\end{figure*}

\begin{figure*}
\centering
\includegraphics[width=0.82\textwidth]{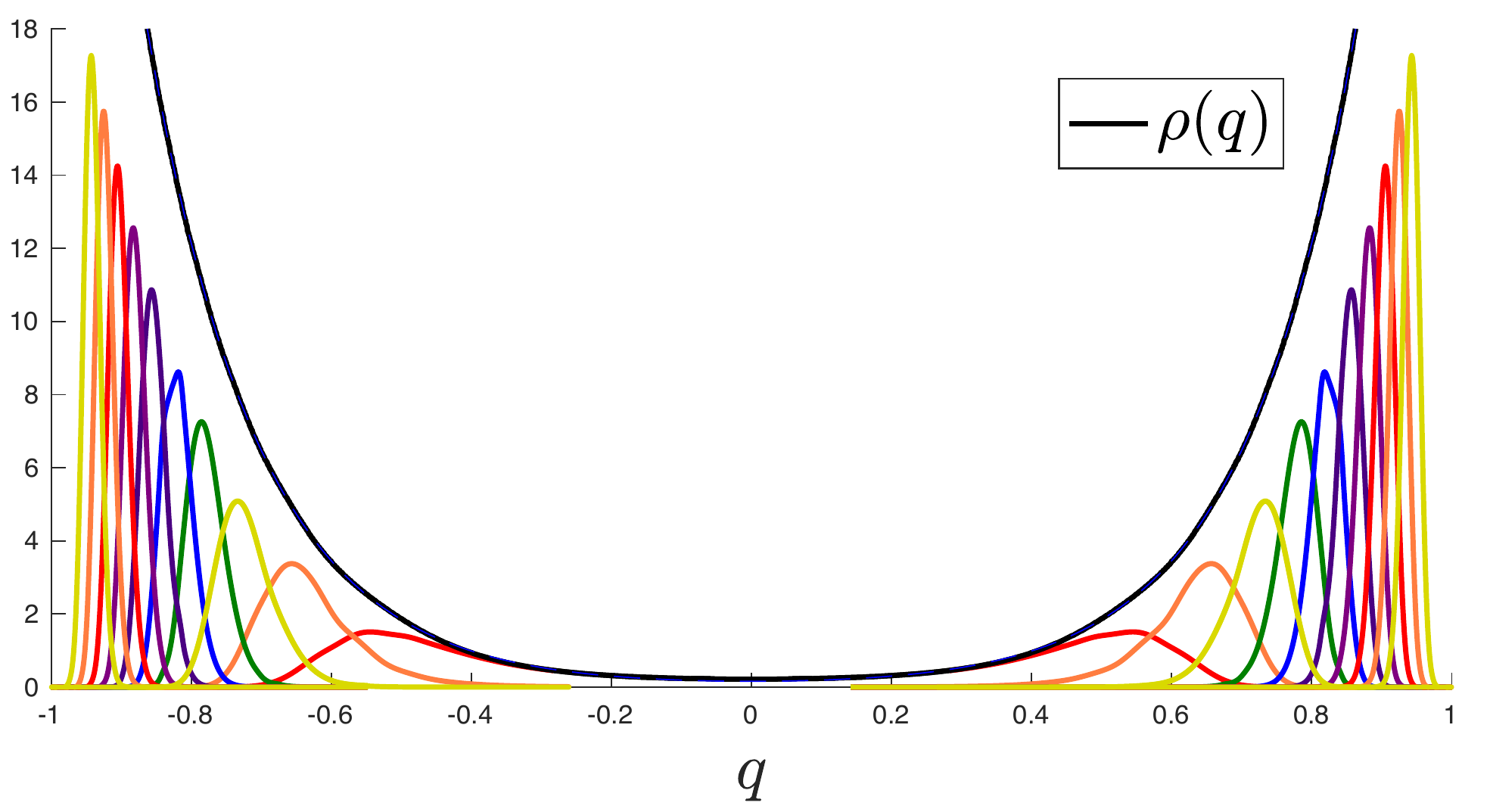}
\caption{\label{fig:JT_supergravity_0B_microstates_10_Q} A two-cut merger: The statistical distribution of the first ten energy levels (and their mirrors) for   0B JT supergravity (in the $q$ basis) shown alongside the leading ($\rho_0(q)$, practically invisible) and the full ($\rho(q)$) spectral density. Here $\hbar{=}1$.}
\end{figure*}

Things are rather more subtle for the Type~0B case, since it is made from combining two sectors and hence there are two sets of wavefunctions, and correspondingly two Fermi surface wavefunctions $\psi(E,\mu)_\pm$, whose WKB forms are of the form given in equation~(\ref{eq:WKB-JT-supergravity}), but with $\Gamma=\pm\frac12$. The zeros of each of these are in different places, so it is a natural puzzle as to which one determines the (approximate) location of the peaks, earmarked by the inflection points in the spectral density. The answer is in fact: neither. This is consistent with the observation~\cite{Johnson:2021owr} that in fact the semi-classical form of the spectral density can be derived (using the same methods mentioned around equation~(\ref{eq:WKB2}) for the JT gravity case) is:
\be
\label{eq:semi-classical-np-3}
\rho_{\rm sc}=\rhoo(q) - \sum_{s=\pm1}\frac{s}{4\pi |q|}\sin\left(2\pi\!\int^q_0\!\!\!\rhoo(q^\prime)dq^\prime-\pi C\right)\ , 
\ee
(where $C=\Gamma-\frac{s}{2}$) and $\rhoo(q)$ is (after changing variables {\it via} $E=q^2$):
\be
\rhoo_(q) =\frac{|q|}{2\pi\hbar}\sum_{s=\pm1} \int^1\frac{dx}{\sqrt{q^2-u_0(x)}} =  \frac{1}{2\pi\hbar}\cos(2\pi q)\ .
\ee 
However, since here the sign of $\Gamma$ is correlated with the sign of $s$,  the constant $C$ vanishes, matching the phases. Hence there is a  cancelling of the semi-classical undulations of the spectral density. It  turns out that this persists to the full non-perturbative solution too, quite remarkably. Nevertheless, there are oscillations, but they are significantly smaller, and quite hard to see (although careful analysis was done in ref.~\cite{Johnson:2021owr} to see that they are there). They can be thought of as corresponding to instanton contributions to the Dyson gas physics that cannot be captured by the WKB analysis.

\begin{table*}
\begin{center}
\begin{tabular}{|c|c|c||c|}
\hline
\textbf{energy} &\textbf{value}&\textbf{variance}&\textbf{Wavefunction}\\
\textbf{level} &\textbf{(Fredholm)}&\textbf{(Fredholm)}&\textbf{zeros}\\
\hline\hline
\textrm{${\cal E}_0$}&\textrm{0.230809}&\textrm{0.010163}&\textrm{0.298164}\\
\hline
\textrm{${\cal E}_1$}&\textrm{0.420580}&\textrm{0.004143}&\textrm{0.457373}\\
\hline
\textrm{${\cal E}_2$}&\textrm{0.533524}&\textrm{0.002166}&\textrm{0.558383}\\
\hline 
\textrm{${\cal E}_3$}&\textrm{0.615025}&\textrm{0.001374}&\textrm{0.633965}\\
\hline
\textrm{${\cal E}_4$}&\textrm{0.679535}&\textrm{0.000968}&\textrm{0.694944}\\
\hline
\textrm{${\cal E}_5$}&\textrm{0.733330}&\textrm{0.000728}&\textrm{0.746333}\\
\hline
\textrm{${\cal E}_6$}&\textrm{0.779616}&\textrm{0.000572}&\textrm{0.790931}\\
\hline
\textrm{${\cal E}_7$}&\textrm{0.820346}&\textrm{0.000467}&\textrm{0.830414}\\
\hline
\textrm{${\cal E}_8$}&\textrm{0.857010}&\textrm{0.000357}&\textrm{0.865924}\\
\hline
\textrm{${\cal E}_9$}&\textrm{0.891344}&\textrm{0.000060}&\textrm{0.898235}\\
\hline
\end{tabular}
\hskip1cm
\begin{tabular}{|c|c|c||c|}
\hline
\textbf{energy} &\textbf{value}&\textbf{variance}&\textbf{Wavefunction}\\
\textbf{level} &\textbf{(Fredholm)}&\textbf{(Fredholm)}&\textbf{zeros}\\
\hline\hline
\textrm{${\cal E}_0$}&\textrm{0.196104}&\textrm{0.012680}&\textrm{---}\\
\hline
\textrm{${\cal E}_1$}&\textrm{0.402042}&\textrm{0.006996}&\textrm{---}\\
\hline
\textrm{${\cal E}_2$}&\textrm{0.522545}&\textrm{0.003315}&\textrm{---}\\
\hline 
\textrm{${\cal E}_3$}&\textrm{0.606660}&\textrm{0.001977}&\textrm{---}\\
\hline
\textrm{${\cal E}_4$}&\textrm{0.673021}&\textrm{0.001418}&\textrm{---}\\
\hline
\textrm{${\cal E}_5$}&\textrm{0.728533}&\textrm{0.001038}&\textrm{---}\\
\hline
\textrm{${\cal E}_6$}&\textrm{0.775473}&\textrm{0.000796}&\textrm{---}\\
\hline
\textrm{${\cal E}_7$}&\textrm{0.816634}&\textrm{0.000648}&\textrm{---}\\
\hline
\textrm{${\cal E}_8$}&\textrm{0.853441}&\textrm{0.000545}&\textrm{---}\\
\hline
\textrm{${\cal E}_9$}&\textrm{0.886810}&\textrm{0.000472}&\textrm{---}\\
\hline
\end{tabular}
\includegraphics[width=0.485\textwidth]{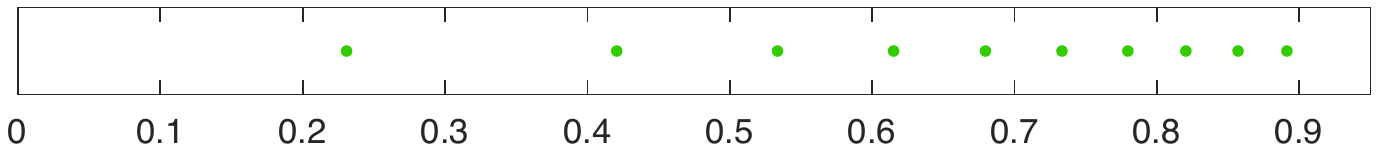}
\includegraphics[width=0.49\textwidth]{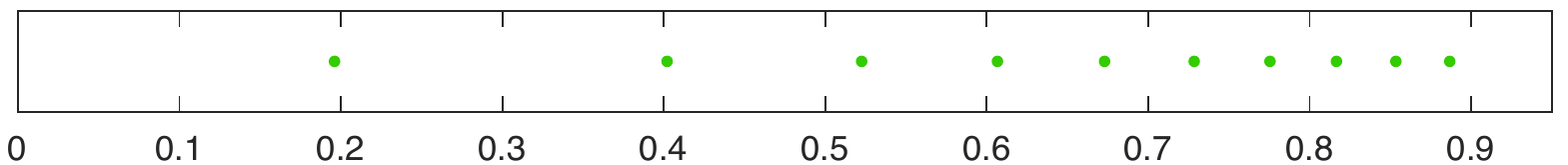}
\end{center}
\caption{The first 10 mean levels ${\cal E}_n=\langle E_n\rangle$ for the (1,2) JT supergravity at $\hbar=1$ (left), computed using  the Fredholm determinant method. On the right is the Type~0B case. The computed values are shown along with the (decreasing with energy) variance. The 4th column  shows the vanishing of the wavefunction~(\ref{eq:WKB-JT-supergravity}), giving an approximation to the peaks' locations that improves as $n$ grows larger.}
\label{tab:mean-levels-fredholm-JT-supergravity-12}
\end{table*}

\begin{figure*}
\centering
\includegraphics[width=0.45\textwidth]{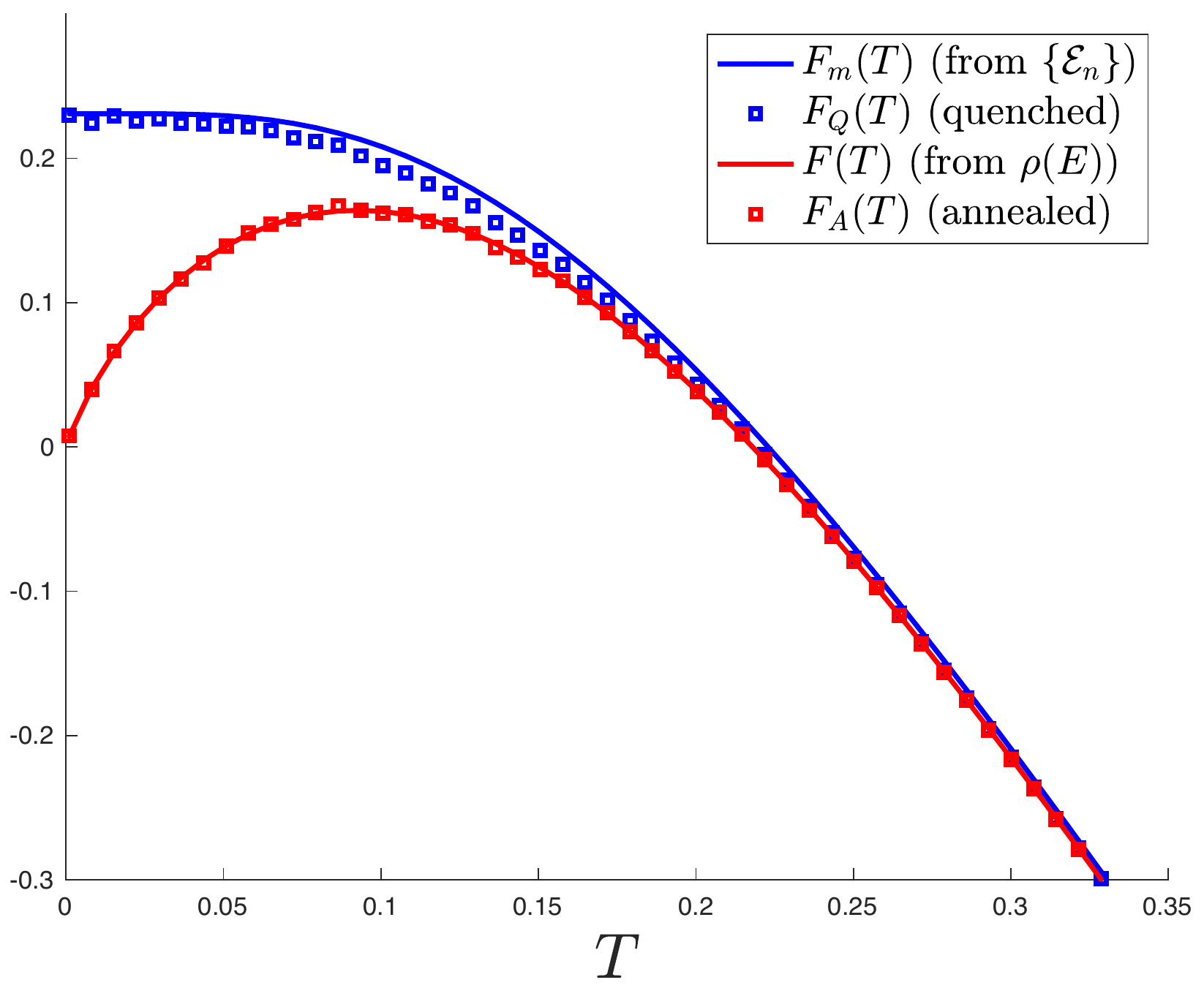}
\hskip1.5 cm
\includegraphics[width=0.45\textwidth]{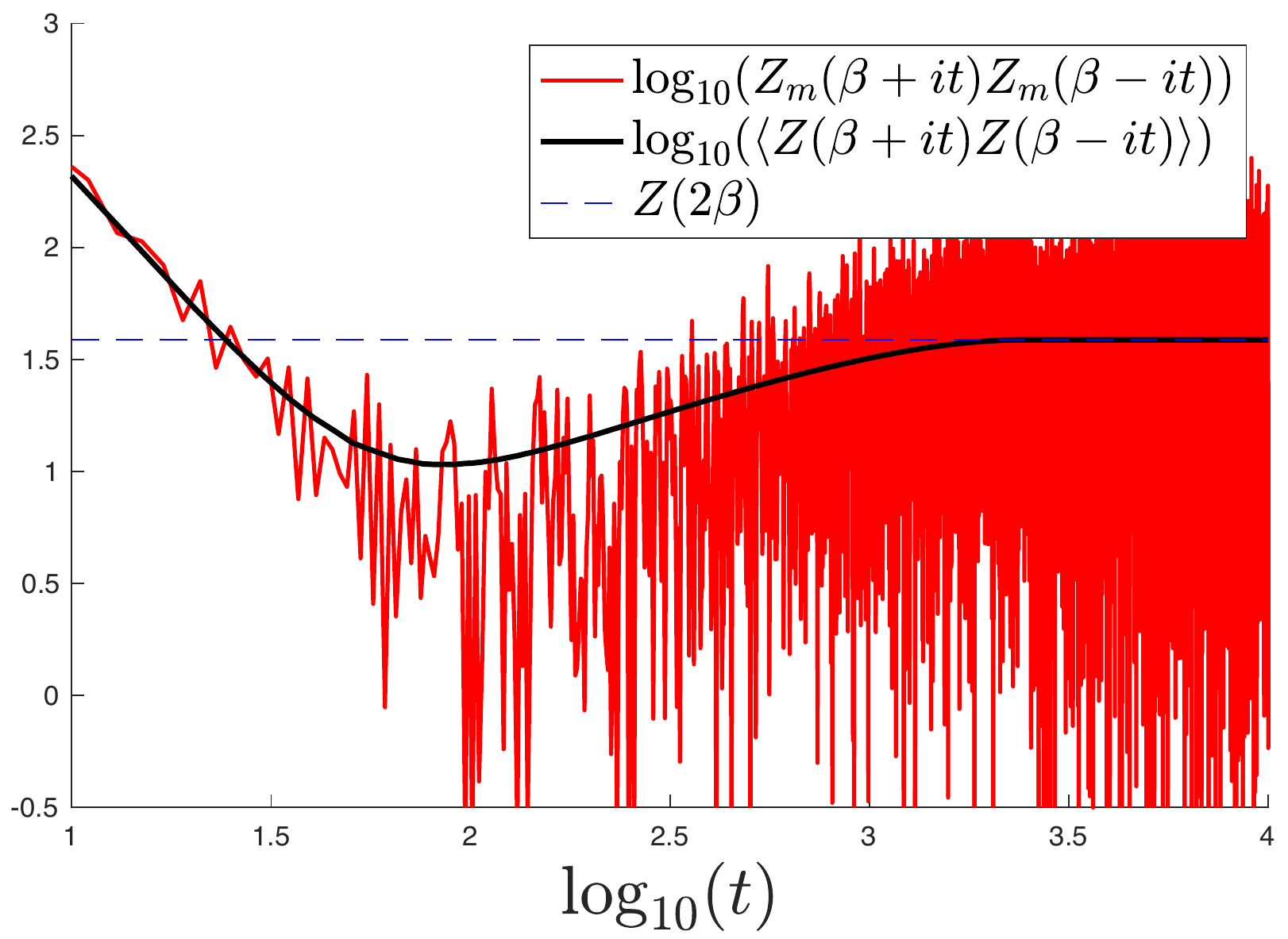}
\caption{\label{fig:JT_supergravity_12_free_energy} The left figure shows the JT supergravity free energy, for $\Gamma {=} 0$ Type 0A (1,2), computed using 34 energy levels (enough for the temperature range explored).  Also shown is the annealed quantity $F_A(T)$, as well as the naive free energy computed using the log of the partition function computed from $\rho(E)$ (integrated up to a similar energy). On the right is the spectral form factor computed for the spectrum, alongside the matrix model wormhole quantity, equivalent to averaging over the entire ensemble of spectra in the same symmetry class.}
\end{figure*}

From the perspective of the deeper understanding of the non-perturbative spectrum of matrix models presented in this paper, it is reasonable to expect that there {\it must be} undulations,  however small. This is {\it still} a model of random matrices (even if it has two merging cuts), with eigenvalues that spread over the line, and there ought to be statistics, and hence peaks. Indeed,  as discussed above, the Fredholm analysis goes through as before, and  in figure~\ref{fig:JT_supergravity_12_microstates_10} the resulting peaks are shown, for both the Type~0A (1,2) case (left) and the Type~0B case (in the~$E$ basis). (It should be noted that there are some notable wobbles on some of the peaks ($p(0;s), p(2;s)$, and $p(4;s)$)  for the Type~0B case, reflecting additional difficulties already encountered with the $\Gamma=-\frac12$ sector, magnified by having to combine them with the $\Gamma=+\frac12$ sector and then do the Fredholm analysis.) Note the relatively slight undulations in $\rho(E)$ for Type~0B as compared to Type~0A~(1,2), or indeed any of the other models. 

Changing to the $q$ variable, the real line over which the two-cut problem naturally lives, figure~\ref{fig:JT_supergravity_0B_microstates_10_Q} shows how the peaks beautifully arrange themselves on either side of the cut-join. In this variable, the non-perturbative oscillations in $\rho(q)$ are barely visible (indeed the blue dashed line of $\rho_0(q)$ is hidden by it), but they are there, and of course line up nicely with the probability peaks.

Just as before, the data for the spectrum of the JT supergravity duals can be extracted by evaluating the means of the peaks, and they are displayed in Tables~\ref{tab:mean-levels-fredholm-JT-supergravity-12}. There are no wavefunction zeros for comparison in the Type~0B case for the reasons already discussed. Consequently, it is difficult to get good estimates of the spectrum at large $E$ (or $q$) in this case (although to be sure it has been shown to exist), and while a free energy and spectral form factor analysis could be done with the ten levels established, it seems unnecessary at this point as the pattern is clear. Those results for the Type~0A (1,2) case are given in figure~\ref{fig:JT_supergravity_12_free_energy}, however.


\section{Concluding Remarks}
\label{sec:concluding-remarks}
Recall that a common view in  the community is that rather than have an holographic dual with a single Hamiltonian, the  dual of JT gravity is instead given by   the {\it whole} random matrix ensemble of Hamiltonians (with a similar statement for variants and deformations already recalled earlier). It is an intriguing and attractive idea. Many interesting attempts to try to learn new things about gravity in higher dimensions have stemmed from it (see {\it e.g.} refs.~\cite{Maloney:2020nni,Cotler:2020ugk,Afkhami-Jeddi:2020ezh,Bousso:2020kmy,Benjamin:2021wzr,Heckman:2021vzx,Cotler:2021cqa,Collier:2022emf}). On the other hand, a host of uncomfortable consequences reverberate outward from it,  starting with the idea that  this implies that the theory fundamentally fails to factorize. However, the best understood examples of  holography in higher dimensions (AdS/CFT) seem to provide definite duals such as Yang-Mills theories, and a vast amount of evidence has been accumulated for AdS/CFT in its traditional form. So this presents a  puzzle. Various proposals for avoiding the puzzle have resulted, including attempts to tinker with the matrix model to somehow restore factorization while at the same time preserving gravity.

 The proposals of this paper are quite a departure from the standard  view of things, and so will initially seem hard to accept. The driving force behind them comes from  going back to the roots of holography (at least as found in the string theory context) which started  with D-branes in a well-defined  theory of quantum gravity and simply taking well-motivated limits, arriving at AdS/CFT~\cite{Maldacena:1997re}. This defined the holographic dual in a manner that first makes sense in a {\it Lorentzian} approach  and { then}  the duality was subsequently further explored in the {\it Euclidean} approach, and seen to accommodate gravity on other smooth geometries with more general topologies. 
 
 It is difficult to see how to take a limit of a D-brane system and recover that gravity as fundamentally dual to an  ensemble. 
 So the logical conclusion is that the same successful route to holographic duals (starting with gravity on a surface with {\it trivial} topology, in the sense that it can be Wick rotated back to Lorentzian signature) must apply  in the 2D context. This urged a closer examination of what the matrix model is really doing, and ultimately a re-think of whether there might be a different interpretation of the matrix model results. The conclusion (with all due respect to those who have worked in this area) is that the full extent of how matrix models work for 2D gravity has  not been appreciated, not just for JT gravity, but even going back to the 1990s. (Although in that period it was not as sharp a problem because Euclidean 2D gravity on all topologies was just what is needed to describe string world sheets, and holography had not yet come along.)
 
The picture that has emerged from this work supports this idea of carefully contrasting Lorentzian {\it vs} Euclidean interpretations  in order to understand the ensemble. It seems clear that the ensemble is indeed a natural part of the story but it is part of the Euclidean machinery of the gravity calculus. It should not be take over wholesale to the Lorentzian picture. The point is that the matrix model contains both pictures, with instructions as to how to move between them. On the Euclidean side,  it was observed that the random matrix model contains two mechanisms (one 't Hooftian, the other Wignerian) for  building smooth Euclidean geometries of any topology. On the other hand it is proposed that in the  Lorentzian completion of the story there is a single Hamiltonian for the gravity model and its holographic dual, just like in higher dimensions. The matrix model, when examined closely, characterizes very specifically how this can work by providing a new type of spectrum (characterized in this paper) that is discrete, but which reproduces the leading Schwarzian spectral density in precisely the regime where the Schwarzian analysis applies ($E\gg\e^{-S_0}$, where $S_0$ is the extremal entropy), {\it i.e.,} the  spectral  density  of the Schwarzian (continuous in the field theory approach) is only an approximation to a discrete structure. This means that the original concern that there could be no sensible quantum mechanical dual is addressed, and having an ensemble dual (with its continuous density) is therefore unnecessary. 
 
 In fact, as pointed out in the Introduction, there is entirely another way of thinking about the origins of  the matrix model that is Lorentzian in spirit,  following the approach Wigner~\cite{10.2307/1970079} made for the problem of characterizing spectra of the Hamiltonians of complex nuclei in the 1950s. (See figure~\ref{fig:paths_to_spectrum} in the Introduction and the nearby discussion.) In fact, the problem is almost exactly the same: Try to determine the typical properties of the spectrum of a system with many states, subject to some general input characteristics. In this case, the key characteristic property is that the leading spectral density is the Schwarzian form. The matrix model that results from this, if studied fully non-perturbatively (as done here) yields very specific results about what that {\it single} spectrum is for the system in hand. Apart from the specific use of the Schwarzian result as a starting constraint, this is entirely a Lorentzian approach, which should have a sensible quantum-mechanical  answer: Wigner did not assume that a given nucleus was fundamentally described  by the ensemble, instead of some quantum mechanical Hamiltonian! Having characterized what the spectrum of the single Hamiltonian most likely looks like (as done here) one can {\it also} take this same model and apply the techniques of the 1970s and beyond~\cite{'tHooft:1973jz,Brezin:1978sv,Bessis:1980ss}: The matrix model can be expanded in  Feynman diagrams and, {\it via}  the $\e^{-S_0}\sim 1/N$ expansion, discovered to be {\it also} encoding the sum, over all geometries, of a family of smooth bordered two dimensional {\it Euclidean} hyperbolic surfaces of all possible topologies~\cite{Mirzakhani:2006fta,Eynard:2007fi}. In other words, the Euclidean approach of ref.~\cite{Saad:2019lba} necessarily also emerges! 
 
  This is in fact quite remarkable. The random matrix model (in the double-scaled limit)  naturally contains the data of both the Lorentzian and Euclidean approaches to quantum gravity, and there is a clear means of moving between  between the two pictures. Moreover this applies readily to the many other 2D gravity models that can be captured by such matrix models, and would seem to imply  several new physics results, including for minimal strings. This is worth exploring. While the Wignerian perspective just outlined implies that a single holographic dual exists, with a great deal of data for what the spectrum most probably looks like (given by the Fredholm peaks discussed in this paper) more guidance as to how to read the specific spectrum  would be welcome. Since the minimal strings are built from simple models (such as the Ising model)   coupled to gravity, models which presumably in Lorentzian signature have definite Hamiltonians, their further study in the light of these results should yield more clues as to how to read off the definite spectrum from the Fredholm data, in case the prescription used here (the mean energy of the peaks at each level, or instead the most frequent energy at each level; see footnote~\ref{fn:peak-subtleties}) is too simple.
 
 This interpretation of the matrix models solves many puzzles about 2D gravity holography and removes the factorization puzzle, since 2D holography is then much more akin to higher dimensional AdS/CFT. This also implies that the lessons learned here can teach many things about how to interpret Euclidean quantum gravity computations in higher dimensions.   For example, it is entirely natural that analogues of ensemble averaging  appear in the Euclidean approach in higher dimensions (see earlier references mentioned above). It really is just an aspect of incorporating all possibilities in the Euclidean sum, and in fact it was seen in the matrix models that ensembles help guarantee that the surfaces in the sum are smooth: The core lesson is that ensembles appearing in the Euclidean approach does {\it not} mean that the Lorentzian theory requires an ensemble definition in any fundamental sense: {\it Holography,  AdS/CFT and factorization are just fine.}

 It is possible that matrix models alone cannot fix the precise form of the spectrum. For that, the specific Hamiltonian that yields the Schwarzian at leading order might need to be known. So matrix models show how it is possible to have a specific Hamiltonian, and predict the likely structure it ought to have (through the Fredholm peaks computed in the paper), but perhaps different frameworks are needed to make further progress in order to get the precise form of the spectrum (denoted  $\{{\cal E}_n\}$ in the body of the paper). Revisiting the study of brane configurations that yield near-horizon geometries with AdS$_2$ factors and the Hamiltonians on their world-volumes would be interesting in this regard (see footnote~\ref{fn:d-brane-wrapping}). The clear suggestion is that they have spectra of the form uncovered  in this paper.\footnote{Perhaps also the  approach in the recently appeared  ref.~\cite{Post:2022dfi} might have some relevance to this question.} 

As an aside that could be relevant, recall the observation that the spectra represent the ``locations'' (boundary cosmological constants) of D-branes in the $N{\times}N$ matrix model and that on the other hand the FZZT D-brane partition function's zeros give the locations of D-branes in the $(N-1){\times}(N-1)$ matrix model. (This does not seem to have been reported in prior literature, incidentally.) So the FZZT D-brane probes don't quite get the right answer (the Fredholm determinants are needed) but for high enough energies it was observed that  those zeros do begin to furnish a good approximation to the correct locations.\footnote{This follows from how Dyson gases spread out with $N$, combined with the fact that the double scaling limit zooms into an endpoint of the gas.} In the simplest example, the Airy model, the D-brane partition function is simply the Airy function. It is possible that there are already  known systems with Hamiltonians that have  spectra that are given by the zeros of special functions like Airy. Those may well be the kinds of systems sought here. The generalization to the more general $\psi(E,\mu)$ of the JT systems would then be interesting to explore.
 
Incidentally, notice that this picture of JT gravity having a definite dual Hamiltonian also furnishes a specific model of the microstates of higher-dimensional near-extremal black holes in the near-horizon limit (from which JT arises by dimensional reduction~\cite{Achucarro:1993fd,Nayak:2018qej,Kolekar:2018sba,Ghosh:2019rcj}), where~$S_0$ is their extremal entropy. The fact that the gaps in the spectrum disappear in the large black hole limit $S_0{\to}\infty$ is consistent with that. It therefore also provides a very clean description of the black hole thermodynamics all the way down to extremality, providing the kind of new model needed to appreciate the low $T$ thermodynamics~\cite{Preskill:1991tb}.

 Finally, it is worth repeating that even if the Reader is not entirely familiar with the random matrix model techniques beyond what are  typically used in the 2D gravity community (which tend to favour smooth objects with a geometrical interpretation, following the 't~Hooftian/Euclidean approach),   it is hoped that it is clear  that the observed discrete structures in the matrix model spectrum, (see {\it e.g.,} equation~(\ref{eq:fattening})) that have  a firm understanding in other areas of application of  random matrix models,  are indisputably present, and really seem to have no satisfactory role or explanation in the standard ``fundamental ensemble'' interpretation of the matrix model. There, the continuous matrix model spectral density $\rho(E)$ is the fundamental object, and its decomposition into the discrete spectrum, with peaks $p(n;E)$, currently has  no significance. Perhaps that is just the way it is, but this seems unlikely. 
 
 It was through the task of trying to answer the question as to the meaning of the peaks $p(n;E)$ within the current framework of JT gravity that the idea emerged to reexamine the entire narrative. This led to thinking more about how gravity really emerges from the matrix model in 2D, what it even is, and how holography, topology and factorization could possibly all play well together.

It is hoped that the resulting alternative picture   suggested   in  this paper solves more problems than it creates.

\phantom{rwgmgrignrigrqngirqngrignr2'ignr2'ignr2g2goigrngirngir;gnrg;rng;irngrignrigrngirgnr2ig;nr2girngir2ngi}

 \begin{acknowledgments}
 
CVJ  thanks  Robbert Dijkgraaf,  Juan Maldacena,  Felipe Rosso,  Joaquin Turiaci, Herman Verlinde, and Edward Witten for questions, comments, and discussions. After this manuscript appeared on the arXiv, further helpful questions and comments from a number of  people (Ahmed Almheiri, Ibrahima Bah, Panos Betzios, Lorenz Eberhardt, Matthew Heydeman, Henry Maxfield, Massimo Porrati, and Edgar Shaghoulian) were very useful in identifying places where clarifying statements could be made to improve presentation of the ideas.  Thanks also go to the  US Department of Energy for support under grant  \protect{DE-SC} 0011687, and, especially during the pandemic,  Amelia for her support and patience.    
\end{acknowledgments}

\begin{widetext}
\bigskip
\bigskip
\bigskip

{\bf Note Added:} Seven weeks after this paper's release on the arXiv, ref.~\cite{Blommaert:2022ucs} appeared. Using techniques similar to those used in this paper (but only solved in a semi-classical approximation), they present a {\it modification} of the matrix model that, it is argued, produces a localization on to a spectrum that is given by the zeros of the FZZT wavefunction of JT gravity, denoted $\psi(E,\mu)$ here. This was offered as the holographic dual spectrum. Of course, it does (as already shown here) have the property that it reproduces  the Schwarzian density at large $E$ and so can be thought of as a discretization of it. However there are several points worth noting. One is that (even if their approach was enhanced to yield the full non-perturbative $\psi(E,\mu)$, as is done here), it is already clear from this paper's results  that $\psi(E,\mu)$'s zeros are manifestly {\it not} the  double-scaled matrix model's mean spectrum, but only an approximation to it. This  is puzzling since the construction is argued to localize the matrix model onto a definite spectrum.  (This matter was communicated to the authors before their paper appeared.) The problematic mismatch is quite clear from the equivalent of their construction for the $k{=}1$ model, where the exact results are well known: The mean of the first peak, the Tracy-Widom distribution, is at ${\sim}1.771$, while the first zero of the Airy function is at ${\sim}2.338$, as listed in Table~\ref{tab:mean-levels-fredholm}. The large discrepancy (and its meaning)  is clear, and already explained in Section~\ref{sec:return-to-gravity}.  The second point is that while their construction is interesting, there should be no need to modify the JT gravity matrix model by adding any additional terms in order to understand how to resolve factorization and see that there is a specific holographic dual spectrum. The answer most likely lies in  fully non-perturbatively exploring the  content of the matrix model as it stands, as presented in this paper,  and finding the correct interpretation of the results.  
\bigskip

\end{widetext}

\bibliographystyle{apsrev4-1}
\bibliography{Fredholm_super_JT_gravity1,Fredholm_super_JT_gravity2}

\end{document}